\documentclass[letterpaper,12pt]{article}
\usepackage{floatrow}
\usepackage{xr-hyper}
\usepackage{hyperref}
\usepackage{fullpage}
\usepackage[ansinew]{inputenc}
\usepackage{verbatim}
\usepackage{graphicx}
\usepackage{tabularx}
\usepackage{array}
\usepackage{delarray}
\usepackage{dcolumn}
\usepackage{float}
\usepackage{amsmath}
\usepackage{psfrag}
\usepackage{booktabs}
\usepackage{multirow,hhline}
\usepackage{amstext}
\usepackage{amssymb}
\usepackage{fancyhdr}

\usepackage{setspace}

\usepackage{xparse} 
\usepackage[margin=1in]{geometry}

\usepackage{titletoc}

\usepackage{amsthm}
\usepackage{makeidx}
\usepackage{tikz}
\usepackage{lipsum}
\usepackage{url}
\usepackage{adjustbox}
\usepackage{enumitem}
\usepackage{mathabx}
\usepackage{authblk}
\usepackage{appendix} 

\newcommand{\hidefromtoc}{%
  \setcounter{oldtocdepth}{\value{tocdepth}}%
  \addtocontents{toc}{\protect\setcounter{tocdepth}{-10}}%
}
\newcommand{\unhidefromtoc}{%
  \addtocontents{toc}{\protect\setcounter{tocdepth}{\value{oldtocdepth}}}%
}
\newcounter{oldtocdepth}

\newcommand{\appendixtoc}{\makeatletter\@starttoc{app}\makeatother}
\newcommand{\appcontentsline}[2]{\addtocontents{app}{\protect\contentsline{section}{\protect\numberline{#1}#2}{\thepage}{section.\thesection}}}

\usepackage{siunitx}  

\RequirePackage[round,authoryear,longnamesfirst]{natbib}
\usetikzlibrary{arrows,calc}
\hypersetup{citecolor=true,colorlinks=true,allcolors=blue}
\allowdisplaybreaks
\numberwithin{equation}{section}
\newtheorem{lem}{Lemma}[section]
\newtheorem{thm}{Theorem}[section]

\newtheorem{ass}{Assumption}
\newenvironment{assprime}[1]
  {\renewcommand{\theass}{#1$^\prime$}\ass}
  {\endass}
\newtheorem{prop}{Proposition}[section]
\theoremstyle{definition}

\newtheorem{rem}{Remark}[section]
\renewcommand{\citep}[1]{\citeauthor{#1}, \citeyear{#1}}

\newcommand{\convP}{\stackrel{p}{\longrightarrow}}
\newcommand{\convD}{\rightsquigarrow}
\newcommand{\Op}{O_p}
\newcommand{\op}{o_p}
\newcommand{\N}{\mathcal{N}}
\newcommand{\eps}{\varepsilon}
\newcommand{\tr}{\text{trace}}
\newcommand{\rank}{\text{rank}}
\renewcommand{\epsilon}{\varepsilon}

\DeclareMathOperator*{\argmin}{arg\,min}
\makeatletter
\newcommand*{\rom}
[1]{\expandafter\@slowromancap\romannumeral #1@}
\makeatother

\title{An Improved Inference for IV Regressions\thanks{This version is dated on \today. We are grateful to Michal Kol\'esar for helpful comments. 
Dou acknowledges financial support from the National Natural Science Foundation of China through Grants 72133005 and 72573149. 
Wang acknowledges the support from the Singapore Ministry of Education Tier
1 Grant RG51/24.
Zhang acknowledges support from the Ministry of Education, Singapore under its MOE Academic Research Fund Tier 2 (Project ID\@: MOE-000767-00). Any opinions, findings and conclusions or recommendations expressed in this material are those of the authors and do not reflect the views of the Ministry of Education, Singapore. 
Any and all errors are our own. \vspace{1mm}} 
	\\ \vspace{1mm}
}

\author[a]{Liyu Dou}
\author[a]{Pengjin Min}
\author[b]{Wenjie Wang}
\author[a]{Yichong Zhang}
\affil[a]{\small \emph{Singapore Management University, Singapore}}
\affil[b]{\small \emph{Nanyang Technological University, Singapore}}

\date{}

\let\oldfootnote\footnote 
\RenewDocumentCommand{\footnote}{ m }{%
    \oldfootnote{\doublespacing#1}%
}

\begin{document}

\maketitle

\begin{abstract}

Empirical instrumental variables (IV) studies often report separate results based on low-dimensional instruments and many base instruments. This paper proposes a combination test that integrates these commonly reported statistics. The test linearly combines a cluster-robust Wald statistic based on low-dimensional IVs with leave-one-cluster-out Lagrangian multiplier (LM) and Anderson-Rubin (AR) statistics constructed from many IVs. 
We establish joint asymptotic normality and asymptotic optimality of the proposed test. The procedure yields costless efficiency improvements, automatically adapts to weak identification of many instruments, and is accompanied by a practical rule of thumb for assessing efficiency gains.


\end{abstract}


\vspace{2mm}

\newpage

\hidefromtoc

\section{Introduction}

Empirical applications of instrumental variables (IV) regressions in economics often involve multiple sets of candidate instruments, some with dimensionality proportional to the sample size and others more parsimonious. A canonical example is the influential study by \cite{Angrist-Krueger(1991)}, which estimates the causal effect of schooling on wages using three quarter-of-birth (QoB) dummies as instruments, as well as their interactions with state- and year-of-birth dummies, yielding a total of 1,530 instruments. A different but related class of examples arises in the context of shift-share IVs,\footnote{See, for instance, \cite{bartik1991}, \cite{blanchard1992}, \cite{adao2019shift}, \cite{Goldsmith(2020)}, \cite{borusyak2022quasi}, \cite{borusyak2025practical}, and references therein.} which, as noted by \cite{Goldsmith(2020)}, can be interpreted as a particular way of aggregating many underlying base instruments. It is common practice to report results based on a one-dimensional shift-share IV and then supplement them with results obtained using the full set of base IVs.



This naturally raises the question of whether one can take the low-dimensional IV regression as the benchmark specification and combine it with its many-IV counterpart in a way that enhances the power of IV inference, for free, in the sense that no additional identification restrictions are imposed on the many-IV specification. In this paper, we provide a constructive solution.


Specifically, we consider an optimal combined inference procedure based on three commonly reported test statistics for IV regressions in a clustered setting, where the data consist of many clusters of bounded size. The three component statistics are: (i) a standard cluster-robust Wald statistic constructed from the low-dimensional IVs; (ii) a leave-one-cluster-out jackknife Lagrange multiplier (LM) statistic; and (iii) a leave-one-cluster-out jackknife Anderson--Rubin (AR) statistic. The LM and AR statistics are constructed using the many-IV specification, and the leave-one-cluster-out design removes the many-IV bias in the presence of within-cluster error dependence.  

We first show that, under the null hypothesis and local alternatives, the Wald, LM, and AR statistics are jointly asymptotically normal, provided that the low-dimensional IVs strongly identify the parameter of interest, while allowing the many-IV specification to be weakly identified. Standard optimal testing theory then implies that, in the corresponding Gaussian limit experiment, the uniformly most powerful unbiased (UMPU) test rejects for large absolute values of an appropriate linear combination of the three limiting Gaussian statistics.\footnote{See, for instance, Section 4.2 of \cite{Lehmann-Romano(2006)} and Lemma 2.2 of \cite{LWZ24} for formal arguments.} Our proposed test, as a function of the three statistics, is asymptotically equivalent to this UMPU test and is therefore weakly more powerful than each of the Wald, LM, and AR tests.

Importantly, the combination test adapts automatically to the identification strength of the many-IV specification. When the parameter of interest is weakly identified under many IVs in the sense of \cite{MS22}, the optimal linear combination places asymptotically negligible weight on the LM and AR components, so that the resulting test reduces to the Wald test. In this case, the procedure remains valid and retains the same asymptotic power as the standalone Wald test.

The notion of optimality in our study is defined relative to the class of tests based on the Wald, LM, and AR statistics and serves primarily as a device for improving inference within this class. We do not attempt to derive a globally optimal test by searching over all possible combinations of a given set of base IVs, nor do we aggregate results across alternative specifications that employ different sets of base IVs or different low-dimensional IVs. Rather, our objective is to take the low-dimensional specification as the benchmark and combine its associated test statistic with those constructed from a given set of many IVs, thereby strengthening inference while remaining agnostic about the identification strength of the many-IV specification. Nonetheless, we attribute a notion of asymptotic optimality to the resulting combination test in the sense of \cite{mueller2011}.

The confidence interval implied by the combination test has the usual ``estimator plus and minus a standard error times a critical value" form. Its center is an estimator that linearly combines a standard GMM estimator based on the low-dimensional IVs with a leave-one-cluster-out jackknife IV estimator based on many IVs and the AR statistic, using weights that capture both the relative identification strength from the two IV sets and the UMPU weights. The efficiency gain manifests itself as an almost surely shorter confidence interval. We measure this gain by the percentage reduction in the length of the resulting confidence interval relative to that of the Wald test based on the low-dimensional IVs. As an illustration, in the immigrant enclave application of \citeauthor{card(2009)} (\citeyear{card(2009)}), the combination procedure reduces the length by between 5\% and 17\%. This reduction depends mainly on the identification strengths of the low-dimensional and many IVs, together with the limiting correlations between the Wald and LM statistics, and between the LM and AR statistics. In Section \ref{sec: illu}, we translate these relationships into a practical rule of thumb based solely on the ratio of the standard errors of the low-dimensional and many-IV estimators, which can be directly read from reported regression tables prior to implementing our combination procedure. We illustrate this rule using the application in \cite{card(2009)}.

\vspace{0.1in}

\noindent
\textbf{Relation to the literature.} This paper contributes to the large literature on many (weak) instruments,\footnote{See, for instance, \cite{Kunitomo1980}, \cite{morimune1983}, \cite{Bekker(1994)}, \cite{donald2001}, \cite{Chao-Swanson(2005)}, \cite{stock2005}, \cite{han2006}, \cite{Andrews-Stock(2007)}, \cite{Hansen-Hausman-Newey(2008)}, \cite{Newey-Windmeijer(2009)}, \cite{anderson2010}, \cite{kuersteiner2010}, \cite{anatolyev2011}, \cite{okui2011}, \cite{belloni2012}, \cite{carrasco2012}, 
\cite{Chao(2012)}, \cite{Haus2012}, \cite{K13}, \cite{hansen2014}, \cite{carrasco2015}, \cite{Wang_Kaffo_2016}, 
\cite{kolesar2018}, \cite{EK18}, \cite{solvsten2020}, \cite{CNT23}, \cite{LWZ24}, \cite{BN24}, \cite{Yap24}, \cite{lim2024dimension}, among others.} 
and is particularly related to \cite{LWZ24}. Building on \cite{Andrews(2016)}, they propose a jackknife conditional linear combination (CLC) test for independent data that is robust to weak identification, many instruments, and heteroskedasticity, combining jackknife AR, LM, and orthogonalized LM statistics to achieve good power across identification regimes. Their analysis, however, is confined to many-IV settings and does not directly apply to our framework. In contrast, we explicitly incorporate low-dimensional IVs within a clustered environment and develop a unified procedure that efficiently combines estimation and inference from both low- and high-dimensional IV specifications. Our primary objective is to construct a theoretically grounded combination test that improves upon the conventional Wald test based solely on low-dimensional IVs. We also note that, in a different context, \cite{jiang2025adjustments} proposes an optimal linear combination of adjusted and unadjusted estimators for the average treatment effect under covariate-adaptive randomization.

Our paper is also related to the literature on inference with many IVs and clustered data. \cite{FLM23} study leave-one-cluster-out jackknife versions of the IV estimator, and \cite{ligtenberg2023inference} develop weak-identification-robust procedures in clustered environments. We depart from this literature by not focusing solely on estimation or inference within the many-IV specification itself. Instead, we leverage the many-IV specification to enhance the power of low-dimensional Wald inference through an optimal combination of test statistics. \cite{CNT23} and \cite{kolesar2026cluster} consider settings with both many IVs and many control variables, which introduce additional challenges for cluster-robust inference. In this paper, we restrict attention to models with a fixed number of control variables and leave the case with many IVs and many controls for future research.

\vspace{0.1in}

\noindent
\textbf{Structure of the paper:} 
Section \ref{sec: illu} details the rationale behind the proposed rule of thumb and demonstrates its use in an empirical application; practitioners mainly interested in applications may focus on this section directly. Section \ref{sec: setup} introduces the model and key preliminaries, while Section \ref{sec: comb} develops the large-sample theory for the combination test and formalizes its theoretical properties. Section \ref{sec: simu} presents the practical implementation of the combination test in an empirically relevant setting and employs simulations to evaluate its power properties, and Section \ref{sec: conclu} concludes. An additional case with weak low-dimensional IVs alongside strong many IVs, as well as all proofs, is presented in the Online Appendices.

\vspace{0.1in}

\noindent
\textbf{Notation.} We write $[n] \equiv \{1,\ldots,n\}$ and $[G] \equiv \{1,\ldots,G\}$. Let $A$ be an $n\times m$ matrix and let $\{n_g\}_{g\in [G]}$ be positive integers with $\sum_{g=1}^G n_g = n$. We denote by $A_{[g]}$ the $g$-th row-wise block of $A$, of dimension $n_g \times m$. When $A$ is an $n \times n$ square matrix, we denote by $A_{[g,h]}$ the $(g,h)$-th block of $A$. For a positive semi-definite square matrix $A$, denote its largest and smallest eigenvalues by $\lambda_{\max}(A)$ and $\lambda_{\min}(A)$, respectively. Let $C$ be a generic positive constant independent of $n$, whose value may change from line to line. For brevity, we write $\sum_{g,h \in [G]^2, g\neq h} := \sum_{g \in [G]}\sum_{h \in [G], h \neq g}$.

\section{Rule of Thumb and Empirical Illustration} \label{sec: illu}

In this section, we develop a practical rule of thumb that can be directly applied to reported estimates and standard errors from regressions using low-dimensional IVs and from regressions employing many IVs. We illustrate its empirical relevance using the application in \citet[Section VII]{Goldsmith(2020)}, which builds on \cite{card(2009)}.

We quantify the efficiency gain from the combination test by the percentage reduction in the length of its confidence interval, relative to that of the Wald test, in large samples. Section~\ref{sec: efficiency_gain} formally derives an analytical expression for this reduction as a function of the identification strengths of both the low-dimensional and many instrumental variables (IVs), as well as the limiting correlations between the Wald and LM statistics and between the LM and AR statistics. Because the percentage reduction is monotonically increasing in the absolute correlation between the LM and AR statistics, we further obtain a lower bound on this reduction by setting this correlation to zero. This yields a lower bound on the efficiency gain that depends only on the correlation between the Wald and LM statistics (denoted as $\rho_1$) and on the ratio of the standard deviations of the GMM estimator based on low-dimensional IVs to the leave-one-cluster-out jackknife IV estimator (JIVE) based on many IVs (or, equivalently, on the relative strength of the many IVs to the low-dimensional IVs).



\begin{figure}[H]
    \caption{Theoretical lower bounds for percentage reduction in confidence interval length.}
    \centering
    \includegraphics[width=0.9\textwidth]
    {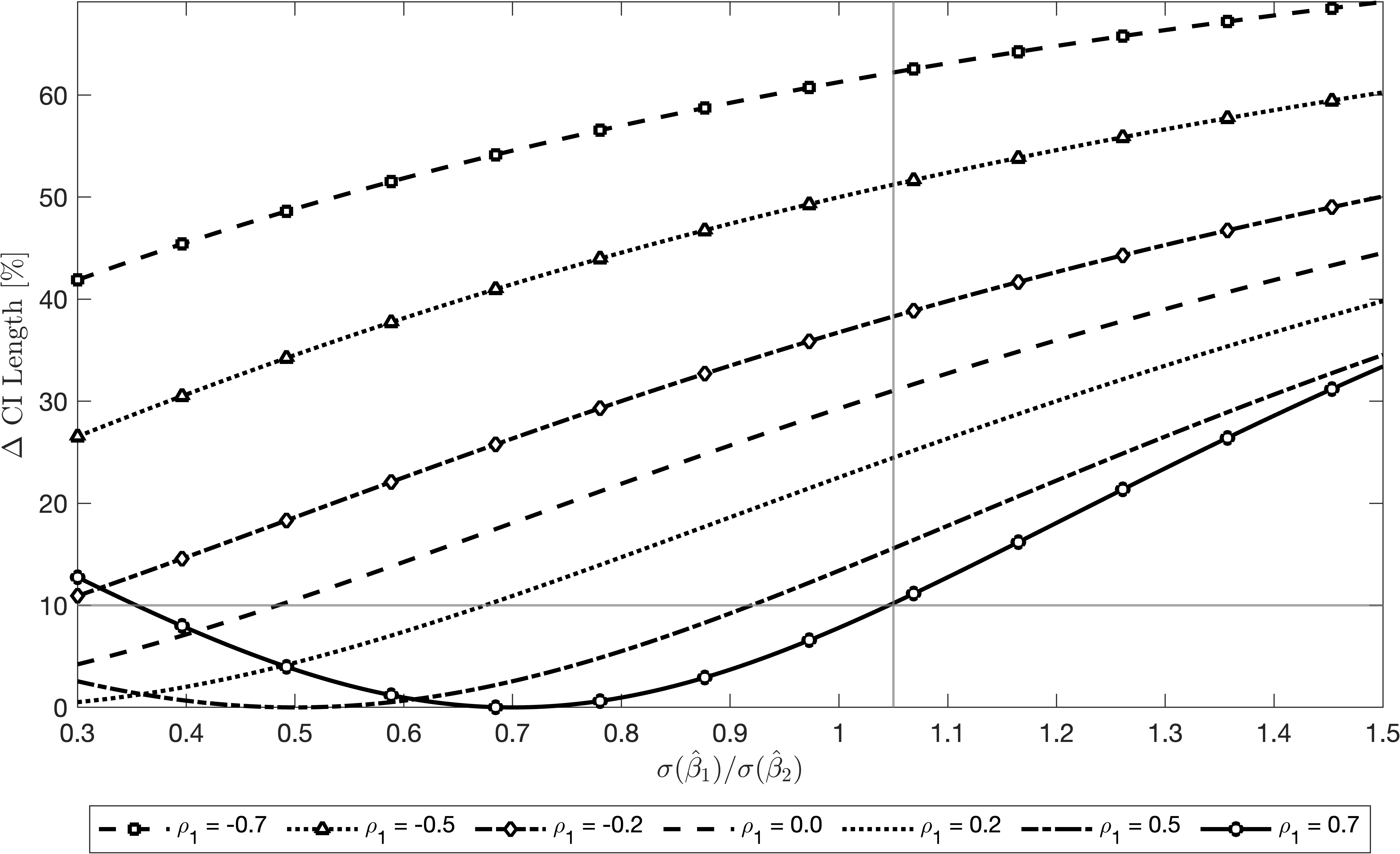}
    \floatfoot{\doublespacing \textit{Notes}: This figure plots the lower bound of the efficiency gain, given by (\ref{eq: efficiency_gain_bound}), as a function of the standard deviation ratio, 
    $\sigma(\hat\beta_1)/\sigma(\hat\beta_2)$, for various values of $\rho_1$, the limiting correlation between the Wald and (leave-one-cluster-out jacknife) LM statistics. The horizontal axis is the ratio of standard deviation of $\hat \beta_1$, the standard GMM estimator using low-dimensional IVs, and standard deviation of $\hat \beta_2$, the leave-one-cluster-out estimator using the many base IVs. The vertical axis is the reduction in the length of confidence interval in percentage points.}
    \label{fig:CI_theory}
\end{figure}

Figure~\ref{fig:CI_theory} plots the lower bound as a function of the standard deviation ratio for different values of $\rho_1$, the limiting correlation between the Wald and LM statistics. Two observations emerge. First, for any fixed $\rho_1$, we show theoretically that whenever the standard deviation ratio exceeds $\rho_1$, the lower bound on efficiency gains increases with the ratio, reflecting the fact that the combination test exploits the additional precision provided by the LM statistic. Second, once the standard deviation ratio exceeds one, the lower bound decreases as $\rho_1$ increases, because the LM statistic then contributes relatively little additional information beyond the highly correlated Wald statistic. As a simple rule of thumb that only requires a back of envelope calculation based on the reported standard errors, we propose: for empirically plausible values of $\rho_1$ (between $-0.7$ and $0.7$), whenever the standard error from the regression with low-dimensional IVs divided by that from the regression with many IVs is greater than $1.05$, the corresponding confidence interval is reduced by at least $10\%$.\footnote{In additional (unreported) plots for $\rho_1 \in [-0.99, 0.99]$, the confidence interval is at least $10\%$ shorter whenever the standard deviation ratio exceeds $1.11$ (with thresholds $1.05$ for $5\%$ and $1.25$ for $20\%$).} We further argue in Section \ref{sec: efficiency_gain} that the same rule-of-thumb can be applied to other many-IV estimators such as HLIML and HFUL developed by \cite{Haus2012}. 

To illustrate the empirical relevance of efficiency gains and the associated rule of thumb, we implement the combination test in an empirical application that estimates the (negative) inverse elasticity of substitution between immigrants and natives, following \cite{card(2009)}. As in \citet[Section VII]{Goldsmith(2020)}, we examine two separate sets of results by skill group: high school equivalent workers and college equivalent workers. The analysis is based on cross-sectional regressions for each skill group in 124 cities in 2000. The dependent variable is the residual log wage gap between immigrant and native men, and the main regressor of interest is the log ratio of immigrant-to-native hours for both men and women within the same skill group. Because a positive labor demand shock to immigrants can simultaneously increase their earnings and labor supply relative to natives, this introduces potential endogeneity. To construct the one-dimensional Bartik instrument (i.e., shift-share instrument), immigration shares from 38 origin countries (groups) in $1980$ are used as the base instruments, and the final instrument is formed as a weighted average of these country-specific shares, where the weights are given by the number of arrivals to the United States between 1990 and 2000 by origin country group and skill group (see \cite{Goldsmith(2020)} for further details). As argued in \cite{card(2009)}, the rationale for the IV is that existing immigrant enclaves are likely to attract additional immigrant labor through social and cultural channels unrelated to labor market outcomes. 

Table \ref{tab:Card_results_rotated_midpoint} reports the point estimates obtained from the two-stage least squares (TSLS) estimator using the Bartik instrument, $\hat \beta_1$, along with those from the leave-one-cluster-out estimator, $\hat \beta_2$, which relies on all the 38 base IVs. We present results separately for specifications that include and exclude city-level controls. As expected and consistent with the findings in \cite{Goldsmith(2020)}, the estimates are broadly similar within each skill group. However, in every specification, the confidence intervals constructed from $\hat \beta_1$ (Wald CI) differ to some extent from those based on $\hat \beta_2$ (LM CI). This discrepancy is partly due to the different standard errors of the two estimators, which we exploit in our combination test to obtain strictly shorter confidence intervals across all specifications. For example, for workers with college equivalent skills, our confidence intervals are roughly $7\%$ and $17\%$ shorter, respectively, than the Wald CIs in the specifications with and without controls. 

\begin{figure}[H]
    \centering
    \includegraphics[width=0.9\textwidth]{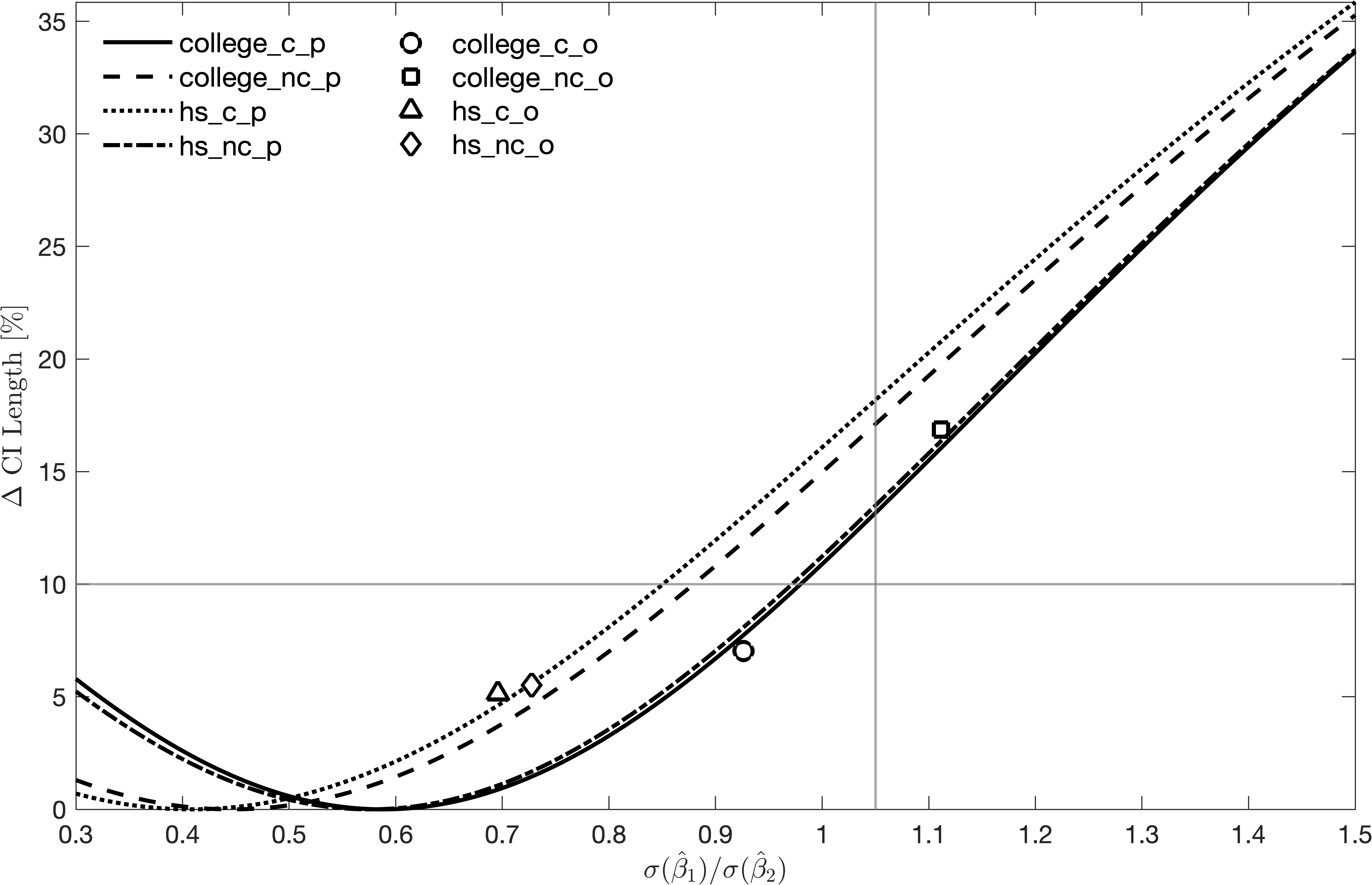}
    \caption{Realized percentage reduction in confidence interval length: Immigrant enclave.}
    \floatfoot{\doublespacing \textit{Notes}: This figure shows, for each specification in the immigrant enclave example, the observed percentage decrease in confidence interval length (Combined CI versus Wald CI, as in Table \ref{tab:Card_results_rotated_midpoint}, and indicated by ``o" in figure legends) plotted as a point against the standard error ratio ($\hat \sigma(\hat \beta_1) / \hat \sigma(\hat \beta_2)$ in Table \ref{tab:Card_results_rotated_midpoint}). Also shown is the theoretical lower bound for the reduction (indicated by ``p" in figure legends), analogous to Figure \ref{fig:CI_theory}, but now computed using the specification-specific estimate $\hat\rho_1$, as reported in Table \ref{tab:Card_results_rotated_midpoint}. Here, ``college" refers to the specifications for college equivalent workers, and ``hs" refers to the specifications for high school equivalent workers. ``c" indicates that controls are included, whereas ``nc" indicates that controls are not included. The horizontal axis is the ratio of standard deviations (errors) of $\hat \beta_1$ and $\hat \beta_2$. The vertical axis is the reduction in the length of confidence interval in percentage points. As a final remark, note that the actual numerical values of the relevant quantities in Table \ref{tab:Card_results_rotated_midpoint}, rather than the rounded values shown there, are used to produce Figure \ref{fig:CI_observed_Card_full}.}
    \label{fig:CI_observed_Card_full}
\end{figure}

Figure \ref{fig:CI_observed_Card_full} displays the realized percentage reduction in confidence interval length achieved by our combination test, together with the asymptotic lower bounds for that reduction, analogous to Figure \ref{fig:CI_theory}, but calculated using the specification-specific estimate $\hat \rho_1$ of the limiting correlation between the Wald and LM statistics. Two observations are worth noting. First, up to finite-sample estimation error and across different specifications, the actual percentage reductions generally fall near or above the corresponding lower bounds, in line with our theoretical predictions. Note also that these bounds are derived solely from the Wald and LM statistics, indicating that, in this empirical setting, incorporating the AR statistic yields little additional efficiency gain. This aligns with our earlier theoretical discussion, which posits that efficiency gains increase monotonically with the absolute correlation between LM and AR statistics, and, in fact, the corresponding consistent estimates $\hat\rho_2$ are not particularly large in almost every specification. 

Second, Figure \ref{fig:CI_observed_Card_full} clearly illustrates our proposed rule of thumb. All estimated $\hat \rho_1$ values fall within $[-0.7, 0.7]$. When the standard error ratio exceeds $1.05$, the actual reduction in the length of the confidence interval is $17\%$ (specification without controls for college equivalent workers), much above the rule-of-thumb benchmark $10\%$ for its lower bound. In contrast, when the standard error ratio remains at or below $1.05$, the improvements are modest, reflecting the converse of our rule of thumb. Nevertheless, even in these cases, our combination test can still deliver confidence intervals that are about $7\%$ (specification with controls for college equivalent workers), $6\%$ (specification without controls for high school equivalent workers), and $5\%$ (specification with controls for high school equivalent workers) shorter. In Supplementary Appendix F, we further consider the return to education application of \cite{Angrist-Krueger(1991)}, which illustrates even more notable efficiency gains provided by our approach.

\section{Model and Preliminaries}\label{sec: setup}

\subsection{Setup} \label{sec: model_setup}

We consider a clustered dataset with $G$ clusters and denote the size of the $g$-th cluster as $n_g$ for $g \in [G]$. We index observations by units followed by clusters. Denote $I_g = [N_{g-1}+1,\cdots,N_{g}]$, where $N_0 = 0$, $N_g = \sum_{g'=0}^g n_{g'}$, and $N_G =n$. Then, $\{I_g\}_{g \in [G]}$ forms a partition of $[n]$, and if $i \in I_g$, this means that the $i$-th observation belongs to the $g$-th cluster. We then consider a linear IV regression with clustered data:
\begin{eqnarray}\label{eq: IV-model}
\tilde Y_{i,g} = \tilde X_{i,g} \beta + W^{\top}_{i,g} \gamma + \tilde e_{i,g},
\end{eqnarray}
where we denote 
$\tilde Y_{i,g} \in \Re$, $\tilde X_{i,g} \in \Re$, and $W_{i,g} \in \Re^{d_w}$ as an outcome variable, an endogenous regressor, and exogenous regressors, respectively. Further denote $\tilde Z_{i,g} \in \Re^{K}$ as the IVs for $\tilde X_{i,g}$. The first-stage equation can be written as
\begin{eqnarray}\label{eq:first-stage}
    \tilde    X_{i,g} = \tilde \Pi_{i,g} + \tilde V_{i,g}, 
\end{eqnarray}
where $\tilde \Pi_{i,g} = \mathbb E (\tilde X_{i,g}|\{\tilde Z_{j,g},W_{j,g}\}_{j \in I_g})$ is not assumed to be linear in $\tilde Z_{i,g}$ and $W_{i,g}$. We assume that $\mathbb E \tilde e_{i,g} =0$ and $\mathbb E \tilde V_{i,g} = 0$, and $\left\{ \tilde e_{i,g}, \tilde V_{i,g} \right\}_{i \in I_g, g \in [G]}$ are independent between clusters, but allow them to have a general dependence structure within each cluster.  Throughout the paper, the dimension $d_w$ of $W_{i,g}$ is assumed to be fixed. If researchers want to include cluster fixed effects in the model, they can obtain \eqref{eq: IV-model} by first demeaning the data (outcome, endogenous regressor, controls, and instruments) at the cluster level. 
We assume that $K$, the dimension of $\tilde Z_{i,g}$, diverges to infinity with the sample size. 

Let $\tilde Y$, $\tilde X$, $\tilde \Pi$, $W$, $\tilde Z$ be $n \times 1$, $n \times 1$, $n\times 1$, $n \times d_w$, and $n \times K$-dimensional vectors and matrices formed by $\tilde Y_{i,g}$, $\tilde X_{i,g}$, $\tilde \Pi_{i,g}$, $W_{i,g}$, and $\tilde Z_{i,g}$, respectively. More specifically, $\tilde Y$ is constructed by stacking up $\tilde Y_{i,g}$ across $i \in I_g$ followed by $g \in [G]$, and similarly for $\tilde X$, $\tilde \Pi$, $W$ and $\tilde Z$. We then partial out $W$ from $\tilde Y$, $\tilde X$, and $\tilde Z$, 
so that the model in \eqref{eq: IV-model}-\eqref{eq:first-stage} can be written in a vector form as
\begin{align}
Y = X \beta + e, \quad  X = \Pi + V,
\label{eq:model}
\end{align}
where $Y = M_W \tilde Y$, $X = M_W \tilde X$, $\Pi = M_W \tilde \Pi$, $e = M_W \tilde e$, $V = M_W \tilde V$, $M_W = I_n - P_W$, $P_W = W(W^\top W)^{-1}W^\top$, and $I_n$ denotes an $n \times n$ identity matrix. We further denote $Z = M_W \tilde Z$. 

In addition, besides the $K$-dimensional many IVs $\tilde Z_{i,g} \in \Re^K$, we assume that there is another set of low-dimensional IVs 
\begin{align*}
    \tilde{z}_{i,g} = f_{i,g}(\tilde Z,W) \in \Re^{d_z}, 
\end{align*}
where $\{f_{i,g}(\cdot)\}_{i \in I_g, g\in [G]}$ is a list of known nonstochastic functions of $d_z$ dimension. 
Specifically, as illustrated by the example of \cite{Angrist-Krueger(1991)} in the Introduction, researchers may begin with certain low-dimensional base IVs $\tilde{z}_{i,g}$, such as the three QoB dummies, and construct a large number of new IVs by taking the interaction between $\tilde z_{i,g}$ and control variables $W_{i,g}$ (e.g., state- and year-of-birth dummies in \cite{Angrist-Krueger(1991)}). Then, $\tilde{z}_{i,g}$  is a subset of the $K$-dimensional many IVs $\tilde Z_{i,g}$ for the model in \eqref{eq: IV-model}-\eqref{eq:first-stage}, which include both the low-dimensional base IVs and interacted IVs. The second example of $\tilde z_{i,g}$ is the widely used shift-share IV. As pointed out by \cite{Goldsmith(2020)}, under their identification strategy that treats the shares as exogenous, the one-dimensional shift-share IV can be regarded as a weighted average of many base IVs.  For instance, in the canonical setting of estimating the inverse elasticity of labor supply (e.g., see Sections I and VI of \cite{Goldsmith(2020)}), the observations are typically clustered at the location level, such as the US commuting zone level, with a short panel dataset of $T$ time periods. The structural equation of interest can thus be written as
\begin{align*}
    \tilde Y_{g,t} = \tilde X_{g,t} \beta + W_{g,t}^{\top}\gamma + \tilde e_{g,t}, 
\end{align*}
where $g$ indexes a location, $t$ a time period, $\tilde Y_{g,t}$ is wage growth, $\tilde X_{g,t}$ is employment growth, and $W_{g,t}$ is a vector of controls which could include location and time fixed effects.
Then, according to our notation, $G$ is equal to the number of locations, and $n_g = T$ for all $g \in [G]$.
The shift-share IV is an inner product of the initial industry-location shares and the industry-period growth rates, i.e., $\tilde z_{g,t} = \sum_{k=1}^K \tilde s_{0,g,k} h_{k,t}$, where the employment share of industry $k$ in the location $g$ in a certain initial period corresponds to the base IV $\tilde s_{0,g,k}$, the growth rate of industry $k$ in the period $t$ corresponds to the weight $h_{k,t}$, and $K$ is the number of industries. 

Additionally, let $\tilde z$ be the $n \times d_z$-dimensional matrix formed by $\tilde z_{i,g}$, and denote $z = M_W \tilde z$. In many empirical applications, the dimension $d_z$ is just one (e.g., the shift-share IV), but our setup also allows for $d_z>1$, while maintaining the requirement that $d_z$ is fixed with respect to the sample size $n$. 

The null and alternative hypotheses studied are $\mathcal{H}_0: \beta = \beta_0 \  \text{against}\  \mathcal{H}_{1}:  \beta \neq \beta_0$. We focus on the model with a scalar endogenous variable for two reasons. First, in many empirical applications of IV regressions, there is only one endogenous variable.\footnote{For example, 101 out of 230 specifications in \cite{andrews2019weak}'s sample and 1,087 out of 1,359 in \cite{young2022consistency}'s sample feature one endogenous regressor and one IV. Similarly, \cite{lee2022valid} find that 61 out of 123 IV papers published in \textit{AER} between 2013 and 2019 use a single IV. Our setting further allows for the overidentified case with one endogenous regressor and multiple IVs. 
In general, empirical researchers can generate many IVs by using polynomials or interactions based on their low-dimensional base IVs and control variables, in the same spirit as \cite{Angrist-Krueger(1991)}. Then, it is possible to achieve efficiency improvement using our combination procedure.} Second, if we assume that at least the low-dimensional IVs provide strong identification, our results can be extended to testing of scalar restrictions with multiple endogenous variables by applying standard subvector inference methods, without appealing to a projection-based weak-identification-robust inference approach.\footnote{For weak-identification-robust subvector inference, in general, one may use a projection approach (\citep{Dufour-Taamouti(2005)}) after implementing inference on the whole vector of endogenous variables. However, the projection approach typically leads to conservative inference. Alternative subvector inference methods for IV regressions (e.g., see \cite{GKMC(2012)}, \cite{Andrews(2017)}, \cite{GKM(2019), GKM(2021)}, and \cite{Wang-Doko(2018)}) provide a power improvement over under a fixed number of instruments (some of these methods further require conditional homoskedasticity). However, whether they can be applied to the setting of many weak instruments is unclear.} 
In addition, our method could potentially be extended to the case in which the identification strength provided by many IVs is mixed for the endogenous variables, by following the approach of \citet{Chao(2012)}. We leave these extensions for future research.



\subsection{Test Statistics}\label{subsec:test_stat}

Under our setting, it is possible to conduct inference directly based on the low-dimensional IVs. Specifically, given a $d_z \times d_z$ positive definite weighting matrix $\hat A_n$, the generalized method of moments (GMM) estimator can be written as 
\begin{align}
    \hat \beta_1 = (X^\top z \hat A_n z^\top X)^{-1}(X^\top z \hat A_n z^\top Y). \label{eq:beta1hat}
\end{align}

It is also possible to construct test statistics using the $K$-dimensional many IVs. Denote $P = Z(Z^\top Z)^{-1}Z^\top$ as the projection matrix of $Z$. 
Then, the leave-one-cluster-out jackknife IV estimator of $\beta$ is denoted as $\hat \beta_2$ and defined as 
\begin{align}
    \hat \beta_2 & = \left(\sum_{g,h \in [G]^2, g \neq h} X_{[g]}^\top P_{[g,h]} X_{[h]} \right)^{-1} \left(\sum_{g,h \in [G]^2, g \neq h} X_{[g]}^\top P_{[g,h]} Y_{[h]} \right) \nonumber \\
    &= \left( X^\top (P - \bar P) X \right)^{-1} \left(X^\top (P - \bar P) Y \right), \label{eq:beta2hat} 
\end{align}
where $\bar P$ is the block diagonal matrix corresponding to $P$ such that the $g$-th block on its diagonal is $P_{[g,g]}$.
Note that under independent data, $\hat \beta_2$ reduces to the JIVE estimator in \cite{angrist1999}, \cite{Chao(2012)}, and \cite{MS22}. 

Given $\hat \beta_1$ and $\hat \beta_2$, we define the estimator $\hat \beta$ as
\begin{align}\label{eq: beta_hat}
    \hat \beta = \frac{\ddot \Phi_2^{1/2}}{\dot \Phi_1^{1/2} + \ddot \Phi_2^{1/2}} \times \hat \beta_1 + \frac{\dot \Phi_1^{1/2}}{\dot \Phi_1^{1/2} + \ddot \Phi_2^{1/2}} \times \hat \beta_2,
\end{align}
where $\dot \Phi_1$ and $\ddot \Phi_2$ are the variance estimators for $\hat \beta_1$ and $\hat \beta_2$, respectively, to be defined later in Section \ref{subsec:assumptions}.
We show that \(\hat{\beta}\) is consistent whenever either the low-dimensional IV estimator \(\hat{\beta}_1\) or the many-IV estimator \(\hat{\beta}_2\) is consistent (i.e., either the low-dimensional IVs or the many IVs provide strong identification for $\beta$). 
Because researchers do not need to know which of the two estimators is consistent when constructing \(\hat{\beta}\), the estimator is doubly robust.


We then use the doubly robust estimator \(\hat{\beta}\) to re-estimate the variances associated with \(\hat{\beta}_1\) and \(X^{\top}(P-\bar P)e\), denoted by \(\widehat{\Phi}_1\) and \(\hat{\Sigma}\), respectively, and defined in Section \ref{subsec:assumptions}. These variance estimates are used to construct the Wald statistic (with low-dimensional IVs) and the leave-one-cluster-out jackknife LM statistic (with many IVs):
\begin{align}
T(\beta_0)
&= \frac{(X^{\top} z \hat A_n z^{\top} X)^{-1} X^{\top} z \hat A_n z^{\top} e(\beta_0)}{\sqrt{\widehat{\Phi}_1}}
= \frac{\hat{\beta}_1 - \beta_0}{\sqrt{\widehat{\Phi}_1}}, \label{eq:Waldstatistic}
\\[6pt]
LM(\beta_0)
&= \frac{X^{\top}(P-\bar P)e(\beta_0)}{\sqrt{\hat{\Sigma}}}, \label{eq:LMstatistic}
\end{align}
where \(e(\beta_0) = Y - X\beta_0\).

Lastly, as pointed out by \cite{Haus2012}, \cite{LWZ24}, and \cite{MS24}, it is possible to use the jackknife AR statistic to further improve the efficiency of the jackknife LM statistic. In the current setting with clustered data, we define the leave-one-cluster-out jackknife AR statistic as 
\begin{align}
    AR = \frac{\hat e^\top (P - \bar P) \hat e}{\sqrt{\hat \Upsilon}}, \label{eq:AR}
\end{align}
where $\hat \Upsilon$ is a consistent variance estimator for the numerator defined later, and $\hat e = Y -X \hat \beta$.

We use the consistent estimator $\hat \beta$ to construct our tests for two reasons. First, using the null value $\beta_0$ to form the AR statistic and then combine it with the LM statistic can produce a non-monotonic power curve (i.e., a power ditch against certain alternatives; see \cite{Andrews(2016)} and \cite{LWZ24} for related discussions). Constructing the AR statistic with $\hat \beta$ avoids this issue. Consequently, the AR statistic in \eqref{eq:AR} does not depend on $\beta_0$ but serves as a normalized estimator of zero, used solely to improve the efficiency of our procedure. Second, the combination test relies on correlations between $T(\beta_0)$ and $LM(\beta_0)$ and between $LM(\beta_0)$ and $AR$, whose consistent estimation requires a consistent $\beta$ estimator. This consistency is crucial to ensure proper size control, especially when the many-IV specification is weakly identified.


In the next section, we illustrate how to optimally combine the three test statistics \(T(\beta_0)\), \(LM(\beta_0)\), and \(AR\). The corresponding variance estimators \((\dot{\Phi}_1, \ddot{\Phi}_2, \widehat{\Phi}_1, \hat{\Sigma}, \hat{\Upsilon})\) are introduced later in Section~\ref{subsec:assumptions}.

\subsection{Combination Test}

Given the three test statistics $(T(\beta_0), LM(\beta_0), AR)$, we seek to combine them in a theoretically justified way that can improve on the Wald test based only on the low-dimensional IVs. The key insight of our paper is that under certain local alternative $\beta - \beta_0 = \delta d_n$ with some deterministic sequence $d_n \downarrow 0$, we have the following joint limiting distribution:
\begin{align}
\begin{pmatrix}
    T(\beta_0) \\
    LM(\beta_0) \\
    AR
\end{pmatrix} \convD \N \left( \begin{pmatrix}
    a_1 \delta \\
    a_2 \delta \\
    0
\end{pmatrix}, \begin{pmatrix}
    1 &  \rho_1 & 0 \\
    \rho_1 & 1 & \rho_2 \\
    0 & \rho_2 & 1
\end{pmatrix} \right) \label{eq:weak_convergence}
\end{align} 
for some $a_1$, $a_2$, $\rho_1$ and $\rho_2$ to be defined later. In this limiting problem, the UMPU level-$\alpha$ test for the default null hypothesis $\delta=0$ against two-sided alternatives, which are solely based on the limiting three-dimensional normal random vector, can be obtained by invoking standard hypothesis-testing results (see, for example, Section 4.2 of \cite{Lehmann-Romano(2006)}), and is stated in Proposition \ref{prop:limit} below.

\begin{prop}

Suppose that one observes $(\N_1,\N_2,\N_3)$, which follows the limiting distribution in (\ref{eq:weak_convergence})
with $\rho_{1}^2 + \rho_{2}^2 < 1$ and wants to test $\mathcal H_0: \delta = 0$ against $\mathcal H_1: \delta \neq 0$ for known values of $(a_1,a_2,\rho_1,\rho_2)$, then the UMPU level-$\alpha$ test rejects if 
\begin{align*}
   \left( \frac{b_1 \tilde \N_1 + b_2 \tilde \N_2 + b_3 \tilde \N_3}{\sqrt{b_1^2 + b_2^2 + b_3^2 }}\right)^2 \geqslant \mathbb C_\alpha,
\end{align*}
where $\mathbb C_\alpha$ is the $(1-\alpha)$ percentile of a chi-squared random variable with one degree of freedom,
\begin{align*}
\begin{pmatrix}
    \tilde \N_1 \\
    \tilde \N_2 \\
    \tilde \N_3
\end{pmatrix} = \begin{pmatrix}
    1 &  \rho_1 & 0 \\
    \rho_1 & 1 & \rho_2 \\
    0 & \rho_2 & 1
\end{pmatrix}^{-1/2} 
\begin{pmatrix}
    \N_1 \\
    \N_2 \\
    \N_3
\end{pmatrix},\ and\   
    \begin{pmatrix}
        b_1 \\
        b_2 \\
        b_3
    \end{pmatrix}= \begin{pmatrix}
    1 &  \rho_1 & 0 \\
    \rho_1 & 1 & \rho_2 \\
    0 & \rho_2 & 1
\end{pmatrix}^{-1/2} 
\begin{pmatrix}
    a_1 \\
    a_2 \\
    0
\end{pmatrix}.
\end{align*} 
The corresponding power function for the UMPU test is 
\begin{align*}
   \mathbb P\left( \chi^2_1 \left(\delta^2 \frac{(1-\rho_2^2)a_1^2-2\rho_1 a_1 a_2 + a_2^2}{1-\rho_1^2-\rho_2^2} \right)  \geq \mathbb C_\alpha \right),
\end{align*}
where $\chi^2_1(\lambda)$ is a noncentral chi-squared with noncentrality $\lambda$ and one degree of freedom. 
\label{prop:limit}
\end{prop}

In light of this optimal testing result in the limiting problem, one may wish to propose implementing the following test:
\begin{align}
        & \phi^{o}_n  = \mathbf 1 \left\{ \left(\omega_1  T(\beta_0) + \omega_2 LM(\beta_0) +  \omega_3 AR \right)^2 \geqslant \mathbb C_\alpha \right\}\equiv \phi^*\left(T(\beta_0),LM(\beta_0),AR\right), \label{eq:opttest_oracle} \\
& \text{where}\quad    \begin{pmatrix}
        \omega_1 \\
        \omega_2 \\
        \omega_3
    \end{pmatrix} = \frac{1}{\sqrt{ b_1^2 +  b_2^2 +  b_3^2}}\begin{pmatrix}
    1 &  \rho_1 & 0 \\
    \rho_1 & 1 & \rho_2 \\
    0 & \rho_2 & 1
\end{pmatrix}^{-1/2} 
\begin{pmatrix}
    b_1 \\
    b_2 \\
    b_3
\end{pmatrix},\label{eq:omega}
\end{align}
and then investigate its asymptotic justification. 

However, the parameters $a_1,a_2,\rho_1,\rho_2$ are usually unknown and need to be estimated. In addition, it turns out that the weights $(\omega_1,\omega_2,\omega_3)$ are invariant to the scale normalization of $(b_1,b_2,b_3)$, and thus, $(a_1,a_2)$. Therefore, to construct the UMPU test, it suffices to consistently estimate $\alpha_1 = a_1/\sqrt{a_1^2+a_2^2}$ and $\alpha_2 = a_2/\sqrt{a_1^2+a_2^2}$ along with $\rho_1$ and $\rho_2$. 

Given the consistent estimators $(\hat \alpha_1,\hat \alpha_2,\hat \rho_1,\hat \rho_2)$ for $( \alpha_1, \alpha_2, \rho_1, \rho_2)$ specified in Section \ref{subsec:assumptions}, we then implement the feasible version of the combination test: 
\begin{align}
        \phi^{*}_n = \mathbf 1 \left\{ \left(\hat \omega_1  T(\beta_0) + \hat \omega_2 LM(\beta_0) +  \hat \omega_3 AR \right)^2 \geqslant \mathbb C_\alpha \right\},  \label{eq:opttest}
\end{align}
\begin{align}\label{eq:omegahat}
    \text{where}    \begin{pmatrix}
    \hat \omega_1 \\
    \hat \omega_2 \\  
    \hat \omega_3
    \end{pmatrix}
    = \frac{1}{\sqrt{\hat b_1^2 + \hat b_2^2 + \hat b_3^2}} 
    \times \begin{pmatrix}
    1 & \hat \rho_1 & 0  \\
    \hat \rho_1 & 1 & \hat \rho_2  \\
    0 & \hat \rho_2 & 1
    \end{pmatrix}^{-1/2} 
    \begin{pmatrix}
            \hat b_1 \\
            \hat b_2 \\
            \hat b_3
    \end{pmatrix}, \begin{pmatrix}
    \hat b_1 \\
    \hat b_2 \\
    \hat b_3        
    \end{pmatrix}
    = \begin{pmatrix}
    1 & \hat \rho_1 & 0 \\
    \hat \rho_1 & 1 & \hat \rho_2 \\
    0 & \hat \rho_2 & 1
    \end{pmatrix}^{-1/2} 
    \begin{pmatrix}
            \hat \alpha_1 \\
            \hat \alpha_2 \\
            0
    \end{pmatrix}.
\end{align}

\section{Large-Sample Theory} \label{sec: comb}

In this section, we investigate the asymptotic behavior of our combination test. We begin by stating and discussing general assumptions about the data-generating process and the identification strength of both the low-dimensional and many IVs. We then establish the asymptotic efficiency properties of the combination test and, finally, compare its efficiency to that of the conventional Wald test based solely on low-dimensional IVs via the limiting length ratio of their confidence intervals.

\subsection{General Assumptions} \label{subsec:assumptions}

As in \cite{Chao(2012)}, we treat $\tilde Z$ and $W$ as fixed. This is equivalent to treating them as random and repeating all the analyses in the paper by conditioning on them. For the data-generating process, we impose the following assumptions. 
\begin{ass} 
The following conditions hold when $n$ is sufficiently large: 
\begin{enumerate}[label=\arabic*., ref=\theass.\arabic*]
    \item \label{ass:reg_moment} $\max_{i \in I_g, g \in [G]} \mathbb E (\tilde e_{i,g}^4 + \tilde V_{i,g}^4) \leqslant C < \infty$;
    \item \label{ass:reg_cluster_size} $\max_{1 \leq g \leq G} n_g \leqslant C < \infty$;
    \item \label{ass:reg_cluster_dependence} Let $$\Omega_g = \mathbb E \left[  \begin{pmatrix} \tilde e_{[g]} \tilde e^\top_{[g]} & \tilde e_{[g]} \tilde V^\top_{[g]} \\ \tilde V_{[g]} \tilde e^\top_{[g]} & \tilde V_{[g]} \tilde V^\top_{[g]} \end{pmatrix}\right], \quad 1 \leq g \leq G,$$
    then $$0 < \frac{1}{C} \leqslant \min_{1\leq g \leq G} \lambda_{\min}\left(\Omega_g \right) \leqslant \max_{1\leq g \leq G} \lambda_{\max} \left(\Omega_g \right) \leqslant C < \infty;$$
    \item \label{ass:reg_design}
    \begin{align*}
    0 < \frac{1}{C} \leqslant \lambda_{\min}\left(\frac{1}{n}\sum_{i \in I_g, g \in [G]} z_{i,g} z_{i,g}^\top \right) \leqslant \lambda_{\max}\left(\frac{1}{n}\sum_{i \in I_g, g \in [G]} z_{i,g} z_{i,g}^\top \right) \leqslant C < \infty, \\ 
    0 < \frac{1}{C} \leqslant \lambda_{\min}\left(\frac{1}{n}\sum_{i \in I_g, g \in [G]}  W_{i,g} W_{i,g}^\top \right) \leqslant \lambda_{\max}\left(\frac{1}{n}\sum_{i \in I_g, g \in [G]} W_{i,g} W_{i,g}^\top \right) \leqslant C < \infty,
    \end{align*} and $$\max_{i \in I_g, g \in [G]} \left\{\left| \tilde \Pi_{i,g} \right| + \left|\Pi_{i,g} \right|  \right\} \leqslant C < \infty, \;\; \max_{i \in I_g, g \in [G]} \left\{\left\Vert z_{i,g} \right\Vert_2 + \left\Vert W_{i,g} \right\Vert_2 \right\} = o(\sqrt{n});$$
    \item \label{ass:low_An} There exists a sequence of non-random positive definite matrices $A_n$ 
    such that
    \begin{align*}
     A_n^{-1/2} \hat A_n A_n^{-1/2} \convP I_{d_z},
    \end{align*}
and $\lambda_{\max} (A_n) / \lambda_{\min} (A_n) \leqslant C < \infty$. 
\end{enumerate}\label{ass:reg}
\end{ass}

\begin{rem}
    Assumption~\ref{ass:reg_moment} is a standard condition on the moments of error terms. Assumption~\ref{ass:reg_cluster_size} restricts the cluster size to be bounded, which incorporates cross-sectional and short-panel data structures. It is also possible to extend our analysis to the case with divergent cluster sizes, especially when the within-cluster dependence is weak. However, when the cluster size is allowed to diverge and there is strong within-cluster dependence, the convergence rates of various IV estimators depend on the identification strength, the within-cluster dependence, the cluster sizes, and the number of clusters in a complicated way. We leave the investigation in this direction for future research. Assumption~\ref{ass:reg_cluster_dependence} ensures that the error covariance matrix is non-singular for each cluster. 
    Assumption~\ref{ass:reg_design} is a mild condition for the design matrix. Assumption~\ref{ass:low_An} states that the weighing matrix $\hat{A}_n$ converges in probability to some non-random positive definite matrix, which is standard in the GMM setup. We note that for two-stage least squares (TSLS), Assumption~\ref{ass:low_An} is actually implied by Assumption~\ref{ass:reg_design}.
\end{rem}

For the low-dimensional IVs, we focus on the case of strong identification strength in the main text. We further investigate the case in which the low-dimensional IVs have weak identification strength in Supplementary Appendix A, and show that, under certain conditions, when the many IVs provide strong identification, our optimal combination test still asymptotically controls size. 

With strong identification, the asymptotic variance of $\hat \beta_1$ is
\begin{align*}
    \Phi_1 = (\Pi^\top \acute \Pi)^{-1} \Psi(\Pi^\top \acute \Pi)^{-1}, 
\end{align*}
where $\Omega = \mathbb E \sum_{g \in [G]} \left(z_{[g]}^\top \tilde e_{[g]}\right) \left(z_{[g]}^\top \tilde e_{[g]}\right)^\top$, $\acute \Pi = z A_n z^\top \Pi$ and $\Psi=\Pi^\top z A_n \Omega A_n z^\top \Pi$. The cluster-robust variance estimator $\widehat \Phi_1$ for the Wald statistic with low-dimensional IVs is 
\begin{align}
 \widehat \Phi_1 =  (X^\top \acute X)^{-1} \hat \Psi (X^\top \acute X)^{-1}, \label{eq:hatPhi_1}
\end{align}
where $\hat \Omega  = \sum_{g \in [G]} \left(z_{[g]}^\top \hat e_{[g]}\right) \left(z_{[g]}^\top \hat e_{[g]}\right)^\top$, $\acute X = z \hat A_n z^\top X$ and $\hat \Psi = X^\top z \hat A_n \hat \Omega \hat A_n z^\top X$. The initial estimator $\dot{\Phi}_1$ for $\Phi_1$ used in the computation of $\hat{\beta}$ in \eqref{eq: beta_hat} is defined in the same way as $\widehat{\Phi}_1$, except that $\hat e_{[g]} = Y_{[g]} - X_{[g]} \hat{\beta}$ is replaced by $\dot e_{[g]} = Y_{[g]} - X_{[g]} \hat{\beta}_1$.

We make the following assumptions regarding the inference with low-dimensional IVs.
\begin{ass} \label{ass:low_id}
     Let $r_n = \left\Vert z^\top \Pi \right\Vert_2$, then $\sqrt{n} / r_n \rightarrow 0$.
\end{ass}

\begin{rem}
Assumption~\ref{ass:low_id} ensures that $z^\top \Pi$, the deterministic component of $z^\top X$, dominates $z^\top V$, its random component. This condition is therefore key for the consistency of $\hat \beta_1$ for $\beta$, that is, it ensures strong identification by the low-dimensional IVs. Throughout the paper, we focus on this strong identification case for the low-dimensional IVs while allowing the many-IV specification to be either weakly or strongly identified. In Section~\ref{subsec: asy_eff}, we show that the optimal combination test adapts to the identification strength of the many IVs. In particular, it controls size asymptotically and retains nontrivial power even when $\hat \beta_2$ is inconsistent. In Section A of the Appendix, we consider the opposite scenario in which the low-dimensional IV specification is weakly identified while the many-IV specification is strongly identified. We provide mild conditions under which the exact inference procedure proposed in the paper still controls asymptotic size under the null. 
\end{rem}

If the many-IV-based identification is strong, similar to \cite{Chao(2012)}, we can show that the asymptotic variance of $\hat \beta_2$ is
\begin{align*}
    \Phi_2 &= \left( \Pi^\top (P - \bar P) \Pi \right)^{-1} \Sigma \left( \Pi^\top (P - \bar P) \Pi \right)^{-1},
\end{align*}
where
\begin{align*}
    \Sigma = \mathbb E \left(\sum_{g,h \in [G]^2, g \neq h} \Pi_{[g]}^\top P_{[g,h]} \left(\sum_{k \in [G]} M_{W,[h,k]} \tilde e_{[k]}\right) \right)^2 + \mathbb E \left(\sum_{g,h \in [G]^2, g \neq h} \tilde V_{[g]}^\top P_{[g,h]} \tilde e_{[h]} \right)^2
\end{align*}
is the asymptotic variance of $X^\top (P - \bar P) e$. A natural estimator for $\Phi_2$ is thus 
\begin{align}
    \widehat \Phi_2 = \left( X^\top (P - \bar P) X \right)^{-1} \hat \Sigma \left( X^\top (P - \bar P) X \right)^{-1}, \label{eq:hatPhi_2}
\end{align}
where $\hat \Sigma$ is a consistent estimator of $\Sigma$. Such an estimator is proposed in \cite{Chao(2012)} for the case with independent data.  
Here, in addition to extending to clustered data, we need to account for the fact that $W$ has already been partialled out, whereas in \cite{Chao(2012)} the coefficients for $W$ are also estimated. Therefore, some adjustments are required, as in \cite{matsushita2024}. For that purpose, define $Q = M_W (P - \bar P) M_W$, and let $\bar Q$ be the block diagonal matrix corresponding to $Q$ such that the $g$-th block on its diagonal is $Q_{[g,g]}$. Our variance estimator is similar to the one in \cite{Chao(2012)} but with $P-\bar P$ replaced by $Q- \bar Q$, i.e., 
\begin{align}
    \hat \Sigma = \sum_{g\in [G]} \left(\sum_{h \in [G], h \neq g} \tilde X_{[h]}^\top Q_{[h,g]} \hat e_{[g]} \right)^2 + \sum_{g,h \in [G]^2, g \neq h} \left( \tilde X_{[g]}^\top Q_{[g,h]} \hat e_{[h]} \right) \left( \tilde X_{[h]}^\top Q_{[h,g]} \hat e_{[g]} \right). \label{eq:LMvarianceestimator}
\end{align} 
The initial estimator $\ddot{\Phi}_2$ for $\Phi_2$ used in the computation of $\hat{\beta}$ is defined in the same way as $\widehat{\Phi}_2$, except that $\hat e_{[g]} = Y_{[g]} - X_{[g]} \hat{\beta}$ is replaced by $\ddot e_{[g]} = Y_{[g]} - X_{[g]} \hat{\beta}_2$.

Last, for the jackknife AR statistic, the variance estimator is given by
\begin{align}
    \hat \Upsilon = 2 \sum_{g,h \in [G]^2, g \neq h} \left( \hat e_{[g]}^\top P_{[g,h]} \hat e_{[h]} \right)^2, \label{eq:ARvariianceestimator}
\end{align}
which is consistent for the asymptotic variance of $\hat e^\top (P - \bar P) \hat e$, given by
\begin{align*}
    \Upsilon =  \mathbb E \left( \sum_{g,h \in [G]^2, g \neq h}  \tilde e_{[g]}^\top P_{[g,h]} \tilde e_{[h]} \right)^2.
\end{align*}

We make the following assumptions regarding the inference with many IVs.
\begin{ass}
\begin{enumerate}[label=\arabic*., ref=\theass.\arabic*]
    \item \label{ass:high_K} $K \rightarrow \infty$ as $n \rightarrow \infty$ such that $\limsup_{n \rightarrow \infty}{K/n} \leqslant C < 1$;
    \item \label{ass:high_P} $\rank(P) = K$ and $\max_{1 \leq g \leq G} \lambda_{\max} \left(P_{[g,g]} \right) \leqslant C < 1$;
    \item \label{ass:high_upper} Let $\hat \Pi = M_W (P - \bar P) \Pi = Q \tilde \Pi$ and $\bar \Pi = (Q - \bar Q) \tilde \Pi$, then 
    \begin{align*}
       \max_{i \in I_g, g \in [G]} \left\{\left| \hat \Pi_{i,g} \right| + \left|\bar \Pi_{i,g} \right|  \right\} \leqslant C < \infty,
    \end{align*}
    and 
    \begin{gather*}
        \tilde \Pi^\top \tilde \Pi \leq C \Pi^\top \Pi, \;\;
        \hat \Pi^\top \hat \Pi \geq \Pi^\top \Pi / C, \;\; 
       \Pi^\top (P - \bar P) \Pi \geq \Pi^\top \Pi/C, 
    \end{gather*} 
    when $n$ is large enough;
    \item \label{ass:high_lower} For all sufficiently large $n$,
    \begin{align*}
    \left|corr \left(\sum_{g,h \in [G]^2, g \neq h} \tilde V_{[g]}^\top P_{[g,h]} \tilde V_{[h]}, \sum_{g,h \in [G]^2, g \neq h} \tilde V_{[g]}^\top P_{[g,h]} \tilde e_{[h]} \right) \right| \leqslant C < 1.
    \end{align*}
    
\end{enumerate} \label{ass:high}
\end{ass}

\begin{rem}
    Assumption~\ref{ass:high_K} allows the dimension of many IVs $K$ to be proportional to the sample size $n$. Assumption~\ref{ass:high_P} is similar to the standard condition that $\max_{1 \leq i \leq n} P_{ii} \leqslant C < 1$ in the literature on many instruments, and the restriction that $\rank(P) = K$ will exclude redundant columns from $Z$. Assumption~\ref{ass:high_upper} holds in general if $\tilde \Pi^\top \tilde \Pi$, $\Pi^\top\Pi$ and $\Pi^\top P \Pi$ are of the same order. Assumption~\ref{ass:high_lower} excludes perfect correlations between the two quadratic forms. Note that we do not impose any restriction on the identification strength of many IVs, i.e., we allow $\Pi^\top \Pi/\sqrt{K}$ to be bounded. 
\end{rem}

\subsection{Asymptotic Efficiency Properties of Combination Test}\label{subsec: asy_eff}

We now investigate the asymptotical properties of $\phi^*_n$ when the low-dimensional IVs are strong, in the sense that Assumption \ref{ass:low_id} holds. This allows us to define the local alternative according to the asymptotic variance of $\hat \beta_1$ and $\hat \beta_2$ and the limiting covariance structure of the component statistics, from which the joint limiting distribution of $(T(\beta_0), LM(\beta_0), AR)$ can be derived. The results with weak low-dimensional IVs are given in Supplementary Appendix A.  The formal regularity condition is stated as follows. 

\begin{ass}
   The following limits exist:
\begin{align*}
    \rho_1 &= \lim_{n \rightarrow \infty} \frac{1}{\sqrt{\Psi \Sigma}} \sum_{g \in [G]} \mathbb E \left[ \left( \acute \Pi_{[g]}^\top \tilde e_{[g]} \right) \left( \hat \Pi_{[g]}^\top \tilde e_{[g]} \right) \right], \\
    \rho_2 &= \lim_{n \rightarrow \infty} \frac{2}{\sqrt{\Sigma \Upsilon}}  \sum_{g,h \in [G]^2, g \neq h} \mathbb E \left[ \left(\tilde V_{[g]}^\top P_{[g,h]} \tilde e_{[h]} \right) \left( \tilde e_{[g]}^\top P_{[g,h]} \tilde e_{[h]} \right) \right],
\end{align*} 
with $\rho_1^2 + \rho_2^2 < 1$. 
\label{ass:local_alternative_and_covariance}
\end{ass}

\begin{rem}
Under many instruments, the asymptotic expansion of $LM(\beta_0)$ includes both linear and quadratic functions of the errors $(\tilde e, \tilde V)$, whereas $AR$ depends only on a quadratic function of $\tilde e$. The linear and quadratic components are asymptotically normal and uncorrelated, and therefore asymptotically independent. Since the Wald statistic $T(\beta_0)$ involves only linear functions of the errors, it is asymptotically uncorrelated with $AR$. Finally, $\rho_1$ and $\rho_2$ denote, respectively, the correlation between the linear components of $T(\beta_0)$ and $LM(\beta_0)$, and the correlation between the quadratic components of $LM(\beta_0)$ and $AR$.
\end{rem}

The following theorem establishes the joint distribution of the three test statistics above under the local alternative. 
\begin{thm} 
Under Assumptions \ref{ass:reg}--\ref{ass:local_alternative_and_covariance}, suppose that there exists a deterministic sequence $d_n \downarrow 0$ such that $d_n \Phi_1^{-1/2} \rightarrow a_1$, $d_n \Phi_2^{-1/2} \rightarrow a_2$, and $\beta - \beta_0 = \delta d_n$ for some fixed $\delta$, where $a_1 \geq 0$, $a_2 \geq 0$ and $a_1^2 + a_2^2 > 0$. Then we have the joint limiting distribution (\ref{eq:weak_convergence}) for $\left(T(\beta_0),LM(\beta_0),AR\right)^\top$.
\label{thm:limit_str_lcl}
\end{thm}

\begin{rem}\label{rem:dn}
The existence of the sequence $d_n$ is ensured by Assumption~\ref{ass:low_id}. In particular, we may define $d_n = \min\!\left(\Phi_1^{1/2},\, \Phi_2^{1/2}\right).$ Under the strong identification of the low-dimensional IVs in Assumption~\ref{ass:low_id}, we have $\Phi_1^{1/2} = O\!\left(\frac{\sqrt{n}}{r_n}\right) = o(1)$, which immediately implies that $d_n = \min\!\left(\Phi_1^{1/2},\, \Phi_2^{1/2}\right) = o(1),
$ regardless of the order of $\Phi_2$.
\end{rem}

\begin{rem}\label{rem:AN}
The joint normality established in Theorem~\ref{thm:limit_str_lcl}  holds even when the many-IV specification is weakly identified in the sense of \cite{MS22}, that is, when $\Pi^\top \Pi / \sqrt{K}$ is bounded. This result follows from two observations. First, the estimator $\hat \beta$ used to construct the variance estimators $\hat \Phi_2$ and $\hat \Upsilon$ for the LM and AR statistics remains consistent due to the strong identification of the low-dimensional IVs and the double robustness of $\hat \beta$. Second, under weak many-IV identification, the quadratic components of the LM and AR statistics dominate their asymptotic behavior and yield asymptotic normality as long as $K \to \infty$. In this regime, we have $\rho_1 = 0$, $d_n = \sqrt{n}/r_n$, and $\Phi_2^{-1} = O(1)$, which further imply that $a_2 = 0$ under Assumption \ref{ass:low_id}.
\end{rem}

To implement the optimal test $\phi_n^*$ defined in \eqref{eq:opttest}, we still need to estimate $\alpha_1$, $\alpha_2$, $\rho_1$, and $\rho_2$. For $\alpha_1$ and $\alpha_2$, we propose the following estimators:
\begin{align}
    \hat \alpha_1 = \frac{\sqrt{\widehat \Phi_2}}{\sqrt{\widehat \Phi_1 + \widehat \Phi_2}}, \;\; \text{and} \;\;
    \hat \alpha_2 = \frac{\sqrt{\widehat \Phi_1}}{\sqrt{\widehat \Phi_1 + \widehat \Phi_2}}. \label{eq:alphahats}
\end{align}
In addition, let $\hat X = M_W (P - \bar P) X$. For $\rho_1$ and $\rho_2$, we propose the following estimators: 
\begin{align}
    \hat \rho_1 &=\frac{1}{\sqrt{\hat \Psi \hat \Sigma}} \sum_{g \in [G]} \left[ \left( \acute X_{[g]}^\top \hat e_{[g]} \right) \left( \hat X_{[g]}^\top \hat e_{[g]} \right) \right], \label{eq:hatrho_1}\;\; \text{and} \\
    \hat \rho_2 &= \frac{2}{\sqrt{\hat \Sigma \hat \Upsilon}}  \sum_{g,h \in [G]^2, g \neq h} \left[ \left( X_{[g]}^\top P_{[g,h]} \hat e_{[h]} \right) \left( \hat e_{[g]}^\top P_{[g,h]} \hat e_{[h]} \right) \right]. \label{eq:hatrho_2}
\end{align}

By combining Proposition \ref{prop:limit} and Theorem \ref{thm:limit_str_lcl}, and invoking the approach developed by \cite{mueller2011}, we obtain a precise sense of asymptotic optimality for our proposed test $\phi^*_n$, which is formalized in the following Theorem \ref{thm:opt_test_asypeff}.

\begin{thm}


Let $\mathcal{M}$ denote the set of data generating processes $m$ that satisfy the conditions of Theorem \ref{thm:limit_str_lcl} pointwise for all $\delta\in\Re$. Suppose that one wants to test $\mathcal H_0: \delta = 0$ against $\mathcal H_1: \delta \neq 0$. Then, for the class $\mathfrak{C}$ of tests $\phi_n$ satisfying that
\begin{align}
    \lim_{n \rightarrow \infty}\mathbb E \left[\phi_n  \right] \leqslant \alpha & \text{\quad for all\ }m\in\mathcal{M}, \delta=0\label{eq:asymp_valid}, \\
    \liminf_{n \rightarrow \infty}\mathbb E \left[\phi_n  \right] \geqslant \alpha & \text{\quad for all\ }m\in\mathcal{M}, \delta\neq 0\label{eq:asymp_unbiased},
\end{align}
we have $\phi^*_n\in\mathfrak{C}$, and, for any $\delta_1\neq 0$ and any $\phi_n\in\mathfrak{C}$,
\begin{align}
    \lim_{n \rightarrow \infty}\mathbb E \left[\phi_n  \right] \leqslant \lim_{n \rightarrow \infty}\mathbb E \left[\phi^*_n  \right] \text{\quad for all\ }m\in\mathcal{M}, \delta=\delta_1. \label{eq:asymp_umpu}
\end{align}
Moreover, for the test $\tilde \phi_n = \mathbf 1 \left\{ T^2(\beta_0) \geqslant \mathbb C_\alpha \right\}$, we have $\tilde \phi_n\in\mathfrak{C}$, and for any $\delta$ and all $m\in\mathcal{M}$,
\begin{align*}
        \lim_{n \rightarrow \infty}\mathbb E \left[\tilde \phi_n  \right] = \lim_{n \rightarrow \infty}\mathbb E \left[ \phi^*_n  \right] \text{\quad if and only if \quad $a_2 = \rho_1 a_1$.}
\end{align*}
\label{thm:opt_test_asypeff}
\end{thm}
\begin{rem}
Theorem~\ref{thm:opt_test_asypeff} shows that, under local alternatives, $\phi_n^*$ attains the asymptotic efficiency bound within the class of tests that remain asymptotically unbiased and valid for all data generating processes inducing the same weak limit for $(T(\beta_0), LM(\beta_0), AR)^\top$ as in Theorem~\ref{thm:limit_str_lcl}. This class of tests includes, in particular, the Wald and jackknife LM tests based solely on the low-dimensional and many IVs, respectively. It also includes the HLIM and HFUL-based tests of \citeauthor{Haus2012} (\citeyear{Haus2012}), both of which are asymptotically equivalent to linear combinations of the LM and AR tests under many IVs.\footnote{E.g., see the discussions below Theorem 4.2 in \cite{LWZ24}.} 
Furthermore, as we will discuss more in detail in Section \ref{sec: efficiency_gain}, the weights in \eqref{eq:omega} also yield an optimally efficient combined estimator of $(\hat \beta_1, \hat \beta_2, AR)$ that achieves the minimal asymptotic variance.

However, our optimal test $\phi_n^*$ does not dominate tests that cannot be expressed directly as functions of $(T(\beta_0), LM(\beta_0), AR)$, such as the sup-score test of \citeauthor{belloni2012} (\citeyear{belloni2012}) and the ridge-regularized AR test of \citeauthor{Dovietal2024} (\citeyear{Dovietal2024}). Indeed, it is possible to construct data generating processes under which either the optimal combination test, the sup-score test, or the ridge-regularized AR test achieves the highest power. We establish the notion of optimality in Theorem \ref{thm:opt_test_asypeff} primarily to provide guidance for constructing tests that improve upon the conventional Wald test, rather than to identify a globally optimal procedure. That said, one could potentially combine our $\phi_n^*$ test with alternative tests such as the sup-score test, in the spirit of \cite{navjeevan2024identificationdimensionalityrobusttest}, to obtain more powerful inference.
    
\end{rem}

\begin{rem}
    Theorem \ref{thm:opt_test_asypeff} also clarifies the necessary and sufficient condition under which the combination test does not deliver a strict power gain over the Wald test $T(\beta_0)$, namely $a_2 = \rho_1 a_1$. Recall that $a_1$ and $a_2$ represent the orders of the concentration parameters for the low-dimensional and many IVs, respectively. As shown below, even when we allow either of them to be zero---thereby covering situations where one IV estimator dominates the other in terms of convergence rate---the condition ($a_2 = \rho_1 a_1$) is still seldom met. Put differently, one should generally anticipate strictly more powerful inference when using our test.
    
     In particular, when $a_1 = 0$ and $a_2 > 0$, corresponding to the case when the identification strength of many IVs is larger than that of the low-dimensional IVs (i.e., the convergence rate of $\hat \beta_2$ is faster than that of $\hat \beta_1$), we obtain a strict power improvement for all values of $\rho_1$ and $\rho_2$ satisfying $\rho_1^2 + \rho_2^2 < 1$. Conversely, when $a_1 > 0$ and $a_2 = 0$, meaning that low-dimensional IVs provide stronger identification than many IVs, we still achieve a strict power gain as long as $\rho_1 \neq 0$. When $a_1 > 0$ and $a_2 > 0$, that is, when the two sets of IVs have identification strengths of the same order, strict power improvement is ensured, provided that $\rho_1 \neq a_2 / a_1$. Indeed, at $\rho_1 = a_2 / a_1$, the sufficient statistic for $\delta$ derived from the joint limiting distribution of the three component statistics (cf. (\ref{eq:weak_convergence})) becomes independent of the limiting Gaussian observations associated with the LM and AR statistics, so it is not surprising that combining the Wald statistic with them does not yield a more powerful inference. 

\end{rem}
\begin{rem} \label{rem:lowIV} As long as the low-dimensional IVs provide strong identification, the optimal combination test $\phi_n^*$ does not lose asymptotic power for any degree of identification strength of the many IVs. In this sense, the efficiency gains delivered by the combination test are essentially a ``free lunch''. Under local alternatives, the weak convergence result in Theorem~\ref{thm:limit_str_lcl} holds uniformly, regardless of whether the many IVs are strong or weak. In particular, when the many IVs are weak so that $a_2=\rho_1=0$ (as the quadratic term in $LM(\beta_0)$ dominates the linear term), the second part of Theorem~\ref{thm:opt_test_asypeff} shows that the combination test asymptotically reduces to the Wald test, implying no efficiency loss from combining. Moreover, as shown below, the combination test remains consistent against fixed alternatives irrespective of the identification strength of the many IVs.


\end{rem}

Finally, for any fixed alternative, both $T(\beta_0)$ and $LM(\beta_0)$ are consistent and, by construction, avoid the issue of non-monotonic power (noted at the end of Section \ref{subsec:test_stat}) when their corresponding set of IVs is strong. Hence, it is reasonable to anticipate that our combined test will retain these desirable properties, a result that we formalize in the theorem below. However, we emphasize once more that these results remain valid regardless of the strength of the many IVs.

\begin{thm}
    Suppose that Assumptions \ref{ass:reg}-\ref{ass:local_alternative_and_covariance} hold. Then, under $\beta - \beta_0 = \delta$ for some fixed $\delta \neq 0$, we have $\lim_{n \rightarrow \infty} \mathbb E \left[ \phi^*_n  \right] = 1$.
\label{thm:fixed_alter}
\end{thm}



\subsection{Quantifying Efficiency Improvement in Large Samples} \label{sec: efficiency_gain}

We measure the efficiency improvement of the combination test over the conventional Wald test that uses only low-dimensional IVs by the percentage reduction in the asymptotic length of the resulting confidence interval. Recall from Remark \ref{rem:lowIV} that, under weak many instruments, the combination test is asymptotically equivalent to the Wald test. Consequently, the associated confidence interval takes the usual ``estimator plus and minus a standard error times a critical value" form: $\bigl[\hat\beta_1-\sqrt{\widehat\Phi_1}\times\sqrt{\mathbb C_{\alpha}},\;\hat\beta_1+\sqrt{\widehat\Phi_1}\times\sqrt{\mathbb C_{\alpha}}\bigr]$, where $\hat\beta_1$ and $\hat\Phi_1$ are as defined above, and $\sqrt{\mathbb C_{\alpha}}$ is the standard normal critical value. In this case, the asymptotic efficiency gain over the conventional Wald test is zero.

When the many IVs provide strong identification, the confidence interval associated with our combination test can also be expressed in the familiar ``estimator plus and minus a standard error times a critical value" form. To see this, observe that under strong identification by many IVs, the component LM statistic can be represented as follows:
\begin{align*}
    LM(\beta_0) = \frac{X^\top (P - \bar P) e(\beta_0)}{\sqrt{\hat \Sigma}}= \frac{\hat \beta_2 - \beta_0}{\sqrt{\widehat \Phi_2}}+o_P(1),
\end{align*}
where the $o_P(1)$ term comes from the fact that under strong identification, 
\begin{align*}
\text{sign}(X^\top (P - \bar P)X) \convP 1.
\end{align*}
Inserting this into (\ref{eq:opttest}) gives the following form of our combination test:
\begin{align*}
        \phi^{*}_n = \mathbf 1 \left\{ \left(\hat \omega_1  \frac{\hat \beta_1 - \beta_0}{\sqrt{\widehat \Phi_1}} + \hat \omega_2 \frac{\hat \beta_2 - \beta_0}{\sqrt{\widehat \Phi_2}}+ \hat\omega_2 o_P(1) +  \hat \omega_3 AR \right)^2 \geqslant \mathbb C_\alpha \right\}.
\end{align*}
The resulting confidence interval is asymptotically equivalent to
\begin{align}
CI^* =    \left[\hat\beta^*-\frac{1}{\left(\hat \omega_1/\sqrt{\widehat \Phi_1} + \hat \omega_2 / \sqrt{\widehat \Phi_2} \right)} \sqrt{\mathbb C_{\alpha}},\;\hat\beta^*+\frac{1}{\left(\hat \omega_1/\sqrt{\widehat \Phi_1} + \hat \omega_2 / \sqrt{\widehat \Phi_2} \right)}\sqrt{\mathbb C_{\alpha}}\right], \label{eq:impliedCI}
\end{align}
where $\hat\beta^*$ is a combined estimator of $\beta$,
\begin{align*}
    \hat \beta^* = \frac{\hat \omega_1/\sqrt{\widehat \Phi_1}}{\left(\hat \omega_1/\sqrt{\widehat \Phi_1} + \hat \omega_2 / \sqrt{\widehat \Phi_2} \right)} \hat \beta_1 + \frac{\hat \omega_2/\sqrt{\widehat \Phi_2}}{\left(\hat \omega_1/\sqrt{\widehat \Phi_1} + \hat \omega_2 / \sqrt{\widehat \Phi_2} \right)} \hat \beta_2 + \frac{\hat \omega_3 AR }{\left(\hat \omega_1/\sqrt{\widehat \Phi_1} + \hat \omega_2 / \sqrt{\widehat \Phi_2} \right)}.
\end{align*}

The intuition behind the combined estimator is fundamentally efficiency-driven. First, $\widehat \Phi_1$ and $\widehat \Phi_2$ estimate the asymptotic variances of $\hat \beta_1$ and $\hat \beta_2$, respectively, so for given weights $(\hat\omega_1,\hat\omega_2)$ the combined estimator $\hat\beta^*$ assigns greater weight to the estimator with 
the smaller asymptotic variance. Second, although the AR statistic is asymptotically centered at zero and therefore does not affect the location of the combined estimator, its correlation with the many-IV-based estimator $\hat \beta_2$ allows it to reduce the variance of the combined estimator. This mechanism parallels the construction of the HLIML and HFUL estimators in \cite{Haus2012}, which can be more efficient than the JIVE estimator. Third, the estimated weights $(\hat \omega_1,\hat \omega_2,\hat \omega_3)$ are consistent for the population weights $(\omega_1,\omega_2,\omega_3)$ in \eqref{eq:omega} associated with the UMPU test. Finally, the optimal weights defined in \eqref{eq:omega} solve the following problem:
\begin{align*}
    \min_{\omega_1,\omega_2,\omega_3} \frac{1}{(a_1\omega_1+a_2\omega_2)^2}
    \quad \text{s.t.} \quad
(\omega_1,\omega_2,\omega_3)
\begin{pmatrix}
1 & \rho_1 & 0 \\
\rho_1 & 1 & \rho_2 \\
0 & \rho_2 & 1
\end{pmatrix}
\begin{pmatrix}
\omega_1 \\
\omega_2 \\
\omega_3
\end{pmatrix}
= 1 .
\end{align*}
The quadratic constraint ensures that $CI^*$ attains the correct asymptotic coverage, while the objective corresponds to the asymptotic variance of the combined estimator $\hat\beta^*$, since
\begin{align*}
\frac{1}{d_n^2\big(\omega_1/\sqrt{\Phi_1}+\omega_2/\sqrt{\Phi_2}\big)^2}
\rightarrow 
\frac{1}{(a_1\omega_1+a_2\omega_2)^2}.
\end{align*}
Thus, the optimal weights in \eqref{eq:omega}, originally motivated by the UMPU testing problem, also yield an optimally efficient combined estimator of $(\hat \beta_1, \hat \beta_2, AR)$ that achieves the minimal asymptotic variance.

A direct implication of the confidence interval in (\ref{eq:impliedCI}) is that, asymptotically, the percentage reduction in its length relative to the confidence interval based on the conventional Wald test can be derived analytically as follows,
\begin{align}
    & \operatorname*{plim}_{n\rightarrow\infty}1- \frac{1/\left(\hat \omega_1/\sqrt{\widehat \Phi_1} + \hat \omega_2 / \sqrt{\widehat \Phi_2} \right)}{\sqrt{\widehat\Phi_1}} \nonumber \\
    = & 1- \sqrt{\frac{(1-\rho_1^2-\rho_2^2)}{(1-\rho_2^2)-2 \rho_1 a_2/a_1 + (a_2/a_1)^2}} \label{eq: efficiency_gain} \\
    \geqslant & 1- \sqrt{\frac{(1-\rho_1^2)a_1^2}{a_1^2-2 \rho_1 a_1a_2 + a_2^2}} = 1- \sqrt{\frac{(1-\rho_1^2)}{(1-\rho_1^2) + (\rho_1  - a_2/a_1)^2}}. \label{eq: efficiency_gain_bound}
\end{align}
As equation (\ref{eq: efficiency_gain}) shows, the efficiency gain is primarily driven by the relative identification strength of the low-dimensional and many IVs, summarized by
\[
\frac{a_2}{a_1} = \lim_{n \to \infty} \sqrt{\frac{\Phi_1}{\Phi_2}},
\]
as well as by the limiting correlations between the Wald and LM statistics, denoted by $\rho_1$, and between the LM and AR statistics, denoted by $\rho_2$. In particular, consistent with Theorem \ref{thm:opt_test_asypeff}, when $a_2/a_1 = \rho_1$, the combination test $\phi_n^*$ yields no efficiency improvement, and its confidence interval is asymptotically of the same length as that based on $\tilde \phi_n$. This case includes the weak many-IV scenario, where $\rho_1 = a_2 = 0$. 

By contrast, whenever $a_2/a_1 \neq \rho_1$, the resulting confidence interval is strictly shorter, implying improved efficiency. Moreover, the efficiency gain in (\ref{eq: efficiency_gain}) is monotonically increasing in $|\rho_2|$, which leads to the lower bound reported in (\ref{eq: efficiency_gain_bound}) by setting $\rho_2 = 0$. This lower bound can also be interpreted as the CI length reduction achieved by optimally combine $\hat \beta_1$ and $\hat \beta_2$ only.\footnote{The optimal reduction in this case is $\lim_{n \rightarrow \infty} 1- \frac{1/\left(\tilde \omega_1/\sqrt{ \Phi_1} + \tilde \omega_2 / \sqrt{ \Phi_2} \right)}{\sqrt{ \Phi_1}} = 1- \frac{1}{\left(\tilde \omega_1 + \tilde \omega_2 a_2/ a_1 \right)}$, where the optimal weights are computed by \begin{align*}
 (\tilde \omega_1,\tilde \omega_2) =  \argmin_{\omega_1,\omega_2} \frac{1}{(a_1\omega_1+a_2\omega_2)^2}
    \quad \text{s.t.} \quad
(\omega_1,\omega_2)
\begin{pmatrix}
1 & \rho_1  \\
\rho_1 & 1 \\
\end{pmatrix}
\begin{pmatrix}
\omega_1 \\
\omega_2 
\end{pmatrix}
= 1 .
\end{align*}
By direct calculation, we can show that 
\begin{align*}
 1- \frac{1}{\left(\tilde \omega_1 + \tilde \omega_2 a_2/ a_1 \right)} =     1- \sqrt{\frac{(1-\rho_1^2)}{(1-\rho_1^2) + (\rho_1  - a_2/a_1)^2}}.
\end{align*}}

Figure \ref{fig:CI_theory} plots this bound as a function of $a_2/a_1$, the ratio of the standard deviations of $\hat\beta_1$ and $\hat\beta_2$, for various values of $\rho_1$. 

In practice, $a_2/a_1$ can be estimated by the ratio of any consistent estimator of $\Phi_1$ and $\Phi_2$ under strong identification.\footnote{If the many-IV specification is weakly identified, then $\Phi_1 = o(1)$ and $1/\Phi_2 = O(1)$, so that $a_2/a_1 = 0$. Suppose we estimate $\Phi_1$ and $\Phi_2$ by $\dot \Phi_1$ and $\ddot \Phi_2$, respectively. Although $\ddot \Phi_2$ is not consistent under weak many IVs, Section C.8 shows that $\ddot \Phi_2^{-1/2} = O_P(1)$ and $\dot \Phi_1^{1/2} = o_P(1)$, implying $\sqrt{\dot \Phi_1 / \ddot \Phi_2} \convP 0 = a_2/a_1$. Hence, the plug-in estimator of $a_2/a_1$ remains consistent even under weak identification of the many-IV specification.} This yields a simple rule of thumb, discussed in Section \ref{sec: illu}: for empirically plausible values of $\rho_1$ between $-0.7$ and $0.7$, if the ratio of the reported standard errors $\sqrt{\dot \Phi_1}$ and $\sqrt{\ddot \Phi_2}$ exceeds $1.05$, then the associated confidence interval shortens by at least $10\%$.

In some applications (e.g., the replication of \cite{card(2009)} by \cite{Goldsmith(2020)}), researchers report alternative many-IV estimators (denoted $\tilde \beta_2$ with variance $\tilde \Phi_2$), such as HFUL, instead of the JIVE estimator ($\hat \beta_2$) considered above. Let $\tilde \rho_1$ denote the correlation between the GMM estimator $\hat \beta_1$ and $\tilde \beta_2$. By the same argument, if $\tilde \rho_1 \in [-0.7,0.7]$ (or $\tilde \rho_1 \in [-0.99,0.99]$), the confidence interval $\widetilde{CI}$ based on the optimal combination of $\hat \beta_1$ and $\tilde \beta_2$ is at least $10\%$ shorter than the conventional Wald interval whenever the ratio $\sqrt{\Phi_1/\tilde \Phi_2}$ exceeds $1.05$ (or $1.1$). Moreover, if $\tilde \beta_2$ is asymptotically equivalent to a combination of the JIVE estimator $\hat \beta_2$ and the $AR$ statistic, as is the case for HLIML and HFUL, then our optimal confidence interval ($CI^*$) is weakly shorter than $\widetilde{CI}$ and therefore at least $10\%$ shorter than the conventional Wald interval. Hence, the rule of thumb is not specific to the JIVE estimator $\hat \beta_2$, but applies more broadly to many-IV estimators asymptotically equivalent to a combination of JIVE and the $AR$ statistic.

\section{Practical Implementation and Simulation Study} \label{sec: simu}
In this section, we first synthesize the preceding discussions to describe how to practically implement our combination test in an empirically relevant model and then apply it in simulations to assess its finite-sample power performance. In particular, we consider the following model with clustered data,
\begin{align}
        \bar Y_{i,g} &= \bar X_{i,g}\beta + \bar W_{i,g}^\top\gamma + \alpha_g + \bar e_{i,g}, \label{eq:simu_eqn1}\\
        \bar X_{i,g} &= \bar Z_{i,g}^\top\pi + \bar W_{i,g}^\top\tau + \xi_g + \bar V_{i,g}, \label{eq:simu_eqn2}
    \end{align}
where $\alpha_g$ and $\xi_g$ denote cluster-specific fixed effects, and $\bar X_{i,g}$, $\bar W_{i,g}$, and $\bar Z_{i,g}$ represent, respectively, the endogenous regressor, exogenous regressors, and (potentially many) base instruments. By demeaning at the cluster level, we partial out the fixed effects and obtain $\tilde Y_{i,g}$, $\tilde X_{i,g}$, $W_{i,g}$, $\tilde Z_{i,g}$, $\tilde V_{i,g}$, and $\tilde e_{i,g}$, echoing the notation in our initial model setup in \eqref{eq: IV-model}-\eqref{eq:first-stage}. The setting we consider in this paper is one in which practitioners are often interested in testing $\mathcal{H}_0: \beta = \beta_0 \ \text{against}\ \mathcal{H}_{1}: \beta \neq \beta_0$ using an alternative set of low-dimensional IVs, $\tilde{z}_{i,g} = f_{i,g}(\tilde Z,W)$. These instruments can be formed, for example, by selecting a subset of the original base IVs $\tilde Z_{i,g}$, or by taking a weighted or simply an unweighted average of them.

\subsection{Practical Implementation}
Our implementation procedure starts by partialling out the covariates $\tilde W$ from $\tilde Y$, $\tilde X$, $\tilde Z$, and $\tilde z$. This produces $Y$, $X$, $Z$, and $z$, which are consistent with the notation used in Section \ref{sec: model_setup}. These transformed variables serve as the effective observations in all subsequent estimation and inference procedures. We outline these subsequent steps below in sequential order.
\begin{enumerate}
    \item \textit{Preliminary estimates of $\beta$}. 
    \begin{enumerate}
        \item Use $Y$, $X$, the low-dimensional IVs $z$, and a given weighting matrix $\hat A_n$ to obtain a preliminary GMM estimate $\hat\beta_1$ by \eqref{eq:beta1hat};
        \item Use $Y$, $X$, base IVs $Z$, and the projection matrix of $Z$ to obtain a preliminary leave-one-cluster-out jackknife estimate $\hat\beta_2$ by \eqref{eq:beta2hat}.
    \end{enumerate} 
    \item \textit{Preliminary variance estimates}. (See Section \ref{subsec:assumptions} for explicit formulas of the asymptotic variances $\Phi_1$ and $\Phi_2$, and their corresponding estimators.)
    \begin{enumerate}
        \item Using the residuals from the structural equation \eqref{eq:model}, computed with $\hat\beta_1$, obtain a preliminary estimate $\dot{\Phi}_1$ for the asymptotic variance $\Phi_1$ of $\hat\beta_1$.
        \item Using the residuals from the structural equation \eqref{eq:model}, computed with $\hat\beta_2$, obtain a preliminary estimate $\ddot{\Phi}_2$ for the asymptotic variance $\Phi_2$ of $\hat\beta_2$. Pay special attention to the adjustments made in \eqref{eq:LMvarianceestimator} to account for the fact that $W$ has already been partialled out. 
    \end{enumerate}
    \item \textit{A doubly robust estimate of $\beta$ and the resulting robust variance estimates}. (See Section \ref{subsec:assumptions} for explicit formulas of $\hat \Psi$, $\hat \Omega$, $\widehat\Phi_1$, $\hat\Sigma$, $\widehat\Phi_2$ and $\hat\Upsilon$.)
    \begin{enumerate}
        \item Use $\hat\beta_1$, $\hat\beta_2$, $\dot{\Phi}_1$, and $\ddot{\Phi}_2$ to obtain a doubly robust estimate $\hat \beta$ of $\beta$ by \eqref{eq: beta_hat}.
        \item Using the residuals from \eqref{eq:model}, computed with $\hat\beta$, obtain the robust variance estimates $\hat \Psi$, $\hat \Omega$ and thus $\widehat\Phi_1$ (cf. \eqref{eq:hatPhi_1}), $\hat\Sigma$ (cf. \eqref{eq:LMvarianceestimator}) and thus $\widehat\Phi_2$ (cf. \eqref{eq:hatPhi_2}), and finally $\hat\Upsilon$ (cf. \eqref{eq:ARvariianceestimator}).
    \end{enumerate}
    \item \textit{Component statistics and their estimated weights}.
    \begin{enumerate}
        \item Using the hypothesized $\beta_0$ and the outputs from Step 3 above, compute the three component statistics---$T(\beta_0)$ (cf. \eqref{eq:Waldstatistic},) $LM(\beta_0)$ (cf. \eqref{eq:LMstatistic}) and $AR$ (cf. \eqref{eq:AR}).
        \item Using the outputs from Step 3 above, compute $\hat\alpha_1$ and $\hat\alpha_2$ (cf. \eqref{eq:alphahats}) and $\hat\rho_1$ and $\hat\rho_2$ (cf. \eqref{eq:hatrho_1} and \eqref{eq:hatrho_2}), and thus the weights $(\hat\omega_1,\hat\omega_2,\hat\omega_3)$ (cf. \eqref{eq:omegahat}).
    \end{enumerate}
    \item Using outputs from Step 4, obtain the feasible combination test $\phi^*_n$ by \eqref{eq:opttest}.
\end{enumerate}

\subsection{Simulation Study}
All of our simulations are based on \eqref{eq:simu_eqn1} and \eqref{eq:simu_eqn2}. The fixed effects are generated by $\alpha_g = u_{1g} + g/G$ and $\xi_g = u_{2g} + g/G$, $g=1, \dots, G$, where $u_{1g}$ and $u_{2g}$ are independent standard normal random variables. 
The control variables in $\bar W_{i,g}$ are generated by the standard normal distribution, and the dimension of $\bar W$ is fixed at $d_w = 10$. The instruments in $\bar Z_{i,g}$ are normally distributed with mean $0$ and cluster-level dependence: within each cluster $g$, the covariance matrix is given by $$\Omega_{1g}= \begin{bmatrix}
1 & \theta_1 & \cdots & \theta_1 \\
\theta_1 & 1 & \cdots & \theta_1 \\
\vdots & \vdots & \ddots & \vdots \\
\theta_1 & \theta_1 & \cdots & 1
\end{bmatrix}_{n_g \times n_g}, \quad g=1, \dots, G,$$ and between clusters these instruments are independent of each other; we set $\theta_1 = 0.5$ in our simulations. 
Finally, to obtain an arguably complex error structure, we first generate $\acute e_{i,g} = \rho \eps_{i,g} + \sqrt{1-\rho^2} \sigma_{i,g} v_{g}$ and $\acute V_{i,g} = \rho \eta_{i,g} + \sqrt{1-\rho^2} \sigma_{i,g} v_{g}$, where $\sigma_{i,g} = \sqrt{\left(0.2+(\bar W_{i,g}^\top\tau)^2\right)/2.4}$. Here, $\eps_{i,g}$, $\eta_{i,g}$, and $v_{g}$ are mutually independent standard normal random variables, $\rho$ governs the degree of endogeneity, $\tau$ is specified below, and we fix $\rho = 0.5$ in all simulations. Next, within each cluster, we premultiply the vectors $\acute e_{i,g}$ and $\acute V_{i,g}$ by
$$
\Omega_{2g}= \begin{bmatrix}
1 & 0 & \cdots & 0 \\
\theta_2 & 1 & \cdots & 0 \\
\vdots & \vdots & \ddots & \vdots \\
\theta_2^{n_g-1} & \theta_2^{n_g-2} & \cdots & 1
\end{bmatrix}_{n_g \times n_g}, \quad g=1, \dots, G,
$$
thereby generating $\bar e_{i,g}$ and $\bar V_{i,g}$. In our simulations, we set $\theta_2 = 0.7$. 

For the parameters, we set $\beta = 0.3$, $\gamma = \tau = (1/\sqrt{d_w}) \times \iota_{d_w}$, where $\iota_{d_w}$ is a $d_w \times 1$ vector of ones. We specify geometrically decaying coefficients of IVs as $\pi = \left( \phi^0, \phi^1, \ldots, \phi^{K-1} \right)$, where $K$ denotes the number of (many) base IVs and $\phi$ controls the relative weight assigned to each instrument. It is noted that $\phi = 0$ represents the case in which only the first instrument has identification strength, while $\phi = 1$ corresponds to the case in which each instrument has the same identification strength. We further normalize $\pi$ to have $\left\Vert \pi \right\Vert_2 = \sqrt{\psi \sqrt{K}/n}$, where $\psi$ controls the identification strength of the many IVs. The one-dimensional IV is constructed by taking the average of the many IVs; as $\phi$ approaches one, the identification strength of the low-dimensional IV becomes stronger since it is closer to the optimal instrument. We set the sample size at $n = 2{,}000$ and the number of clusters at $G = 500$, and then generate heterogeneous cluster sizes following a procedure similar to that in \cite{Djogbenou-Mackinnon-Nielsen-2019}. Specifically, for $g=1, \cdots,G-1$, we set $n_g = \max\left\{1, n \exp(2 g / G) / (1+\sum_{g=1}^{G-1} \exp(2 g / G)) \right\}$ and then the size of the last cluster as $n_{G} = \max\left\{1, n - \sum_{g=1}^{G-1}n_g \right\}$. For the dimension of $\bar Z$, we consider $K = 100$ and $K = 500$, respectively. All the results below are based on $5,000$ simulations.

\begin{figure}[H]
\centering

\textbf{Panel A: $K=100$ and $\psi=30$}\\[-0.2em]
\includegraphics[width=0.9\textwidth]{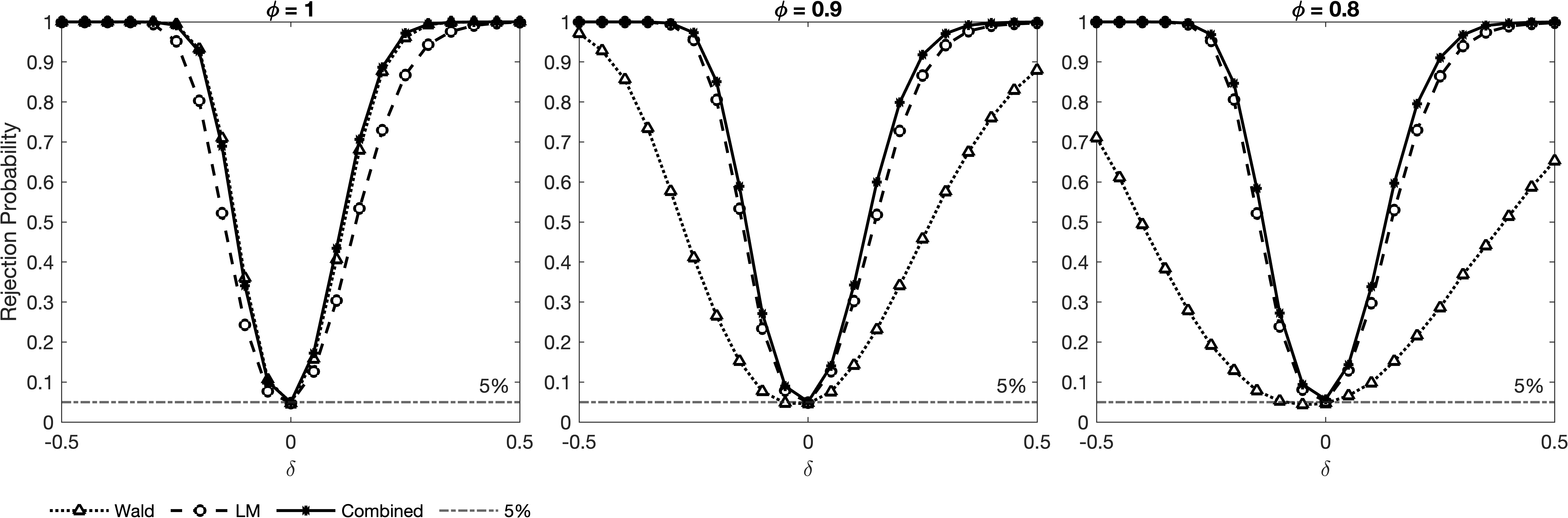}

\vspace{0.3cm}

\textbf{Panel B: $K=500$ and $\psi=30$}\\[-0.2em]
\includegraphics[width=0.9\textwidth]{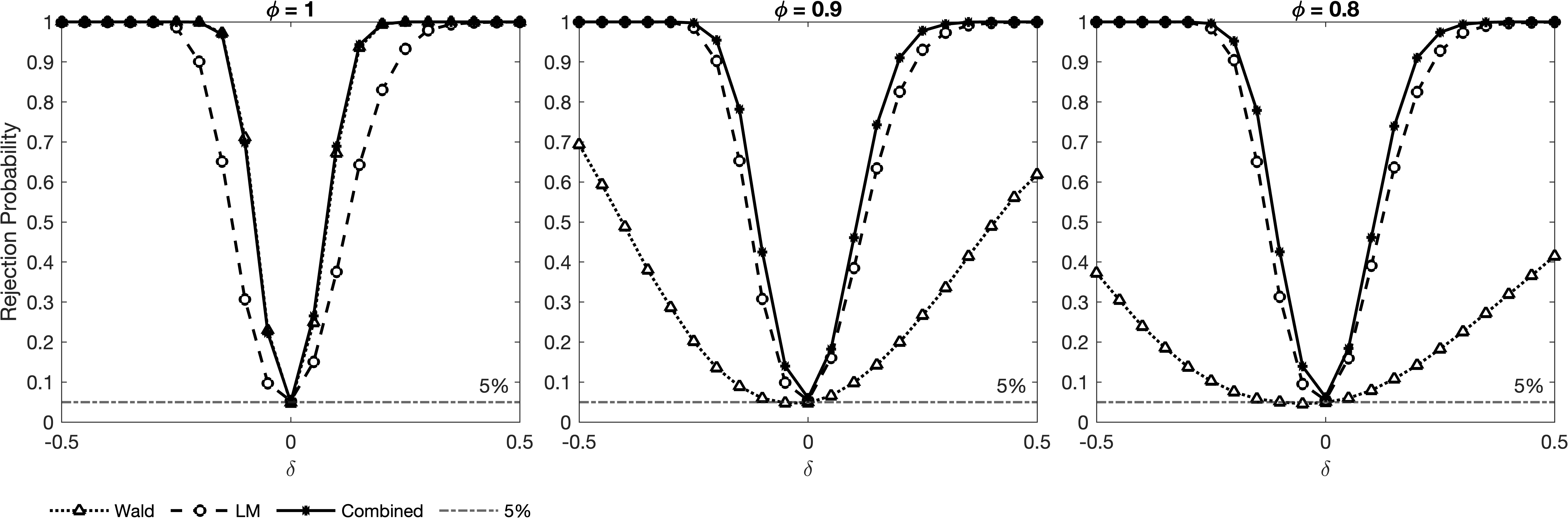}

\vspace{0.3cm}

\textbf{Panel C: $K=500$ and $\phi=1$}\\[-0.2em]
\includegraphics[width=0.9\textwidth]{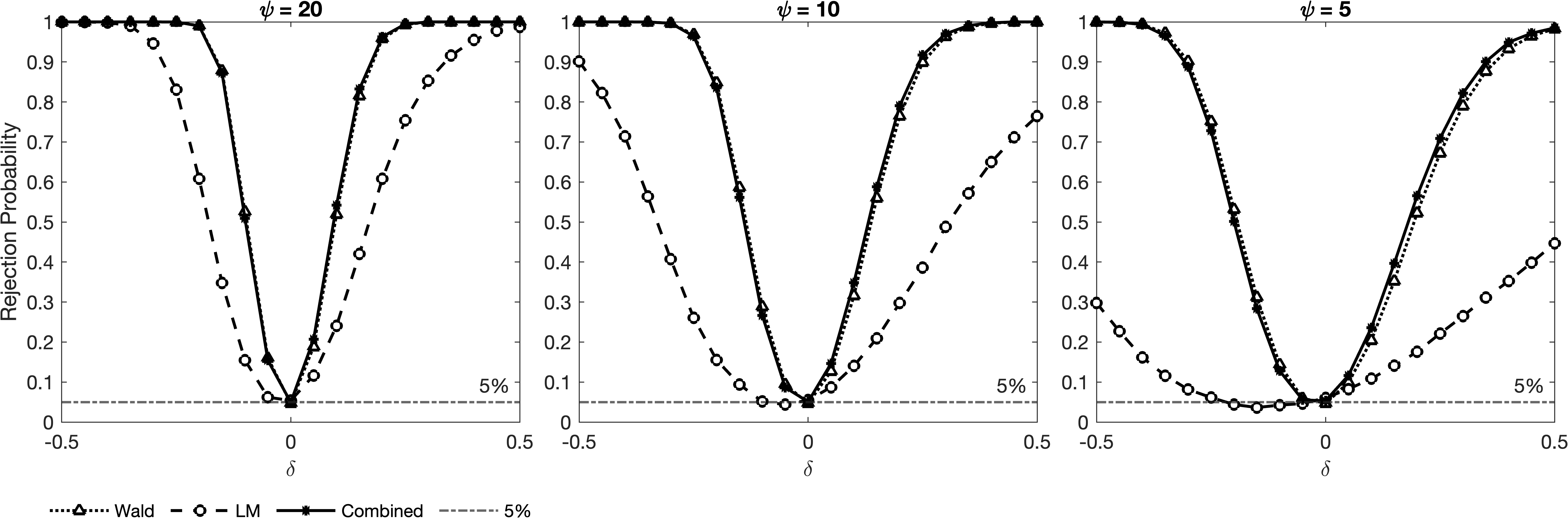}

\caption{Power curves of the combination, Wald, and LM tests.}
\floatfoot{\doublespacing \textit{Notes}: This figure displays the power curves for our combination test $\phi_n^*$ along with those for the component Wald and LM tests, at different values of $K$ (the dimension of the many IVs), $\psi$ (which governs the identification strength of the many IVs), and $\phi$ (which controls the identification strength of the one-dimensional IV relative to the many IVs). The horizontal axis represents the deviations in the parameter of interest from the maintained hypothesis, that is, we test $\mathcal{H}_0: \beta = \beta_0$ against $\mathcal{H}_{1}: \beta \neq \beta_0$, where $\delta = \beta - \beta_0$. See Section \ref{sec: simu} for details of the simulation setup. All results are based on $5,000$ simulations.}
\label{fig:simulations-stacked}
\end{figure}

Figure \ref{fig:simulations-stacked} displays the power curves for our combination test $\phi_n^*$ along with those for the component Wald and jackknife LM tests, at different values of $K$ (the dimension of the many IVs), $\psi$ (which governs the identification strength of the many IVs), and $\phi$ (which controls the identification strength of the one-dimensional IV relative to the many IVs). We identified three main observations, each of which aligns with our large-sample theory. First, in every scenario, the combination test $\phi^*_n$ attains the correct size and is more powerful than each of the other two tests. In particular, as shown in Panel C, the power curve $\phi^*_n$ is never dominated by that of the Wald test, regardless of the strength of the many IVs, thereby underscoring the ``free lunch" efficiency gains delivered by our combination test. Second, within Panels A and B of Figure \ref{fig:simulations-stacked}, we observe that for fixed $K$ and $\psi$, the power improvement of $\phi_n^*$ over the Wald test becomes more substantial as the identification strength of the one-dimensional IV weakens relative to that of the many IVs (i.e., as $\phi$ decreases). This is reflected in the widening gap between the power curves of $\phi_n^*$ and the Wald test. It emphasizes how many-IV-based LM and AR statistics contribute to the power enhancement. Third, in contrast to the cases in which the one-dimensional IV dominates many IVs in strength (first figure in Panel A or B) and the power curve of $\phi_n^*$ coincides with that of the Wald test, noticeable gaps persist between the power curves of $\phi_n^*$ and the LM test in the flipped cases (second and third figures in Panel A or B). These gaps highlight how the AR component contributes to power enhancement through its correlation with the LM statistic.

\section{Conclusion}\label{sec: conclu}
This paper proposes an inference approach that improves conventional estimation and inference in instrumental variables regressions using low-dimensional instruments (e.g., aggregated shift-share IVs) and their underlying high-dimensional base instruments. The procedure requires strong identification only for the low-dimensional IV regression while allowing the many-instrument specification to be weakly identified. We also provide a practical rule of thumb for when inference improves by at least $10\%$, based solely on the ratio of variances of the low- and high-dimensional IV estimators commonly reported in applications. Extensions to settings with many controls or diverging cluster sizes, alternative bootstrap procedures, and combinations with tests such as the sup-score test are left for future work.

\newpage

\newpage


\begin{table}[H]
\centering
\caption{Point estimates and confidence intervals: Immigrant enclave.}
\label{tab:Card_results_rotated_midpoint}
\begin{tabular}{lcccc}
\hline\hline
 & \multicolumn{2}{c}{College Equivalent Workers}
 & \multicolumn{2}{c}{High School Equivalent Workers} \\
\cline{2-3} \cline{4-5}
 & {Yes} & {No} & {Yes} & {No} \\
\hline
$\hat \rho_1$
& 0.588 & 0.446 & 0.408 & 0.576 \\
$\hat \rho_2$
& 0.103 & 0.120 & 0.129 & 0.165 \\
$\hat \sigma(\hat \beta_1) / \hat \sigma(\hat \beta_2)$
& 0.926 & 1.112 & 0.696 & 0.727 \\
$\hat \beta_1$
& -0.078 & -0.080 & -0.037 & -0.024 \\
Wald CI
& {(-0.103, -0.053)} & {(-0.107, -0.052)}
& {(-0.051, -0.023)} & {(-0.037, -0.011)} \\
$\hat \beta_2$
& -0.066 & -0.058 & -0.043 & -0.030 \\
LM CI
& {(-0.093, -0.039)} & {(-0.083, -0.033)}
& {(-0.063, -0.024)} & {(-0.048, -0.012)} \\
$\hat \beta^*$
& -0.072 & -0.064 & -0.039 & -0.025 \\
Comb. CI
& {(-0.095, -0.049)} & {(-0.087, -0.041)}
& {(-0.052, -0.026)} & {(-0.038, -0.013)} \\
\hline\hline
\end{tabular}
    \floatfoot{\doublespacing \textit{Notes}: This table reports the estimation and inference results for the immigrant enclave example using the \cite{card(2009)} dataset, shown separately for college equivalent workers and high school equivalent workers. Columns with ``Yes" contain city-level controls, while columns with ``No" do not. The point estimates are obtained from the standard two-stage least squares (TSLS) estimator with the one-dimensional Bartik instrument, $\hat\beta_1$, and, in addition, from the leave-one-cluster-out estimator, $\hat\beta_2$, which makes use of all base IVs. Wald CI and LM CI denote the confidence intervals based on $\hat \beta_1$ and $\hat \beta_2$, respectively. The estimator $\hat\beta^*$ is the combined estimator for $\beta$, defined in Section \ref{sec: efficiency_gain}. It is essentially the midpoint of the confidence interval in (\ref{eq:impliedCI}), which is obtained from our combination test and labeled as ``Comb. CI" in the table. In addition, $\hat\rho_1$ and $\hat\rho_2$ denote estimates of the asymptotic correlation between the Wald and LM statistics, and between the LM and AR statistics, respectively. Finally, $\hat \sigma(\hat \beta_1) / \hat \sigma(\hat \beta_2)$ denotes the ratio of standard errors of $\hat \beta_1$ and $\hat \beta_2$. All displayed numbers are rounded to three decimal places.}
\end{table}

\newpage

\newpage

\newpage

\bibliographystyle{ecta}
\bibliography{DMWZ}


\newpage
\appendix
\noindent{\huge\textbf{Appendix}}

\unhidefromtoc 
\tableofcontents

\newpage

\section{Combination Test with Weak Low-Dimensional IVs} \label{sec:low_weak}
\appcontentsline{A}{Combination Test with Weak Low-Dimensional IVs}

In this section, we study the property of our combination test when Assumption \ref{ass:low_id} is violated. 
Specifically, we assume that the identification strength provided by the low-dimensional IVs is weak, rendering the Wald test invalid. 
On the other hand, we assume that the identification strength provided by the many IVs is strong.  
To characterize the limiting behavior of the low-dimensional IVs under weak identification, we introduce the following assumptions.

\begin{assprime}{2}  \label{ass:low_weak}

The following conditions hold: 
    \begin{enumerate}[label=\arabic*., ref=\theass.\arabic*]
        \item  For $r_n = \left\Vert z^\top \Pi \right\Vert_2$, $r_n/\sqrt{n} = O(1)$; \label{ass:low_weak_id}
        \item For all $n$ large enough,
        \begin{align*}
            \frac{1}{n}
             \begin{pmatrix}
                \sum_{g \in [G]} \mathbb E \left(z_{[g]}^\top \tilde e_{[g]} \right)\left(z_{[g]}^\top \tilde e_{[g]} \right)^\top &
                \sum_{g \in [G]} \mathbb E \left(z_{[g]}^\top \tilde e_{[g]} \right)\left( z_{[g]}^\top \tilde V_{[g]} \right)^\top \\
                \sum_{g \in [G]} \mathbb E \left(z_{[g]}^\top \tilde V_{[g]} \right)\left(z_{[g]}^\top \tilde e_{[g]} \right)^\top &
                \sum_{g \in [G]} \mathbb E \left(z_{[g]}^\top \tilde V_{[g]} \right)\left(z_{[g]}^\top \tilde V_{[g]} \right)^\top
            \end{pmatrix} > 0,
        \end{align*}
        in the matrix sense.
        \label{ass:low_weak_lower}
    \end{enumerate}
\end{assprime}

\begin{rem}
    Compared to Assumption \ref{ass:low_id}, the main difference here is that we have $r_n/\sqrt{n} = O(1)$. This corresponds to weak identification of the parameter of interest $\beta$ under the low-dimensional IVs, since in this case the deterministic part and the random part of $z^\top X$ are of the same order. This setting is similar to the weak-IV asymptotics considered in the literature,  where $z^\top X/\sqrt{n}$ converges to a random limit instead of diverging to infinity (the latter would happen under a standard asymptotics where $z^\top X/n$ is assumed to converge to a non-zero fixed limit). In addition, perfect correlations between the two linear forms are excluded.
\end{rem}

In the following, we first present the results for the case with $d_z=1$. Note that it is the most important case for empirical applications of IV regressions. For instance, 101 out of 230 specifications in \cite{andrews2019weak}'s sample and 1,087 out of 1,359 in \cite{young2022consistency}'s sample feature one endogenous regressor and one IV. Similarly, \cite{lee2022valid} find that 61 out of 123 IV papers published in \textit{AER} between 2013 and 2019 use one endogenous regressor and one IV. For these applications, empirical researchers can generate many IVs by using polynomials or interactions based on their one-dimensional base IV and control variables. Then, it is possible to achieve efficiency improvement using our combination procedure. 
Furthermore, the widely used shift-share or Bartik instrument is also one-dimensional. 
It turns out that in the case with $d_z=1$, the local asymptotic power function of our combination test is equal to that of the asymptotically optimal test based on $z^\top e(\beta_0)/\sqrt{\Omega}$, $LM(\beta_0)$, and $AR$, where the first statistic corresponds to the conventional cluster-robust AR test using $z$ as an instrument.

\begin{thm} 

Suppose that $d_z = 1$. Assume that the following limits exist:
\begin{align*}
    \bar \rho_1 &= \lim_{n \rightarrow \infty} \frac{1}{\sqrt{\Omega \Sigma}} \sum_{g \in [G]} \mathbb E \left[ \left( z_{[g]}^\top \tilde e_{[g]} \right) \left( \hat \Pi_{[g]}^\top \tilde e_{[g]} \right) \right], \\
    \bar \rho_2 &= \lim_{n \rightarrow \infty} \frac{2}{\sqrt{\Sigma \Upsilon}}  \sum_{g,h \in [G]^2, g \neq h} \mathbb E \left[ \left(\tilde V_{[g]}^\top P_{[g,h]} \tilde e_{[h]} \right) \left( \tilde e_{[g]}^\top P_{[g,h]} \tilde e_{[h]} \right) \right],
\end{align*} 
and $\bar \rho_1^2 + \bar \rho_2^2 < 1$. Under Assumptions \ref{ass:reg}, \ref{ass:low_weak} and \ref{ass:high}, and assuming that $\Pi^\top \Pi / \sqrt{K} \rightarrow \infty$, we have:
\begin{itemize}
    \item [1.] Suppose that there exists a deterministic sequence $d_n \downarrow 0$ such that 
    \begin{align*}
    d_n \Phi_2^{-1/2} \rightarrow a > 0, \quad \text{and}\quad  \beta - \beta_0 = \delta d_n,
    \end{align*}    
    for some fixed $\delta$, then
  \begin{align*}
   \lim_{n \rightarrow \infty}\mathbb E \left[ \phi^*_n  \right] & = \mathbb P \left( \left(-\frac{\bar \rho_1}{\sqrt{1- \bar \rho_1^2 - \bar \rho_2^2}} \N_1 + \frac{1}{\sqrt{1 - \bar \rho_1^2 - \bar \rho_2^2}} \N_2 -  \frac{\bar \rho_2}{\sqrt{1 - \bar \rho_1^2 - \bar \rho_2^2}} \N_3 \right)^2 \geqslant \mathbb C_\alpha \right) \\ 
   & =    \mathbb P\left( \chi^2_1 \left(\delta^2 \frac{a^2}{1-\bar \rho_1^2 - \bar \rho_2^2} \right)  \geq \mathbb C_\alpha \right) ,
       \end{align*}
where $\begin{pmatrix}
        \N_1 \\
        \N_2 \\
        \N_3 \\
    \end{pmatrix}  \stackrel{d}{=} \N \left( \begin{pmatrix}
        0 \\
        a \delta \\
        0 \\
    \end{pmatrix}, \begin{pmatrix}
        1 & \bar \rho_1 & 0  \\
        \bar \rho_1 & 1 & \bar \rho_2  \\
        0 & \bar \rho_2 & 1 \\
    \end{pmatrix} \right).$

    \item [2.] Suppose that $\beta - \beta_0 = \delta$ for some fixed $\delta \neq 0$, then $\lim_{n \rightarrow \infty}\mathbb E \left[ \phi_n^*  \right] = 1$.
\end{itemize}

\label{thm:low_weak_scalar}
\end{thm}

Theorem \ref{thm:low_weak_scalar} implies that, in the scalar case ($d_z=1$), our approach is equivalent to an optimal combination of the low-dimensional AR, jackknife LM, and jackknife AR. A key scalar-specific feature is that the discrepancy between the low-dimensional AR and low-dimensional Wald (i.e., $T(\beta_0)$) effectively reduces to a random sign that cannot be consistently estimated when the IV $z$ is weak; correspondingly, $\hat \rho_1$ is inconsistent, and the low-dimensional Wald is non-normal. However, by appropriately choosing combination weights, this unidentifiable sign component cancels out, preserving the validity of our combination test in the important scalar setting.

When $d_z > 1$, it is no longer possible to cancel the random sign. Nevertheless, if the correlation between $T(\beta_0)$ and $LM(\beta_0)$ is asymptotically negligible, our combined test remains robust to weak low-dimensional IVs, as shown in the next result.

\begin{thm}\label{thm: weak-low-2} 

Assume that the following limit exists:
\begin{align*}
    \rho = \lim_{n \rightarrow \infty} \frac{2}{\sqrt{\Sigma \Upsilon}}  \sum_{g,h \in [G]^2, g \neq h} \mathbb E \left[ \left(\tilde V_{[g]}^\top P_{[g,h]} \tilde e_{[h]} \right) \left( \tilde e_{[g]}^\top P_{[g,h]} \tilde e_{[h]} \right) \right],
\end{align*}
with $\rho^2 < 1$. Under Assumptions \ref{ass:reg}, \ref{ass:low_weak} and \ref{ass:high}, and assuming that $\Pi^\top \Pi / \sqrt{K} \rightarrow \infty$ with $\Pi^\top \Pi/ K \rightarrow 0$, we have:
\begin{itemize}
    \item [1.] Suppose that there exists a deterministic sequence $d_n \downarrow 0$ such that 
    \begin{align*}
    d_n \Phi_2^{-1/2} \rightarrow a > 0, \quad \text{and}\quad  \beta - \beta_0 = \delta d_n,
    \end{align*}    
    for some fixed $\delta$, then
    \begin{align*}
    & \lim_{n \rightarrow \infty}\mathbb E \left[ \phi^*_n  \right]  = \mathbb P \left( \left( \frac{1}{\sqrt{1 - \rho^2}} \N_1 -  \frac{\rho}{\sqrt{1 - \rho^2}} \N_2 \right)^2 \geqslant \mathbb C_\alpha \right)   =  \mathbb P\left( \chi^2_1 \left(\delta^2 \frac{a^2}{1-\rho^2} \right)  \geq \mathbb C_\alpha \right) ,  
    \end{align*}
    where $ \begin{pmatrix}
        \N_1 \\
        \N_2 \\
    \end{pmatrix}  \stackrel{d}{=} \N \left( \begin{pmatrix}
        a \delta \\
        0 \\
    \end{pmatrix}, \begin{pmatrix}
        1 & \rho  \\
        \rho & 1  \\
    \end{pmatrix} \right).$
    \item [2.] Suppose that $\beta - \beta_0 = \delta$ for some fixed $\delta \neq 0$, then $\lim_{n \rightarrow \infty}\mathbb E \left[ \phi_n^*  \right] = 1$.
\end{itemize}

\label{thm:low_weak_multi}
\end{thm}

Theorem \ref{thm: weak-low-2} shows that under the local alternative $\beta - \beta_0 = \delta d_n$, $\lim_{n \rightarrow \infty}\mathbb E \left[ \phi^*_n  \right]$ coincides exactly with the local asymptotic power function of the asymptotically optimal test based on $LM(\beta_0)$ and $AR$ in \cite{LWZ24}. The assumption that $\Pi^\top \Pi / K \rightarrow 0$ is similar to that imposed in \citet[Theorem 4]{MS22}. As pointed out by \cite{MS22}, this condition is quite weak as it still covers both weakly and strongly identified cases with many IVs. In addition, we note that under fixed alternatives, the combination test remains consistent even when the identification strength of the low-dimensional IVs is weak.

Note also that instead of using the low-dimensional Wald statistic, it is possible to combine the low-dimensional AR, jackknife LM, and jackknife AR directly, provided that the many IVs are strong. In this way, the combination test is robust against weak low-dimensional IVs when $d_z > 1$ without additional assumptions (e.g., $\Pi^{\top}\Pi/K \rightarrow 0$). In this paper, however, we emphasize Wald-based inference for the low-dimensional IV setting because it is the workhorse in empirical applications, and when low-dimensional IVs are strong, Wald is more powerful than AR in the overidentified case.

Finally, if both low-dimensional IVs and many IVs may be weak, then using null-imposed variance estimators for all component statistics, the combination test that combines the low-dimensional AR, jackknife LM, and jackknife AR is robust to arbitrary weak identification. It is also possible to consider a broader combination that includes low-dimensional AR and LM tests, along with jackknife LM and AR tests. This is beyond the scope of the current paper. 


\section{Technical Lemmas} \label{sec:lemma}

We use the following notation throughout Sections \ref{sec:lemma}-\ref{sec:proof_thm}. Recall the definition of $\tilde Y$, $\tilde X$, $\tilde Z$, $\tilde z$, $W$, $\tilde \Pi$, $\tilde V$ and $\tilde e$. Denote $P_W$ as the projection matrix of $W$, $M_W = I_n - P_W$, and let $Y = M_W \tilde Y$, $X = M_W \tilde X$, $Z = M_W \tilde Z$, $z = M_W \tilde z$, $\Pi = M_W \tilde \Pi$, $V = M_W \tilde V$ and $e = M_W \tilde e$. Denote $P$ as the projection matrix of $Z$, $M = I_n - P$, and $Q = M_W (P-\bar P) M_W$, where $\bar P$ is the block diagonal matrix corresponding to $P$ such that the $g$-th block on its diagonal is $P_{[g,g]}$; also denote $\bar Q$ as the block diagonal matrix corresponding to $Q$. Let $\acute \Pi = z A_n z^\top X$, $\hat \Pi = M_W (P - \bar P) \Pi = Q \tilde \Pi$, $\bar \Pi = (Q - \bar Q) \tilde \Pi$, $\acute X = z \hat A_n z^\top X$ and $\hat X = M_W (P - \bar P) X$. Finally, we use $\Omega_{\tilde e}$ to denote the block diagonal matrix with $g$-th block $\Omega_g^{\tilde e, \tilde e}$, and $\Omega_{\tilde V}$, $\Omega_{\tilde e, \tilde V}$ and $\Omega_{\tilde V, \tilde e}$ are defined similarly. 

\begin{lem} \label{lem:4th_moment}
    Under Assumption \ref{ass:reg}, we have
    \begin{align*}
        &\max_{g \in [G]}  \left(\tilde \Pi_{[g]}^\top \tilde \Pi_{[g]} \right)^2 + \max_{g \in [G]} \mathbb E \left( \tilde V_{[g]}^\top \tilde V_{[g]} \right)^2 + \max_{g \in [G]} \mathbb E \left(\tilde e_{[g]}^\top \tilde e_{[g]} \right)^2 \leqslant C, \\
        &\max_{g \in [G]}  \left( \Pi_{[g]}^\top \Pi_{[g]} \right)^2 + \max_{g \in [G]} \mathbb E \left( V_{[g]}^\top V_{[g]} \right)^2 + \max_{g \in [G]} \mathbb E \left( e_{[g]}^\top e_{[g]} \right)^2 \leqslant C,
    \end{align*}
    for some constant $C < \infty$. In addition, let $\hat \gamma_{\tilde e} = (W^\top W)^{-1} W^\top \tilde e$ and $\hat \gamma_{\tilde V} = (W^\top W)^{-1} W^\top \tilde V$, we have
    \begin{align*}
        \max_{1 \leq g \leq G} \left\Vert W_{[g]} \hat \gamma_{\tilde e} \right\Vert_2 = o_P(1), \quad \max_{1 \leq g \leq G} \left\Vert W_{[g]} \hat \gamma_{\tilde V} \right\Vert_2 = o_P(1).
    \end{align*}
\end{lem}

\begin{lem} \label{lem:PandQ}
    If Assumptions \ref{ass:reg} and \ref{ass:high} hold, then both $P$ and $Q$ are symmetric, and satisfy
    \begin{align*}
        \left\Vert P \right\Vert_{op} = O(1), \quad \left\Vert \bar P \right\Vert_{op} = O(1), \quad \left\Vert P \right\Vert_{F} = O(\sqrt{K}), \\
        \left\Vert Q \right\Vert_{op} = O(1), \quad \left\Vert \bar Q \right\Vert_{op} = o(1), \quad \left\Vert Q \right\Vert_{F} = O(\sqrt{K}).
    \end{align*}
    In addition, let $\tilde P$ be the block lower triangular matrix corresponding to $P - \bar P$ (i.e. $\tilde P_{[g,h]} = P_{[g,h]}$ for $g > h$ and $\tilde P_{[g,h]} = 0_{n_g \times n_h}$ otherwise), then \begin{align*}
        \left\Vert \tilde P \tilde P^{\top} \right\Vert_F = O(\sqrt{K}).
    \end{align*}
\end{lem}

\begin{lem} \label{lem:quad_form_1}
    Under Assumptions \ref{ass:reg} and \ref{ass:high}, we have
    \begin{align*}
        \tilde u^\top \bar P P_W \tilde v = O_P(1), \quad \tilde u^\top P_W \bar P \tilde v = O_P(1), \quad \tilde u^\top \bar P P_W \bar P \tilde v = O_P(1),
    \end{align*}
    for $(\tilde u, \tilde v) \in \{\tilde V, \tilde e\} \times \{\tilde V, \tilde e\}$.
\end{lem}

\begin{lem} \label{lem:quad_form_2}
    Under Assumptions \ref{ass:reg} and \ref{ass:high}, we have
    \begin{align*}
        \frac{1}{K} \sum_{g \in [G]} \left( \sum_{h \in [G], h \neq g} u_{[h]}^{\top} P_{[h,g]} v_{[g]} \right)^2 &= \frac{1}{K} \sum_{g \in [G]} \left( \sum_{h \in [G], h \neq g} \tilde u_{[h]}^{\top} P_{[h,g]} \tilde v_{[g]} \right)^2 + o_P(1) \\
        & = \frac{1}{K} \sum_{g \in [G]} \mathbb E \left( \sum_{h \in [G], h \neq g} \tilde u_{[h]}^{\top} P_{[h,g]} \tilde v_{[g]} \right)^2 + o_P(1), \\
        \frac{1}{K} \sum_{g,h \in [G]^2, g \neq h}\left( u_{[h]}^\top P_{[h,g]} v_{[g]} \right) \left( u_{[g]}^\top P_{[g,h]} v_{[h]} \right)  &= \frac{1}{K} \sum_{g,h \in [G]^2, g \neq h} \left( \tilde u_{[h]}^\top P_{[h,g]} \tilde v_{[g]} \right)\left(\tilde u_{[g]}^\top P_{[g,h]} \tilde v_{[h]} \right) + o_P(1) \\
        &= \frac{1}{K} \sum_{g,h \in [G]^2, g \neq h} \mathbb E \left( \tilde u_{[h]}^\top P_{[h,g]} \tilde v_{[g]} \right) \left(\tilde u_{[g]}^\top P_{[g,h]} \tilde v_{[h]} \right)  + o_P(1),
    \end{align*}
    for $(u,v) \in \{ V, e\} \times \{ V, e\}$, and the same results hold if we replace $P$ with $Q$. In addition, we have
    \begin{align*}
        & \frac{1}{K} \sum_{g, h \in [G]^2, g \neq h} \mathbb E \left( \tilde u_{[h]}^{\top} P_{[h,g]} \tilde v_{[g]} \right)^2 = \frac{1}{K} \sum_{g, h \in [G]^2, g \neq h} \mathbb E \left( \tilde u_{[h]}^{\top} Q_{[h,g]} \tilde v_{[g]} \right)^2 + o(1), \\
        & \frac{1}{K} \sum_{g, h \in [G]^2, g \neq h} \mathbb E \left( \tilde u_{[h]}^{\top} P_{[h,g]} \tilde v_{[g]} \right)\left( \tilde u_{[g]}^{\top} P_{[g,h]} \tilde v_{[h]} \right)  \\
        &= \frac{1}{K} \sum_{g, h \in [G]^2, g \neq h} \mathbb E \left( \tilde u_{[h]}^{\top} Q_{[h,g]} \tilde v_{[g]} \right) \left( \tilde u_{[g]}^{\top} Q_{[g,h]} \tilde v_{[h]} \right) + o(1),
    \end{align*}
    for $(\tilde u, \tilde v) \in \{\tilde V, \tilde e\} \times \{\tilde V, \tilde e\}$.
\end{lem}

\begin{lem} \label{lem:linear_quad_form_1}

Under Assumptions \ref{ass:reg} and \ref{ass:high}, we have $\Sigma \geq C (\Pi^\top \Pi + K)$ for some constant $C>0$, and
\begin{align*}
&\quad \frac{1}{\Sigma} \sum_{g\in [G]} \left(\sum_{h \in [G], h \neq g} \tilde X_{[h]}^\top Q_{[h,g]} \tilde e_{[g]} \right)^2 \\&= \frac{1}{\Sigma} \sum_{g\in [G]} \mathbb E \left(\sum_{h \in [G], h \neq g} \tilde X_{[h]}^\top Q_{[h,g]} \tilde e_{[g]} \right)^2 + o_P(1), \\
&= \frac{1}{\Sigma} \sum_{g\in [G]} \mathbb E \left(\bar \Pi_{[g]}^\top \tilde e_{[g]} \right)^2 + \frac{1}{\Sigma} \sum_{g, h\in [G]^2, g \neq h} \mathbb E \left( \tilde V_{[h]}^\top Q_{[h,g]} \tilde e_{[g]} \right)^2 + o_P(1),
\end{align*}
\begin{align*}
&\quad \frac{1}{\Sigma} \sum_{g\in [G]} \left(\sum_{h \in [G], h \neq g} \tilde X_{[h]}^\top Q_{[h,g]}  e_{[g]} \right)^2 \\
&= \frac{1}{\Sigma} \sum_{g\in [G]} \left(\sum_{h \in [G], h \neq g} \tilde X_{[h]}^\top Q_{[h,g]} \tilde e_{[g]} \right)^2 + o_P(1), 
\end{align*} 
\begin{align*}
&\quad \frac{1}{\Sigma}  \sum_{g\in [G]} \left(\sum_{h \in [G], h \neq g} \tilde X_{[h]}^\top Q_{[h,g]} X_{[g]} \right)^2 \\
& = \frac{1}{\Sigma} \sum_{g\in [G]} \left(\bar \Pi_{[g]}^\top \Pi_{[g]} \right)^2 + \frac{1}{\Sigma} \sum_{g\in [G]} \mathbb E \left(\bar \Pi_{[g]}^\top \tilde V_{[g]} \right)^2 \\
& + \frac{1}{\Sigma} \sum_{g\in [G]} \mathbb E \left(\sum_{h \in [G], h \neq g} \tilde V_{[h]}^\top Q_{[h,g]} \Pi_{[g]} \right)^2 \\
& + \frac{1}{\Sigma} \sum_{g\in [G]} \mathbb E\left(\sum_{h \in [G], h \neq g} \tilde V_{[h]}^\top Q_{[h,g]} \tilde V_{[g]} \right)^2 + o_P(1), 
\end{align*}
and 
\begin{align*}
&\quad \frac{1}{\Sigma}  \sum_{g\in [G]} \left(\sum_{h \in [G], h \neq g} \tilde X_{[h]}^\top Q_{[h,g]} X_{[g]} \right) \left(\sum_{k \in [G], k \neq g} \tilde X_{[k]}^\top Q_{[k,g]} e_{[g]} \right) \\
& = \frac{1}{\Sigma} \sum_{g \in [G]} \mathbb E \left(\bar \Pi_{[g]}^\top \tilde V_{[g]} \right) \left(\bar \Pi_{[g]}^\top \tilde e_{[g]} \right) \\
&+ \frac{1}{\Sigma}  \sum_{g\in [G]} \mathbb E \left(\sum_{h \in [G], h \neq g} \tilde V_{[h]}^\top Q_{[h,g]} \tilde V_{[g]} \right) \left(\sum_{k \in [G], k \neq g} \tilde V_{[k]}^\top Q_{[k,g]} \tilde e_{[g]} \right) + o_P(1).
\end{align*}
\end{lem}

\begin{lem} \label{lem:linear_quad_form_2}
    Under Assumptions \ref{ass:reg} and \ref{ass:high}, we have
\begin{align*}
    &\quad \frac{1}{\Sigma} \sum_{g,h \in [G]^2, g \neq h} \left( \tilde X_{[g]}^\top Q_{[g,h]} \tilde e_{[h]} \right) \left( \tilde X_{[h]}^\top Q_{[h,g]} \tilde e_{[g]} \right) \\
    & = \frac{1}{\Sigma} \sum_{g,h \in [G]^2, g \neq h} \mathbb E\left( \tilde V_{[g]}^\top Q_{[g,h]} \tilde e_{[h]} \right) \left( \tilde V_{[h]}^\top Q_{[h,g]} \tilde e_{[g]} \right) + o_P(1), 
\end{align*} 
\begin{align*}
    &\quad \frac{1}{\Sigma} \sum_{g,h \in [G]^2, g \neq h} \left( \tilde X_{[g]}^\top Q_{[g,h]} e_{[h]} \right) \left( \tilde X_{[h]}^\top Q_{[h,g]} e_{[g]} \right) \\
    & = \frac{1}{\Sigma} \sum_{g,h \in [G]^2, g \neq h} \left( \tilde X_{[g]}^\top Q_{[g,h]} \tilde e_{[h]} \right) \left( \tilde X_{[h]}^\top Q_{[h,g]} \tilde e_{[g]} \right) + o_P(1),     
\end{align*}    
\begin{align*}
    &\quad \frac{1}{\Sigma} \sum_{g,h \in [G]^2, g \neq h} \left( \tilde X_{[g]}^\top Q_{[g,h]} X_{[h]} \right) \left( \tilde X_{[h]}^\top Q_{[h,g]} X_{[g]} \right) \\
    & = \frac{1}{\Sigma} \sum_{g,h \in [G]^2, g \neq h} \mathbb E \left( \tilde V_{[g]}^\top Q_{[g,h]} \tilde V_{[h]} \right) \left( \tilde V_{[h]}^\top Q_{[h,g]} \tilde V_{[g]} \right)\\
        &+ \frac{1}{\Sigma} \sum_{g,h \in [G]^2, g \neq h} \left( \tilde \Pi_{[g]}^\top Q_{[g,h]} \Pi_{[h]} \right) \left( \tilde \Pi_{[h]}^\top Q_{[h,g]} \Pi_{[g]} \right) \\
        & + \frac{2}{\Sigma} \sum_{g,h \in [G]^2, g \neq h} \mathbb E \left( \tilde V_{[g]}^\top Q_{[g,h]} \Pi_{[h]} \right) \left( \tilde \Pi_{[h]}^\top Q_{[h,g]} \tilde V_{[g]} \right) + o_P(1),    
\end{align*}    
and    
    \begin{align*}
         &\quad \frac{1}{\Sigma} \sum_{g,h \in [G]^2, g \neq h} \left( \tilde X_{[g]}^\top Q_{[g,h]} X_{[h]} \right) \left( \tilde X_{[h]}^\top Q_{[h,g]} e_{[g]} \right) \\
         & = \frac{1}{\Sigma} \sum_{g,h \in [G]^2, g \neq h} \mathbb E \left( \tilde V_{[g]}^\top Q_{[g,h]} \tilde V_{[h]} \right) \left( \tilde V_{[h]}^\top Q_{[h,g]} \tilde e_{[g]} \right) \\
        &+ \frac{1}{\Sigma} \sum_{g,h \in [G]^2, g \neq h} \mathbb E \left( \tilde V_{[g]}^\top Q_{[g,h]} \Pi_{[h]} \right) \left( \tilde \Pi_{[h]}^\top Q_{[h,g]} \tilde e_{[g]} \right) + o_P(1).
    \end{align*}
\end{lem}

\begin{lem}\label{lem:var_est}
    Let $\acute \beta$ be a generic estimator of $\beta$.
Further define 
\begin{align*}
& \acute \Psi = X^\top z \hat A_n \acute \Omega \hat A_n z^\top X, \\
& \acute \Sigma  = \sum_{g\in [G]} \left(\sum_{h \in [G], h \neq g} \tilde X_{[h]}^\top Q_{[h,g]} \acute e_{[g]} \right)^2 + \sum_{g,h \in [G]^2, g \neq h} \left( \tilde X_{[g]}^\top Q_{[g,h]} \acute e_{[h]} \right) \left( \tilde X_{[h]}^\top Q_{[h,g]} \acute e_{[g]} \right),\\
& \acute \Upsilon =  2 \sum_{g,h \in [G]^2, g \neq h} \left( \acute e_{[g]}^\top P_{[g,h]} \acute e_{[h]} \right)^2 ,
\end{align*}
where $\acute e = Y - X \acute \beta$ and 
\begin{align*}
\acute \Omega & = \sum_{g \in [G]} \left(z_{[g]}^\top \acute e_{[g]}\right) \left(z_{[g]}^\top \acute e_{[g]}\right)^\top.
\end{align*}

Suppose that $\acute \beta \convP \beta$, then the following holds. 
    \begin{enumerate}
        \item If Assumption \ref{ass:reg} holds, then 
            \begin{align*}
                \hat \Omega^{-1/2} \Omega^{1/2} = I_{d_z} + o_P(1).
            \end{align*}
        \item If Assumptions \ref{ass:reg} and \ref{ass:low_id} hold, then
        \begin{align*}
            \frac{\acute \Psi}{\Psi} &= 1 + o_P(1).
        \end{align*}
        \item If Assumptions \ref{ass:reg} and \ref{ass:high} hold, then
        \begin{align*}
            \frac{\acute \Sigma}{\Sigma} &= 1 + o_P(1), \\
            \frac{\acute \Upsilon}{\Upsilon} &= 1 + o_P(1).
        \end{align*}
    \end{enumerate}
\end{lem}

\begin{lem}\label{lem:betahat}
 Under Assumptions \ref{ass:reg}-\ref{ass:high}, we have $\hat \beta \convP \beta$ and $(\hat \beta - \beta)^2\Pi^\top \Pi /\sqrt{K} = o_P(1)$. Alternatively, if Assumptions \ref{ass:reg}, \ref{ass:low_weak} and \ref{ass:high} hold and $\Pi^\top \Pi / \sqrt{K} \rightarrow \infty$, we have $\hat \beta \convP \beta$ and $(\hat \beta - \beta)^2\Pi^\top \Pi /\sqrt{K} = o_P(1)$. 
\end{lem}

\begin{lem} \label{lem:linear_form}
    Under Assumptions \ref{ass:reg}-\ref{ass:local_alternative_and_covariance}, if the assumptions for $a_1$ and $a_2$ in Theorem \ref{thm:limit_str_lcl} hold, then
    \begin{align*}
        \begin{pmatrix}
            T(\beta_0) \\
            LM(\beta_0) \\
            AR
        \end{pmatrix} = 
        \begin{pmatrix}
            \frac{1}{\sqrt{\Psi}} \sum_{g=1}^G \acute \Pi_{[g]}^\top \tilde e_{[g]} \\
            \frac{1}{\sqrt{\Sigma}} \left( \sum_{g \in [G]} \hat \Pi_{[g]}^\top \tilde e_{[g]} + \sum_{g,h \in [G]^2, g \neq h} \tilde V_{[g]}^\top P_{[g,h]} \tilde e_{[h]} \right)  \\
            \frac{1}{\sqrt{\Upsilon}} \sum_{g,h \in [G]^2, g \neq h} \tilde e_{[g]}^\top P_{[g,h]} \tilde e_{[h]}
        \end{pmatrix} + 
        \begin{pmatrix}
            a_1 \delta \\
            a_2 \delta \\
            0
        \end{pmatrix} + o_P(1).
    \end{align*}
\end{lem}

\begin{lem}\label{lem:limit_dist}
    If Assumptions \ref{ass:reg}-\ref{ass:local_alternative_and_covariance} hold, then
    \begin{align*}
    \begin{pmatrix}
        \frac{1}{\sqrt{\Psi}} \sum_{g=1}^G \acute \Pi_{[g]}^\top \tilde e_{[g]} \\
        \frac{1}{\sqrt{\Sigma}} \left( \sum_{g \in [G]} \hat \Pi_{[g]}^\top \tilde e_{[g]} + \sum_{g,h \in [G]^2, g \neq h} \tilde V_{[g]}^\top P_{[g,h]} \tilde e_{[h]} \right)  \\
        \frac{1}{\sqrt{\Upsilon}} \sum_{g,h \in [G]^2, g \neq h} \tilde e_{[g]}^\top P_{[g,h]} \tilde e_{[h]}
    \end{pmatrix} \convD \N \left( \begin{pmatrix}
    0 \\
    0 \\
    0
\end{pmatrix}, \begin{pmatrix}
    1 &  \rho_1 & 0 \\
    \rho_1 & 1 & \rho_2 \\
    0 & \rho_2 & 1
\end{pmatrix} \right).
\end{align*}
\end{lem}

\begin{lem} \label{lem:cov_est}
    Under Assumptions \ref{ass:reg} and \ref{ass:high}, we have
    \begin{align*}
        \frac{1}{\sqrt{\Sigma}} \Omega^{-1/2}   \sum_{g \in [G]} \left[ \left( z_{[g]}^\top \hat e_{[g]} \right) \left( \hat X_{[g]}^\top \hat e_{[g]} \right) \right] &= \frac{1}{\sqrt{\Sigma}} \Omega^{-1/2}   \sum_{g \in [G]} \mathbb E\left[ \left( z_{[g]}^\top \tilde e_{[g]} \right) \left( \hat \Pi_{[g]}^\top \tilde e_{[g]} \right) \right] + o_P(1), 
    \end{align*}
    and
    \begin{align*}
        &\quad \frac{1}{\sqrt{\Sigma \Upsilon}} \sum_{g,h \in [G]^2, g \neq h} \left(X_{[g]}^\top P_{[g,h]} \hat e_{[h]} \right) \left( \hat e_{[g]}^\top P_{[g,h]} \hat e_{[h]} \right) \\
        &= \frac{1}{\sqrt{\Sigma \Upsilon}}  \sum_{g,h \in [G]^2, g \neq h} \mathbb E \left(\tilde V_{[g]}^\top P_{[g,h]} \tilde e_{[h]} \right) \left( \tilde e_{[g]}^\top P_{[g,h]} \tilde e_{[h]} \right) + o_P(1).
    \end{align*}
\end{lem}

\begin{lem}\label{lem:nuisance_parameter}
    Under Assumptions \ref{ass:reg}--\ref{ass:local_alternative_and_covariance}, we have
    \begin{align*}
        \hat \rho_1 \convP \rho_1, \quad \hat \rho_2 \convP \rho_2. 
    \end{align*}
    If in addition the assumptions for $a_1$ and $a_2$ in Theorem \ref{thm:limit_str_lcl} hold, then
        \begin{align*}
            \hat \alpha_1 \convP \alpha_1,  \quad 
            \hat \alpha_2 \convP \alpha_2.
        \end{align*} 
\end{lem}

\section{Proofs of Technical Lemmas} \label{sec:proof_lemma}

\subsection{Proof of Lemma \ref{lem:4th_moment}}

The first result follows readily from Assumption \ref{ass:reg}. For the second result, note that for any $g \in [G]$, we have
\begin{align*}
    \left( \Pi_{[g]}^\top \Pi_{[g]} \right)^2 = \left( \sum_{i \in I_g} \Pi_{i,g}^2 \right)^2 \leqslant C \sum_{i \in I_g} \Pi_{i,g}^4 \leqslant C.
\end{align*}
In addition, by an abuse of notation, for any $i \in [n]$ and $g \in [G]$, we denote $M_W^{(i)}$ as the $i$-th column of $M_W$, and $M_{W,[g]}^{(i)} \in \Re^{n_g}$ is vector that collects all elements in the $n$-dimensional vector  $M_W^{(i)}$ that belong to the $g$-th cluster. Then, we have
\begin{align*}
    \mathbb E V_{i,g}^4 &= \mathbb E \left( \sum_{g \in [G]} M_{W,[g]}^{(i),\top} \tilde V_{[g]} \right)^4 \\
    &\leqslant C \left( \sum_{g \in [G]} \left( M_{W,[g]}^{(i),\top} M_{W,[g]}^{(i)} \right)^2 + \sum_{g,h \in [G]^2, h \neq g} \left( M_{W,[g]}^{(i),\top} M_{W,[g]} \right) \left( M_{W,[h]}^{(i),\top} M_{W,[h]} \right) \right) \\
    &\leqslant C,
\end{align*}
by the first result and the fact that
\begin{align*}
    \sum_{g \in [G]} \left( M_{W,[g]}^{(i),\top} M_{W,[g]}^{(i)} \right) = M_W^{(i),\top} M_W^{(i)} = M_{W,ii} \leqslant C.
\end{align*}
It follows that
\begin{align*}
    \mathbb E \left( V_{[g]}^\top V_{[g]} \right)^2 = \mathbb E \left( \sum_{i \in I_g} V_{i,g}^2 \right)^2 \leqslant C \sum_{i \in I_g} \mathbb E V_{i,g}^4 \leqslant C. 
\end{align*}
Using the same argument, we also have
\begin{align*}
    \mathbb E \left( e_{[g]}^\top e_{[g]} \right)^2 \leqslant C,
\end{align*}
and the desired result follows. Finally, for the last result, by Assumption \ref{ass:reg}, we have $ \hat \gamma_{\tilde e} = O_P(1/\sqrt{n})$ and $ \hat \gamma_{\tilde V} = O_P(1/\sqrt{n})$, and thus 
\begin{align*}
    \max_{1 \leq g \leq G} \left\Vert W_{[g]} \hat \gamma_{\tilde e} \right\Vert_2^2 &\leqslant \max_{1 \leq g \leq G} n_g \times \max_{i \in I_g, g \in [G]} \left\Vert W_{i,g} \right\Vert_2^2 \times \left\Vert \hat \gamma_{\tilde e} \right\Vert_2^2 = o_P(1), \\
    \max_{1 \leq g \leq G} \left\Vert W_{[g]} \hat \gamma_{\tilde V} \right\Vert_2^2 &\leqslant \max_{1 \leq g \leq G} n_g \times \max_{i \in I_g, g \in [G]} \left\Vert W_{i,g} \right\Vert_2^2 \times \left\Vert \hat \gamma_{\tilde V} \right\Vert_2^2 = o_P(1).
\end{align*}

$\hfill\qedsymbol$

\subsection{Proof of Lemma \ref{lem:PandQ}}

For the first part of Lemma \ref{lem:PandQ}, the results for $P$ are standard for projection matrix, so we focus on the results for $Q$. We have
\begin{align*}
    \left\Vert Q \right\Vert_{op} = \left\Vert M_W (P-\bar P) M_W \right\Vert_{op} \leqslant \left\Vert M_W \right\Vert_{op}^2 \left\Vert P - \bar P \right\Vert_{op} = O(1),
\end{align*}
and 
\begin{align*}
    \left\Vert Q \right\Vert_{F} = \sqrt{\tr(M_W(P-\bar P) M_W (P -\bar P) M_W)} \leqslant C \left\Vert P - \bar P \right\Vert_{F} = O(\sqrt{K}).
\end{align*}
In addition, we note that 
\begin{align*}
    \bar Q = \bar P_W \bar P + \bar P \bar P_W - \hat P,
\end{align*}
where $\bar P_W$ is the block diagonal matrix corresponding to $P_W$ and $\hat P$ is a block diagonal matrix such that the $g$-th block on its diagonal is $\sum_{h=1}^G P_{W,[g,h]}P_{[h,h]}P_{W,[h,g]}$ (corresponding to the block diagonals of $P_W \bar P P_W$). By Assumption \ref{ass:reg}, we have 
\begin{align*}
    \max_{1 \leq i \leq n} P_{W, ii} \leq \max_{i \in I_g, g \in [G]} \left\Vert W_{i,g} \right\Vert_2^2 \times \lambda_{\max}\left( (W^\top W)^{-1} \right) = o(1),
\end{align*}
and thus 
\begin{align*}
    \max_{1 \leq g \leq G} \lambda_{\max} \left(P_{W,[g,g]} \right) \leq \max_{1 \leq g \leq G} n_g \times \max_{1 \leq i \leq n} P_{W, ii} = o(1),
\end{align*}
which implies that $\lambda_{\max}  \left( \bar P_W \right) = o(1)$. It follows that
\begin{align*}
    \left\Vert \bar Q \right\Vert_{op} &\leqslant \left\Vert \bar P_W \bar P \right\Vert_{op} +  \left\Vert \bar P \bar P_W \right\Vert_{op} + \left\Vert \hat P \right\Vert_{op} \\
    &\leqslant 2\left\Vert \bar P_W  \right\Vert_{op} \left\Vert \bar P \right\Vert_{op} + \max_{1 \leq g \leq G} \left\Vert \sum_{h \in [G]} P_{W,[g,h]} P_{[h,h]} P_{W,[h,g]} \right\Vert_{op} \\
    & \leq C\max_{1 \leq g,h\leq G}  \left\Vert P_{W,[g,h]}P_{W,[h,g]} \right\Vert_{op} + o(1) \\
    &\leqslant C\max_{1 \leq g \leq G}  \left\Vert P_{W,[g,g]} \right\Vert_{op} + o(1) \\
    &= o(1).
\end{align*}

For the second part of Lemma \ref{lem:PandQ}, we shall use an argument similar to \cite{Chao(2012)}. A closer inspection of their proof suggests that, all the equalities in the proof of their Lemma B.2. remain unchanged if we replace $P_{ij}$ with $P_{[g,h]}$ and keep the trace operator; note also that we have $\tr\left((P - \bar P)^{4}\right) = O(K)$ so that (i) of Lemma B.2. still holds. To obtain (iii) of Lemma B.2. we establish results similar to their Lemma B.1.: for any subset $\mathcal I_2$ of the set $\{g,h \}_{g,h=1}^G$, we have
\begin{align*}
\tr\left(\sum_{\mathcal I_2} P_{[g,h]}P_{[h,g]}P_{[g,h]}P_{[h,g]} \right) & \leqslant C \sum_{\mathcal I_2} \tr \left(P_{[g,h]} P_{[h,g]}\right)\\
& \leqslant C\sum_{g,h \in [G]^2} \tr\left(P_{[g,h]}  P_{[h,g]}\right) \\
&= O(K),
\end{align*}
and similarly for any subset $\mathcal I_3$ of the set $\{g,h,k\}_{g,h,k=1}^G$, we have \begin{align*}
\tr \left(\sum_{\mathcal I_3} P_{[g,h]}P_{[h,k]}P_{[k,h]}P_{[h,g]}\right) &= O(K), \\
\tr \left(\sum_{\mathcal I_3} P_{[g,h]}P_{[h,g]}P_{[g,k]}P_{[k,g]}\right) &= O(K), 
\end{align*} and then it is easy to see that (iii) of Lemma B.2. holds. To obtain (ii) of Lemma B.2. we define, as in their paper, the following random variables 
\begin{align*}
\Delta_1 &= \sum_{g < h < k} \left( \xi_{[h]}^\top P_{[h,g]}P_{[g,k]}\xi_{[k]} + \xi_{[g]}^\top P_{[g,h]}P_{[h,k]}\xi_{[k]} + \xi_{[g]}^\top P_{[g,k]}P_{[k,h]}\xi_{[h]}\right) \\
\Delta_2 &= \sum_{g < h < k} \left( \xi_{[h]}^\top P_{[h,g]}P_{[g,k]}\xi_{[k]} + \xi_{[g]}^\top P_{[g,h]}P_{[h,k]}\xi_{[k]}\right) \\ 
\Delta_3 &= \sum_{g < h < k} \left( \xi_{[g]}^\top P_{[g,k]}P_{[k,h]}\xi_{[h]}\right)
\end{align*} where $\{\xi_i \}_{i=1}^n$ is a sequence of i.i.d. random variables with mean $0$ and variance $1$, and independent of $\tilde{Z}, W$ (note that $\xi_i$ are not only independent across clusters but also within clusters), and then it is straightforward to verify that (ii) of Lemma B.2. also holds. These results, together with a similar argument as in the proof of their Lemma B.3., allow us to conclude that $\left\Vert \tilde P \tilde P^{\top} \right\Vert_F = O(\sqrt{K})$. This concludes the proof. $\hfill\qedsymbol$

\subsection{Proof of Lemma \ref{lem:quad_form_1}}

We focus on the case when $\tilde u = \tilde V$ and $\tilde v = \tilde e$. We have
\begin{align*}
    \left| \tilde V^\top P_W \bar P \tilde e \right| &\leqslant \sqrt{\tilde V^\top P_W \tilde V} \times \sqrt{\tilde e^\top \bar P P_W \bar P \tilde e} = O_P(1),
\end{align*}
because
\begin{align*}
    \mathbb E \left( \tilde V^\top P_W \tilde V \right) &= \tr \left(P_W \Omega_{\tilde V} \right) \leq \lambda_{\max}(\Omega_{\tilde V}) \tr \left(P_W  \right) \leqslant C, \\
    \mathbb E \left( \tilde e^\top \bar P P_W \bar P \tilde e \right) &= \tr \left(\bar P P_W \bar P \Omega_{\tilde e} \right) \leq \lambda_{\max}(\Omega_{\tilde e}) \lambda_{\max}(\bar P^2) \tr \left(P_W  \right) \leqslant C,
\end{align*}
since $d_w$ is fixed. The other two terms can be handled similarly. $\hfill\qedsymbol$

\subsection{Proof of Lemma \ref{lem:quad_form_2}}

For the first result, we focus on the case when $u = V$ and $v = e$. To show
\begin{align}\label{eq:Vpe1}
    \frac{1}{K} \sum_{g \in [G]} \left( \sum_{h \in [G], h \neq g} V_{[h]}^{\top} P_{[h,g]} e_{[g]} \right)^2  - \frac{1}{K} \sum_{g \in [G]}  \left( \sum_{h \in [G], h \neq g} \tilde V_{[h]}^{\top} P_{[h,g]} \tilde e_{[g]} \right)^2 = o_P(1),
\end{align}
we note that
\begin{align*}
    &\quad \frac{1}{K} \sum_{g \in [G]} \left( \sum_{h \in [G], h \neq g} V_{[h]}^{\top} P_{[h,g]} e_{[g]} \right)^2 \\
    &= \frac{1}{K} \sum_{g \in [G]} \left( \sum_{h \in [G], h \neq g} (\tilde V_{[h]} - W_{[h]} \hat \gamma_{\tilde V} )^{\top} P_{[h,g]} (\tilde e_{[g]} - W_{[g]} \hat \gamma_{\tilde e}) \right)^2 \\
    &= \frac{1}{K} \sum_{g \in [G]} \left( \sum_{h \in [G], h \neq g} \begin{pmatrix}
        & \tilde V_{[h]} P_{[h,g]} \tilde e_{[g]} - (W_{[h]} \hat \gamma_{\tilde V})^\top P_{[h,g]} \tilde e_{[g]} \\
        & - \tilde V_{[h]}^\top P_{[h,g]} (W_{[g]} \hat \gamma_{\tilde e}) + (W_{[h]} \hat \gamma_{\tilde V})^\top P_{[h,g]} (W_{[g]} \hat \gamma_{\tilde e}) 
    \end{pmatrix}  \right)^2.
\end{align*}
We have
\begin{align*}
    \frac{1}{K} \sum_{g \in [G]} \left( \sum_{h \in [G], h \neq g} (W_{[h]} \hat \gamma_{\tilde V})^\top P_{[h,g]} \tilde e_{[g]} \right)^2 &= \frac{1}{K} \sum_{g \in [G]} \left( (W_{[g]} \hat \gamma_{\tilde V})^\top P_{[g,g]} \tilde e_{[g]} \right)^2 \\
    &\leq \frac{C \max_{1 \leq g \leq G} \left\Vert W_{[g]} \hat \gamma_{\tilde V} \right\Vert_2^2}{K} \sum_{g \in [G]} \left\Vert P_{[g,g]} \tilde e_{[g]} \right\Vert_2^2 \\
    &= o_P(1),
\end{align*}
by Lemma \ref{lem:4th_moment}, where we use the fact that $W^\top P = 0_{n \times n}$ and
\begin{align*}
     \mathbb E \frac{1}{K} \sum_{g \in [G]} \left\Vert P_{[g,g]} \tilde e_{[g]} \right\Vert_2^2 \leq \frac{C}{K} \sum_{g \in [G]} \tr(P_{[g,g]}) = O(1).
\end{align*}
Similarly, we have
\begin{align*}
    \frac{1}{K} \sum_{g \in [G]} \left( \sum_{h \in [G], h \neq g} \tilde V_{[h]}^\top P_{[h,g]} (W_{[g]} \hat \gamma_{\tilde e}) \right)^2 &\leq \frac{\max_{1 \leq g \leq G} \left\Vert W_{[g]} \hat \gamma_{\tilde e} \right\Vert_2^2}{K} \sum_{g \in [G]} \left\Vert \sum_{h \in [G], h \neq g} P_{[g,h]} \tilde V_{[h]} \right\Vert_2^2 \\
    &= o_P(1),
\end{align*}
since
\begin{align*}
    \mathbb E \frac{1}{K} \sum_{g \in [G]} \left\Vert \sum_{h \in [G], h \neq g} P_{[g,h]} \tilde V_{[h]} \right\Vert_2^2 &= \frac{1}{K} \sum_{g,h \in [G]^2, g \neq h} \mathbb E \left\Vert P_{[g,h]} \tilde V_{[h]} \right\Vert_2^2 \\
    &\leq \frac{C}{K} \sum_{g,h \in [G]^2, g \neq h} \tr(P_{[g,h]} P_{[h,g]}) \\
    &= O(1).
\end{align*}
Finally, we have
\begin{align*}
  &\quad \frac{1}{K} \sum_{g \in [G]} \left( \sum_{h \in [G], h \neq g} (W_{[h]} \hat \gamma_{\tilde V})^\top P_{[h,g]} (W_{[g]} \hat \gamma_{\tilde e}) \right)^2 \\
  &= \frac{1}{K} \sum_{g \in [G]} \left((W_{[g]} \hat \gamma_{\tilde V})^\top P_{[g,g]} (W_{[g]} \hat \gamma_{\tilde e}) \right)^2 \\
    &\leq \frac{C \max_{1 \leq g \leq G} \left\Vert W_{[g]} \hat \gamma_{\tilde V} \right\Vert_2^2 \max_{1 \leq g \leq G} \left\Vert W_{[g]} \hat \gamma_{\tilde e} \right\Vert_2^2}{K} \sum_{g \in [G]} \tr(P_{[g,g]}) \\
    &= o_P(1).
\end{align*}
Combining the above results with the triangle inequality, we have 
\begin{align*}
\left(    \frac{1}{K} \sum_{g \in [G]} \left( \sum_{h \in [G], h \neq g} V_{[h]}^{\top} P_{[h,g]} e_{[g]} \right)^2 \right)^{1/2} = \left( \frac{1}{K} \sum_{g \in [G]} \left( \sum_{h \in [G], h \neq g} \tilde V_{[h]}^{\top} P_{[h,g]} \tilde e_{[g]} \right)^2 \right)^{1/2} + o_P(1).
\end{align*}

In addition, we have
\begin{align*}
    \mathbb E \frac{1}{K} \sum_{g \in [G]} \left( \sum_{h \in [G], h \neq g} \tilde V_{[h]} P_{[h,g]} \tilde e_{[g]} \right)^2 = \frac{1}{K} \sum_{g,h \in [G]^2, g \neq h} \mathbb E \left( \tilde V_{[h]} P_{[h,g]} \tilde e_{[g]} \right)^2 \leq C ||P-\bar P||_F^2/K =  O(1),
\end{align*}
which implies
\begin{align*}
\frac{1}{K} \sum_{g \in [G]} \left( \sum_{h \in [G], h \neq g} \tilde V_{[h]} P_{[h,g]} \tilde e_{[g]} \right)^2   = O_P(1).  
\end{align*}

Therefore, we have
\begin{align*}
&\quad \left\vert \frac{1}{K} \sum_{g \in [G]} \left( \sum_{h \in [G], h \neq g} V_{[h]}^{\top} P_{[h,g]} e_{[g]} \right)^2 - \frac{1}{K} \sum_{g \in [G]} \left( \sum_{h \in [G], h \neq g} \tilde V_{[h]}^{\top} P_{[h,g]} \tilde e_{[g]} \right)^2 \right\vert \\
& = \left\vert \left(    \frac{1}{K} \sum_{g \in [G]} \left( \sum_{h \in [G], h \neq g} V_{[h]}^{\top} P_{[h,g]} e_{[g]} \right)^2 \right)^{1/2} - \left( \frac{1}{K} \sum_{g \in [G]} \left( \sum_{h \in [G], h \neq g} \tilde V_{[h]}^{\top} P_{[h,g]} \tilde e_{[g]} \right)^2 \right)^{1/2} \right\vert \\
& \times \left\vert \left(    \frac{1}{K} \sum_{g \in [G]} \left( \sum_{h \in [G], h \neq g} V_{[h]}^{\top} P_{[h,g]} e_{[g]} \right)^2 \right)^{1/2} + \left( \frac{1}{K} \sum_{g \in [G]} \left( \sum_{h \in [G], h \neq g} \tilde V_{[h]}^{\top} P_{[h,g]} \tilde e_{[g]} \right)^2 \right)^{1/2} \right\vert = o_P(1). 
\end{align*}

Next, we show that
\begin{align}\label{eq:Vpe2}
    \frac{1}{K} \sum_{g \in [G]} \left( \sum_{h \in [G], h \neq g} \tilde V_{[h]}^{\top} P_{[h,g]} \tilde e_{[g]} \right)^2  - \frac{1}{K} \sum_{g \in [G]} \mathbb E \left( \sum_{h \in [G], h \neq g} \tilde V_{[h]}^{\top} P_{[h,g]} \tilde e_{[g]} \right)^2 = o_P(1).
\end{align}
By Markov inequality, it suffices to show that the RHS of the  following display is $o(1)$
\begin{align}\label{eq:Vpe_var}
    &\quad \mathbb E \left( \frac{1}{K} \left(\sum_{g\in [G]} \left(\sum_{h \in [G], h \neq g} \tilde V_{[h]}^\top P_{[h,g]} \tilde e_{[g]} \right)^2 -  \sum_{g\in [G]} \mathbb E \left(\sum_{h \in [G], h \neq g} \tilde V_{[h]}^\top P_{[h,g]} \tilde e_{[g]} \right)^2 \right)  \right)^2 \notag \\
    &= \frac{1}{K^2} \mathbb V \left( \sum_{g,h,k\in [G]^3, h \neq g, k \neq g}  \tilde V_{[h]}^\top P_{[h,g]} \tilde e_{[g]} \tilde V_{[k]}^\top P_{[k,g]} \tilde e_{[g]} \right) \notag \\
    &\leqslant \frac{C}{K^2} \mathbb V \left( \sum_{g,h\in [G]^2, g \neq h}  \left(\tilde V_{[h]}^\top P_{[h,g]} \tilde e_{[g]} \right)^2 \right) + \frac{C}{K^2} \mathbb V \left( \sum_{g,h,k\in [G]^3, g \neq h \neq k}  \tilde V_{[h]}^\top P_{[h,g]} \tilde e_{[g]} \tilde V_{[k]}^\top P_{[k,g]} \tilde e_{[g]} \right).
\end{align}
For the first term on the RHS of \eqref{eq:Vpe_var}, we have
\begin{align*}
    &\quad \frac{1}{K^2} \mathbb V \left( \sum_{g,h\in [G]^2, g \neq h}  \left(\tilde V_{[h]}^\top P_{[h,g]} \tilde e_{[g]} \right)^2 \right) \\
    &= \frac{1}{K^2}  \mathbb E \left( \sum_{g,h\in [G]^2, g \neq h}  \left(\tilde V_{[h]}^\top P_{[h,g]} \tilde e_{[g]} \right)^2 - \mathbb E \left(\tilde V_{[h]}^\top P_{[h,g]} \tilde e_{[g]} \right)^2 \right)^2 \\
    &\leqslant \frac{C}{K^2} \mathbb E \left( \sum_{g,h\in [G]^2, g \neq h} \tr \left( P_{[g,h]} \left( \tilde V_{[h]} \tilde V_{[h]}^\top - \Omega_h^{\tilde V, \tilde V} \right)  P_{[h,g]} \left( \tilde e_{[g]} \tilde e_{[g]}^\top - \Omega_g^{\tilde e, \tilde e} \right) \right) \right)^2 \\
    &+ \frac{C}{K^2} \mathbb E \left( \sum_{g,h\in [G]^2, g \neq h} \tr \left( P_{[g,h]} \Omega_h^{\tilde V, \tilde V}  P_{[h,g]} \left( \tilde e_{[g]} \tilde e_{[g]}^\top - \Omega_g^{\tilde e, \tilde e} \right) \right) \right)^2 \\
    &+ \frac{C}{K^2} \mathbb E \left( \sum_{g,h\in [G]^2, g \neq h} \tr \left( \left( \tilde V_{[h]} \tilde V_{[h]}^\top - \Omega_h^{\tilde V, \tilde V} \right)  P_{[h,g]} \Omega_g^{\tilde e, \tilde e} P_{[g,h]}  \right)  \right)^2.
\end{align*}
In addition, we have
\begin{align*}
    &\quad \frac{1}{K^2} \mathbb E \left( \sum_{g,h\in [G]^2, g \neq h} \tr \left( P_{[g,h]} \left( \tilde V_{[h]} \tilde V_{[h]}^\top - \Omega_h^{\tilde V, \tilde V} \right)  P_{[h,g]} \left( \tilde e_{[g]} \tilde e_{[g]}^\top - \Omega_g^{\tilde e, \tilde e} \right) \right) \right)^2 \\
    &\leqslant \frac{C}{K^2} \sum_{g,h\in [G]^2, g \neq h}  \mathbb E \left( \tr \left( P_{[g,h]} \left( \tilde V_{[h]} \tilde V_{[h]}^\top - \Omega_h^{\tilde V, \tilde V} \right)  P_{[h,g]} \left( \tilde e_{[g]} \tilde e_{[g]}^\top - \Omega_g^{\tilde e, \tilde e} \right) \right) \right)^2 \\
    & \leqslant \frac{C}{K^2} \sum_{g,h\in [G]^2, g \neq h} \tr \left( P_{[g,h]} \mathbb E \left( \tilde V_{[h]} \tilde V_{[h]}^\top - \Omega_h^{\tilde V, \tilde V} \right)^2  P_{[h,g]} \right) \\
    &\qquad \qquad \qquad \quad \times \tr \left( P_{[h,g]} \mathbb E \left( \tilde e_{[g]} \tilde e_{[g]}^\top - \Omega_g^{\tilde e, \tilde e} \right)^2  P_{[g,h]} \right) \\
    &\leqslant \frac{C}{K^2} \sum_{g,h\in [G]^2, g \neq h} \tr \left( P_{[g,h]} P_{[h,g]} \right) \\
    &\leqslant \frac{C}{K^2} \tr \left((P - \bar P)^2 \right) \\
    &= o(1), 
\end{align*}
where the second inequality is by the trace Cauchy-Schwartz inequality.  Similarly, we have
\begin{align*}
    &\quad \frac{1}{K^2} \mathbb E \left( \sum_{g,h\in [G]^2, g \neq h} \tr \left( P_{[g,h]} \Omega_h^{\tilde V, \tilde V}  P_{[h,g]} \left( \tilde e_{[g]} \tilde e_{[g]}^\top - \Omega_g^{\tilde e, \tilde e} \right) \right) \right)^2 \\
    &= \frac{1}{K^2} \mathbb E \left( \sum_{g\in [G]} \tr \left( \left( \sum_{h \neq g} P_{[g,h]} \Omega_h^{\tilde V, \tilde V}  P_{[h,g]}\right) \left( \tilde e_{[g]} \tilde e_{[g]}^\top - \Omega_g^{\tilde e, \tilde e} \right) \right) \right)^2 \\
    &=\frac{1}{K^2} \sum_{g\in [G]} \mathbb E \left( \tr \left( \left( \sum_{h \neq g} P_{[g,h]} \Omega_h^{\tilde V, \tilde V}  P_{[h,g]}\right) \left( \tilde e_{[g]} \tilde e_{[g]}^\top - \Omega_g^{\tilde e, \tilde e} \right) \right) \right)^2 \\
    &\leqslant \frac{1}{K^2} \sum_{g\in [G]} \tr \left( \left( \sum_{h \neq g} P_{[g,h]} \Omega_h^{\tilde V, \tilde V}  P_{[h,g]}\right)^2 \right) \tr \left( \mathbb E \left( \tilde e_{[g]} \tilde e_{[g]}^\top - \Omega_g^{\tilde e, \tilde e} \right)^2 \right) \\
    &\leqslant \frac{C}{K^2} \sum_{g\in [G]} \tr \left( \sum_{h \neq g} P_{[g,h]} \Omega_h^{\tilde V, \tilde V}  P_{[h,g]} \right) \\
    &\leqslant \frac{C}{K^2} \sum_{g,h \in [G]^2, g \neq h} \tr \left(P_{[g,h]} P_{[h,g]} \right) \\
    &= o(1),
\end{align*}
where the first inequality is by the trace Cauchy-Schwartz inequality, and, following the same argument,  
\begin{align*}
\frac{1}{K^2} \mathbb E \left( \sum_{g,h\in [G]^2, g \neq h} \tr \left( \left( \tilde V_{[h]} \tilde V_{[h]}^\top - \Omega_h^{\tilde V, \tilde V} \right)  P_{[h,g]} \Omega_g^{\tilde e, \tilde e} P_{[g,h]}  \right)  \right)^2 = o(1).
\end{align*}
Combining these bounds with \eqref{eq:Vpe_var}, we have
\begin{align}\label{eq:Vpe_var1}
\frac{1}{K^2} \mathbb V \left( \sum_{g,h\in [G]^2, g \neq h}  \left(\tilde V_{[h]}^\top P_{[h,g]} \tilde e_{[g]} \right)^2 \right) = o(1).
\end{align}

Now consider the second term on the RHS of \eqref{eq:Vpe_var}. We have
\begin{align*}
    &\quad \frac{1}{K^2} \mathbb V \left( \sum_{g,h,k\in [G]^3, g \neq h \neq k}  \tilde V_{[h]}^\top P_{[h,g]} \tilde e_{[g]} \tilde V_{[k]}^\top P_{[k,g]} \tilde e_{[g]} \right) \\
    &= \frac{1}{K^2} \mathbb E \left( \sum_{g,h,k\in [G]^3, g \neq h \neq k}  \tilde V_{[h]}^\top P_{[h,g]} \tilde e_{[g]} \tilde V_{[k]}^\top P_{[k,g]} \tilde e_{[g]} \right)^2 \\
    &\leqslant \frac{C}{K^2} \mathbb E \left( \sum_{g,h,k\in [G]^3, g \neq h \neq k}  \tilde V_{[h]}^\top P_{[h,g]} \left( \tilde e_{[g]} \tilde e_{[g]}^\top - \Omega_g^{\tilde e, \tilde e}  \right) P_{[g,k]} \tilde V_{[k]}  \right)^2 \\
    &+ \frac{C}{K^2} \mathbb E \left( \sum_{g,h,k\in [G]^3, g \neq h \neq k}  \tilde V_{[h]}^\top P_{[h,g]} \Omega_g^{\tilde e, \tilde e} P_{[g,k]} \tilde V_{[k]}  \right)^2, 
\end{align*}
where
\begin{align*}
    &\quad\frac{1}{K^2} \mathbb E \left( \sum_{g,h,k\in [G]^3, g \neq h \neq k}  \tilde V_{[h]}^\top P_{[h,g]} \left( \tilde e_{[g]} \tilde e_{[g]}^\top - \Omega_g^{\tilde e, \tilde e}  \right) P_{[g,k]} \tilde V_{[k]}  \right)^2 \\
    &\leqslant \frac{C}{K^2} \sum_{g,h,k\in [G]^3, g \neq h \neq k} \mathbb E \left( \tilde V_{[h]}^\top P_{[h,g]} \left( \tilde e_{[g]} \tilde e_{[g]}^\top - \Omega_g^{\tilde e, \tilde e}  \right) P_{[g,k]} \tilde V_{[k]}  \right)^2 \\
    &\leqslant \frac{C}{K^2} \sum_{g,h\in [G]^2, g \neq h} \tr \left( P_{[g,h]} P_{[h,g]} \right) \\
    &= o(1),
\end{align*}
and
\begin{align*}
    &\quad \frac{1}{K^2} \mathbb E \left( \sum_{g,h,k\in [G]^3, g \neq h \neq k}  \tilde V_{[h]}^\top P_{[h,g]} \Omega_g^{\tilde e, \tilde e} P_{[g,k]} \tilde V_{[k]}  \right)^2 \\
    &= \frac{1}{K^2} \mathbb E \left( \sum_{h,k\in [G]^2, h \neq k}  \tilde V_{[h]}^\top \left( \sum_{g \neq h \neq k} P_{[h,g]} \Omega_g^{\tilde e, \tilde e} P_{[g,k]} \right) \tilde V_{[k]}  \right)^2 \\
    &\leqslant \frac{C}{K^2}\sum_{h,k\in [G]^2, h \neq k}  \mathbb E \left(  \tilde V_{[h]}^\top \left( \sum_{g \neq h \neq k} P_{[h,g]} \Omega_g^{\tilde e, \tilde e} P_{[g,k]} \right) \tilde V_{[k]}  \right)^2 \\
    &\leqslant \frac{C}{K^2}\sum_{h,k\in [G]^2, h \neq k} \tr \left( \left( (P - \bar P) \Omega_{\tilde e} (P - \bar P) \right)_{[h,k]} \left( (P - \bar P) \Omega_{\tilde e} (P - \bar P) \right)_{[k,h]} \right) \\
    &\leqslant \frac{C}{K^2} \tr \left( (P - \bar P) \Omega_{\tilde e} (P - \bar P) (P - \bar P) \Omega_{\tilde e} (P - \bar P) \right) \\
    &= o(1).
\end{align*} 
This implies that
\begin{align}\label{eq:Vpe_var2}
\frac{1}{K^2} \mathbb V \left( \sum_{g,h,k\in [G]^3, g \neq h \neq k}  \tilde V_{[h]}^\top P_{[h,g]} \tilde e_{[g]} \tilde V_{[k]}^\top P_{[k,g]} \tilde e_{[g]} \right) = o(1).
\end{align}
Combining \eqref{eq:Vpe_var}--\eqref{eq:Vpe_var2}, we have established \eqref{eq:Vpe2}, which further implies the desired result that 
\begin{align*}
\frac{1}{K} \sum_{g \in [G]} \left( \sum_{h \in [G], h \neq g} V_{[h]}^{\top} P_{[h,g]} e_{[g]} \right)^2 = \frac{1}{K} \sum_{g \in [G]} \mathbb E \left( \sum_{h \in [G], h \neq g} \tilde V_{[h]}^{\top} P_{[h,g]} \tilde e_{[g]} \right)^2 + o_P(1).
\end{align*}
Note that, by Lemma \ref{lem:PandQ} and the fact that $W^\top Q = 0$, we can show 
\begin{align*}
\frac{1}{K} \sum_{g \in [G]} \left( \sum_{h \in [G], h \neq g} V_{[h]}^{\top} Q_{[h,g]} e_{[g]} \right)^2 = \frac{1}{K} \sum_{g \in [G]} \mathbb E \left( \sum_{h \in [G], h \neq g} \tilde V_{[h]}^{\top} Q_{[h,g]} \tilde e_{[g]} \right)^2 + o_P(1)    
\end{align*}
in the same manner by replacing $P$ by $Q$. 

Next, we note that
\begin{align*}
    &\quad \frac{1}{K} \sum_{g,h \in [G]^2, g \neq h}\left( V_{[h]}^\top P_{[h,g]} e_{[g]} \right) \left( V_{[g]}^\top P_{[g,h]} e_{[h]} \right) \\
    &= \frac{1}{K} \sum_{g,h \in [G]^2, g \neq h}\left(\underbrace{\tilde V_{[h]}^\top P_{[h,g]} \tilde e_{[g]}}_{U_{gh}^{(1)}} -  \underbrace{(W_{[h]} \hat \gamma_{\tilde V})^\top P_{[h,g]} \tilde e_{[g]}}_{U_{gh}^{(2)}} - \underbrace{\tilde V_{[h]}^\top P_{[h,g]} (W_{[g]} \hat \gamma_{\tilde e})}_{U_{gh}^{(3)}} + \underbrace{(W_{[h]} \hat \gamma_{\tilde V})^\top P_{[h,g]} (W_{[g]} \hat \gamma_{\tilde e})}_{U_{gh}^{(4)}} \right) \\
    &\qquad \qquad \qquad \times \left(\underbrace{\tilde V_{[g]}^\top P_{[g,h]} \tilde e_{[h]}}_{U_{hg}^{(1)}} -  \underbrace{(W_{[g]} \hat \gamma_{\tilde V})^\top P_{[g,h]} \tilde e_{[h]}}_{U_{hg}^{(2)}} - \underbrace{\tilde V_{[g]}^\top P_{[g,h]} (W_{[h]} \hat \gamma_{\tilde e})}_{U_{hg}^{(3)}} + \underbrace{(W_{[g]} \hat \gamma_{\tilde V})^\top P_{[g,h]} (W_{[h]} \hat \gamma_{\tilde e})}_{U_{hg}^{(4)}} \right),
\end{align*}
and by using a similar argument as in the proof for (\ref{eq:Vpe1}), we have
\begin{align*}
    \frac{1}{K} \sum_{g,h \in [G]^2, g \neq h} U_{gh}^{(1),2} &= \frac{1}{K} \sum_{g,h \in [G]^2, g \neq h} U_{hg}^{(1),2} = O_P(1), \\
    \frac{1}{K} \sum_{g,h \in [G]^2, g \neq h} U_{gh}^{(2),2} &= \frac{1}{K} \sum_{g,h \in [G]^2, g \neq h} U_{hg}^{(2),2} = o_P(1), \\
    \frac{1}{K} \sum_{g,h \in [G]^2, g \neq h} U_{gh}^{(3),2} &= \frac{1}{K} \sum_{g,h \in [G]^2, g \neq h} U_{hg}^{(3),2} = o_P(1), \\
    \frac{1}{K} \sum_{g,h \in [G]^2, g \neq h} U_{gh}^{(4),2} &= \frac{1}{K} \sum_{g,h \in [G]^2, g \neq h} U_{hg}^{(4),2} = o_P(1).
\end{align*}
Then by repeatedly applying Cauchy-Schwarz inequality, we have
\begin{align*}
    \left| \frac{1}{K} \sum_{g,h \in [G]^2, g \neq h} U_{gh}^{(s_1)} U_{hg}^{(s_2)} \right| \leqslant \left(\frac{1}{K} \sum_{g,h \in [G]^2, g \neq h} U_{gh}^{(s_1),2} \right)^{1/2} \left(\frac{1}{K} \sum_{g,h \in [G]^2, g \neq h} U_{gh}^{(s_2),2} \right)^{1/2} = o_P(1)
\end{align*}
for $s_1 \neq 1$ or $s_2 \neq 1$, whence
\begin{align*}
    &\quad \frac{1}{K} \sum_{g,h \in [G]^2, g \neq h}\left( V_{[h]}^\top P_{[h,g]} e_{[g]} \right) \left( V_{[g]}^\top P_{[g,h]} e_{[h]} \right) \\
    &= \frac{1}{K} \sum_{g,h \in [G]^2, g \neq h}\left(\tilde V_{[h]}^\top P_{[h,g]} \tilde e_{[g]} \right) \left( \tilde V_{[g]}^\top P_{[g,h]} \tilde e_{[h]} \right) + o_P(1).
\end{align*}

Next, we note that
\begin{align*}
    &\quad \mathbb V \left(\frac{1}{K} \sum_{g,h \in [G]^2, g \neq h} \left( \tilde V_{[h]}^\top P_{[h,g]} \tilde e_{[g]} \right)\left( \tilde V_{[g]}^\top P_{[g,h]} \tilde e_{[h]} \right)  \right) \\
    &= \frac{1}{K^2} \mathbb E \left( \sum_{g,h \in [G]^2, g \neq h}\left( \tilde V_{[h]}^\top P_{[h,g]} \tilde e_{[g]} \right) \left( \tilde V_{[g]}^\top P_{[g,h]} \tilde e_{[h]} \right) \right.\\
    & \left. \qquad \qquad \quad-\sum_{g,h \in [G]^2, g \neq h} \mathbb E \left( \tilde V_{[h]}^\top P_{[h,g]} \tilde e_{[g]} \right) \left( \tilde V_{[g]}^\top P_{[g,h]} \tilde e_{[h]} \right) \right)^2 \\
    &\leqslant \frac{C}{K^2} \mathbb E \left( \sum_{g,h \in [G]^2, g \neq h}  \tr \left(P_{[g,h]} \left(\tilde e_{[h]} \tilde V_{[h]}^\top - \Omega_h^{\tilde e, \tilde V} \right) P_{[h,g]} \left(\tilde e_{[g]} \tilde V_{[g]}^\top - \Omega_g^{\tilde e, \tilde V} \right) \right) \right)^2 \\
    &+ \frac{C}{K^2} \mathbb E \left( \sum_{g,h \in [G]^2, g \neq h}  \tr \left(P_{[g,h]} \Omega_h^{\tilde e, \tilde V} P_{[h,g]} \left(\tilde e_{[g]} \tilde V_{[g]}^\top - \Omega_g^{\tilde e, \tilde V} \right) \right) \right)^2 \\
    &+ \frac{C}{K^2} \mathbb E \left( \sum_{g,h \in [G]^2, g \neq h}  \tr \left( \left(\tilde e_{[h]} \tilde V_{[h]}^\top - \Omega_h^{\tilde e, \tilde V} \right) P_{[h,g]} \Omega_g^{\tilde e, \tilde V} P_{[g,h]} \right) \right)^2 \\
    & = o(1), 
\end{align*}
where the last equality holds because
\begin{align*}
    &\quad \frac{1}{K^2} \mathbb E \left( \sum_{g,h \in [G]^2, g \neq h}  \tr \left(P_{[g,h]} \left(\tilde e_{[h]} \tilde V_{[h]}^\top - \Omega_h^{\tilde e, \tilde V} \right) P_{[h,g]} \left(\tilde e_{[g]} \tilde V_{[g]}^\top - \Omega_g^{\tilde e, \tilde V} \right) \right) \right)^2 \\
    &\leqslant \frac{C}{K^2} \sum_{g,h \in [G]^2, g \neq h} \mathbb E \left( \tr \left(P_{[g,h]} \left(\tilde e_{[h]} \tilde V_{[h]}^\top - \Omega_h^{\tilde e, \tilde V} \right) P_{[h,g]} \left(\tilde e_{[g]} \tilde V_{[g]}^\top - \Omega_g^{\tilde e, \tilde V} \right) \right) \right)^2 \\
    &\leqslant \frac{C}{K^2} \sum_{g,h \in [G]^2, g \neq h} \tr \left(P_{[g,h]} \mathbb E \left(\tilde e_{[h]} \tilde V_{[h]}^\top - \Omega_h^{\tilde e, \tilde V} \right)  \left(\tilde V_{[h]} \tilde e_{[h]}^\top - \Omega_h^{\tilde V, \tilde e} \right) P_{[h,g]} \right) \\
    &\qquad\qquad\qquad\quad\times \tr \left(P_{[h,g]} \mathbb E \left(\tilde e_{[g]} \tilde V_{[g]}^\top - \Omega_g^{\tilde e, \tilde V} \right)  \left(\tilde V_{[g]} \tilde e_{[g]}^\top - \Omega_g^{\tilde V, \tilde e} \right) P_{[g,h]} \right) \\
    &\leqslant \frac{C}{K^2} \sum_{g,h \in [G]^2, g \neq h} \tr \left( P_{[g,h]} P_{[h,g]} \right) \\
    &= o(1),
\end{align*}
and 
\begin{align*}
    &\quad \frac{1}{K^2} \mathbb E \left( \sum_{g,h \in [G]^2, g \neq h}  \tr \left(P_{[g,h]} \Omega_h^{\tilde e, \tilde V} P_{[h,g]} \left(\tilde e_{[g]} \tilde V_{[g]}^\top - \Omega_g^{\tilde e, \tilde V} \right) \right) \right)^2 \\
    &= \frac{1}{K^2} \mathbb E \left( \sum_{g \in [G]}  \tr \left( \left( \sum_{h \neq g} P_{[g,h]} \Omega_h^{\tilde e, \tilde V} P_{[h,g]} \right) \left(\tilde e_{[g]} \tilde V_{[g]}^\top - \Omega_g^{\tilde e, \tilde V} \right) \right) \right)^2 \\
    &= \frac{1}{K^2} \sum_{g \in [G]} \mathbb E \left( \tr \left( \left( \sum_{h \neq g} P_{[g,h]} \Omega_h^{\tilde e, \tilde V} P_{[h,g]} \right) \left(\tilde e_{[g]} \tilde V_{[g]}^\top - \Omega_g^{\tilde e, \tilde V} \right) \right) \right)^2 \\
    &\leqslant \frac{1}{K^2} \sum_{g \in [G]} \tr \left( \left( \sum_{h \neq g} P_{[g,h]} \Omega_h^{\tilde e, \tilde V} P_{[h,g]} \right)  \left( \sum_{k \neq g} P_{[g,k]} \Omega_k^{\tilde V, \tilde e} P_{[k,g]} \right) \right) \\
    &\qquad \qquad \quad \times \tr \left( \mathbb E \left(\tilde e_{[g]} \tilde V_{[g]}^\top - \Omega_g^{\tilde e, \tilde V} \right) \left(\tilde V_{[g]} \tilde e_{[g]}^\top - \Omega_g^{\tilde V, \tilde e} \right) \right) \\
    &\leqslant \frac{C}{K^2} \sum_{g,h,k \in [G]^2, h \neq g, k \neq g}  \tr \left( P_{[g,h]} \Omega_h^{\tilde e, \tilde V} P_{[h,g]} P_{[g,k]} \Omega_k^{\tilde V, \tilde e} P_{[k,g]} \right)  \\ 
    &\leqslant \frac{C}{K^2} \sum_{g,h,k \in [G]^2, h \neq g, k \neq g} \left( \tr \left(P_{[k,g]}  P_{[g,h]} \Omega_h^{\tilde e, \tilde V} \Omega_h^{\tilde V, \tilde e} P_{[h,g]} P_{[g,k]} \right) \right)^{1/2} \\
    & \qquad\qquad\qquad\qquad\quad\times \left(\tr \left(P_{[h,g]}  P_{[g,k]} \Omega_k^{\tilde V, \tilde e} \Omega_k^{\tilde e, \tilde V} P_{[k,g]} P_{[g,h]} \right) \right)^{1/2} \\
    &\leqslant \frac{C}{K^2} \sum_{g,h,k \in [G]^2, h \neq g, k \neq g} \tr \left(P_{[k,g]}  P_{[g,h]} P_{[h,g]} P_{[g,k]} \right) \\
    &\leqslant \frac{C}{K^2} \sum_{g,h \in [G]^2, g \neq h} \tr \left( P_{[g,h]} P_{[h,g]} \right) \\
    &= o(1),
\end{align*}
and, following the same argument, 
\begin{align*}
\frac{1}{K^2} \mathbb E \left( \sum_{g,h \in [G]^2, g \neq h}  \tr \left( \left(\tilde e_{[h]} \tilde V_{[h]}^\top - \Omega_h^{\tilde e, \tilde V} \right) P_{[h,g]} \Omega_g^{\tilde e, \tilde V} P_{[g,h]} \right) \right)^2 = o(1).
\end{align*}
This implies that
\begin{align*}
    &\quad \frac{1}{K} \sum_{g,h \in [G]^2, g \neq h} \left( \tilde V_{[h]}^\top P_{[h,g]} \tilde e_{[g]} \right)\left( \tilde V_{[g]}^\top P_{[g,h]} \tilde e_{[h]} \right) \\
    &= \frac{1}{K} \sum_{g,h \in [G]^2, g \neq h} \mathbb E \left( \tilde V_{[h]}^\top P_{[h,g]} \tilde e_{[g]} \right)\left( \tilde V_{[g]}^\top P_{[g,h]} \tilde e_{[h]} \right) + o_P(1).
\end{align*}
Note also that we can replace $P$ by $Q$, as in the proof for (\ref{eq:Vpe1}) and (\ref{eq:Vpe2}), and this concludes the proof for the first result.

For the second result, we focus on the case when $\tilde u = \tilde V$ and $\tilde v = \tilde e$. Recall that
\begin{align*}
    Q = M_W (P - \bar P)M_W = P - \bar P + P_W \bar P + \bar P P_W - P_W \bar P P_W,
\end{align*}
which implies that
\begin{align*}
    Q_{[h,g]} = P_{[h,g]} + P_{W,[h,g]} P_{[g,g]} + P_{[h,h]} P_{W,[h,g]} - \sum_{k \in [G]} P_{W,[h,k]} P_{[k,k]} P_{W,[k,g]}, \quad g \neq h.
\end{align*}
Therefore, we have
\begin{align*}
    &\quad \frac{1}{K} \sum_{g,h \in [G]^2, g \neq h} \mathbb E \left( \tilde V_{[h]}^{\top} Q_{[h,g]} \tilde e_{[g]} \right)^2 \\
    &= \frac{1}{K} \sum_{g,h \in [G]^2, g \neq h} \mathbb E \left( \tilde V_{[h]}^{\top} \left( P_{[h,g]} + P_{W,[h,g]} P_{[g,g]} + P_{[h,h]} P_{W,[h,g]} - \sum_{k \in [G]} P_{W,[h,k]} P_{[k,k]} P_{W,[k,g]} \right) \tilde e_{[g]} \right)^2,
\end{align*}
where
\begin{align*}
    \frac{1}{K} \sum_{g,h \in [G]^2, g \neq h} \mathbb E \left( \tilde V_{[h]}^\top P_{W, [h,g]} P_{[g,g]} \tilde e_{[g]} \right)^2 \leqslant \frac{C}{K} \sum_{g,h \in [G]^2, g \neq h} \tr \left( P_{W, [g,h]} P_{W, [h,g]} \right) = o(1),
\end{align*}
since $d_w$ is fixed, and
\begin{align*}
    \frac{1}{K} \sum_{g,h \in [G]^2, g \neq h} \mathbb E \left( \tilde V_{[g]}^{\top}  P_{[g,g]} P_{W,[g,h]}  \tilde e_{[g]} \right)^2 = o(1)
\end{align*}
by the same argument as above, and
\begin{align*}
    &\quad \frac{1}{K} \sum_{g,h \in [G]^2, g \neq h} \mathbb E \left(\tilde V_{[h]}^\top \left(\sum_{k \in [G]} P_{W,[h,k]} P_{[k,k]} P_{W,[k,g]} \right) \tilde e_{[g]} \right)^2 \\
    &\leqslant \frac{C}{K} \sum_{g,h \in [G]^2, g \neq h} \tr \left( \left(P_W \bar P P_W\right)_{[g,h]} \left(P_W \bar P P_W \right)_{[h,g]} \right) \\
    &\leqslant \frac{C}{K} \tr \left( P_W \bar P P_W P_W \bar P P_W \right) \\
    &= o(1).
\end{align*}
It follows that
\begin{align*}
    \frac{1}{K} \sum_{g \in [G]} \mathbb E \left( \sum_{h \in [G], h \neq g} \tilde V_{[h]}^{\top} Q_{[h,g]} \tilde e_{[g]} \right)^2 = \frac{1}{K} \sum_{g \in [G]} \mathbb E \left( \sum_{h \in [G], h \neq g} \tilde V_{[h]}^{\top} P_{[h,g]} \tilde e_{[g]} \right)^2 + o(1).
\end{align*}
Similarly, we can show that 
\begin{align*}
& \quad \frac{1}{K} \sum_{g, h \in [G]^2, g \neq h} \mathbb E \left( \tilde V_{[h]}^{\top} P_{[h,g]} \tilde e_{[g]} \right)\left( \tilde V_{[g]}^{\top} P_{[g,h]} \tilde e_{[h]} \right) \\
& = \frac{1}{K} \sum_{g, h \in [G]^2, g \neq h} \mathbb E \left( \tilde V_{[h]}^{\top} Q_{[h,g]} \tilde e_{[g]} \right) \left( \tilde V_{[g]}^{\top} Q_{[g,h]} \tilde e_{[h]} \right) + o(1). 
\end{align*}
This concludes the proof. $\hfill\qedsymbol$

\subsection{Proof of Lemma \ref{lem:linear_quad_form_1}}

We prove each result in turn. To begin with, we note that
\begin{align*}
    \Sigma = \mathbb V \left( \sum_{g \in [G]} \hat \Pi_{[g]}^\top \tilde e_{[g]}\right) + \mathbb V \left( \sum_{g,h \in [G]^2, g \neq h} \tilde V_{[g]}^\top P_{[g,h]} \tilde e_{[h]} \right),
\end{align*}
where
\begin{align*}
\mathbb V \left( \sum_{g \in [G]} \hat \Pi_{[g]}^\top \tilde e_{[g]}\right) = \mathbb E \left(\hat \Pi^\top \tilde e \right)^2 & \geqslant \frac{1}{C} \hat \Pi^\top \hat \Pi \geqslant \frac{1}{C} \Pi^\top \Pi,
\end{align*}
and 
\begin{align*}
    &\quad \mathbb E \left(\sum_{g,h \in [G]^2, g \neq h} \tilde V_{[g]}^\top P_{[g,h]} \tilde e_{[h]} \right)^2 \\
    &=\sum_{g,h \in [G]^2, g \neq h} \mathbb E \left(\tilde V_{[g]}^\top P_{[g,h]} \tilde e_{[h]} \right)^2 + \sum_{g,h \in [G]^2, g \neq h} \mathbb E \left(\tilde V_{[g]}^\top P_{[g,h]} \tilde e_{[h]} \right) \left(\tilde V_{[h]}^\top P_{[h,g]} \tilde e_{[g]} \right) \\
    &= \frac{1}{2} \sum_{g,h \in [G]^2, g \neq h} \mathbb E \left[\left(\tilde V_{[h]}^\top P_{[h,g]} \tilde e_{[g]} \right) + \left(\tilde V_{[g]}^\top P_{[g,h]} \tilde e_{[h]} \right) \right]^2 \\
    &= \frac{1}{2} \sum_{g,h \in [G]^2, g \neq h} \mathbb E \left[\begin{pmatrix}
        \tilde e_{[g]} & \tilde V_{[g]} 
    \end{pmatrix} \begin{pmatrix}
        0_{n_g \times n_h} & P_{[g,h]} \\ P_{[g,h]} & 0_{n_g \times n_h} 
    \end{pmatrix} \begin{pmatrix}
        \tilde e_{[h]} \\ \tilde V_{[h]} 
    \end{pmatrix} \right]^2 \\
    &\geqslant \frac{1}{2C} \sum_{g,h \in [G]^2, g \neq h} \tr \left[\begin{pmatrix}
        0_{n_g \times n_h} & P_{[g,h]} \\ P_{[g,h]} & 0_{n_g \times n_h} 
    \end{pmatrix} \begin{pmatrix}
        0_{n_h \times n_g} & P_{[h,g]} \\ P_{[h,g]} & 0_{n_h \times n_g} 
    \end{pmatrix} \right] \\
    &= \frac{1}{C} \sum_{g,h \in [G]^2, g \neq h} \tr \left[P_{[g,h]} P_{[h,g]} \right] \\
    &= \frac{1}{C} \sum_{g \in [G]} \tr \left[P_{[g,g]} - P_{[g,g]}^2 \right] \\
    &\geqslant \frac{1}{C} \sum_{g \in [G]} \left(1 - \lambda_{\max}(P_{[g,g]}) \right) \tr \left[P_{[g,g]}\right] \\
    &\geqslant \frac{1}{C} K.
\end{align*}
These two lower bounds imply the desired result that 
\begin{align*}
\Sigma \geq (\Pi^\top \Pi + K) / C. 
\end{align*}

Next, we note that
\begin{align}\label{eq:XtildeQetilde}
    &\quad \frac{1}{\Sigma} \sum_{g\in [G]} \left(\sum_{h \in [G], h \neq g} \tilde X_{[h]}^\top Q_{[h,g]} \tilde e_{[g]} \right)^2 \notag \\
    &= \frac{1}{\Sigma} \sum_{g\in [G]} \left(\sum_{h \in [G], h \neq g} \tilde \Pi_{[h]}^\top Q_{[h,g]} \tilde e_{[g]} \right)^2 \notag  \\
    &+ \frac{1}{\Sigma} \sum_{g\in [G]} \left(\sum_{h \in [G], h \neq g} \tilde V_{[h]}^\top Q_{[h,g]} \tilde e_{[g]} \right)^2 \notag  \\
    &+ \frac{2}{\Sigma} \sum_{g\in [G]} \left(\sum_{h \in [G], h \neq g} \tilde V_{[h]}^\top Q_{[h,g]} \tilde e_{[g]}\right) \left(\sum_{k \in [G], k \neq g} \tilde \Pi_{[k]}^\top Q_{[k,g]} \tilde e_{[g]} \right).
\end{align}
For the first term on the RHS of \eqref{eq:XtildeQetilde}, we have
\begin{align*}
    \mathbb E \frac{1}{\Sigma} \left(\sum_{g\in [G]} \left(\sum_{h \in [G], h \neq g} \tilde \Pi_{[h]}^\top Q_{[h,g]} \tilde e_{[g]} \right)^2 -  \sum_{g\in [G]} \mathbb E \left(\sum_{h \in [G], h \neq g} \tilde \Pi_{[h]}^\top Q_{[h,g]} \tilde e_{[g]} \right)^2 \right) = 0,
\end{align*}
and by Assumption \ref{ass:high}
\begin{align*}
    &\quad \mathbb V \left(\frac{1}{\Sigma}\sum_{g\in [G]} \left(\sum_{h \in [G], h \neq g} \tilde \Pi_{[h]}^\top Q_{[h,g]} \tilde e_{[g]} \right)^2 \right) \\
    &= \frac{1}{\Sigma^2} \mathbb V \left( \sum_{g\in [G]} \left(\bar \Pi_{[g]}^\top \tilde e_{[g]} \right)^2 \right) \\
    &\leqslant \frac{1}{\Sigma^2} \sum_{g\in [G]} \mathbb E \left(\bar \Pi_{[g]}^\top \tilde e_{[g]} \right)^4 \\
    &\leqslant \frac{C}{\Sigma^2} \sum_{g\in [G]} \left(\bar \Pi_{[g]}^\top \bar \Pi_{[g]} \right)^2 \\
    &\leqslant \frac{C\max_{1 \leq g \leq G} \left\Vert\bar \Pi_{[g]} \right\Vert_2^2 \bar \Pi^\top \bar \Pi}{\left( \Pi^\top \Pi +K \right)^2} \\
    & = o(1).
\end{align*}
Therefore, we have
\begin{align}\label{eq:XtildeQetilde1}
    \frac{1}{\Sigma} \sum_{g\in [G]} \left(\sum_{h \in [G], h \neq g} \tilde \Pi_{[h]}^\top Q_{[h,g]} \tilde e_{[g]} \right)^2 = \frac{1}{\Sigma} \sum_{g\in [G]} \mathbb E\left(\sum_{h \in [G], h \neq g} \tilde \Pi_{[h]}^\top Q_{[h,g]} \tilde e_{[g]} \right)^2 + o_P(1).
\end{align}
For the second term on the RHS of \eqref{eq:XtildeQetilde}, by Lemma \ref{lem:quad_form_2} and the fact that $K/\Sigma = O(1)$, we have
\begin{align}\label{eq:XtildeQetilde2}
    \frac{1}{\Sigma} \sum_{g\in [G]} \left(\sum_{h \in [G], h \neq g} \tilde V_{[h]}^\top Q_{[h,g]} \tilde e_{[g]} \right)^2 = \frac{1}{\Sigma} \sum_{g\in [G]} \mathbb E \left(\sum_{h \in [G], h \neq g} \tilde V_{[h]}^\top Q_{[h,g]} \tilde e_{[g]} \right)^2 + o_P(1).
\end{align}
For the last term on the RHS of \eqref{eq:XtildeQetilde}, we have
\begin{align*}
    \mathbb E \frac{1}{\Sigma} \sum_{g\in [G]} \left(\sum_{k \in [G], k \neq g} \tilde \Pi_{[k]}^\top Q_{[k,g]} \tilde e_{[g]} \right)\left(\sum_{h \in [G], h \neq g} \tilde V_{[h]}^\top Q_{[h,g]} \tilde e_{[g]}\right) = 0,
\end{align*}
and 
\begin{align*}
    &\quad \mathbb V \left(\frac{1}{\Sigma} \sum_{g\in [G]} \left(\sum_{k \in [G], k \neq g} \tilde \Pi_{[k]}^\top Q_{[k,g]} \tilde e_{[g]} \right) \left(\sum_{h \in [G], h \neq g} \tilde V_{[h]}^\top Q_{[h,g]} \tilde e_{[g]} \right) \right) \\
    &= \frac{1}{\Sigma^2} \mathbb E \left(\sum_{g,h\in [G]^2, g \neq h} \bar \Pi_{[g]}^\top \tilde e_{[g]} \tilde V_{[h]}^\top Q_{[h,g]} \tilde e_{[g]}\right)^2 \\
    &\leqslant \frac{C}{\Sigma^2} \mathbb E \left(\sum_{g,h\in [G]^2, g \neq h} \bar \Pi_{[g]}^\top \left( \tilde e_{[g]} \tilde e_{[g]}^\top - \Omega_g^{\tilde e, \tilde e} \right) Q_{[g,h]} \tilde V_{[h]}  \right)^2 \\
    &+ \frac{C}{\Sigma^2} \mathbb E \left(\sum_{g,h\in [G]^2, g \neq h} \bar \Pi_{[g]}^\top \Omega_g^{\tilde e, \tilde e} Q_{[g,h]} \tilde V_{[h]}\right)^2, 
\end{align*}
where
\begin{align*}
    &\quad \frac{1}{\Sigma^2} \mathbb E \left(\sum_{g,h\in [G]^2, g \neq h} \bar \Pi_{[g]}^\top \left( \tilde e_{[g]} \tilde e_{[g]}^\top - \Omega_g^{\tilde e, \tilde e} \right) Q_{[g,h]} \tilde V_{[h]}  \right)^2 \\
    &\leqslant \frac{C}{\Sigma^2}\sum_{g,h\in [G]^2, g \neq h}  \mathbb E \left( \bar \Pi_{[g]}^\top \left( \tilde e_{[g]} \tilde e_{[g]}^\top - \Omega_g^{\tilde e, \tilde e} \right) Q_{[g,h]} \tilde V_{[h]}  \right)^2 \\
    &\leqslant \frac{C\max_{1\leq g\leq G} \left\Vert \bar \Pi_{[g]} \right\Vert_2^2}{\Sigma^2} \sum_{g,h\in [G]^2, g \neq h} \tr \left( Q_{[g,h]} Q_{[h,g]} \right) \\
    &= o(1),
\end{align*}
and
\begin{align*}
    &\quad \frac{1}{\Sigma^2} \mathbb E \left(\sum_{g,h\in [G]^2, g \neq h} \bar \Pi_{[g]}^\top \Omega_g^{\tilde e, \tilde e} Q_{[g,h]} \tilde V_{[h]} \right)^2 \\
    &= \frac{1}{\Sigma^2} \mathbb E \left( \bar \Pi^\top \Omega_{\tilde e} (Q - \bar Q) \tilde V \right)^2 \\
    &\leqslant \frac{C \bar \Pi^\top \bar \Pi}{\left( \Pi^\top \Pi + K \right)^2} \\
    & = o(1).
\end{align*}
Therefore, we have
\begin{align}\label{eq:XtildeQetilde3}
\frac{2}{\Sigma} \sum_{g\in [G]} \left(\sum_{h \in [G], h \neq g} \tilde V_{[h]}^\top Q_{[h,g]} \tilde e_{[g]}\right) \left(\sum_{k \in [G], k \neq g} \tilde \Pi_{[k]}^\top Q_{[k,g]} \tilde e_{[g]} \right) = o_P(1).
\end{align}
Combining \eqref{eq:XtildeQetilde}--\eqref{eq:XtildeQetilde3}, we have the desired result that
\begin{align*}
    &\quad \frac{1}{\Sigma} \sum_{g\in [G]} \left(\sum_{h \in [G], h \neq g} \tilde X_{[h]}^\top Q_{[h,g]} \tilde e_{[g]} \right)^2 \\
    &= \frac{1}{\Sigma} \sum_{g\in [G]} \left(\sum_{h \in [G], h \neq g} \tilde \Pi_{[h]}^\top Q_{[h,g]} \tilde e_{[g]} \right)^2 + \frac{1}{\Sigma} \sum_{g\in [G]} \left(\sum_{h \in [G], h \neq g} \tilde V_{[h]}^\top Q_{[h,g]} \tilde e_{[g]} \right)^2 + o_P(1) \\
    &= \frac{1}{\Sigma} \sum_{g\in [G]} \mathbb E\left(\sum_{h \in [G], h \neq g} \tilde \Pi_{[h]}^\top Q_{[h,g]} \tilde e_{[g]} \right)^2 + \frac{1}{\Sigma} \sum_{g\in [G]} \mathbb E\left(\sum_{h \in [G], h \neq g} \tilde V_{[h]}^\top Q_{[h,g]} \tilde e_{[g]} \right)^2 + o_P(1) \\
    &= \frac{1}{\Sigma} \sum_{g\in [G]} \mathbb E \left(\sum_{h \in [G], h \neq g} \tilde X_{[h]}^\top Q_{[h,g]} \tilde e_{[g]} \right)^2 + o_P(1).
\end{align*}

Next, we note that
\begin{align}\label{eq:XtildeQe}
    &\quad\frac{1}{\Sigma} \left( \sum_{g\in [G]} \left(\sum_{h \in [G], h \neq g} \tilde X_{[h]}^\top Q_{[h,g]} e_{[g]} \right)^2 -\sum_{g\in [G]} \left(\sum_{h \in [G], h \neq g} \tilde X_{[h]}^\top Q_{[h,g]} \tilde e_{[g]} \right)^2 \right) \notag \\
    &=  \frac{1}{\Sigma} \sum_{g\in [G]} \left(\sum_{h \in [G], h \neq g} \tilde X_{[h]}^\top Q_{[h,g]} W_{[g]} \hat \gamma_{\tilde e} \right)^2 \\
    &  - \frac{2}{\Sigma} \sum_{g\in [G]} \left(\sum_{h \in [G], h \neq g} \tilde X_{[h]}^\top Q_{[h,g]} W_{[g]} \hat \gamma_{\tilde e} \right) \left(\sum_{k \in [G], k \neq g} \tilde X_{[k]}^\top Q_{[k,g]} \tilde e_{[g]} \right).
\end{align}
For the first term on the RHS of \eqref{eq:XtildeQe}, we have
\begin{align*}
    &\quad \frac{1}{\Sigma} \sum_{g\in [G]} \left(\sum_{h \in [G], h \neq g} \tilde X_{[h]}^\top Q_{[h,g]} W_{[g]} \hat \gamma_{\tilde e} \right)^2\\
    &\leqslant \max_{1 \leq g \leq G} \left\Vert W_{[g]} \hat \gamma_{\tilde e} \right\Vert_2^2 \times \frac{1}{\Sigma} \sum_{g\in [G]} \left\Vert \sum_{h \in [G], h \neq g} Q_{[g,h]} \tilde X_{[h]} \right\Vert_2^2 \\
    &\leqslant \max_{1 \leq g \leq G} \left\Vert W_{[g]} \hat \gamma_{\tilde e} \right\Vert_2^2 \times \frac{C}{\Sigma} \sum_{g\in [G]} \left(\left\Vert \bar \Pi_{[g]}\right\Vert_2^2 + \left\Vert \sum_{h \in [G], h \neq g} Q_{[g,h]} \tilde V_{[h]} \right\Vert_2^2 \right) \\
    &= o_P(1),
\end{align*}
by Lemma \ref{lem:4th_moment}. For the second term on the RHS of \eqref{eq:XtildeQe}, we have
\begin{align*}
    &\quad \left| \frac{1}{\Sigma} \sum_{g\in [G]} \left(\sum_{h \in [G], h \neq g} \tilde X_{[h]}^\top Q_{[h,g]} W_{[g]} \hat \gamma_{\tilde e} \right) \left(\sum_{k \in [G], k \neq g} \tilde X_{[k]}^\top Q_{[k,g]} \tilde e_{[g]} \right) \right| \\
    &\leqslant \left(\frac{1}{\Sigma} \sum_{g\in [G]} \left(\sum_{h \in [G], h \neq g} \tilde X_{[h]}^\top Q_{[h,g]} W_{[g]} \hat \gamma_{\tilde e} \right)^2 \right)^{1/2} \times \left(\frac{1}{\Sigma} \sum_{g\in [G]} \left(\sum_{k \in [G], k \neq g} \tilde X_{[k]}^\top Q_{[k,g]} \tilde e_{[g]} \right)^2 \right)^{1/2} \\
    &= o_P(1).
\end{align*}
Combining the two bounds above, we have the desired result that 
\begin{align*}
    \frac{1}{\Sigma} \sum_{g\in [G]} \left(\sum_{h \in [G], h \neq g} \tilde X_{[h]}^\top Q_{[h,g]} e_{[g]} \right)^2 = \frac{1}{\Sigma} \sum_{g\in [G]} \left(\sum_{h \in [G], h \neq g} \tilde X_{[h]}^\top Q_{[h,g]} \tilde e_{[g]} \right)^2 + o_P(1).
\end{align*}

Next, we note that
\begin{align}\label{eq:XtildeQX}
    &\quad \frac{1}{\Sigma} \sum_{g\in [G]} \left(\sum_{h \in [G], h \neq g} \tilde X_{[h]}^\top Q_{[h,g]} X_{[g]} \right)^2 \notag \\
    &= \underbrace{\frac{1}{\Sigma} \sum_{g\in [G]} \left(\bar \Pi_{[g]}^\top X_{[g]} \right)^2}_{R_1}  \notag \\
    &+ \underbrace{\frac{1}{\Sigma} \sum_{g\in [G]} \left(\sum_{h \in [G], h \neq g} \tilde V_{[h]}^\top Q_{[h,g]} X_{[g]} \right)^2}_{R_2} \notag  \\
    &+ 2\times \underbrace{\frac{1}{\Sigma} \sum_{g, h \in [G]^2, g \neq h}  \bar \Pi_{[g]}^\top X_{[g]}  \tilde V_{[h]}^\top Q_{[h,g]} X_{[g]}}_{R_3}. 
\end{align}
For $R_1$, we have
\begin{align*}
    R_1 &= \frac{1}{\Sigma} \sum_{g\in [G]} \left(\bar \Pi_{[g]}^\top \Pi_{[g]} \right)^2 + \frac{1}{\Sigma} \sum_{g\in [G]} \left(\bar \Pi_{[g]}^\top V_{[g]} \right)^2 + \frac{2}{\Sigma} \sum_{g\in [G]} \left(\bar \Pi_{[g]}^\top \Pi_{[g]} \right) \left(\bar \Pi_{[g]}^\top V_{[g]} \right) \\
    &= \frac{1}{\Sigma} \sum_{g\in [G]} \left(\bar \Pi_{[g]}^\top \Pi_{[g]} \right)^2 + \frac{1}{\Sigma} \sum_{g\in [G]} \left(\bar \Pi_{[g]}^\top \tilde V_{[g]} \right)^2 + \frac{2}{\Sigma} \sum_{g\in [G]} \left(\bar \Pi_{[g]}^\top \Pi_{[g]} \right) \left(\bar \Pi_{[g]}^\top \tilde V_{[g]} \right) + o_P(1) \\
    &= \frac{1}{\Sigma} \sum_{g\in [G]} \left(\bar \Pi_{[g]}^\top \Pi_{[g]} \right)^2 + \frac{1}{\Sigma} \sum_{g\in [G]} \mathbb E \left(\bar \Pi_{[g]}^\top \tilde V_{[g]} \right)^2 + o_P(1),
\end{align*}
where the second equality holds by using $V_{[g]} = \tilde V_{[g]} - W_{[g]} \hat \gamma_{\tilde V}$ and Lemma \ref{lem:4th_moment}, and the last equality holds because
\begin{align*}
    &\quad \mathbb V \left( \frac{1}{\Sigma} \sum_{g\in [G]} \left(\bar \Pi_{[g]}^\top \tilde V_{[g]} \right)^2 \right) \\
    &\leq \frac{1}{\Sigma^2} \sum_{g\in [G]} \mathbb E \left(\bar \Pi_{[g]}^\top \tilde V_{[g]} \right)^4 \\
    &\leq \frac{C}{\Sigma^2} \sum_{g\in [G]} \left(\bar \Pi_{[g]}^\top \bar \Pi_{[g]} \right)^2 \\
    &\leq \frac{C \max_{g \in [G]} \left\Vert \bar \Pi_{[g]} \right\Vert_2^2 \bar \Pi^\top \bar \Pi}{(\Pi^\top \Pi + K)^2} \\
    &= o(1),
\end{align*}
and
\begin{align*}
    &\quad \mathbb V \left( \frac{1}{\Sigma} \sum_{g\in [G]} \left(\bar \Pi_{[g]}^\top \Pi_{[g]} \right) \left(\bar \Pi_{[g]}^\top \tilde V_{[g]} \right) \right) \\
    &= \frac{1}{\Sigma^2} \sum_{g\in [G]} \mathbb E \left( \left(\bar \Pi_{[g]}^\top \Pi_{[g]} \right) \left(\bar \Pi_{[g]}^\top \tilde V_{[g]} \right) \right)^2 \\
    &\leq \frac{C}{\Sigma^2} \sum_{g\in [G]} \left(\bar \Pi_{[g]}^\top \Pi_{[g]} \right)^2 \left(\bar \Pi_{[g]}^\top \bar \Pi_{[g]} \right) \\
    &\leq \frac{C \max_{g \in [G]} \left\Vert \bar \Pi_{[g]} \right\Vert_2^2 \bar \Pi^\top \bar \Pi}{(\Pi^\top \Pi + K)^2} \\
    &= o(1).
\end{align*}

For $R_2$, we have
\begin{align*}
    R_2 &= \underbrace{\frac{1}{\Sigma} \sum_{g\in [G]} \left(\sum_{h \in [G], h \neq g} \tilde V_{[h]}^\top Q_{[h,g]} \Pi_{[g]} \right)^2}_{R_{2,1}} \\
    &+ \underbrace{\frac{1}{\Sigma} \sum_{g\in [G]} \left(\sum_{h \in [G], h \neq g} \tilde V_{[h]}^\top Q_{[h,g]} V_{[g]} \right)^2}_{R_{2,2}} \\
    &+ 2 \times \underbrace{\frac{1}{\Sigma} \sum_{g\in [G]} \left(\sum_{h \in [G], h \neq g} \tilde V_{[h]}^\top Q_{[h,g]} \Pi_{[g]} \right)\left(\sum_{k \in [G], k \neq g} \tilde V_{[k]}^\top Q_{[k,g]} V_{[g]} \right)}_{R_{2,3}}.
\end{align*}
For $R_{2,1}$, we have
\begin{align*}
    \mathbb V \left( R_{2,1} \right) &\leqslant \frac{C}{\Sigma^2}  \mathbb V \left( \sum_{g,h \in [G]^2, g \neq h} \left( \tilde V_{[h]}^\top Q_{[h,g]} \Pi_{[g]} \right)^2  \right) \\
    &+ \frac{C}{\Sigma^2}  \mathbb V \left( \sum_{g,h,k \in [G]^3, g \neq h \neq k} \left( \tilde V_{[h]}^\top Q_{[h,g]} \Pi_{[g]} \right) \left( \tilde V_{[k]}^\top Q_{[k,g]} \Pi_{[g]} \right)  \right),
\end{align*}
where
\begin{align*}
    &\quad \frac{1}{\Sigma^2}  \mathbb V \left( \sum_{g,h \in [G]^2, g \neq h} \left( \tilde V_{[h]}^\top Q_{[h,g]} \Pi_{[g]} \right)^2  \right) \\
    &= \frac{1}{\Sigma^2} \mathbb V \left(  \sum_{h \in [G]} \tilde V_{[h]}^\top \left(\sum_{g \in [G], g \neq h} Q_{[h,g]} \Pi_{[g]} \Pi_{[g]}^\top Q_{[g,h]} \right) \tilde V_{[h]} \right)^2 \\
    &\leq \frac{C}{\Sigma^2} \sum_{h \in [G]} \left\Vert \sum_{g \in [G], g \neq h} Q_{[h,g]} \Pi_{[g]} \Pi_{[g]}^\top Q_{[g,h]} \right\Vert_{op}^2 \\
    &\leq \frac{C}{\Sigma^2} \sum_{h \in [G]} \tr \left( \sum_{g \in [G], g \neq h} Q_{[h,g]} \Pi_{[g]} \Pi_{[g]}^\top Q_{[g,h]} \right) \\
    &\leq \frac{C}{\Sigma^2} \sum_{g,h \in [G]^2, g \neq h} \tr \left( Q_{[h,g]}  Q_{[g,h]} \right) \\
    &= o(1),
\end{align*}
and
\begin{align*}
    &\quad \frac{1}{\Sigma^2}  \mathbb V \left( \sum_{g,h,k \in [G]^3, g \neq h \neq k} \left( \tilde V_{[h]}^\top Q_{[h,g]} \Pi_{[g]} \right) \left( \tilde V_{[k]}^\top Q_{[k,g]} \Pi_{[g]} \right)  \right) \\
    &= \frac{1}{\Sigma^2} \mathbb E \left(  \sum_{g,h,k \in [G]^3, g \neq h \neq k} \left( \tilde V_{[h]}^\top Q_{[h,g]} \Pi_{[g]} \right) \left( \tilde V_{[k]}^\top Q_{[k,g]} \Pi_{[g]} \right)  \right)^2 \\
    &\leq \frac{C}{\Sigma^2} \sum_{g,h,k \in [G]^3, g \neq h \neq k} \mathbb E \left( \left( \tilde V_{[h]}^\top Q_{[h,g]} \Pi_{[g]} \right) \left( \tilde V_{[k]}^\top Q_{[k,g]} \Pi_{[g]} \right)  \right)^2 \\
    &\leq \frac{C}{\Sigma^2} \sum_{g,h,k \in [G]^3, g \neq h \neq k} \ \tr \left( Q_{[h,g]}  Q_{[g,h]} \right) \tr \left( Q_{[k,g]}  Q_{[g,k]} \right) \\
    &= o(1).
\end{align*}
This implies that
\begin{align*}
    R_{2,1} = \frac{1}{\Sigma} \sum_{g\in [G]} \mathbb E \left(\sum_{h \in [G], h \neq g} \tilde V_{[h]}^\top Q_{[h,g]} \Pi_{[g]} \right)^2 + o_P(1).
\end{align*}
For $R_{2,2}$, since $K/\Sigma = O(1)$, by Lemma \ref{lem:linear_quad_form_2}, we have
\begin{align*}
    R_{2,2} = \frac{1}{\Sigma} \sum_{g\in [G]} \mathbb E\left(\sum_{h \in [G], h \neq g} \tilde V_{[h]}^\top Q_{[h,g]} \tilde V_{[g]} \right)^2 + o_P(1).
\end{align*}
For $R_{2,3}$, we have
\begin{align*}
    R_{2,3} &= \frac{1}{\Sigma} \sum_{g\in [G]} \left(\sum_{h \in [G], h \neq g} \tilde V_{[h]}^\top Q_{[h,g]} \Pi_{[g]} \right)\left(\sum_{k \in [G], k \neq g} \tilde V_{[k]}^\top Q_{[k,g]} \tilde V_{[g]} \right) + o_P(1) \\
    &= o_P(1),
\end{align*}
where the second equality holds because
\begin{align*}
    \mathbb E \frac{1}{\Sigma} \sum_{g\in [G]} \left(\sum_{h \in [G], h \neq g} \tilde V_{[h]}^\top Q_{[h,g]} \Pi_{[g]} \right)\left(\sum_{k \in [G], k \neq g} \tilde V_{[k]}^\top Q_{[k,g]} \tilde V_{[g]} \right)  = 0,
\end{align*}
\begin{align*}
    &\quad \mathbb V \left( \frac{1}{\Sigma} \sum_{g\in [G]} \left(\sum_{h \in [G], h \neq g} \tilde V_{[h]}^\top Q_{[h,g]} \Pi_{[g]} \right)\left(\sum_{k \in [G], k \neq g} \tilde V_{[k]}^\top Q_{[k,g]} \tilde V_{[g]} \right) \right) \\
    &\leq \frac{C}{\Sigma^2} \mathbb V \left( \sum_{g,h \in [G]^2, g \neq h} \tilde V_{[h]}^\top Q_{[h,g]} \Pi_{[g]} \tilde V_{[h]}^\top Q_{[h,g]} \tilde V_{[g]} \right) \\
    &+ \frac{C}{\Sigma^2} \mathbb V \left( \sum_{g,h,k \in [G]^3, g \neq h \neq k} \tilde V_{[h]}^\top Q_{[h,g]} \Pi_{[g]} \tilde V_{[k]}^\top Q_{[k,g]} \tilde V_{[g]} \right),
\end{align*}
where
\begin{align*}
    &\quad \frac{1}{\Sigma^2} \mathbb V \left( \sum_{g,h \in [G]^2, g \neq h} \tilde V_{[h]}^\top Q_{[h,g]} \Pi_{[g]} \tilde V_{[h]}^\top Q_{[h,g]} \tilde V_{[g]} \right) \\
    &\leq \frac{C}{\Sigma^2} \mathbb V \left( \sum_{g,h \in [G]^2, g \neq h} \Pi_{[g]}^\top Q_{[g,h]} \Omega_h^{\tilde V, \tilde V} Q_{[h,g]} \tilde V_{[g]} \right)^2 \\
    &+ \frac{C}{\Sigma^2} \mathbb V \left( \sum_{g,h \in [G]^2, g \neq h} \Pi_{[g]}^\top Q_{[g,h]} (\tilde V_{[h]} \tilde V_{[h]}^\top - \Omega_h^{\tilde V, \tilde V}) Q_{[h,g]} \tilde V_{[g]} \right)^2\\
    &\leq \frac{C}{\Sigma^2} \sum_{g \in [G]} \left\Vert \left(\sum_{h \in [G], h \neq g} Q_{[g,h]} \Omega_h^{\tilde V, \tilde V} Q_{[h,g]} \right) \Pi_{[g]} \right\Vert_2^2 \\
    &+ \frac{C}{\Sigma^2}  \sum_{g,h \in [G]^2, g \neq h} \mathbb E \left( \tilde V_{[h]}^\top Q_{[h,g]} \Pi_{[g]} \tilde V_{[h]}^\top Q_{[h,g]} \tilde V_{[g]} \right)^2 \\
    &\leq \frac{C}{\Sigma^2} \sum_{g,h \in [G]^2, g \neq h} \tr \left( Q_{[h,g]}  Q_{[g,h]} \right) \\
    &= o(1),
\end{align*}
and
\begin{align*}
    &\quad \frac{1}{\Sigma^2} \mathbb V \left( \sum_{g,h,k \in [G]^3, g \neq h \neq k} \tilde V_{[h]}^\top Q_{[h,g]} \Pi_{[g]} \tilde V_{[k]}^\top Q_{[k,g]} \tilde V_{[g]} \right) \\
    &= \frac{1}{\Sigma^2} \mathbb E \left( \sum_{g,h,k \in [G]^3, g \neq h \neq k} \tilde V_{[h]}^\top Q_{[h,g]} \Pi_{[g]} \tilde V_{[k]}^\top Q_{[k,g]} \tilde V_{[g]} \right)^2 \\
    &\leq \frac{C}{\Sigma^2} \sum_{g,h,k \in [G]^3, g \neq h \neq k} \mathbb E \left( \tilde V_{[h]}^\top Q_{[h,g]} \Pi_{[g]} \tilde V_{[k]}^\top Q_{[k,g]} \tilde V_{[g]} \right)^2 \\
    &\leq \frac{C}{\Sigma^2} \sum_{g,h,k \in [G]^3, g \neq h \neq k} \tr \left( Q_{[h,g]}  Q_{[g,h]} \right) \tr \left( Q_{[k,g]} Q_{[g,k]} \right) \\
    &= o(1).
\end{align*}
Combining the results above, we have
\begin{align*}
    R_2 = \frac{1}{\Sigma} \sum_{g\in [G]} \mathbb E \left(\sum_{h \in [G], h \neq g} \tilde V_{[h]}^\top Q_{[h,g]} \Pi_{[g]} \right)^2 + \frac{1}{\Sigma} \sum_{g\in [G]} \mathbb E\left(\sum_{h \in [G], h \neq g} \tilde V_{[h]}^\top Q_{[h,g]} \tilde V_{[g]} \right)^2 + o_P(1).
\end{align*}
For $R_3$, we have
\begin{align*}
    R_3 &= \underbrace{\frac{1}{\Sigma} \sum_{g, h \in [G]^2, g \neq h}  \bar \Pi_{[g]}^\top \Pi_{[g]}  \tilde V_{[h]}^\top Q_{[h,g]} \Pi_{[g]}}_{R_{3,1}} \\
    &+ \underbrace{\frac{1}{\Sigma} \sum_{g, h \in [G]^2, g \neq h}  \bar \Pi_{[g]}^\top \Pi_{[g]}  \tilde V_{[h]}^\top Q_{[h,g]} V_{[g]}}_{R_{3,2}} \\
    &+ \underbrace{\frac{1}{\Sigma} \sum_{g, h \in [G]^2, g \neq h}  \bar \Pi_{[g]}^\top V_{[g]}  \tilde V_{[h]}^\top Q_{[h,g]} \Pi_{[g]}}_{R_{3,3}} \\
    &+ \underbrace{\frac{1}{\Sigma} \sum_{g, h \in [G]^2, g \neq h}  \bar \Pi_{[g]}^\top V_{[g]}  \tilde V_{[h]}^\top Q_{[h,g]} V_{[g]}}_{R_{3,4}}.
\end{align*}
For $R_{3,1}$, it has mean zero and
\begin{align*}
    \mathbb V \left( R_{3,1} \right) & = \frac{1}{\Sigma^2} \mathbb E \left( \sum_{g, h \in [G]^2, g \neq h}  \bar \Pi_{[g]}^\top \Pi_{[g]}  \tilde V_{[h]}^\top Q_{[h,g]} \Pi_{[g]} \right)^2 \\
    &\leq \frac{C}{\Sigma^2} \sum_{h \in [G]} \left\Vert \sum_{g \in [G], g \neq h} Q_{[h,g]} \Pi_{[g]} \bar \Pi_{[g]}^\top \Pi_{[g]} \right\Vert_2^2 \\
    &\leq \frac{C}{\Sigma^2} \sum_{h \in [G]}  \left\Vert \sum_{g \in [G]} Q_{[h,g]} \Pi_{[g]} \bar \Pi_{[g]}^\top \Pi_{[g]} \right\Vert_2^2 + \frac{C}{\Sigma^2} \sum_{h \in [G]}  \left\Vert  Q_{[h,h]} \Pi_{[h]} \bar \Pi_{[h]}^\top \Pi_{[h]} \right\Vert_2^2 \\
    &\leq \frac{C}{\Sigma^2} \sum_{h \in [G]}  \left\Vert \sum_{g \in [G]} Q_{[h,g]} \Pi_{[g]} \bar \Pi_{[g]}^\top \Pi_{[g]} \right\Vert_2^2 + o(1) \\
    &= o(1),
\end{align*}
where the last equality holds because 
\begin{align*}
    &\quad \frac{1}{\Sigma^2} \sum_{h \in [G]}  \left\Vert \sum_{g \in [G]} Q_{[h,g]} \Pi_{[g]} \bar \Pi_{[g]}^\top \Pi_{[g]} \right\Vert_2^2 \\
    &= \frac{1}{\Sigma^2}  \sum_{g, g' \in [G]^2} (\Pi_{[g]} \bar \Pi_{[g]}^\top \Pi_{[g]})^\top \left( \sum_{h \in [G]} Q_{[g,h]} Q_{[h,g']} \right) \Pi_{[g']} \bar \Pi_{[g']}^\top \Pi_{[g']} \\
    & \leq \frac{1}{\Sigma^2}  \sum_{g \in [G]}\left\Vert \Pi_{[g]} \bar \Pi_{[g]}^\top \Pi_{[g]} \right\Vert_2^2 \\
    &= o(1).
\end{align*}
For $R_{3,2}$, we have
\begin{align*}
    R_{3,2} = \frac{1}{\Sigma} \sum_{g, h \in [G]^2, g \neq h}  \bar \Pi_{[g]}^\top \Pi_{[g]}  \tilde V_{[h]}^\top Q_{[h,g]} \tilde V_{[g]} + o_P(1),
\end{align*}
where the first term has mean zero and
\begin{align*}
    &\quad \mathbb V \left(\frac{1}{\Sigma} \sum_{g, h \in [G]^2, g \neq h}  \bar \Pi_{[g]}^\top \Pi_{[g]}  \tilde V_{[h]}^\top Q_{[h,g]} \tilde V_{[g]} \right) \\
    &= \frac{1}{\Sigma^2} \mathbb E \left( \sum_{g, h \in [G]^2, g \neq h}  \bar \Pi_{[g]}^\top \Pi_{[g]}  \tilde V_{[h]}^\top Q_{[h,g]} \tilde V_{[g]} \right)^2 \\
    &\leq \frac{C}{\Sigma^2} \sum_{g, h \in [G]^2, g \neq h} \mathbb E \left( \bar \Pi_{[g]}^\top \Pi_{[g]}  \tilde V_{[h]}^\top Q_{[h,g]} \tilde V_{[g]} \right)^2 \\
    &\leq \frac{C\max_{g \in [G]}\left\Vert \bar \Pi_{[g]} \right\Vert_2^2}{\Sigma^2} \sum_{g,h \in [G]^2, g \neq h} \tr \left( Q_{[h,g]}  Q_{[g,h]} \right) \\
    &= o(1).
\end{align*}
By using the same argument, we also have $R_{3,3} = o_P(1)$. For $R_{3,4}$, we have
\begin{align*}
    R_{3,4} = \frac{1}{\Sigma} \sum_{g, h \in [G]^2, g \neq h}  \bar \Pi_{[g]}^\top \tilde V_{[g]}  \tilde V_{[h]}^\top Q_{[h,g]} \tilde V_{[g]} + o_P(1),
\end{align*}
where the first term has mean zero and
\begin{align*}
    &\quad \mathbb V \left(\frac{1}{\Sigma} \sum_{g, h \in [G]^2, g \neq h}  \bar \Pi_{[g]}^\top \tilde V_{[g]}  \tilde V_{[h]}^\top Q_{[h,g]} \tilde V_{[g]} \right) \\
    &\leq \frac{C}{\Sigma^2} \mathbb V \left(\sum_{g, h \in [G]^2, g \neq h}  \bar \Pi_{[g]}^\top \Omega_g^{\tilde V, \tilde V} Q_{[g,h]} \tilde V_{[h]}\right) \\
    &+ \frac{C}{\Sigma^2} \mathbb V \left(\sum_{g, h \in [G]^2, g \neq h}  \bar \Pi_{[g]}^\top (\tilde V_{[g]} \tilde V_{[g]}^\top - \Omega_g^{\tilde V, \tilde V}) Q_{[g,h]} \tilde V_{[h]}\right) \\
    &\leq \frac{C}{\Sigma^2} \mathbb E \left( \bar \Pi^\top \Omega_{\tilde V} (Q - \bar Q) \tilde V \right)^2 \\
    &+ \frac{C}{\Sigma^2} \sum_{g, h \in [G]^2, g \neq h} \mathbb E \left( \bar \Pi_{[g]}^\top \tilde V_{[g]}  \tilde V_{[h]}^\top Q_{[h,g]} \tilde V_{[g]} \right)^2 \\
    &\leq \frac{C\bar \Pi^\top \bar \Pi}{\Sigma^2} + \frac{C\max_{g \in [G]}\left\Vert \bar \Pi_{[g]} \right\Vert_2^2}{\Sigma^2} \sum_{g,h \in [G]^2, g \neq h} \tr \left( Q_{[h,g]}  Q_{[g,h]} \right) \\
    &= o(1).
\end{align*}
Combining the results above, we have $R_3 = o_P(1)$. Therefore, by combining \eqref{eq:XtildeQX} with the calculations about for terms $R_1$ to $R_3$, we have
\begin{align*}
    \frac{1}{\Sigma} \sum_{g\in [G]} \left(\sum_{h \in [G], h \neq g} \tilde X_{[h]}^\top Q_{[h,g]} X_{[g]} \right)^2 &= \frac{1}{\Sigma} \sum_{g\in [G]} \left(\bar \Pi_{[g]}^\top \Pi_{[g]} \right)^2 + \frac{1}{\Sigma} \sum_{g\in [G]} \mathbb E \left(\bar \Pi_{[g]}^\top \tilde V_{[g]} \right)^2 \\
    &+ \frac{1}{\Sigma} \sum_{g\in [G]} \mathbb E \left(\sum_{h \in [G], h \neq g} \tilde V_{[h]}^\top Q_{[h,g]} \Pi_{[g]} \right)^2 \\
    &+ \frac{1}{\Sigma} \sum_{g\in [G]} \mathbb E\left(\sum_{h \in [G], h \neq g} \tilde V_{[h]}^\top Q_{[h,g]} \tilde V_{[g]} \right)^2 + o_P(1).
\end{align*}

Finally, we note that
\begin{align}\label{eq:XtildeQXXtildeQe}
    &\quad \frac{1}{\Sigma}  \sum_{g\in [G]} \left(\sum_{h \in [G], h \neq g} \tilde X_{[h]}^\top Q_{[h,g]} X_{[g]} \right) \left(\sum_{k \in [G], k \neq g} \tilde X_{[k]}^\top Q_{[k,g]} e_{[g]} \right) \notag \\
    &= \underbrace{\frac{1}{\Sigma}  \sum_{g\in [G]} \left(\sum_{h \in [G], h \neq g} \tilde X_{[h]}^\top Q_{[h,g]} \Pi_{[g]} \right) \left(\sum_{k \in [G], k \neq g} \tilde X_{[k]}^\top Q_{[k,g]} e_{[g]} \right)}_{R_4} \notag \\
    &+ \underbrace{\frac{1}{\Sigma}  \sum_{g\in [G]} \left(\sum_{h \in [G], h \neq g} \tilde X_{[h]}^\top Q_{[h,g]} V_{[g]} \right) \left(\sum_{k \in [G], k \neq g} \tilde X_{[k]}^\top Q_{[k,g]} e_{[g]} \right)}_{R_5}.
\end{align}

For $R_4$, we have
\begin{align*}
    R_4 &= \frac{1}{\Sigma}  \sum_{g\in [G]} \left(\sum_{h \in [G], h \neq g} \tilde X_{[h]}^\top Q_{[h,g]} \Pi_{[g]} \right) \left(\sum_{k \in [G], k \neq g} \tilde X_{[k]}^\top Q_{[k,g]} \tilde e_{[g]} \right) + o_P(1) \\
    &= \underbrace{\frac{1}{\Sigma} \sum_{g \in [G]} \left(\bar \Pi_{[g]}^\top \Pi_{[g]} \right) \left(\bar \Pi_{[g]}^\top \tilde e_{[g]} \right)}_{R_{4,1}} \\
    &+ \underbrace{\frac{1}{\Sigma} \sum_{g,h \in [G]^2, g \neq h} \left(\bar \Pi_{[g]}^\top \Pi_{[g]} \right) \left(\tilde V_{[h]}^\top Q_{[h,g]} \tilde e_{[g]} \right)}_{R_{4,2}} \\
    &+ \underbrace{\frac{1}{\Sigma} \sum_{g, h \in [G]^2, g \neq h} \left(\tilde V_{[h]}^\top Q_{[h,g]} \Pi_{[g]} \right) \left(\bar \Pi_{[g]}^\top \tilde e_{[g]} \right)}_{R_{4,3}} \\
    &+ \underbrace{\frac{1}{\Sigma}  \sum_{g\in [G]} \left(\sum_{h \in [G], h \neq g} \tilde V_{[h]}^\top Q_{[h,g]} \Pi_{[g]} \right) \left(\sum_{k \in [G], k \neq g} \tilde V_{[k]}^\top Q_{[k,g]} \tilde e_{[g]} \right)}_{R_{4,4}} + o_P(1).
\end{align*}
By using a similar argument as in the proof for $R_1$, we have $R_{4,1} = o_P(1)$. For $R_{4,2}$, it has mean zero and
\begin{align*}
    \mathbb V \left( R_{4,2} \right) &= \frac{1}{\Sigma^2} \mathbb E \left( \sum_{g,h \in [G]^2, g \neq h} \left(\bar \Pi_{[g]}^\top \Pi_{[g]} \right) \left(\tilde V_{[h]}^\top Q_{[h,g]} \tilde e_{[g]} \right)\right)^2 \\
    &\leq \frac{C}{\Sigma^2} \sum_{g, h \in [G]^2, g \neq h} \left(\bar \Pi_{[g]}^\top \Pi_{[g]} \right)^2 \mathbb E \left(\tilde V_{[h]}^\top Q_{[h,g]} \tilde e_{[g]} \right)^2 \\
    &\leq \frac{C\max_{g \in [G]}\left\Vert \bar \Pi_{[g]} \right\Vert_2^2}{\Sigma^2} \sum_{g,h \in [G]^2, g \neq h} \tr \left( Q_{[h,g]}  Q_{[g,h]} \right) \\
    &= o(1).
\end{align*}
Therefore, we have $R_{4,2} = o_P(1)$. Using the same argument, we also have $R_{4,3} = o_P(1)$. In addition, by using a similar argument as in the proof for $R_{2,3}$, we have $R_{4,4} = o_P(1)$, which implies $R_4 = o_P(1)$.

For $R_5$, we have
\begin{align*}
    R_5 &= \frac{1}{\Sigma}  \sum_{g\in [G]} \left(\sum_{h \in [G], h \neq g} \tilde X_{[h]}^\top Q_{[h,g]} \tilde V_{[g]} \right) \left(\sum_{k \in [G], k \neq g} \tilde X_{[k]}^\top Q_{[k,g]} 
    \tilde e_{[g]} \right) + o_P(1) \\
    &= \underbrace{\frac{1}{\Sigma} \sum_{g \in [G]} \left(\bar \Pi_{[g]}^\top \tilde V_{[g]} \right) \left(\bar \Pi_{[g]}^\top \tilde e_{[g]} \right)}_{R_{5,1}} \\
    &+ \underbrace{\frac{1}{\Sigma} \sum_{g,h \in [G]^2, g \neq h} \left(\bar \Pi_{[g]}^\top \tilde V_{[g]} \right) \left(\tilde V_{[h]}^\top Q_{[h,g]} \tilde e_{[g]} \right)}_{R_{5,2}} \\
    &+ \underbrace{\frac{1}{\Sigma} \sum_{g, h \in [G]^2, g \neq h} \left(\tilde V_{[h]}^\top Q_{[h,g]} \tilde V_{[g]} \right) \left(\bar \Pi_{[g]}^\top \tilde e_{[g]} \right)}_{R_{5,3}} \\
    &+ \underbrace{\frac{1}{\Sigma}  \sum_{g\in [G]} \left(\sum_{h \in [G], h \neq g} \tilde V_{[h]}^\top Q_{[h,g]} \tilde V_{[g]} \right) \left(\sum_{k \in [G], k \neq g} \tilde V_{[k]}^\top Q_{[k,g]} \tilde e_{[g]} \right)}_{R_{5,4}} + o_P(1).
\end{align*}
For $R_{5,1}$, by using a similar argument as in the proof for $R_1$, we have 
\begin{align*}
    R_{5,1} = \frac{1}{\Sigma} \sum_{g \in [G]} \mathbb E \left(\bar \Pi_{[g]}^\top \tilde V_{[g]} \right) \left(\bar \Pi_{[g]}^\top \tilde e_{[g]} \right) + o_P(1).
\end{align*}
In addition, by using a similar argument as in the proof for $R_{3,4}$, we have $R_{5,2} = o_P(1)$ and $R_{5,3} = o_P(1)$. Lastly, by using a similar argument as in the proof for $R_{2,2}$, we have 
\begin{align*}
    R_{5,4} = \frac{1}{\Sigma}  \sum_{g\in [G]} \mathbb E \left(\sum_{h \in [G], h \neq g} \tilde V_{[h]}^\top Q_{[h,g]} \tilde V_{[g]} \right) \left(\sum_{k \in [G], k \neq g} \tilde V_{[k]}^\top Q_{[k,g]} \tilde e_{[g]} \right) + o_P(1).
\end{align*}
Combining the results with \eqref{eq:XtildeQXXtildeQe}, we have
\begin{align*}
    &\quad \frac{1}{\Sigma}  \sum_{g\in [G]} \left(\sum_{h \in [G], h \neq g} \tilde X_{[h]}^\top Q_{[h,g]} X_{[g]} \right) \left(\sum_{k \in [G], k \neq g} \tilde X_{[k]}^\top Q_{[k,g]} e_{[g]} \right) \\ 
    &= \frac{1}{\Sigma} \sum_{g \in [G]} \mathbb E \left(\bar \Pi_{[g]}^\top \tilde V_{[g]} \right) \left(\bar \Pi_{[g]}^\top \tilde e_{[g]} \right) \\
    &+ \frac{1}{\Sigma}  \sum_{g\in [G]} \mathbb E \left(\sum_{h \in [G], h \neq g} \tilde V_{[h]}^\top Q_{[h,g]} \tilde V_{[g]} \right) \left(\sum_{k \in [G], k \neq g} \tilde V_{[k]}^\top Q_{[k,g]} \tilde e_{[g]} \right) + o_P(1).
\end{align*}
This concludes the proof. $\hfill\qedsymbol$

\subsection{Proof of Lemma \ref{lem:linear_quad_form_2}}

For the first result in Lemma \ref{lem:linear_quad_form_2}, we note that
\begin{align*}
    &\quad \frac{1}{\Sigma}  \sum_{g,h \in [G]^2, g \neq h} \left( \tilde X_{[g]}^\top Q_{[g,h]} \tilde e_{[h]} \right) \left( \tilde X_{[h]}^\top Q_{[h,g]} \tilde e_{[g]} \right) \\
    &= \underbrace{\frac{1}{\Sigma}  \sum_{g,h \in [G]^2, g \neq h} \left( \tilde \Pi_{[g]}^\top Q_{[g,h]} \tilde e_{[h]} \right) \left( \tilde \Pi_{[h]}^\top Q_{[h,g]} \tilde e_{[g]} \right)}_{R_6} \\
    &+ \underbrace{\frac{1}{\Sigma}  \sum_{g,h \in [G]^2, g \neq h} \left( \tilde V_{[g]}^\top Q_{[g,h]} \tilde e_{[h]} \right) \left( \tilde V_{[h]}^\top Q_{[h,g]} \tilde e_{[g]} \right)}_{R_7} \\
    &+ 2 \times \underbrace{\frac{1}{\Sigma}  \sum_{g,h \in [G]^2, g \neq h} \left( \tilde \Pi_{[g]}^\top Q_{[g,h]} \tilde e_{[h]} \right) \left( \tilde X_{[h]}^\top Q_{[h,g]} \tilde e_{[g]} \right)}_{R_8}.
\end{align*}
For $R_6$, it has mean zero and 
\begin{align*}
    \mathbb V \left( R_6 \right) &= \frac{1}{\Sigma^2} \mathbb E \left( \sum_{g,h \in [G]^2, g \neq h} \left( \tilde \Pi_{[g]}^\top Q_{[g,h]} \tilde e_{[h]} \right) \left( \tilde \Pi_{[h]}^\top Q_{[h,g]} \tilde e_{[g]} \right) \right)^2 \\
    &\leqslant \frac{C}{\Sigma^2} \sum_{g,h \in [G]^2, g \neq h} \mathbb E \left(\tilde \Pi_{[g]}^\top Q_{[g,h]} \tilde e_{[h]} \tilde \Pi_{[h]}^\top Q_{[h,g]} \tilde e_{[g]} \right)^2 \\
    &\leqslant \frac{C}{\Sigma^2} \sum_{g,h \in [G]^2, g \neq h} \tr \left( Q_{[g,h]} Q_{[h,g]} \right) \\
    &= o(1),
\end{align*}
whence $R_6 = o_P(1)$. For $R_7$, since $K/\Sigma = O(1)$, by an application of Lemma \ref{lem:quad_form_2}, we have
\begin{align*}
    R_7 = \frac{1}{\Sigma}  \sum_{g,h \in [G]^2, g \neq h} \mathbb E \left( \tilde V_{[g]}^\top Q_{[g,h]} \tilde e_{[h]} \right) \left( \tilde V_{[h]}^\top Q_{[h,g]} \tilde e_{[g]} \right) + o_P(1).
\end{align*}
For $R_8$, we have
\begin{align*}
    \mathbb V \left( R_8 \right) &= \frac{1}{\Sigma^2} \mathbb E \left( \sum_{g,h \in [G]^2, g \neq h} \left( \tilde \Pi_{[g]}^\top Q_{[g,h]} \tilde e_{[h]} \right) \left( \tilde V_{[h]}^\top Q_{[h,g]} \tilde e_{[g]} \right) \right)^2 \\
    &\leqslant \frac{C}{\Sigma^2} \mathbb E \left( \sum_{g,h \in [G]^2, g \neq h} \tilde \Pi_{[g]}^\top Q_{[g,h]} \left( \tilde e_{[h]} \tilde V_{[h]}^\top - \Omega_h^{\tilde e, \tilde V} \right) Q_{[h,g]} \tilde e_{[g]}  \right)^2 \\
    &+ \frac{C}{\Sigma^2} \mathbb E \left( \sum_{g,h \in [G]^2, g \neq h} \tilde \Pi_{[g]}^\top Q_{[g,h]} \Omega_h^{\tilde e, \tilde V} Q_{[h,g]} \tilde e_{[g]}  \right)^2 \\
    & = o(1),
\end{align*}
where the last equality holds because 
\begin{align*}
    &\quad \frac{1}{\Sigma^2} \mathbb E \left( \sum_{g,h \in [G]^2, g \neq h} \tilde \Pi_{[g]}^\top Q_{[g,h]} \left( \tilde e_{[h]} \tilde V_{[h]}^\top - \Omega_h^{\tilde e, \tilde V} \right) Q_{[h,g]} \tilde e_{[g]}  \right)^2 \\
    &\leqslant \frac{C}{\Sigma^2} \sum_{g,h \in [G]^2, g \neq h} \mathbb E \left(\tilde \Pi_{[g]}^\top Q_{[g,h]} \left( \tilde e_{[h]} \tilde V_{[h]}^\top - \Omega_h^{\tilde e, \tilde V} \right) Q_{[h,g]} \tilde e_{[g]}  \right)^2 \\
    &\leqslant \frac{C}{\Sigma^2} \sum_{g,h \in [G]^2, g \neq h} \tr \left( Q_{[g,h]} Q_{[h,g]} \right) \\
    &= o(1),
\end{align*}
and
\begin{align*}
    &\quad \frac{1}{\Sigma^2} \mathbb E \left( \sum_{g,h \in [G]^2, g \neq h} \tilde \Pi_{[g]}^\top Q_{[g,h]} \Omega_h^{\tilde e, \tilde V} Q_{[h,g]} \tilde e_{[g]}  \right)^2 \\
    &= \frac{1}{\Sigma^2} \mathbb E \left( \sum_{g \in [G]} \tilde \Pi_{[g]}^\top \left( \sum_{h \neq g} Q_{[g,h]} \Omega_h^{\tilde e, \tilde V} Q_{[h,g]} \right) \tilde e_{[g]}  \right)^2 \\
    &\leqslant \frac{C}{\Sigma^2}\sum_{g \in [G]} \tr \left( \left( \sum_{h \neq g} Q_{[g,h]} \Omega_h^{\tilde e, \tilde V} Q_{[h,g]} \right) \left( \sum_{k \neq g} Q_{[g,k]} \Omega_k^{\tilde V, \tilde e} Q_{[k,g]} \right) \right) \\
    & \leq \frac{C}{\Sigma^2} \tr \left( (Q-\bar Q) \Omega^{\tilde e, \tilde V} (Q-\bar Q)^2 \Omega^{\tilde V, \tilde e} (Q-\bar Q) \right)\\
    &= o(1).
\end{align*}
It follows that $R_8 = o_P(1)$. Combining the results above, we have
\begin{align*}
    &\quad \frac{1}{\Sigma}  \sum_{g,h \in [G]^2, g \neq h} \left( \tilde X_{[g]}^\top Q_{[g,h]} \tilde e_{[h]} \right) \left( \tilde X_{[h]}^\top Q_{[h,g]} \tilde e_{[g]} \right) \\
    &= \frac{1}{\Sigma}  \sum_{g,h \in [G]^2, g \neq h} \mathbb E \left( \tilde V_{[g]}^\top Q_{[g,h]} \tilde e_{[h]} \right) \left( \tilde V_{[h]}^\top Q_{[h,g]} \tilde e_{[g]} \right) + o_P(1).
\end{align*}

Next, we turn to the second result in Lemma \ref{lem:linear_quad_form_2}. We note that
\begin{align*}
    &\quad \frac{1}{\Sigma} \sum_{g,h \in [G]^2, g \neq h} \left( \tilde X_{[g]}^\top Q_{[g,h]} e_{[h]} \right) \left( \tilde X_{[h]}^\top Q_{[h,g]} e_{[g]} \right) \\
    &= \frac{1}{\Sigma} \sum_{g,h \in [G]^2, g \neq h} \left( \tilde X_{[g]}^\top Q_{[g,h]} \tilde e_{[h]} \right) \left( \tilde X_{[h]}^\top Q_{[h,g]} \tilde e_{[g]} \right)  \\
    &+ \frac{1}{\Sigma} \sum_{g,h \in [G]^2, g \neq h} \left( \tilde X_{[g]}^\top Q_{[g,h]} W_{[h]} \hat \gamma_{\tilde e} \right) \left( \tilde X_{[h]}^\top Q_{[h,g]} W_{[g]} \hat \gamma_{\tilde e} \right) \\
    &- \frac{2}{\Sigma} \sum_{g,h \in [G]^2, g \neq h} \left( \tilde X_{[g]}^\top Q_{[g,h]} W_{[h]} \hat \gamma_{\tilde e} \right) \left( \tilde X_{[h]}^\top Q_{[h,g]} \tilde e_{[g]} \right),
\end{align*}
where
\begin{align*}
    &\quad \left| \frac{1}{\Sigma} \sum_{g,h \in [G]^2, g \neq h} \left( \tilde X_{[g]}^\top Q_{[g,h]} W_{[h]} \hat \gamma_{\tilde e} \right) \left( \tilde X_{[h]}^\top Q_{[h,g]} W_{[g]} \hat \gamma_{\tilde e} \right) \right| \\
    &\leqslant \frac{1}{\Sigma} \sum_{g,h \in [G]^2, g \neq h} \left( \tilde X_{[g]}^\top Q_{[g,h]} W_{[h]} \hat \gamma_{\tilde e} \right)^2 \\
    &\leqslant \max_{1 \leq g \leq G} \left\Vert W_{[g]} \hat \gamma_{\tilde e} \right\Vert_2^2 \times \frac{1}{\Sigma} \sum_{g,h \in [G]^2, g \neq h} \left\Vert Q_{[h,g]} \tilde X_{[g]} \right\Vert_2^2 \\
    &= o_P(1),
\end{align*}
by Lemma \ref{lem:4th_moment}, and
\begin{align*}
    &\quad \left| \frac{1}{\Sigma} \sum_{g,h \in [G]^2, g \neq h} \left( \tilde X_{[g]}^\top Q_{[g,h]} W_{[h]} \hat \gamma_{\tilde e} \right) \left( \tilde X_{[h]}^\top Q_{[h,g]} \tilde e_{[g]} \right) \right| \\
    &\leqslant \left(\frac{1}{\Sigma} \sum_{g,h \in [G]^2, g \neq h} \left( \tilde X_{[g]}^\top Q_{[g,h]} W_{[h]} \hat \gamma_{\tilde e} \right)^2\right)^{1/2} \\
    &\times \left(\frac{1}{\Sigma} \sum_{g,h \in [G]^2, g \neq h} \left( \tilde X_{[h]}^\top Q_{[h,g]} \tilde e_{[g]} \right)^2 \right)^{1/2} \\
    &= o_P(1).
\end{align*}
Therefore, we have
\begin{align*}
    &\quad \frac{1}{\Sigma} \sum_{g,h \in [G]^2, g \neq h} \left( \tilde X_{[g]}^\top Q_{[g,h]} e_{[h]} \right) \left( \tilde X_{[h]}^\top Q_{[h,g]} e_{[g]} \right) \\
    &= \frac{1}{\Sigma} \sum_{g,h \in [G]^2, g \neq h} \left( \tilde X_{[g]}^\top Q_{[g,h]} \tilde e_{[h]} \right) \left( \tilde X_{[h]}^\top Q_{[h,g]} \tilde e_{[g]} \right) + o_P(1) \\
    &= \frac{1}{\Sigma} \sum_{g,h \in [G]^2, g \neq h} \mathbb E \left( \tilde X_{[g]}^\top Q_{[g,h]} \tilde e_{[h]} \right) \left( \tilde X_{[h]}^\top Q_{[h,g]} \tilde e_{[g]} \right) + o_P(1).
\end{align*}

Next, we turn to the third result in Lemma \ref{lem:linear_quad_form_2}. Note that
\begin{align*}
    &\quad \frac{1}{\Sigma} \sum_{g,h \in [G]^2, g \neq h} \left( \tilde X_{[g]}^\top Q_{[g,h]} X_{[h]} \right) \left( \tilde X_{[h]}^\top Q_{[h,g]} X_{[g]} \right) \\
    &= \underbrace{\frac{1}{\Sigma} \sum_{g,h \in [G]^2, g \neq h} \left( \tilde X_{[g]}^\top Q_{[g,h]} V_{[h]} \right) \left( \tilde X_{[h]}^\top Q_{[h,g]} V_{[g]} \right)}_{R_9} \\
    &+ \underbrace{\frac{1}{\Sigma} \sum_{g,h \in [G]^2, g \neq h} \left( \tilde X_{[g]}^\top Q_{[g,h]} \Pi_{[h]} \right) \left( \tilde X_{[h]}^\top Q_{[h,g]} \Pi_{[g]} \right)}_{R_{10}} \\
    &+ 2 \times \underbrace{\frac{1}{\Sigma} \sum_{g,h \in [G]^2, g \neq h} \left( \tilde X_{[g]}^\top Q_{[g,h]} \Pi_{[h]} \right) \left( \tilde X_{[h]}^\top Q_{[h,g]} V_{[g]} \right)}_{R_{11}}.
\end{align*}
By using the same argument as in the proof above, we have
\begin{align*}
    R_9 = \frac{1}{\Sigma} \sum_{g,h \in [G]^2, g \neq h} \mathbb E \left( \tilde X_{[g]}^\top Q_{[g,h]} \tilde V_{[h]} \right) \left( \tilde X_{[h]}^\top Q_{[h,g]} 
    \tilde V_{[g]} \right) + o_P(1).
\end{align*}
For $R_{10}$, we have
\begin{align*}
    R_{10} &= \frac{1}{\Sigma} \sum_{g,h \in [G]^2, g \neq h} \left( \tilde \Pi_{[g]}^\top Q_{[g,h]} \Pi_{[h]} \right) \left( \tilde \Pi_{[h]}^\top Q_{[h,g]} \Pi_{[g]} \right) \\
    &+ \underbrace{\frac{1}{\Sigma} \sum_{g,h \in [G]^2, g \neq h} \left( \tilde V_{[g]}^\top Q_{[g,h]} \Pi_{[h]} \right) \left( \tilde V_{[h]}^\top Q_{[h,g]} \Pi_{[g]} \right)}_{R_{10,1}} \\
    &+ 2 \times \underbrace{\frac{1}{\Sigma} \sum_{g,h \in [G]^2, g \neq h} \left( \tilde \Pi_{[g]}^\top Q_{[g,h]} \Pi_{[h]} \right) \left( \tilde V_{[h]}^\top Q_{[h,g]} \Pi_{[g]} \right)}_{R_{10,2}}.
\end{align*}
Note that 
\begin{align*}
    \mathbb V \left( R_{10,1} \right) &= \frac{1}{\Sigma^2} \mathbb E \left( \sum_{g,h \in [G]^2, g \neq h} \left( \tilde V_{[g]}^\top Q_{[g,h]} \Pi_{[h]} \right) \left( \tilde V_{[h]}^\top Q_{[h,g]} \Pi_{[g]} \right) \right)^2 \\
    &\leq \frac{C}{\Sigma^2}  \sum_{g,h \in [G]^2, g \neq h} \mathbb E \left( \left( \tilde V_{[g]}^\top Q_{[g,h]} \Pi_{[h]} \right) \left( \tilde V_{[h]}^\top Q_{[h,g]} \Pi_{[g]} \right) \right)^2 \\
    &\leq \frac{C}{\Sigma^2} \sum_{g,h \in [G]^2, g \neq h} \tr \left( Q_{[h,g]}  Q_{[g,h]} \right) \\
    &= o(1),
\end{align*}
and
\begin{align*}
    \mathbb V \left( R_{10,2} \right) &= \frac{1}{\Sigma^2} \mathbb E \left( \sum_{g,h \in [G]^2, g \neq h} \left( \tilde \Pi_{[g]}^\top Q_{[g,h]} \Pi_{[h]} \right) \left( \tilde V_{[h]}^\top Q_{[h,g]} \Pi_{[g]} \right) \right)^2 \\
    & \leq \frac{C}{\Sigma^2} \sum_{h \in [G]} \left\Vert \sum_{g \in [G], g \neq h} Q_{[h,g]} \Pi_{[g]} \tilde \Pi_{[g]}^\top Q_{[g,h]} \Pi_{[h]}  \right\Vert_{2}^2 \\ 
    &\leq \frac{C}{\Sigma^2} \sum_{h \in [G]} \left\Vert \sum_{g \in [G], g \neq h} Q_{[h,g]} \Pi_{[g]} \tilde \Pi_{[g]}^\top Q_{[g,h]}  \right\Vert_{F}^2 \\
    &\leq \frac{C}{\Sigma^2} \sum_{h \in [G]} \sum_{g,g' \in [G]^2, g, g' \neq h} \tr \left(Q_{[h,g]} \Pi_{[g]} \tilde \Pi_{[g]}^\top Q_{[g,h]} Q_{[h,g']} \tilde \Pi_{[g']} \Pi_{[g']}^\top Q_{[g',h]} \right) \\
    &\leq \frac{C}{\Sigma^2} \sum_{h \in [G]} \sum_{g,g' \in [G]^2, g, g' \neq h} \left(\tilde \Pi_{[g]}^\top Q_{[g,h]} Q_{[h,g']} \tilde \Pi_{[g']} \right)^2 \\
    &+ \frac{C}{\Sigma^2} \sum_{h \in [G]} \sum_{g,g' \in [G]^2, g, g' \neq h} \left(\Pi_{[g]}^\top Q_{[g,h]} Q_{[h,g']} \Pi_{[g']} \right)^2 \\
    &\leq \frac{C}{\Sigma^2} \sum_{h \in [G]} \sum_{g,g' \in [G]^2, g, g' \neq h} \tr \left( Q_{[h,g]}  Q_{[g,h]} \right) \tr \left( Q_{[h,g']}  Q_{[g',h]} \right) \\
    &= o(1).
\end{align*}

Therefore, we have
\begin{align*}
    R_{10} = \frac{1}{\Sigma} \sum_{g,h \in [G]^2, g \neq h} \left( \tilde \Pi_{[g]}^\top Q_{[g,h]} \Pi_{[h]} \right) \left( \tilde \Pi_{[h]}^\top Q_{[h,g]} \Pi_{[g]} \right) + o_P(1).
\end{align*}
For $R_{11}$, we have
\begin{align*}
    R_{11} &= \frac{1}{\Sigma} \sum_{g,h \in [G]^2, g \neq h} \left( \tilde X_{[g]}^\top Q_{[g,h]} \Pi_{[h]} \right) \left( \tilde X_{[h]}^\top Q_{[h,g]} \tilde V_{[g]} \right) + o_P(1) \\
    &= \underbrace{\frac{1}{\Sigma} \sum_{g,h \in [G]^2, g \neq h} \left( \tilde \Pi_{[g]}^\top Q_{[g,h]} \Pi_{[h]} \right) \left( \tilde \Pi_{[h]}^\top Q_{[h,g]} \tilde V_{[g]} \right)}_{R_{11,1}} \\
    &+ \underbrace{\frac{1}{\Sigma} \sum_{g,h \in [G]^2, g \neq h} \left( \tilde \Pi_{[g]}^\top Q_{[g,h]} \Pi_{[h]} \right) \left( \tilde V_{[h]}^\top Q_{[h,g]} \tilde V_{[g]} \right)}_{R_{11,2}} \\
    &+ \underbrace{\frac{1}{\Sigma} \sum_{g,h \in [G]^2, g \neq h} \left( \tilde V_{[g]}^\top Q_{[g,h]} \Pi_{[h]} \right) \left( \tilde \Pi_{[h]}^\top Q_{[h,g]} \tilde V_{[g]} \right)}_{R_{11,3}} \\
    &+ \underbrace{\frac{1}{\Sigma} \sum_{g,h \in [G]^2, g \neq h} \left( \tilde V_{[g]}^\top Q_{[g,h]} \Pi_{[h]} \right) \left( \tilde V_{[h]}^\top Q_{[h,g]} \tilde V_{[g]} \right)}_{R_{11,4}}.
\end{align*}
By using a similar argument as in the proof for $R_{10,1}$ and $R_{10,2}$, we have $R_{11,1} = o_P(1)$ and $R_{11,2} = o_P(1)$. For $R_{11,3}$, we have
\begin{align*}
    \mathbb V \left( R_{11,3} \right) &= \frac{1}{\Sigma^2} \mathbb E \left( \sum_{g,h \in [G]^2, g \neq h} \left( \tilde V_{[g]}^\top Q_{[g,h]} \Pi_{[h]} \right) \left( \tilde \Pi_{[h]}^\top Q_{[h,g]} \tilde V_{[g]} \right) \right)^2 \\
    &\leq \frac{C}{\Sigma^2} \sum_{g \in [G]} \left\Vert \sum_{h \in [G], h \neq g} Q_{[g,h]} \Pi_{[h]} \tilde \Pi_{[h]}^\top Q_{[h,g]} \right\Vert_F^2 \\
    &\leq \frac{C}{\Sigma^2} \sum_{g \in [G]} \sum_{h, h' \in [G]^2, h,h' \neq g} \left( \tilde \Pi_{[h]}^\top Q_{[h,g]} Q_{[g,h']} \tilde \Pi_{[h']} \right)^2 \\
    &+ \frac{C}{\Sigma^2} \sum_{g \in [G]} \sum_{h, h' \in [G]^2, h,h' \neq g} \left( \Pi_{[h]}^\top Q_{[h,g]} Q_{[g,h']} \Pi_{[h']} \right)^2 \\
    &\leq \frac{C}{\Sigma^2} \sum_{g \in [G]} \sum_{h, h' \in [G]^2, h,h' \neq g} \tr \left( Q_{[g,h]} Q_{[h,g]} \right) \tr \left( Q_{[g,h']} Q_{[h',g]} \right) \\
    &= o(1),
\end{align*}
and thus
\begin{align*}
    R_{11,3} = \frac{1}{\Sigma} \sum_{g,h \in [G]^2, g \neq h} \mathbb E \left( \tilde V_{[g]}^\top Q_{[g,h]} \Pi_{[h]} \right) \left( \tilde \Pi_{[h]}^\top Q_{[h,g]} \tilde V_{[g]} \right) + o_P(1).
\end{align*}
For $R_{11,4}$, we have
\begin{align*}
    \mathbb V \left( R_{11,4} \right) &\leq \mathbb V \left( \frac{1}{\Sigma} \sum_{g,h \in [G]^2, g \neq h} \tilde V_{[h]}^\top Q_{[h,g]} \Omega_g^{\tilde V, \tilde V} Q_{[g,h]} \Pi_{[h]} \right) \\
    &+ \mathbb V \left( \frac{1}{\Sigma} \sum_{g,h \in [G]^2, g \neq h} \tilde V_{[h]}^\top Q_{[h,g]} (\tilde V_{[g]} \tilde V_{[g]}^\top - \Omega_g^{\tilde V, \tilde V}) Q_{[g,h]} \Pi_{[h]} \right),
\end{align*}
where
\begin{align*}
    &\quad \mathbb V \left( \frac{1}{\Sigma} \sum_{g,h \in [G]^2, g \neq h} \tilde V_{[h]}^\top Q_{[h,g]} \Omega_g^{\tilde V, \tilde V} Q_{[g,h]} \Pi_{[h]} \right) \\
    &= \frac{1}{\Sigma^2} \mathbb E \left(\sum_{h \in [G]} \tilde V_{[h]}^\top \left(\sum_{g \in [G], g \neq h} Q_{[h,g]} \Omega_g^{\tilde V, \tilde V} Q_{[g,h]}  \right) \Pi_{[h]} \right)^2 \\
    &\leq \frac{C\Pi^\top \Pi}{(\Pi^\top \Pi + K)^2} \\
    &= o(1),
\end{align*}
and 
\begin{align*}
    &\quad \mathbb V \left( \frac{1}{\Sigma} \sum_{g,h \in [G]^2, g \neq h} \tilde V_{[h]}^\top Q_{[h,g]} (\tilde V_{[g]} \tilde V_{[g]}^\top - \Omega_g^{\tilde V, \tilde V}) Q_{[g,h]} \Pi_{[h]} \right) \\
    &= \frac{1}{\Sigma^2} \mathbb E \left( \sum_{g,h \in [G]^2, g \neq h} \tilde V_{[h]}^\top Q_{[h,g]} (\tilde V_{[g]} \tilde V_{[g]}^\top - \Omega_g^{\tilde V, \tilde V}) Q_{[g,h]} \Pi_{[h]} \right)^2 \\
    &\leq \frac{C}{\Sigma^2}  \sum_{g,h \in [G]^2, g \neq h} \mathbb E \left( \tilde V_{[h]}^\top Q_{[h,g]} \tilde V_{[g]} \tilde V_{[g]}^\top Q_{[g,h]} \Pi_{[h]} \right)^2 \\
    &\leq \frac{C}{\Sigma^2} \sum_{g,h \in [G]^2, g \neq h} \tr \left( Q_{[h,g]}  Q_{[g,h]} \right) \\
    &= o(1).
\end{align*}
It follows that $R_{11,4} = o_P(1)$, whence
\begin{align*}
    R_{11} = \frac{1}{\Sigma} \sum_{g,h \in [G]^2, g \neq h} \mathbb E \left( \tilde V_{[g]}^\top Q_{[g,h]} \Pi_{[h]} \right) \left( \tilde \Pi_{[h]}^\top Q_{[h,g]} \tilde V_{[g]} \right) + o_P(1).
\end{align*}
Combining the results above, we have
\begin{align*}
    &\quad \frac{1}{\Sigma} \sum_{g,h \in [G]^2, g \neq h} \left( \tilde X_{[g]}^\top Q_{[g,h]} X_{[h]} \right) \left( \tilde X_{[h]}^\top Q_{[h,g]} X_{[g]} \right) \\
    &= \frac{1}{\Sigma} \sum_{g,h \in [G]^2, g \neq h} \mathbb E \left( \tilde X_{[g]}^\top Q_{[g,h]} \tilde V_{[h]} \right) \left( \tilde X_{[h]}^\top Q_{[h,g]} 
    \tilde V_{[g]} \right) \\
    &+ \frac{1}{\Sigma} \sum_{g,h \in [G]^2, g \neq h} \left( \tilde \Pi_{[g]}^\top Q_{[g,h]} \Pi_{[h]} \right) \left( \tilde \Pi_{[h]}^\top Q_{[h,g]} \Pi_{[g]} \right) \\
    &+ \frac{2}{\Sigma} \sum_{g,h \in [G]^2, g \neq h} \mathbb E \left( \tilde V_{[g]}^\top Q_{[g,h]} \Pi_{[h]} \right) \left( \tilde \Pi_{[h]}^\top Q_{[h,g]} \tilde V_{[g]} \right) + o_P(1).
\end{align*}

Finally, for the last result of Lemma \ref{lem:linear_quad_form_2}, we note that
\begin{align*}
    &\quad \frac{1}{\Sigma} \sum_{g,h \in [G]^2, g \neq h} \left( \tilde X_{[g]}^\top Q_{[g,h]} X_{[h]} \right) \left( \tilde X_{[h]}^\top Q_{[h,g]} e_{[g]} \right) \\
    &= \underbrace{\frac{1}{\Sigma} \sum_{g,h \in [G]^2, g \neq h} \left( \tilde X_{[g]}^\top Q_{[g,h]} V_{[h]} \right) \left( \tilde X_{[h]}^\top Q_{[h,g]} e_{[g]} \right)}_{R_{12}} \\
    &+ \underbrace{\frac{1}{\Sigma} \sum_{g,h \in [G]^2, g \neq h} \left( \tilde X_{[g]}^\top Q_{[g,h]} \Pi_{[h]} \right) \left( \tilde X_{[h]}^\top Q_{[h,g]} e_{[g]} \right)}_{R_{13}}.
\end{align*}
By using the same argument as in the proof for $R_{9}$, we have
\begin{align*}
    R_{12} = \frac{1}{\Sigma} \sum_{g,h \in [G]^2, g \neq h} \mathbb E \left( \tilde X_{[g]}^\top Q_{[g,h]} \tilde V_{[h]} \right) \left( \tilde X_{[h]}^\top Q_{[h,g]} \tilde e_{[g]} \right) + o_P(1).
\end{align*}
In addition, by using the same argument as in the proof for $R_{11}$, we have
\begin{align*}
    R_{13} &= \frac{1}{\Sigma} \sum_{g,h \in [G]^2, g \neq h} \mathbb E \left( \tilde V_{[g]}^\top Q_{[g,h]} \Pi_{[h]} \right) \left( \tilde \Pi_{[h]}^\top Q_{[h,g]} \tilde e_{[g]} \right) + o_P(1).
\end{align*}
It follows that
\begin{align*}
    &\quad \frac{1}{\Sigma} \sum_{g,h \in [G]^2, g \neq h} \left( \tilde X_{[g]}^\top Q_{[g,h]} X_{[h]} \right) \left( \tilde X_{[h]}^\top Q_{[h,g]} e_{[g]} \right) \\ 
    &= \frac{1}{\Sigma} \sum_{g,h \in [G]^2, g \neq h} \mathbb E \left( \tilde X_{[g]}^\top Q_{[g,h]} \tilde V_{[h]} \right) \left( \tilde X_{[h]}^\top Q_{[h,g]} \tilde e_{[g]} \right) \\
    &+ \frac{1}{\Sigma} \sum_{g,h \in [G]^2, g \neq h} \mathbb E \left( \tilde V_{[g]}^\top Q_{[g,h]} \Pi_{[h]} \right) \left( \tilde \Pi_{[h]}^\top Q_{[h,g]} \tilde e_{[g]} \right) + o_P(1).
\end{align*}
This concludes the proof. $\hfill\qedsymbol$

\subsection{Proof of Lemma \ref{lem:var_est}}

Throughout the proof we denote $\acute \Delta = \acute \beta - \beta$, and note that $\acute \Delta = o_P(1)$ by assumption. We divide the proof into four steps.

\vspace{5mm}
\noindent \textbf{Step 1: Consistency of $\acute \Omega$.} Note that $1/C \leq \lambda_{\min}(\Omega/n) \leq \lambda_{\min}(\Omega/n) \leq C$ by Assumption \ref{ass:reg}, and thus it suffices to show that
\begin{align*}
    \frac{1}{n} \acute \Omega - \frac{1}{n} \Omega = o_P(1).
\end{align*}
Let 
\begin{align*}
    \tilde \Omega = \sum_{g \in [G]} \left(z_{[g]}^\top \tilde e_{[g]} \right)\left(z_{[g]}^\top \tilde e_{[g]} \right)^\top \quad \text{and} \quad 
    \bar \Omega = \sum_{g \in [G]} \left(z_{[g]}^\top e_{[g]} \right)\left(z_{[g]}^\top e_{[g]} \right)^\top.
\end{align*}
We aim to show that
\begin{align}
    \frac{1}{n} \tilde \Omega - \frac{1}{n} \Omega &= o_P(1), \label{eq:omegatilde-omega}\\
    \frac{1}{n} \bar \Omega - \frac{1}{n} \tilde \Omega &= o_P(1), \label{eq:omegabar-omegatilde} \\
    \frac{1}{n} \acute \Omega - \frac{1}{n} \bar \Omega &= o_P(1). \label{eq:omegacute-omegabar}
\end{align}

For \eqref{eq:omegatilde-omega}, consider its $(j,k)$-th element for $1 \leq j,k \leq d_z$, given by
\begin{align*}
    \frac{1}{n} \sum_{g \in [G]} \left[ z_{[g], j}^\top \tilde e_{[g]} \tilde e_{[g]}^\top  z_{[g], k} - \mathbb E \left( z_{[g], j}^\top \tilde e_{[g]} \tilde e_{[g]}^\top  z_{[g], k} \right) \right],
\end{align*}
where we use $z_{[g], j}$ ($z_{[g], k}$) to denote the $j$-th ($k$-th) column of $z_{[g]}$; note that it has mean zero and
\begin{align*}
    &\quad \mathbb V \left( \frac{1}{n} \sum_{g \in [G]} \left[ z_{[g], j}^\top \tilde e_{[g]} \tilde e_{[g]}^\top  z_{[g], k} - \mathbb E \left( z_{[g], j}^\top \tilde e_{[g]} \tilde e_{[g]}^\top  z_{[g], k} \right) \right] \right) \\
    &\leqslant \frac{1}{n^2} \sum_{g \in [G]} \mathbb E \left( z_{[g], j}^\top \tilde e_{[g]} \tilde e_{[g]}^\top  z_{[g], k} \right)^2 \\
    &\leqslant \frac{1}{n^2} \sum_{g \in [G]} z_{[g], j}^\top z_{[g], j} z_{[g], k}^\top z_{[g], k}  \mathbb E \left( \tilde e_{[g]}^\top \tilde e_{[g]} \right)^2 \\
    &\leqslant \frac{C \max_{i \in I_g, g \in [G]} \left\Vert z_{i,g} \right\Vert_2^2}{n} \times \frac{1}{n} \sum_{g \in [G]} z_{[g], j}^\top z_{[g], j} \\
    &= o(1)
\end{align*}
by Assumption \ref{ass:reg}, and the result follows since $d_z$ is fixed.

For \eqref{eq:omegabar-omegatilde}, its $(j,k)$-th element can be written as 
\begin{align*}
    &\quad \frac{1}{n} \sum_{g \in [G]} \left( z_{[g], j}^\top  e_{[g]}  e_{[g]}^\top  z_{[g], k} - z_{[g], j}^\top \tilde e_{[g]} \tilde e_{[g]}^\top  z_{[g], k} \right) \\
    &= \frac{1}{n} \sum_{g \in [G]} (z_{[g], j}^\top W_{[g]} \hat \gamma_{\tilde e} ) (z_{[g], k}^\top W_{[g]} \hat \gamma_{\tilde e} ) \\
    &- \frac{1}{n} \sum_{g \in [G]} (z_{[g], j}^\top W_{[g]} \hat \gamma_{\tilde e} ) (z_{[g], k}^\top \tilde e_{[g]} )\\
    &- \frac{1}{n} \sum_{g \in [G]} (z_{[g], k}^\top W_{[g]} \hat \gamma_{\tilde e} ) (z_{[g], j}^\top \tilde e_{[g]} ).
\end{align*}
We have
\begin{align*}
    &\quad \left| \frac{1}{n} \sum_{g \in [G]} (z_{[g], j}^\top W_{[g]} \hat \gamma_{\tilde e} ) (z_{[g], k}^\top W_{[g]} \hat \gamma_{\tilde e} ) \right| \\
    &\leqslant \sqrt{\frac{1}{n} \sum_{g \in [G]} (z_{[g], j}^\top W_{[g]} \hat \gamma_{\tilde e} )^2} \times \sqrt{\frac{1}{n} \sum_{g \in [G]} (z_{[g], k}^\top W_{[g]} \hat \gamma_{\tilde e} )^2} \\
    &\leqslant \max_{1 \leq g \leq G} \left\Vert W_{[g]} \hat \gamma_{\tilde e} \right\Vert_2^2 \times \sqrt{\frac{1}{n} \sum_{g \in [G]} z_{[g], j}^\top z_{[g], j}} \times \sqrt{\frac{1}{n} \sum_{g \in [G]} z_{[g], k}^\top z_{[g], k}} \\
    &\leqslant \max_{1 \leq g \leq G} \left\Vert W_{[g]} \hat \gamma_{\tilde e} \right\Vert_2^2 \times \sqrt{v_j^\top \left(\frac{1}{n} \sum_{i \in I_g, g \in [G]} z_{i,g} z_{i,g}^\top \right) v_j} \times \sqrt{v_k^\top \left(\frac{1}{n} \sum_{i \in I_g, g \in [G]} z_{i,g} z_{i,g}^\top \right) v_k} \\
    &= o_P(1)
\end{align*}
by Assumption \ref{ass:reg} and Lemma \ref{lem:4th_moment}, where we use $v_j$ ($v_k$) to denote the $d_z$-dimensional unit vector with $j$-th ($k$-th) element one and other elements zero; we also have
\begin{align*}
    &\quad \left|\frac{1}{n} \sum_{g \in [G]} (z_{[g], j}^\top W_{[g]} \hat \gamma_{\tilde e} ) (z_{[g], k}^\top \tilde e_{[g]} ) \right| \\
    &\leqslant \max_{1 \leq g \leq G} \left\Vert W_{[g]} \hat \gamma_{\tilde e} \right\Vert_2 \times \sqrt{\frac{1}{n} \sum_{g \in [G]} z_{[g], j}^\top z_{[g], j}} \times \sqrt{\frac{1}{n} \sum_{g \in [G]} z_{[g], k}^\top \tilde e_{[g]} \tilde e_{[g]}^\top z_{[g], k}} \\
    &= o_P(1),
\end{align*}
and by using the same argument,
\begin{align*}
    \left| \frac{1}{n} \sum_{g \in [G]} (z_{[g], k}^\top W_{[g]} \hat \gamma_{\tilde e} ) (z_{[g], j}^\top \tilde e_{[g]} ) \right| = o_P(1).
\end{align*}

For \eqref{eq:omegacute-omegabar}, its $(j,k)$-th element can be written as 
\begin{align*}
    &\quad \frac{1}{n} \sum_{g \in [G]} \left( z_{[g], j}^\top \acute e_{[g]} \acute e_{[g]}^\top  z_{[g], k} - z_{[g], j}^\top e_{[g]} e_{[g]}^\top  z_{[g], k} \right) \\
    &= \frac{\acute \Delta^2}{n} \sum_{g \in [G]} (z_{[g], j}^\top X_{[g]}) (z_{[g], k}^\top X_{[g]}) \\
    &- \frac{\acute \Delta}{n} \sum_{g \in [G]} (z_{[g], j}^\top X_{[g]}) (z_{[g], k}^\top e_{[g]} ) \\
    &- \frac{\acute \Delta}{n} \sum_{g \in [G]} (z_{[g], k}^\top X_{[g]}) (z_{[g], j}^\top e_{[g]} ),
\end{align*}
and note that
\begin{align*}
    \frac{1}{n} \sum_{g \in [G]} (z_{[g], j}^\top X_{[g]}) (z_{[g], k}^\top X_{[g]}) & \leqslant \sqrt{\frac{1}{n} \sum_{g \in [G]} z_{[g], j}^\top z_{[g], j} X_{[g]}^\top X_{[g]}} \times \sqrt{\frac{1}{n} \sum_{g \in [G]} z_{[g], k}^\top z_{[g], k} X_{[g]}^\top X_{[g]}} \\
    & = O_P(1),
\end{align*}
since $\max_{g \in [G]} \mathbb E \left( X_{[g]}^\top X_{[g]} \right) = O(1)$ by Lemma \ref{lem:4th_moment}; the other two terms can be handled similarly.

\vspace{5mm}
\noindent \textbf{Step 2: Consistency of $\acute \Psi$.} By Assumption \ref{ass:low_id}, we have
\begin{align} \label{eq:low_id}
    \frac{1}{r_n} z^\top X = \frac{1}{r_n} z^\top \Pi + \frac{1}{r_n} z^\top V = \frac{1}{r_n} z^\top \Pi + o_P(1),
\end{align} 
and note that
\begin{align} 
    \frac{1}{\lambda_n} \hat A_n &= \frac{1}{\lambda_n} A_n + \frac{1}{\lambda_n} (\hat A_n - A_n) \notag \\
    &= \frac{1}{\lambda_n} A_n + (\frac{1}{\lambda_n} A_n)^{1/2} (A_n^{-1/2} \hat A_n A_n^{-1/2} - I_{d_{z}} ) (\frac{1}{\lambda_n} A_n)^{1/2} \notag \\
    &= \frac{1}{\lambda_n} A_n + o_P(1), \label{eq:low_An}
\end{align}
where $\lambda_n = \lambda_{\max}(A_n)$. In addition, we have
\begin{align*}
\frac{1}{(\frac{1}{r_n} \Pi^\top z) (\frac{1}{\lambda_n} A_n) (\frac{1}{n} \Omega) (\frac{1}{\lambda_n} A_n) (\frac{1}{r_n} z^\top \Pi)} \leq C.
\end{align*}
Therefore, we have
\begin{align*}
    \frac{\acute \Psi}{\Psi} = \frac{X^\top z \hat A_n \acute \Omega \hat A_n z^\top X}{\Pi^\top  z A_n \Omega A_n z^\top \Pi} = \frac{(\frac{1}{r_n}X^\top z) (\frac{1}{\lambda_n} \hat A_n) (\frac{1}{n} \acute \Omega) (\frac{1}{\lambda_n} \hat A_n) (\frac{1}{r_n} z^\top X)}{(\frac{1}{r_n} \Pi^\top z) (\frac{1}{\lambda_n} A_n) (\frac{1}{n} \Omega) (\frac{1}{\lambda_n} A_n) (\frac{1}{r_n} z^\top \Pi)} = 1 + o_P(1)
\end{align*}
by Step 1.

\vspace{5mm}
\noindent \textbf{Step 3: Consistency of $\acute \Sigma$.} 
Define \begin{align*}
    \bar \Sigma = \mathbb E \left( \tilde \Pi^\top (Q - \bar Q) \tilde e \right)^2 + \mathbb E \left( \tilde V^\top (Q - \bar Q) \tilde e \right)^2,
\end{align*}
and recall
\begin{align*}
    \Sigma & = \mathbb E \left(\hat \Pi^\top \tilde e \right)^2 + \mathbb E \left( \tilde V^\top (P - \bar P) \tilde e \right)^2 = \mathbb E \left(\tilde \Pi^\top Q \tilde e \right)^2 + \mathbb E \left( \tilde V^\top (P - \bar P) \tilde e \right)^2.
\end{align*} 
Using the notation $\left\Vert X \right\Vert_{\mathbb P,2} = \left(\mathbb E\left(X^2 \right)\right)^{1/2}$, we have
\begin{align*}
\left\vert \frac{\bar \Sigma - \Sigma}{ \Sigma} \right\vert &\leq \left\vert \frac{\left\Vert \tilde \Pi^\top Q \tilde e \right\Vert_{\mathbb P,2}^2 - \left\Vert \tilde \Pi^\top (Q-\bar Q) \tilde e \right\Vert_{\mathbb P,2}^2}{ \Sigma} \right\vert + \left\vert \frac{\left\Vert \tilde V^\top (P - \bar P) \tilde e \right\Vert_{\mathbb P,2}^2 - \left\Vert \tilde V^\top (Q - \bar Q) \tilde e \right\Vert_{\mathbb P,2}^2}{ \Sigma} \right\vert \\
&\leq \left( \frac{2\left\Vert \tilde \Pi^\top Q \tilde e \right\Vert_{\mathbb P,2} \left\Vert \tilde \Pi^\top \bar Q \tilde e \right\Vert_{\mathbb P,2} + \left\Vert \tilde \Pi^\top \bar Q \tilde e \right\Vert_{\mathbb P,2}^2}{ \Sigma} \right) + \left\vert \frac{\left\Vert \tilde V^\top (P - \bar P) \tilde e \right\Vert_{\mathbb P,2}^2 - \left\Vert \tilde V^\top (Q - \bar Q) \tilde e \right\Vert_{\mathbb P,2}^2}{ \Sigma} \right\vert \\
&\leq \left( \frac{2\left\Vert \tilde \Pi^\top Q \tilde e \right\Vert_{\mathbb P,2} \left\Vert \tilde \Pi^\top \bar Q \tilde e \right\Vert_{\mathbb P,2} + \left\Vert \tilde \Pi^\top \bar Q \tilde e \right\Vert_{\mathbb P,2}^2}{\Pi^\top \Pi} \right) + \left\vert \frac{\left\Vert \tilde V^\top (P - \bar P) \tilde e \right\Vert_{\mathbb P,2}^2 - \left\Vert \tilde V^\top (Q - \bar Q) \tilde e \right\Vert_{\mathbb P,2}^2}{K} \right\vert \\
& = o(1),
\end{align*}
where the first inequality is by triangular inequality, the second inequality is by Cauchy-Schwarz inequality, the third inequality is by Lemma \ref{lem:linear_quad_form_1}, and the last equality holds because 
\begin{align*}
    \mathbb E \left(\tilde \Pi^\top \bar Q \tilde e \right)^2 \leqslant C \left\Vert \bar Q \right\Vert_{op}^2 \tilde \Pi^\top \tilde \Pi = o(\Pi^\top \Pi)
\end{align*}
by Lemma \ref{lem:PandQ}, and
\begin{align*}
    \frac{1}{K} \mathbb E \left( \tilde V^\top (Q - \bar Q) \tilde e \right)^2 &= \frac{1}{K} \sum_{g,h \in [G]^2, g \neq h} \mathbb E \left[ \left( \tilde V_{[g]}^\top Q_{[g,h]} \tilde e_{[h]} \right)^2 + \left( \tilde V_{[g]}^\top Q_{[g,h]} \tilde e_{[h]} \right) \left( \tilde V_{[h]}^\top Q_{[h,g]} \tilde e_{[g]} \right) \right] \\
    &= \frac{1}{K} \sum_{g,h \in [G]^2, g \neq h} \mathbb E \left[ \left( \tilde V_{[g]}^\top P_{[g,h]} \tilde e_{[h]} \right)^2 + \left( \tilde V_{[g]}^\top P_{[g,h]} \tilde e_{[h]} \right) \left( \tilde V_{[h]}^\top P_{[h,g]} \tilde e_{[g]} \right) \right] + o(1) \\
    &= \frac{1}{K} \mathbb E \left( \tilde V^\top (P - \bar P) \tilde e \right)^2 + o(1)
\end{align*}
by Lemma \ref{lem:quad_form_2}. Therefore, we have
\begin{align*}
\frac{ \bar \Sigma }{\Sigma} \rightarrow 1.     
\end{align*}



In addition, note that we can write
\begin{align*}
    \bar \Sigma = \sum_{g\in [G]} \mathbb E \left(\sum_{h \in [G], h \neq g} \tilde X_{[h]}^\top Q_{[h,g]} \tilde e_{[g]} \right)^2 + \sum_{g,h \in [G]^2, g \neq h} \mathbb E \left(\tilde X_{[g]}^\top Q_{[g,h]} \tilde e_{[h]} \right) \left(\tilde X_{[h]}^\top Q_{[h,g]} \tilde e_{[g]} \right),
\end{align*}
as in \cite{Chao(2012)}, and thus
\begin{align*}
    \frac{\acute \Sigma - \bar \Sigma}{\bar \Sigma} &= \underbrace{ \frac{1}{\bar \Sigma} \left(\sum_{g\in [G]} \left(\sum_{h \in [G], h \neq g} \tilde X_{[h]}^\top Q_{[h,g]} \tilde e_{[g]} \right)^2 -  \sum_{g\in [G]} \mathbb E \left(\sum_{h \in [G], h \neq g} \tilde X_{[h]}^\top Q_{[h,g]} \tilde e_{[g]} \right)^2 \right)}_{R_{14}}  \\
    &+  \underbrace{ \frac{1}{\bar \Sigma} \begin{pmatrix}
    &    \sum_{g,h \in [G]^2, g \neq h} \left( \tilde X_{[g]}^\top Q_{[g,h]} \tilde e_{[h]} \right) \left( \tilde X_{[h]}^\top Q_{[h,g]} \tilde e_{[g]} \right)\\
    & - \sum_{g,h \in [G]^2, g \neq h} \mathbb E \left(\tilde X_{[g]}^\top Q_{[g,h]} \tilde e_{[h]} \right) \left(\tilde X_{[h]}^\top Q_{[h,g]} \tilde e_{[g]} \right) 
    \end{pmatrix} }_{R_{15}}\\
    &+  \underbrace{ \frac{1}{\bar \Sigma} \left( \sum_{g\in [G]} \left(\sum_{h \in [G], h \neq g} \tilde X_{[h]}^\top Q_{[h,g]} \acute e_{[g]} \right)^2 -\sum_{g\in [G]} \left(\sum_{h \in [G], h \neq g} \tilde X_{[h]}^\top Q_{[h,g]} \tilde e_{[g]} \right)^2 \right)}_{R_{16}} \\
    &+  \underbrace{ \frac{1}{\bar \Sigma} \begin{pmatrix}
      &  \sum_{g,h \in [G]^2, g \neq h} \left( \tilde X_{[g]}^\top Q_{[g,h]} \acute e_{[h]} \right) \left( \tilde X_{[h]}^\top Q_{[h,g]} \acute e_{[g]} \right) \\
    &  - \sum_{g,h \in [G]^2, g \neq h} \left( \tilde X_{[g]}^\top Q_{[g,h]} \tilde e_{[h]} \right) \left( \tilde X_{[h]}^\top Q_{[h,g]} \tilde e_{[g]} \right) 
    \end{pmatrix}}_{R_{17}}.
\end{align*}
Note also that, by Lemmas \ref{lem:linear_quad_form_1} and \ref{lem:linear_quad_form_2}, we have $R_{14} = o_P(1)$ and $R_{15} = o_P(1)$. 

For $R_{16}$, we have
\begin{align*}
    R_{16} &= \underbrace{ \frac{1}{\bar \Sigma} \left( \sum_{g\in [G]} \left(\sum_{h \in [G], h \neq g} \tilde X_{[h]}^\top Q_{[h,g]} \acute e_{[g]} \right)^2 -\sum_{g\in [G]} \left(\sum_{h \in [G], h \neq g} \tilde X_{[h]}^\top Q_{[h,g]} e_{[g]} \right)^2 \right) }_{R_{16,1}}\\
    &+ \underbrace{ \frac{1}{\bar \Sigma} \left( \sum_{g\in [G]} \left(\sum_{h \in [G], h \neq g} \tilde X_{[h]}^\top Q_{[h,g]} e_{[g]} \right)^2 -\sum_{g\in [G]} \left(\sum_{h \in [G], h \neq g} \tilde X_{[h]}^\top Q_{[h,g]} \tilde e_{[g]} \right)^2 \right) }_{R_{16,2}}.
\end{align*}
For $R_{16,1}$, we have
\begin{align*}
    R_{16,1} &= \frac{\acute \Delta^2}{\bar \Sigma} \sum_{g\in [G]} \left(\sum_{h \in [G], h \neq g} \tilde X_{[h]}^\top Q_{[h,g]} X_{[g]} \right)^2 \\
    & - \frac{2 \acute \Delta}{\bar \Sigma} \sum_{g\in [G]} \left(\sum_{h \in [G], h \neq g} \tilde X_{[h]}^\top Q_{[h,g]} X_{[g]} \right)\left(\sum_{k \in [G], k \neq g} \tilde X_{[k]}^\top Q_{[k,g]} e_{[g]} \right).
\end{align*}
By Lemma \ref{lem:linear_quad_form_1}, we have
\begin{align*}
    \frac{1}{\bar \Sigma} \sum_{g\in [G]} \left(\sum_{h \in [G], h \neq g} \tilde X_{[h]}^\top Q_{[h,g]} X_{[g]} \right)^2 = O_P(1),
\end{align*}
since
\begin{align*}
    &\frac{1}{\bar \Sigma} \sum_{g\in [G]} \left(\bar \Pi_{[g]}^\top \Pi_{[g]} \right)^2 \leqslant \frac{C\bar \Pi^\top \bar \Pi}{\bar \Sigma} = O(1), \\
    &\frac{1}{\bar \Sigma} \sum_{g\in [G]} \mathbb E \left(\bar \Pi_{[g]}^\top \tilde V_{[g]} \right)^2 \leqslant \frac{C\bar \Pi^\top \bar \Pi}{\bar \Sigma} = O(1), \\
    &\frac{1}{\bar \Sigma} \sum_{g\in [G]} \mathbb E \left(\sum_{h \in [G], h \neq g} \tilde V_{[h]}^\top Q_{[h,g]} \Pi_{[g]} \right)^2 \leqslant \frac{C\Pi^\top \Pi}{\bar \Sigma} = O(1), \\
    &\frac{1}{\bar \Sigma} \sum_{g\in [G]} \mathbb E\left(\sum_{h \in [G], h \neq g} \tilde V_{[h]}^\top Q_{[h,g]} \tilde V_{[g]} \right)^2 \leqslant \frac{C}{\bar \Sigma} \sum_{g, h \in [G]^2, g \neq h} \tr \left( Q_{[g,h]} Q_{[h,g]} \right) = O(1).
\end{align*}
By the same lemma, we also have
\begin{align*}
    \frac{1}{\bar \Sigma} \sum_{g\in [G]} \left(\sum_{h \in [G], h \neq g} \tilde X_{[h]}^\top Q_{[h,g]} X_{[g]} \right)\left(\sum_{k \in [G], k \neq g} \tilde X_{[k]}^\top Q_{[k,g]} e_{[g]} \right) = O_P(1),
\end{align*}
since
\begin{align*}
    \left\vert \frac{1}{\bar \Sigma} \sum_{g \in [G]} \mathbb E \left(\bar \Pi_{[g]}^\top \tilde V_{[g]} \right) \left(\bar \Pi_{[g]}^\top \tilde e_{[g]} \right) \right\vert \leqslant \frac{C\bar \Pi^\top \bar \Pi}{\bar \Sigma} = O(1)
\end{align*}
and 
\begin{align*}
    &\quad \left\vert \frac{1}{\bar \Sigma}  \sum_{g\in [G]} \mathbb E \left(\sum_{h \in [G], h \neq g} \tilde V_{[h]}^\top Q_{[h,g]} \tilde V_{[g]} \right) \left(\sum_{k \in [G], k \neq g} \tilde V_{[k]}^\top Q_{[k,g]} \tilde e_{[g]} \right) \right\vert \\
    & \leqslant \frac{C}{\bar \Sigma} \sum_{g, h \in [G]^2, g \neq h} \tr \left( Q_{[g,h]} Q_{[h,g]} \right) = O(1).
\end{align*}
It follows that $R_{16,1} = o_P(1)$, and since $R_{16,2} = o_P(1)$ by Lemma \ref{lem:linear_quad_form_1}, we have $R_{16} = o_P(1)$.

For $R_{17}$, we have
\begin{align*}
    R_{17} &=  \underbrace{ \frac{1}{\bar \Sigma} \begin{pmatrix}
      &  \sum_{g,h \in [G]^2, g \neq h} \left( \tilde X_{[g]}^\top Q_{[g,h]} \acute e_{[h]} \right) \left( \tilde X_{[h]}^\top Q_{[h,g]} \acute e_{[g]} \right) \\
    &  - \sum_{g,h \in [G]^2, g \neq h} \left( \tilde X_{[g]}^\top Q_{[g,h]} e_{[h]} \right) \left( \tilde X_{[h]}^\top Q_{[h,g]}  e_{[g]} \right) 
    \end{pmatrix}}_{R_{17,1}} \\
    &+  \underbrace{ \frac{1}{\bar \Sigma} \begin{pmatrix}
      &  \sum_{g,h \in [G]^2, g \neq h} \left( \tilde X_{[g]}^\top Q_{[g,h]} e_{[h]} \right) \left( \tilde X_{[h]}^\top Q_{[h,g]} e_{[g]} \right) \\
    &  - \sum_{g,h \in [G]^2, g \neq h} \left( \tilde X_{[g]}^\top Q_{[g,h]} \tilde e_{[h]} \right) \left( \tilde X_{[h]}^\top Q_{[h,g]} \tilde e_{[g]} \right) 
    \end{pmatrix}}_{R_{17,2}}.
\end{align*}
For $R_{17,1}$, we have
\begin{align*}
    R_{17,1} &= \frac{\hat \Delta^2}{\bar \Sigma}\sum_{g,h \in [G]^2, g \neq h} \left( \tilde X_{[g]}^\top Q_{[g,h]} X_{[h]} \right) \left( \tilde X_{[h]}^\top Q_{[h,g]} X_{[g]} \right) \\
    &-  \frac{2 \hat \Delta}{\bar \Sigma}\sum_{g,h \in [G]^2, g \neq h} \left( \tilde X_{[g]}^\top Q_{[g,h]} X_{[h]} \right) \left( \tilde X_{[h]}^\top Q_{[h,g]} e_{[g]} \right).
\end{align*}
By Lemma \ref{lem:linear_quad_form_2}, we have
\begin{align*}
    \frac{1}{\bar \Sigma}\sum_{g,h \in [G]^2, g \neq h} \left( \tilde X_{[g]}^\top Q_{[g,h]} X_{[h]} \right) \left( \tilde X_{[h]}^\top Q_{[h,g]} X_{[g]} \right) = O_P(1),
\end{align*}
since
\begin{align*}
    &\left\vert \frac{1}{\bar \Sigma} \sum_{g,h \in [G]^2, g \neq h} \mathbb E \left( \tilde X_{[g]}^\top Q_{[g,h]} \tilde V_{[h]} \right) \left( \tilde X_{[h]}^\top Q_{[h,g]} \tilde V_{[g]} \right) \right\vert \leq \frac{C}{\bar \Sigma} \sum_{g,h \in [G]^2, g \neq h} \tr \left( Q_{[h,g]} Q_{[g,h]} \right) = O(1), \\
    &\left\vert \frac{1}{\bar \Sigma} \sum_{g,h \in [G]^2, g \neq h} \left( \tilde \Pi_{[g]}^\top Q_{[g,h]} \Pi_{[h]} \right) \left( \tilde \Pi_{[h]}^\top Q_{[h,g]} \Pi_{[g]} \right) \right\vert \leq \frac{C \Pi^\top \Pi}{\bar \Sigma} = O(1), \\
    &\left\vert \frac{1}{\bar \Sigma} \sum_{g,h \in [G]^2, g \neq h} \mathbb E \left( \tilde V_{[g]}^\top Q_{[g,h]} \Pi_{[h]} \right) \left( \tilde \Pi_{[h]}^\top Q_{[h,g]} \tilde V_{[g]} \right) \right\vert \leq \frac{C \Pi^\top \Pi}{\bar \Sigma} = O(1).
\end{align*}
By the same lemma, we also have
\begin{align*}
    \frac{1}{\bar \Sigma}\sum_{g,h \in [G]^2, g \neq h} \left( \tilde X_{[g]}^\top Q_{[g,h]} X_{[h]} \right) \left( \tilde X_{[h]}^\top Q_{[h,g]} e_{[g]} \right ) = O_P(1),
\end{align*}
since
\begin{align*}
    &\left\vert \frac{1}{\bar \Sigma} \sum_{g,h \in [G]^2, g \neq h} \mathbb E \left( \tilde X_{[g]}^\top Q_{[g,h]} \tilde V_{[h]} \right) \left( \tilde X_{[h]}^\top Q_{[h,g]} \tilde e_{[g]} \right) \right\vert \leq \frac{C}{\bar \Sigma} \sum_{g,h \in [G]^2, g \neq h} \tr \left( Q_{[h,g]} Q_{[g,h]} \right) = O(1), \\
    &\left\vert \frac{1}{\bar \Sigma} \sum_{g,h \in [G]^2, g \neq h} \mathbb E \left( \tilde V_{[g]}^\top Q_{[g,h]} \Pi_{[h]} \right) \left( \tilde \Pi_{[h]}^\top Q_{[h,g]} \tilde e_{[g]} \right) \right\vert  \leq \frac{C \Pi^\top \Pi}{\bar \Sigma} = O(1).
\end{align*}
It follows that $R_{17,1} = o_P(1)$, and since $R_{17,2} = o_P(1)$ by Lemma \ref{lem:linear_quad_form_2}, we have $R_{17} = o_P(1)$. Combining the results above, we have
\begin{align*}
    \frac{\acute \Sigma - \bar \Sigma}{\bar \Sigma} = o_P(1),
\end{align*}
and the desired result follows from $\Sigma/\bar \Sigma \rightarrow 1$.

\vspace{5mm}
\noindent \textbf{Step 4: Consistency of $\acute \Upsilon$.} We have
\begin{align*}
    \frac{\acute \Upsilon - \Upsilon}{\Upsilon} 
    &= 2 \times \underbrace{\frac{1}{\Upsilon} \left( \sum_{g,h \in [G]^2, g \neq h} \left(\tilde e_{[g]}^\top P_{[g,h]} \tilde e_{[h]} \right)^2 - \sum_{g,h \in [G]^2, g \neq h} \mathbb E \left(\tilde e_{[g]}^\top P_{[g,h]} \tilde e_{[h]} \right)^2 \right)}_{R_{18}} \\
    &+ 2 \times \underbrace{\frac{1}{\Upsilon} \left( \sum_{g,h \in [G]^2, g \neq h} \left( e_{[g]}^\top P_{[g,h]} e_{[h]} \right)^2 - \sum_{g,h \in [G]^2, g \neq h} \left(\tilde e_{[g]}^\top P_{[g,h]} \tilde e_{[h]} \right)^2 \right)}_{R_{19}} \\
    &+ 2 \times \underbrace{\frac{1}{\Upsilon} \left( \sum_{g,h \in [G]^2, g \neq h} \left( \acute e_{[g]}^\top P_{[g,h]} \acute e_{[h]} \right)^2 - \sum_{g,h \in [G]^2, g \neq h} \left( e_{[g]}^\top P_{[g,h]} e_{[h]} \right)^2 \right)}_{R_{20}},
\end{align*}
and note that
\begin{align}
    \Upsilon &= \mathbb V \left( \sum_{g,h \in [G]^2, g \neq h}  \tilde e_{[g]}^\top P_{[g,h]} \tilde e_{[h]} \right) \notag \\
    &=\mathbb E \left( \sum_{g,h \in [G]^2, g \neq h}  \tilde e_{[g]}^\top P_{[g,h]} \tilde e_{[h]} \right)^2 \notag \\
    &= 2\sum_{g,h \in [G]^2, g \neq h} \mathbb E \left( \tilde e_{[g]}^{\top} P_{[g,h]} \tilde e_{[h]} \right)^2 \notag \\
    &= 2\sum_{g,h \in [G]^2, g \neq h} \tr \left[\Omega_g^{\tilde e, \tilde e} P_{[g,h]} \Omega_h^{\tilde e, \tilde e} P_{[h,g]} \right] \notag \\
    &\geqslant \frac{1}{C} \sum_{g,h \in [G]^2, g \neq h} \tr \left[ P_{[g,h]} P_{[h,g]} \right] \notag \\
    &\geqslant \frac{1}{C} K \label{eq:Upsilon}.
\end{align} 
By Lemma \ref{lem:quad_form_2}, we have $R_{18} = o_P(1)$ and $R_{19} = o_P(1)$. For $R_{20}$, it can be written as
\begin{align*}
    R_{20} &=\frac{1}{\Upsilon} \sum_{g,h \in [G]^2, g \neq h} \left( \left( (\acute e_{[g]} - e_{[g]})^\top P_{[g,h]} (\acute e_{[h]} - e_{[h]}) + e_{[g]}^\top P_{[g,h]} (\acute e_{[h]} - e_{[h]}) \right. \right.\\
    &\left. \left. \qquad\qquad\qquad\qquad+ (\acute e_{[g]} - e_{[g]})^\top P_{[g,h]} e_{[h]} + e_{[g]}^\top P_{[g,h]} e_{[h]}\right)^2 - \left(e_{[g]}^\top P_{[g,h]} e_{[h]} \right)^2 \right) \\
    & = \acute \Delta^4 \times \underbrace{\frac{1}{\Upsilon} \sum_{g,h \in [G]^2, g \neq h} \left( X_{[g]}^\top P_{[g,h]} X_{[h]} \right)^2}_{R_{20,1}} \\
    &+ \acute \Delta^2 \times \underbrace{\frac{1}{\Upsilon} \sum_{g,h \in [G]^2, g \neq h} \left( e_{[g]}^\top P_{[g,h]} X_{[h]} \right)^2}_{R_{20,2}} \\
    & + \acute \Delta^2 \times \underbrace{\frac{1}{\Upsilon} \sum_{g,h \in [G]^2, g \neq h} \left( X_{[g]}^\top P_{[g,h]} e_{[h]} \right)^2}_{R_{20,3}} \\ 
    &- 4 \times \acute \Delta^3 \times \underbrace{\frac{1}{\Upsilon} \sum_{g,h \in [G]^2, g \neq h} \left( X_{[g]}^\top P_{[g,h]} X_{[h]} \right) \left( e_{[g]}^\top P_{[g,h]} X_{[h]} \right)}_{R_{20,4}} \\
    &+ 2 \times \acute \Delta^2 \times \underbrace{\frac{1}{\Upsilon} \sum_{g,h \in [G]^2, g \neq h} \left( X_{[g]}^\top P_{[g,h]} X_{[h]} \right) \left(e_{[g]}^\top P_{[g,h]} e_{[h]} \right)}_{R_{20,5}} \\
    & + 2 \times \acute \Delta^2 \times \underbrace{\frac{1}{\Upsilon} \sum_{g,h \in [G]^2, g \neq h} \left( e_{[g]}^\top P_{[g,h]} X_{[h]} \right) \left( X_{[g]}^\top P_{[g,h]} e_{[h]} \right)}_{R_{20,6}} \\
    &- 4 \times \acute \Delta \times \underbrace{\frac{1}{\Upsilon} \sum_{g,h \in [G]^2, g \neq h} \left( e_{[g]}^\top P_{[g,h]} X_{[h]} \right) \left(e_{[g]}^\top P_{[g,h]} e_{[h]} \right)}_{R_{20,7}}.
\end{align*}
By using the same argument as in the proof of Lemma \ref{lem:linear_quad_form_2}, we can show that
\begin{align*}
    R_{20,i} = O_P(1), \quad i = 1, \dots, 7,
\end{align*}
whence $R_{20} = o_P(1)$. Combining the results above, we have
\begin{align*}
    \frac{\acute \Upsilon - \Upsilon}{\Upsilon} = o_P(1).
\end{align*}
This concludes the proof. $\hfill\qedsymbol$

\subsection{Proof of Lemma \ref{lem:betahat}}
\label{sec:pf_lem_8}
We first introduce some notation. Denote
\begin{align*}
    \dot \Phi_1 &=  (X^\top z \hat A_n z^\top X)^{-1} (X^\top z \hat A_n \dot \Omega \hat A_n z^\top X)(X^\top z \hat A_n z^\top X)^{-1}, \\
    \dot \Omega & = \sum_{g \in [G]} \left(z_{[g]}^\top \dot e_{[g]}\right) \left(z_{[g]}^\top \dot e_{[g]}\right)^\top,
\end{align*}
where $\dot e = Y - X \hat \beta_1$, and denote $\hat \Delta_1 = \hat \beta_1 - \beta$. In addition, denote
\begin{align*}
    \ddot \Phi_2 &= \left( X^\top (P - \bar P) X \right)^{-1} \ddot \Sigma \left( X^\top (P - \bar P) X \right)^{-1}, \\
    \ddot \Sigma &= \sum_{g\in [G]} \left(\sum_{h \in [G], h \neq g} \tilde X_{[h]}^\top Q_{[h,g]} \ddot e_{[g]} \right)^2 + \sum_{g,h \in [G]^2, g \neq h} \left( \tilde X_{[g]}^\top Q_{[g,h]} \ddot e_{[h]} \right) \left( \tilde X_{[h]}^\top Q_{[h,g]} \ddot e_{[g]} \right),
\end{align*}
where $\ddot e = Y - X \hat \beta_2$, and denote $\hat \Delta_2 = \hat \beta_2 - \beta$. Finally, denote $\tilde \omega_1 = \ddot \Phi_2^{1/2} / \left( \dot \Phi_1^{1/2} + \ddot \Phi_2^{1/2} \right) $ and $\tilde \omega_2 = \dot \Phi_1^{1/2} / \left( \dot \Phi_1^{1/2} + \ddot \Phi_2^{1/2} \right)$.

We divide the proof into four steps. In the first step, we show that if Assumptions \ref{ass:reg}-\ref{ass:high} hold, then 
\begin{align*}
        \hat \beta_1 \convP \beta, \quad \dot \Phi_1 = o_P(1), \quad \text{and} \quad (\hat \beta_1 - \beta)/\dot \Phi_1^{1/2} = O_P(1). 
\end{align*}

In the second step, we show that if Assumptions \ref{ass:reg} and \ref{ass:high} hold and $\Pi^\top \Pi/\sqrt{K} \rightarrow \infty$, we have
\begin{align*}
  (\hat  \beta_2 - \beta)^2\Pi^\top \Pi /\sqrt{K} = o_P(1)  \quad \text{and} \quad \ddot \Phi_2 \Pi^\top \Pi /\sqrt{K} = o_P(1). 
\end{align*}

In the third step, we show that if Assumptions \ref{ass:reg}-\ref{ass:high} hold and $\Pi^\top \Pi/\sqrt{K} = O(1)$, then 
\begin{align*}
\hat  \beta_2 - \beta = O_P(1)  \quad \text{and} \quad 1/\ddot \Phi_2^{1/2} = O_P(1).   
\end{align*}

In the last step, we show that if Assumptions \ref{ass:reg}, \ref{ass:low_weak} and \ref{ass:high} hold, then 
\begin{align*}
\hat \beta_1 - \beta = O_P(1) \quad \text{and} \quad 1/\dot \Phi_1^{1/2} = O_P(1).     
\end{align*}

Consequently, if Assumptions \ref{ass:reg}-\ref{ass:high} hold and $\Pi^\top \Pi/\sqrt{K} \rightarrow \infty$, by the results in Steps 1 and 2, we have
\begin{align*}
    (\hat \beta - \beta)^2 \Pi^\top \Pi /\sqrt{K} & \leq \left( \frac{\ddot \Phi_2^{1/2}}{\dot \Phi_1^{1/2} + \ddot \Phi_2^{1/2}}\right)^2 \times (\hat \beta_1 - \beta)^2 \Pi^\top \Pi /\sqrt{K} \\
    & + \left( \frac{\dot \Phi_1^{1/2}}{\dot \Phi_1^{1/2} + \ddot \Phi_2^{1/2}}\right)^2 \times (\hat \beta_2 - \beta)^2 \Pi^\top \Pi /\sqrt{K} \\
    & \leq  \frac{(\hat \beta_1 - \beta)^2 }{\dot \Phi_1}  \ddot \Phi_2  \Pi^\top \Pi /\sqrt{K} + o_P(1)
    = o_P(1).
\end{align*}

If Assumptions \ref{ass:reg}-\ref{ass:high} hold and $\Pi^\top \Pi/\sqrt{K} = O(1)$, then by results in Steps 1 and 3, we have
\begin{align*}
    (\hat \beta - \beta)^2 \Pi^\top \Pi /\sqrt{K} & \leq \left( \frac{\ddot \Phi_2^{1/2}}{\dot \Phi_1^{1/2} + \ddot \Phi_2^{1/2}}\right)^2 \times (\hat \beta_1 - \beta)^2 \Pi^\top \Pi /\sqrt{K} \\
    & + \left( \frac{\dot \Phi_1^{1/2}}{\dot \Phi_1^{1/2} + \ddot \Phi_2^{1/2}}\right)^2 \times (\hat \beta_2 - \beta)^2 \Pi^\top \Pi /\sqrt{K} \\
    & = (\hat \beta_1 - \beta)^2  \times O_P(1) + \dot \Phi_1 \times O_P(1)= o_P(1).
\end{align*}

If Assumptions \ref{ass:reg}, \ref{ass:low_weak} and \ref{ass:high} hold and $\Pi^\top \Pi/\sqrt{K} \rightarrow \infty$, by results in Steps 2 and 4, we have
\begin{align*}
   (\hat \beta - \beta)^2 \Pi^\top \Pi /\sqrt{K} & \leq \left( \frac{\ddot \Phi_2^{1/2}}{\dot \Phi_1^{1/2} + \ddot \Phi_2^{1/2}}\right)^2 \times (\hat \beta_1 - \beta)^2 \Pi^\top \Pi /\sqrt{K} \\
    & + \left( \frac{\dot \Phi_1^{1/2}}{\dot \Phi_1^{1/2} + \ddot \Phi_2^{1/2}}\right)^2 \times (\hat \beta_2 - \beta)^2 \Pi^\top \Pi /\sqrt{K} \\
    & \leq \frac{ (\hat \beta_1 - \beta)^2 }{\dot \Phi_1}  \ddot \Phi_2 \Pi^\top \Pi /\sqrt{K} + o_P(1) =o_P(1).
\end{align*}

Therefore, we have established the desired results. Next, we focus on proving results in Steps 1--4.

\vspace{5mm}
\noindent \begin{large} \textbf{Step 1: Assumptions \ref{ass:reg}-\ref{ass:high} hold} \end{large}

We have
\begin{align*}
    \hat \beta_1 - \beta = \frac{X^\top z \hat A_n z^\top e}{X^\top z \hat A_n z^\top X}.
\end{align*}
By (\ref{eq:low_id}) and (\ref{eq:low_An}), we have
\begin{align} \label{eq:low_strong}
    \frac{X^\top z \hat A_n z^\top X}{\Pi^\top z A_n z^\top \Pi} = \frac{(\frac{1}{r_n} X^\top z) (\frac{1}{\lambda_n} \hat A_n) (\frac{1}{r_n} z^\top X)}{(\frac{1}{r_n} \Pi^\top z) (\frac{1}{\lambda_n} A_n) (\frac{1}{r_n} z^\top \Pi)} = 1 + o_P(1).
\end{align}
In addition, we have
\begin{align*}
\frac{X^\top z \hat A_n z^\top e}{\sqrt{\Pi^\top z A_n \Omega A_n z^\top \Pi}} &= \frac{(\frac{1}{r_n}X^\top z) (\frac{1}{\lambda_n} \hat A_n) (\frac{1}{\sqrt n} z^\top e)}{\sqrt{(\frac{1}{r_n} \Pi^\top z) (\frac{1}{\lambda_n} A_n) (\frac{1}{n} \Omega)  (\frac{1}{\lambda_n} A_n) (\frac{1}{r_n} z^\top \Pi)}} \\
&= \frac{(\frac{1}{r_n}\Pi^\top z) (\frac{1}{\lambda_n} A_n) (\frac{1}{\sqrt n}z^\top \tilde e) + o_P(1)}{\sqrt{(\frac{1}{r_n} \Pi^\top z) (\frac{1}{\lambda_n} A_n) (\frac{1}{n} \Omega)  (\frac{1}{\lambda_n} A_n) (\frac{1}{r_n} z^\top \Pi)}}  \\
&= \frac{\Pi^\top z A_n z^\top \tilde e}{\sqrt{\Pi^\top z A_n \Omega A_n z^\top \Pi}} + o_{p}(1),
\end{align*} 
where the last equality holds because by Assumptions \ref{ass:reg} and \ref{ass:low_id},
\begin{align*}
    \left(\frac{1}{r_n} \Pi^\top z\right) \left(\frac{1}{\lambda_n} A_n\right) \left(\frac{1}{n} \Omega\right)  \left(\frac{1}{\lambda_n} A_n\right) \left(\frac{1}{r_n} z^\top \Pi \right) \geqslant \lambda_{\min} \left( \left(\frac{1}{\lambda_n} A_n \right) \left( \frac{1}{n} \Omega \right)  \left( \frac{1}{\lambda_n} A_n \right) \right) \geqslant \frac{1}{C}.
\end{align*}
Combining the above results and recalling
\begin{align*}
\Phi_1 & =  (\Pi^\top z A_n z^\top \Pi)^{-1} (\Pi^\top z A_n \Omega A_n z^\top \Pi)(\Pi^\top z A_n z^\top \Pi)^{-1},
\end{align*}
we have
\begin{align*}
    \frac{\hat \beta_1 - \beta}{\sqrt{\Phi_1}} = \frac{\Pi^\top z A_n z^\top \Pi}{X^\top z \hat A_n z^\top X} \times \frac{X^\top z \hat A_n z^\top e}{\sqrt{\Pi^\top z A_n \Omega A_n z^\top \Pi}} = O_P(1),
\end{align*}
and
\begin{align}
    \Phi_1 = \frac{\Pi^\top z A_n \Omega A_n z^\top \Pi}{\left(\Pi^\top z A_n z^\top \Pi \right)^2} = O\left(\frac{n}{r_n^2}\right) = o(1), \label{eq:Phi1}
\end{align}
which further imply
\begin{align*}
    \hat \beta_1 - \beta &= O_P(\sqrt{\Phi_1}) = o_P(1).
\end{align*}

Given the consistency of $\hat \beta_1$, we can apply Lemma \ref{lem:var_est} to show that $\dot \Phi_1 / \Phi_1 \convP 1$, which further implies that 
\begin{align*}
\dot \Phi_1 \convP 0 \quad \text{and} \quad \frac{(\hat \beta_1 - \beta)^2}{\dot \Phi_1} = O_P(1).    
\end{align*}

\vspace{5mm}
\noindent \begin{large} \textbf{Step 2: Assumptions \ref{ass:reg} and \ref{ass:high} hold, and $\Pi^\top \Pi/\sqrt{K} \rightarrow \infty$} \end{large}

We have
\begin{align*}
    \hat \beta_2 - \beta = \frac{X^\top (P - \bar P) e}{X^\top (P - \bar P) X},
\end{align*}
and
\begin{align}\label{eq:X(P-Pbar)e}
    X^\top (P - \bar P) e = \hat \Pi^\top \tilde e + \tilde V^\top (P - \bar P) \tilde e +  \tilde V^\top P_W \bar P \tilde e + \tilde V^\top \bar P P_W \tilde e - \tilde V^\top P_W \bar P P_W \tilde e,
\end{align}
where we use the fact that $P_W P = P P_W = 0_{n \times n}$ since $Z = M_W \tilde Z$. For the third term on the RHS of \eqref{eq:X(P-Pbar)e}, as $\Sigma \geq C(\Pi^\top \Pi + K)$ by Lemma \ref{lem:linear_quad_form_1}, we have
\begin{align*}
\tilde V^\top P_W \bar P \tilde e = O_P(1) = o_P( \sqrt{\Sigma})
\end{align*}
by Lemma \ref{lem:quad_form_1}. Following the similar argument, we can show that 
\begin{align*}
& \tilde V^\top \bar P P_W \tilde e = o_P( \sqrt{\Sigma}) \quad \text{and} \quad \tilde V^\top P_W \bar P P_W \tilde e   = o_P( \sqrt{\Sigma}). 
\end{align*}
Then, by \eqref{eq:X(P-Pbar)e}, we have
\begin{align}\label{eq:X(P-Pbar)e2}
X^\top (P - \bar P) e = \hat \Pi^\top \tilde e + \tilde V^\top (P - \bar P) \tilde e + o_P(\sqrt{\Sigma}).
\end{align}

In addition, we have 
\begin{align}\label{eq:X(P-Pbar)X}
X^\top (P - \bar P) X & =   \Pi^\top (P - \bar P) \Pi +   2 \hat \Pi^\top \tilde V  + \tilde V^\top M_W (P - \bar P) M_W \tilde V \notag \\
& = \Pi^\top (P - \bar P) \Pi +   2 \hat \Pi^\top \tilde V  + \tilde V^\top (P - \bar P) \tilde V +  2\tilde V^\top P_W (P - \bar P) \tilde V + \tilde V^\top P_W (P - \bar P) P_W \tilde V \notag  \\
& = \Pi^\top (P - \bar P) \Pi +   2 \hat \Pi^\top \tilde V  + \tilde V^\top (P - \bar P) \tilde V -  2\tilde V^\top P_W \bar P \tilde V - \tilde V^\top P_W \bar P P_W \tilde V \notag  \\
& = \Pi^\top (P - \bar P) \Pi +   2 \hat \Pi^\top \tilde V  + \tilde V^\top (P - \bar P) \tilde V  + o_P(\sqrt{\Sigma}) \notag  \\
& =  \Pi^\top (P - \bar P) \Pi  + O_P( \sqrt{\Sigma}),
\end{align}
where the second last equality is by 
\begin{align*}
\tilde V^\top P_W \bar P \tilde V = o_P(\sqrt{\Sigma})  \quad \text{and} \quad \tilde V^\top P_W \bar P P_W \tilde V = o_P(\sqrt{\Sigma})
\end{align*}
due to the same argument above, and the last equality holds because 
\begin{align*}
    \mathbb V \left( \hat \Pi^\top \tilde V \right) = O( \Pi^\top \Pi) = O(\Sigma)
\end{align*}
and 
\begin{align*}
\mathbb V\left( \tilde V^\top (P - \bar P) \tilde V \right) \leq C ||P - \bar P||_F^2 \leq C K = O(\Sigma). 
\end{align*}

Combining \eqref{eq:X(P-Pbar)e2} and \eqref{eq:X(P-Pbar)X}, we have
\begin{align*}
    \hat \beta_2 - \beta = \frac{X^\top (P - \bar P) e}{X^\top (P - \bar P) X} = \frac{\hat \Pi^\top \tilde e + \tilde V^\top (P - \bar P) \tilde e + o_P(\sqrt{\Sigma})}{\Pi^\top (P - \bar P) \Pi + O_P(\sqrt{\Sigma})}.
\end{align*}
In addition, we have
\begin{align}
    \Phi_2 = \frac{\Sigma}{\left(\Pi^\top (P - \bar P) \Pi \right)^2} = O(\frac{\Pi^\top \Pi + K}{\left(\Pi^\top \Pi\right)^2}) = o(1) \label{eq:Phi2}
\end{align}
because as $\Pi^\top \Pi/ \sqrt{K} \rightarrow \infty$, we have $\Pi^\top \Pi \rightarrow \infty$; this also implies that
\begin{align} \label{eq:high_strong}
    \frac{X^\top (P - \bar P) X}{\Pi^\top (P - \bar P) \Pi} = \frac{\Pi^\top (P - \bar P) \Pi  + O_P( \sqrt{\Sigma})}{\Pi^\top (P - \bar P) \Pi} = 1 + o_P(1).
\end{align}

Therefore, we have
\begin{align*}
    \frac{|\hat \beta_2 - \beta|}{\sqrt{\Phi_2}} = \frac{\left(\hat \Pi^\top \tilde e + \tilde V^\top (P - \bar P) \tilde e\right) / \sqrt{\Sigma} + o_P(1)}{1 + O_P\left( \sqrt{\Sigma}/|\Pi^\top (P - \bar P) \Pi | \right)} = O_P(1),
\end{align*}
where we use the fact that 
\begin{align*}
\hat \Pi^\top \tilde e = O_P(\sqrt{\Pi^\top \Pi}) = O_P(\sqrt{\Sigma})
\end{align*}
and 
\begin{align*}
\tilde V^\top (P - \bar P) \tilde e = O_P(\sqrt{K}) = O_P(\sqrt{\Sigma}).     
\end{align*}
This also implies 
\begin{align*}
    (\hat \beta_2 - \beta)^2\Pi^\top \Pi/\sqrt{K}  &= O_P(\Phi_2 \Pi^\top \Pi/\sqrt{K}) \\
    & = O_P\left( \frac{\Pi^\top \Pi + K}{\left(\Pi^\top \Pi\right)^2} \times \frac{\Pi^\top \Pi}{\sqrt{K}}   \right) \\
    & = O_P\left( \frac{1}{\sqrt{K}} + \frac{\sqrt{K}}{\Pi^\top \Pi} \right)=o_P(1).
\end{align*} 

Given the consistency of $\hat \beta_2$, we can apply Lemma \ref{lem:var_est} to show that $\ddot \Phi_2 / \Phi_2 \convP 1$, which further implies that 
\begin{align*}
  \quad \ddot \Phi_2 \Pi^\top \Pi /\sqrt{K} = o_P(1). 
\end{align*}

\vspace{5mm}
\noindent \begin{large} \textbf{Step 3: Assumptions \ref{ass:reg} and \ref{ass:high} hold, and $\Pi^\top \Pi / \sqrt{K}$ is bounded} \end{large}

Note that $\Pi^\top (P - \bar P) \Pi / \sqrt{K}$ is bounded in this case. In addition, let
\begin{align*}
    \Gamma_{\tilde V, \tilde V} &=  \sum_{g,h \in [G]^2, g \neq h} \mathbb E \left[ \left( \tilde V_{[g]}^\top P_{[g,h]} \tilde V_{[h]} \right)^2 + \left( \tilde V_{[g]}^\top P_{[g,h]} \tilde V_{[h]} \right) \left( \tilde V_{[h]}^\top P_{[h,g]} \tilde V_{[g]} \right) \right], \\
    \Gamma_{\tilde V, \tilde e} &= \sum_{g,h \in [G]^2, g \neq h} \mathbb E \left[ \left( \tilde V_{[g]}^\top P_{[g,h]} \tilde e_{[h]} \right)^2 + \left( \tilde V_{[g]}^\top P_{[g,h]} \tilde e_{[h]} \right) \left( \tilde V_{[h]}^\top P_{[h,g]} \tilde e_{[g]} \right) \right], \\
    \Lambda &= \sum_{g,h \in [G]^2, g \neq h} \mathbb E \left[  \left( \tilde V_{[g]}^\top P_{[g,h]} \tilde e_{[h]} \right) \left( \tilde V_{[g]}^\top P_{[g,h]} \tilde V_{[h]} \right) + \left( \tilde V_{[g]}^\top P_{[g,h]} \tilde e_{[h]} \right) \left( \tilde V_{[h]}^\top P_{[h,g]} \tilde V_{[g]} \right)\right],
\end{align*}
then by Assumption \ref{ass:high_lower} we have $1/C \leq \Gamma_{\tilde V, \tilde V}/K \leq C$, $1/C \leq \Gamma_{\tilde V, \tilde e}/K \leq C$ and $|\Lambda|/\sqrt{\Gamma_{\tilde V, \tilde V} \Gamma_{\tilde V, \tilde e}} \leq C < 1$. We shall argue along the subsequence where $\Pi^\top (P - \bar P) \Pi / \sqrt{K} \rightarrow \gamma \in \Re$ and
\begin{align*}
    \frac{1}{K}
    \begin{pmatrix}
        \Gamma_{\tilde V, \tilde V} & \Lambda \\
        \Lambda & \Gamma_{\tilde V, \tilde e}  \\
    \end{pmatrix}
    \rightarrow 
    \begin{pmatrix}
    \Gamma_{11} & \Gamma_{12} \\
    \Gamma_{21} & \Gamma_{22} \\
    \end{pmatrix}
    \equiv \Gamma,
\end{align*}
where $\Gamma > 0$ (in the matrix sense) by $\Gamma_{\tilde V, \tilde V}/K \geq 1/C > 0$, $\Gamma_{\tilde V, \tilde e}/K \geq 1/C > 0$ and $|\Lambda / K| / \sqrt{(\Gamma_{\tilde V, \tilde V}/K) (\Gamma_{\tilde V, \tilde e}/K)} \leq C < 1$. 
    

By Assumption \ref{ass:high}, we have
\begin{align*}
    \begin{pmatrix}
        \frac{1}{\sqrt{K}} \tilde V^\top (P- \bar P) \tilde V  \\
        \frac{1}{\sqrt{K}} \tilde V^\top (P- \bar P) \tilde e 
    \end{pmatrix} \convD \N \left( \begin{pmatrix}
        0 \\ 0
    \end{pmatrix}, \begin{pmatrix}
    \Gamma_{11} & \Gamma_{12} \\
    \Gamma_{21} & \Gamma_{22}
\end{pmatrix} \right),
\end{align*}
which can be proved by following the same steps as in the proof of Lemma \ref{lem:limit_dist}. 

In addition, by \eqref{eq:X(P-Pbar)X}, we have
\begin{align*}
    \frac{1}{\sqrt{K}} X^\top (P - \bar P) X & = \frac{1}{\sqrt{K}} \Pi^\top (P - \bar P) \Pi +   \frac{2}{\sqrt{K}} \hat \Pi^\top \tilde V  \\
    & + \frac{1}{\sqrt{K}} \tilde V^\top (P - \bar P) \tilde V -  \frac{2}{\sqrt{K}} \tilde V^\top P_W \bar P \tilde V - \frac{1}{\sqrt{K}} \tilde V^\top P_W \bar P P_W \tilde V \\
    & \convD \N(\gamma,\Gamma_{11}),
\end{align*}
where we use the facts that 
\begin{align*}
    \mathbb V \left( \frac{1}{\sqrt{K}}\hat \Pi \tilde V \right) = O\left( \frac{\Pi^\top \Pi}{K} \right) = o(1), 
\end{align*}
\begin{align*}
\frac{2}{\sqrt{K}} \tilde V^\top P_W \bar P \tilde V = O_P\left(1/\sqrt{K} \right) = o_P(1),   \quad \text{and} \quad \frac{1}{\sqrt{K}} \tilde V^\top P_W \bar P P_W \tilde V = O_P\left(1/\sqrt{K} \right) = o_P(1).
\end{align*}
This implies 
\begin{align}\label{eq:X(P-Pbar)X_inverse}
  \frac{1}{\frac{1}{\sqrt{K}} X^\top (P - \bar P) X} = O_P(1).   
\end{align}

Similarly, by \eqref{eq:X(P-Pbar)e}, we have
\begin{align*}
\frac{1}{\sqrt{K}} X^\top (P - \bar P) e & = \frac{1}{\sqrt{K}} \hat \Pi^\top \tilde e + \frac{1}{\sqrt{K}} \tilde V^\top (P - \bar P) \tilde e \\
& + \frac{1}{\sqrt{K}} \tilde V^\top P_W \bar P \tilde e + \frac{1}{\sqrt{K}} \tilde V^\top \bar P P_W \tilde e -  \frac{1}{\sqrt{K}} \tilde V^\top P_W \bar P P_W \tilde e \\
&  = \frac{1}{\sqrt{K}} \tilde V^\top (P - \bar P) \tilde e + o_P(1) = O_P(1),
\end{align*}
which further implies 
\begin{align*}
    \hat \Delta_2 = \frac{X^\top (P - \bar P) e}{X^\top (P - \bar P) X} = \frac{\frac{1}{\sqrt{K}}X^\top (P - \bar P) e}{\frac{1}{\sqrt{K}}X^\top (P - \bar P) X} = O_P(1).
\end{align*}

Next, we analyze $\ddot \Phi_2$. Note that 
\begin{align}\label{eq:tildeSigma}
    \frac{1}{K} \ddot \Sigma = \frac{1}{K} \sum_{g\in [G]} \left(\sum_{h \in [G], h \neq g} \tilde X_{[h]}^\top Q_{[h,g]} \ddot e_{[g]} \right)^2 + \frac{1}{K} \sum_{g,h \in [G]^2, g \neq h} \left( \tilde X_{[g]}^\top Q_{[g,h]} \ddot e_{[h]} \right) \left( \tilde X_{[h]}^\top Q_{[h,g]} \ddot e_{[g]} \right),
\end{align}
where $\ddot e = Y - X \hat \beta_2$. For the first term on the RHS of \eqref{eq:tildeSigma}, we have
\begin{align*}
    &\quad \frac{1}{K} \sum_{g\in [G]} \left(\sum_{h \in [G], h \neq g} \tilde X_{[h]}^\top Q_{[h,g]} \ddot e_{[g]} \right)^2 \\
    &= \underbrace{ \frac{1}{K} \sum_{g\in [G]} \left(\sum_{h \in [G], h \neq g} \tilde X_{[h]}^\top Q_{[h,g]} \tilde e_{[g]} \right)^2}_{R_{21}} \\
    & + \underbrace{  \frac{1}{K} \left( \sum_{g\in [G]} \left(\sum_{h \in [G], h \neq g} \tilde X_{[h]}^\top Q_{[h,g]} e_{[g]} \right)^2 - \sum_{g\in [G]} \left(\sum_{h \in [G], h \neq g} \tilde X_{[h]}^\top Q_{[h,g]} \tilde e_{[g]} \right)^2 \right)}_{R_{22}} \\
    & + \underbrace{ \frac{1}{K} \left( \sum_{g\in [G]} \left(\sum_{h \in [G], h \neq g} \tilde X_{[h]}^\top Q_{[h,g]} \ddot e_{[g]} \right)^2 - \sum_{g\in [G]} \left(\sum_{h \in [G], h \neq g} \tilde X_{[h]}^\top Q_{[h,g]} e_{[g]} \right)^2 \right)}_{R_{23}}.
\end{align*}
By the fact that $\Pi^\top \Pi / K = o(1)$, we have
\begin{align*}
    1/C \leq \Sigma/K \leq C
\end{align*}
for some constant $C \in (0,\infty)$, so that we can apply Lemma \ref{lem:linear_quad_form_1} to obtain
\begin{align*}
    R_{21} &= \frac{1}{K} \sum_{g\in [G]} \mathbb E \left(\sum_{h \in [G], h \neq g} \tilde X_{[h]}^\top Q_{[h,g]} \tilde e_{[g]} \right)^2 + o_P(1), \\
    &= \frac{1}{K} \sum_{g\in [G]} \mathbb E \left(\sum_{h \in [G], h \neq g} \tilde V_{[h]}^\top Q_{[h,g]} \tilde e_{[g]} \right)^2 + o_P(1),
\end{align*}
and $R_{22} = o_P(1)$. For $R_{23}$, we have
\begin{align*}
    R_{23} &= \hat \Delta_2^2 \times \underbrace{\frac{1}{K}  \sum_{g\in [G]} \left(\sum_{h \in [G], h \neq g} \tilde X_{[h]}^\top Q_{[h,g]} X_{[g]} \right)^2}_{R_{23,1}} \\
    & - 2 \times \hat \Delta_2 \times \underbrace{\frac{1}{K}  \sum_{g\in [G]} \left(\sum_{h \in [G], h \neq g} \tilde X_{[h]}^\top Q_{[h,g]} X_{[g]} \right) \left(\sum_{k \in [G], k \neq g} \tilde X_{[k]}^\top Q_{[k,g]} e_{[g]} \right)}_{R_{23,2}}.
\end{align*}
By Lemma \ref{lem:linear_quad_form_1} and the fact that $\Pi^\top \Pi / K = o(1)$, we have
\begin{align*}
    R_{23,1} &= \frac{1}{K} \sum_{g\in [G]} \mathbb E\left(\sum_{h \in [G], h \neq g} \tilde V_{[h]}^\top Q_{[h,g]} \tilde V_{[g]} \right)^2 + o_P(1), \\
    R_{23,2} &= \frac{1}{K}  \sum_{g\in [G]} \mathbb E \left(\sum_{h \in [G], h \neq g} \tilde V_{[h]}^\top Q_{[h,g]} \tilde V_{[g]} \right) \left(\sum_{k \in [G], k \neq g} \tilde V_{[k]}^\top Q_{[k,g]} \tilde e_{[g]} \right) + o_P(1).
\end{align*}
It follows that
\begin{align}\label{eq:tildeSigma1}
    &\quad \frac{1}{K} \sum_{g\in [G]} \left(\sum_{h \in [G], h \neq g} \tilde X_{[h]}^\top Q_{[h,g]} \ddot e_{[g]} \right)^2 \notag \\
    &= \frac{1}{K} \sum_{g\in [G]} \mathbb E \left(\sum_{h \in [G], h \neq g} \tilde V_{[h]}^\top Q_{[h,g]} \tilde e_{[g]} \right)^2 \notag  \\
    &+ \frac{\hat \Delta_2^2}{K} \sum_{g\in [G]} \mathbb E\left(\sum_{h \in [G], h \neq g} \tilde V_{[h]}^\top Q_{[h,g]} \tilde V_{[g]} \right)^2 \notag  \\
    &- \frac{2 \hat \Delta_2}{K}  \sum_{g\in [G]} \mathbb E \left(\sum_{h \in [G], h \neq g} \tilde V_{[h]}^\top Q_{[h,g]} \tilde V_{[g]} \right) \left(\sum_{k \in [G], k \neq g} \tilde V_{[k]}^\top Q_{[k,g]} \tilde e_{[g]} \right) + o_P(1),
\end{align}
since $\hat \Delta_2 = O_P(1)$. For the second term on the RHS of \eqref{eq:tildeSigma}, we have
\begin{align*}
     &\quad \frac{1}{K} \sum_{g,h \in [G]^2, g \neq h} \left( \tilde X_{[g]}^\top Q_{[g,h]} \ddot e_{[h]} \right) \left( \tilde X_{[h]}^\top Q_{[h,g]} \ddot e_{[g]} \right) \\
     &= \underbrace{\frac{1}{K} \sum_{g,h \in [G]^2, g \neq h} \left( \tilde X_{[g]}^\top Q_{[g,h]} \tilde e_{[h]} \right) \left( \tilde X_{[h]}^\top Q_{[h,g]} \tilde e_{[g]} \right)}_{R_{24}} \\
     &+ \underbrace{ \frac{1}{K} \begin{pmatrix}
      &  \sum_{g,h \in [G]^2, g \neq h} \left( \tilde X_{[g]}^\top Q_{[g,h]} e_{[h]} \right) \left( \tilde X_{[h]}^\top Q_{[h,g]} e_{[g]} \right) \\
    &  - \sum_{g,h \in [G]^2, g \neq h} \left( \tilde X_{[g]}^\top Q_{[g,h]} \tilde e_{[h]} \right) \left( \tilde X_{[h]}^\top Q_{[h,g]} \tilde e_{[g]} \right) 
    \end{pmatrix}}_{R_{25}} \\
     &+ \underbrace{ \frac{1}{K} \begin{pmatrix}
      &  \sum_{g,h \in [G]^2, g \neq h} \left( \tilde X_{[g]}^\top Q_{[g,h]} \ddot e_{[h]} \right) \left( \tilde X_{[h]}^\top Q_{[h,g]} \ddot e_{[g]} \right) \\
    &  - \sum_{g,h \in [G]^2, g \neq h} \left( \tilde X_{[g]}^\top Q_{[g,h]} e_{[h]} \right) \left( \tilde X_{[h]}^\top Q_{[h,g]}  e_{[g]} \right) 
    \end{pmatrix}}_{R_{26}}.
\end{align*}
By Lemma \ref{lem:linear_quad_form_2} and the fact that $\Pi^\top \Pi / K = o(1)$, we have
\begin{align*}
    R_{24} = \frac{1}{K} \sum_{g,h \in [G]^2, g \neq h} \mathbb E \left( \tilde V_{[g]}^\top Q_{[g,h]} \tilde e_{[h]} \right) \left( \tilde V_{[h]}^\top Q_{[h,g]} \tilde e_{[g]} \right) + o_P(1),
\end{align*}
and $R_{25} = o_P(1)$. For $R_{26}$, we have
\begin{align*}
    R_{26} &= \hat \Delta^2 \times \underbrace{\frac{1}{K}\sum_{g,h \in [G]^2, g \neq h} \left( \tilde X_{[g]}^\top Q_{[g,h]} X_{[h]} \right) \left( \tilde X_{[h]}^\top Q_{[h,g]} X_{[g]} \right)}_{R_{26,1}} \\
    &- 2 \times \hat \Delta \times \underbrace{\frac{1}{K}\sum_{g,h \in [G]^2, g \neq h} \left( \tilde X_{[g]}^\top Q_{[g,h]} X_{[h]} \right) \left( \tilde X_{[h]}^\top Q_{[h,g]} e_{[g]} \right)}_{R_{26,2}}.
\end{align*}
By Lemma \ref{lem:linear_quad_form_2} and the fact that $\Pi^\top \Pi / K = o(1)$, we have
\begin{align*}
    R_{26,1} &= \frac{1}{K} \sum_{g,h \in [G]^2, g \neq h} \mathbb E \left( \tilde X_{[g]}^\top Q_{[g,h]} \tilde V_{[h]} \right) \left( \tilde X_{[h]}^\top Q_{[h,g]} \tilde V_{[g]} \right) + o_P(1) \\
    &= \frac{1}{K} \sum_{g,h \in [G]^2, g \neq h} \mathbb E \left( \tilde V_{[g]}^\top Q_{[g,h]} \tilde V_{[h]} \right) \left( \tilde V_{[h]}^\top Q_{[h,g]} \tilde V_{[g]} \right) + o_P(1),
\end{align*}
and
\begin{align*}
    R_{26,2} &= \frac{1}{K} \sum_{g,h \in [G]^2, g \neq h} \mathbb E \left( \tilde X_{[g]}^\top Q_{[g,h]} \tilde V_{[h]} \right) \left( \tilde X_{[h]}^\top Q_{[h,g]} \tilde e_{[g]} \right) + o_P(1) \\
    &= \frac{1}{K} \sum_{g,h \in [G]^2, g \neq h} \mathbb E \left( \tilde V_{[g]}^\top Q_{[g,h]} \tilde V_{[h]} \right) \left( \tilde V_{[h]}^\top Q_{[h,g]} \tilde e_{[g]} \right) + o_P(1).
\end{align*}
It follows that
\begin{align}\label{eq:tildeSigma2}
    &\quad \frac{1}{K} \sum_{g,h \in [G]^2, g \neq h} \left( \tilde X_{[g]}^\top Q_{[g,h]} \ddot e_{[h]} \right) \left( \tilde X_{[h]}^\top Q_{[h,g]} \ddot e_{[g]} \right) \notag \\
    &= \frac{1}{K} \sum_{g,h \in [G]^2, g \neq h} \mathbb E \left( \tilde V_{[g]}^\top Q_{[g,h]} \tilde e_{[h]} \right) \left( \tilde V_{[h]}^\top Q_{[h,g]} \tilde e_{[g]} \right) \notag  \\
    &+ \frac{\hat \Delta_2^2}{K} \sum_{g,h \in [G]^2, g \neq h} \mathbb E \left( \tilde V_{[g]}^\top Q_{[g,h]} \tilde V_{[h]} \right) \left( \tilde V_{[h]}^\top Q_{[h,g]} \tilde V_{[g]} \right) \notag  \\
    &- \frac{2 \hat \Delta_2}{K} \sum_{g,h \in [G]^2, g \neq h} \mathbb E \left( \tilde V_{[g]}^\top Q_{[g,h]} \tilde V_{[h]} \right) \left( \tilde V_{[h]}^\top Q_{[h,g]} \tilde e_{[g]} \right) + o_P(1),
\end{align}
since $\hat \Delta_2 = O_P(1)$. Combining \eqref{eq:tildeSigma}--\eqref{eq:tildeSigma2}, we  obtain that
\begin{align*}
\frac{1}{K} \ddot \Sigma 
    &= \frac{\hat \Delta_2^2}{K} \left( \sum_{g, h\in [G]^2, g \neq h} \mathbb E \left( \tilde V_{[h]}^\top Q_{[h,g]} \tilde V_{[g]} \right)^2 + \sum_{g,h \in [G]^2, g \neq h} \mathbb E \left( \tilde V_{[g]}^\top Q_{[g,h]} \tilde V_{[h]} \right) \left( \tilde V_{[h]}^\top Q_{[h,g]} \tilde V_{[g]} \right)\right) \\
    &+ \frac{1}{K} \left( \sum_{g, h\in [G]^2, g \neq h} \mathbb E \left( \tilde V_{[h]}^\top Q_{[h,g]} \tilde e_{[g]} \right)^2 + \sum_{g,h \in [G]^2, g \neq h} \mathbb E\left( \tilde V_{[g]}^\top Q_{[g,h]} \tilde e_{[h]} \right) \left( \tilde V_{[h]}^\top Q_{[h,g]} \tilde e_{[g]} \right) \right) \\
    & - \frac{2\hat \Delta_2}{K} \left[ \sum_{g, h \in [G]^2, g \neq h} \mathbb E \left(\tilde V_{[h]}^\top Q_{[h,g]} \tilde V_{[g]} \right) \left(\tilde V_{[h]}^\top Q_{[h,g]} \tilde e_{[g]} \right) \right.\\
    &\left. \qquad\qquad\quad+ \sum_{g,h \in [G]^2, g \neq h} \mathbb E\left( \tilde V_{[g]}^\top Q_{[g,h]} \tilde V_{[h]} \right) \left( \tilde V_{[h]}^\top Q_{[h,g]} \tilde e_{[g]} \right) \right]+ o_P(1)  \\
    &=\frac{\hat \Delta_2^2}{K} \mathbb E \left( \tilde V^\top (Q - \bar Q) \tilde V \right)^2 - \frac{2 \hat \Delta_2}{K} \mathbb E \left( \tilde V^\top (Q - \bar Q) \tilde V \right) \left( \tilde V^\top (Q - \bar Q) \tilde e \right) \\
    & + \frac{1}{K} \mathbb E \left( \tilde V^\top (Q - \bar Q) \tilde e \right)^2 + o_P(1).
\end{align*}


In addition, by Lemma \ref{lem:quad_form_2}, we have
\begin{align*}
    \frac{1}{K} \mathbb E \left( \tilde V^\top (Q - \bar Q) \tilde V \right)^2 &= \frac{1}{K} \mathbb E \left( \tilde V^\top (P - \bar P) \tilde V \right)^2 + o(1)  = \Gamma_{11} + o(1), \\
    \frac{1}{K} \mathbb E \left( \tilde V^\top (Q - \bar Q) \tilde e \right)^2 &= \frac{1}{K} \mathbb E \left( \tilde V^\top (P - \bar P) \tilde e \right)^2 + o(1) =  \Gamma_{22} + o(1), \\
    \frac{1}{K} \mathbb E \left( \tilde V^\top (Q - \bar Q) \tilde V \right) \left( \tilde V^\top (Q - \bar Q) \tilde e \right) &= \frac{1}{K} \mathbb E \left( \tilde V^\top (P - \bar P) \tilde V \right) \left( \tilde V^\top (P - \bar P) \tilde e \right) + o(1) = \Gamma_{12} + o(1).
\end{align*}
Combining the above results, we have
\begin{align}\label{eq:Sigmatilde}
    \frac{1}{K} \ddot \Sigma = \hat \Delta_2^2 \Gamma_{11}  - 2\hat \Delta_2 \Gamma_{12} + \Gamma_{22} + o_P(1).
\end{align}
It follows that
\begin{align*}
    \ddot \Phi_2 &= \left( \frac{1}{\sqrt{K}} X^\top (P-\bar P) X \right)^{-2} \left(\frac{1}{K} \ddot \Sigma \right) \\
    &= \frac{(\frac{1}{\sqrt{K}} X^\top (P-\bar P) e)^2}{(\frac{1}{\sqrt{K}}X^\top (P-\bar P) X)^4} \times \Gamma_{11} - 2 \frac{(\frac{1}{\sqrt{K}}X^\top (P-\bar P) e)}{(\frac{1}{\sqrt{K}}X^\top (P-\bar P)X)^3} \times \Gamma_{12} \\
    & + \frac{1}{(\frac{1}{\sqrt{K}}X^\top (P-\bar P) X)^2} \times \Gamma_{22} + o_P(1) \\
    & \convD \bar \Phi_2,
\end{align*}
where the second equality is by \eqref{eq:X(P-Pbar)X_inverse} and \eqref{eq:Sigmatilde}, and $\bar \Phi_2$ is defined as 
\begin{align*}
    \bar \Phi_2 &= \frac{\eta_{\tilde V, \tilde e}^2}{(\eta_{\tilde V, \tilde V}+\gamma)^4} \Gamma_{11} - \frac{2\eta_{\tilde V, \tilde e}}{(\eta_{\tilde V, \tilde V}+\gamma)^3} \Gamma_{12} +  \frac{1}{(\eta_{\tilde V, \tilde V}+\gamma)^2} \Gamma_{22} \\
    &= \frac{1}{(\eta_{\tilde V, \tilde V}+\gamma)^4} \left(\eta_{\tilde V, \tilde e}^2 \Gamma_{11} - 2 \eta_{\tilde V, \tilde e} (\eta_{\tilde V, \tilde V}+\gamma) \Gamma_{12} + (\eta_{\tilde V, \tilde V}+\gamma)^2 \Gamma_{22} \right) \\
    & = \frac{\begin{pmatrix}
        - \eta_{\tilde V, \tilde e} \\
        \eta_{\tilde V, \tilde V}+\gamma
    \end{pmatrix}^\top 
    \begin{pmatrix}
        \Gamma_{11} & \Gamma_{12} \\
        \Gamma_{21} & \Gamma_{22}
    \end{pmatrix}
    \begin{pmatrix}
        - \eta_{\tilde V, \tilde e} \\
        \eta_{\tilde V, \tilde V}+\gamma
    \end{pmatrix}}{(\eta_{\tilde V, \tilde V}+\gamma)^4},
\end{align*}
with
\begin{align*}
\begin{pmatrix}
        \eta_{\tilde V, \tilde V} \\ \eta_{\tilde V, \tilde e} 
    \end{pmatrix} \stackrel{d}{=} \N \left(\begin{pmatrix}
        0 \\ 0
    \end{pmatrix}, \begin{pmatrix}
    \Gamma_{11} & \Gamma_{12} \\
    \Gamma_{21} & \Gamma_{22}
\end{pmatrix} \right).    
\end{align*}
Note that 
\begin{align*}
    \begin{pmatrix}
        - \eta_{\tilde V, \tilde e} \\
        \eta_{\tilde V, \tilde V}+\gamma
    \end{pmatrix}^\top 
    \begin{pmatrix}
        \Gamma_{11} & \Gamma_{12} \\
        \Gamma_{21} & \Gamma_{22}
    \end{pmatrix}
    \begin{pmatrix}
        - \eta_{\tilde V, \tilde e} \\
        \eta_{\tilde V, \tilde V}+\gamma
    \end{pmatrix}
    \geqslant 0,
\end{align*}
and the equality holds if and only if $\eta_{\tilde V, \tilde e} = 0$ and $\eta_{\tilde V, \tilde V}+\gamma = 0$, which has probability zero since $\eta_{\tilde V, \tilde e}$ has a non-degenerate normal distribution and similarly for $\eta_{\tilde V, \tilde V}$. In addition, the denominator of $\bar \Phi_2$ is positive with probability one. Therefore, we have $\bar \Phi_2 > 0 $ with probability one, which implies that 
\begin{align*}
    1/\ddot \Phi_2^{1/2} = O_P(1). 
\end{align*}

\vspace{5mm}
\noindent \begin{large} \textbf{Step 4: Assumptions \ref{ass:reg}, \ref{ass:low_weak} and \ref{ass:high} hold } \end{large}

By Assumptions \ref{ass:low_weak_id} and \ref{ass:low_weak_lower}, we can argue along the subsequence where $z^\top \Pi/\sqrt{n} \rightarrow \pi \in \Re^{d_z}$,
\begin{align*}
    \frac{1}{n}
    \begin{pmatrix}
         \sum_{g \in [G]} \mathbb E \left(z_{[g]}^\top \tilde e_{[g]} \right)\left(z_{[g]}^\top \tilde e_{[g]} \right)^\top &
        \sum_{g \in [G]} \mathbb E \left(z_{[g]}^\top \tilde e_{[g]} \right)\left( z_{[g]}^\top \tilde V_{[g]} \right)^\top \\
        \sum_{g \in [G]} \mathbb E \left(z_{[g]}^\top \tilde V_{[g]} \right)\left(z_{[g]}^\top \tilde e_{[g]} \right)^\top & 
        \sum_{g \in [G]} \mathbb E \left(z_{[g]}^\top \tilde V_{[g]} \right)\left(z_{[g]}^\top \tilde V_{[g]} \right)^\top
    \end{pmatrix} \rightarrow
        \begin{pmatrix}
            \Omega_z^{\tilde e, \tilde e} & \Omega_z^{\tilde e, \tilde V} \\
            \Omega_z^{\tilde V, \tilde e} & \Omega_z^{\tilde V, \tilde V} 
        \end{pmatrix} > 0,
    \end{align*}
    and
    \begin{align*}
        \frac{1}{n} \sum_{g \in [G]} \left(z_{[g]}^\top \Pi_{[g]} \right)\left(z_{[g]}^\top \Pi_{[g]} \right)^\top \rightarrow \Omega_z^{\Pi, \Pi} \geqslant 0,
    \end{align*}
in the matrix sense. In addition, note that since Assumption~\ref{ass:low_An} holds, $A_n/\lambda_n$ has eigenvalues bounded and bounded away from zero, where $\lambda_n = \lambda_{\max}(A_n)$. Therefore, without loss of generality, we assume  that $A_n/ \lambda_n \rightarrow A$ for some non-random positive definite matrix $A$ with eigenvalues bounded and bounded away from zero (otherwise we argue along a further subsequence). 


By Assumption \ref{ass:low_weak}, we have
\begin{align*}
    \begin{pmatrix}
        \frac{1}{\sqrt{n}} z^\top \tilde e \\
        \frac{1}{\sqrt{n}} z^\top \tilde V
    \end{pmatrix} \convD \N \left( 0, \begin{pmatrix}
                \Omega_z^{\tilde e, \tilde e} & \Omega_z^{\tilde e, \tilde V} \\
                \Omega_z^{\tilde V, \tilde e} & \Omega_z^{\tilde V, \tilde V}
            \end{pmatrix}  \right),
\end{align*}
which can be proved by following the same steps as in the proof of Lemma \ref{lem:limit_dist} (see also \cite{Hansen-Lee-2019} and \cite{Djogbenou-Mackinnon-Nielsen-2019}). This also implies that 
\begin{align*}
    \frac{1}{\sqrt{n}} z^\top X \convD \N \left( \pi, \Omega_z^{\tilde V, \tilde V} \right).
\end{align*}
We have
\begin{align*}
    \hat \Delta_1 = \frac{X^\top z \hat A_n z^\top e}{X^\top z \hat A_n z^\top X} = \frac{(\frac{1}{\sqrt{n}}X^\top z) (\frac{1}{\tilde \lambda_n} \hat A_n) (\frac{1}{\sqrt{n}} z^\top \tilde e)}{(\frac{1}{\sqrt{n}}X^\top z) (\frac{1}{\tilde \lambda_n}\hat A_n) (\frac{1}{\sqrt{n}}z^\top X)}
\end{align*}
and
\begin{align*}
    (\frac{1}{\sqrt{n}}X^\top z) (\frac{1}{\tilde \lambda_n} \hat A_n) (\frac{1}{\sqrt{n}} z^\top \tilde e) = O_P(1).
\end{align*}
We also have
\begin{align*}
    \frac{1}{(\frac{1}{\sqrt{n}}X^\top z) (\frac{1}{\tilde \lambda_n}\hat A_n) (\frac{1}{\sqrt{n}}z^\top X)} \leqslant \frac{1}{\xi_n^2},
\end{align*}
where $\xi_n$ is the first element of $(\hat A_n/\tilde \lambda_n)^{1/2} (z^\top X/\sqrt{n})$, and note that $\xi_n \convD \xi$ where $\xi$ has a non-degenerate normal distribution. 
This implies that 
\begin{align*}
    \frac{1}{(\frac{1}{\sqrt{n}}X^\top z) (\frac{1}{\tilde \lambda_n}\hat A_n) (\frac{1}{\sqrt{n}}z^\top X)} = O_P(1),
\end{align*}
whence $\hat \Delta_1 = O_P(1)$.


Next, we analyze $\dot \Phi_1$. Note that
\begin{align*}
    \frac{1}{n} \dot \Omega &= \left( \frac{1}{n} \dot \Omega - \frac{1}{n} \bar \Omega \right) + \left( \frac{1}{n} \bar \Omega - \frac{1}{n} \tilde \Omega \right) + \left( \frac{1}{n} \tilde \Omega - \frac{1}{n} \Omega \right) + \frac{1}{n} \Omega \\
    &= \left( \frac{1}{n} \dot \Omega - \frac{1}{n} \bar \Omega \right) + \Omega_z^{\tilde e, \tilde e} + o_P(1),
\end{align*}
as in the proof of Lemma \ref{lem:var_est}, where $\dot \Omega$ is just $\acute \Omega$ with $\acute \beta = \hat \beta_1$. We have
\begin{align*}
    \frac{1}{n} \tilde \Omega - \frac{1}{n} \bar \Omega 
    &= \hat \Delta_1^2 \times \underbrace{\frac{1}{n} \sum_{g \in [G]} (z_{[g]}^\top X_{[g]}) (X_{[g]}^\top z_{[g]})}_{R_{14}} \\
    &- \hat \Delta_1 \times \underbrace{\frac{1}{n} \sum_{g \in [G]} (z_{[g]}^\top X_{[g]}) e_{[g]}^\top z_{[g]}}_{R_{15}} \\
    &- \hat \Delta_1 \times \underbrace{\frac{1}{n} \sum_{g \in [G]} (z_{[g]}^\top e_{[g]}) (X_{[g]}^\top z_{[g]})}_{R_{16}}.
\end{align*}
By using the same argument as in the proof of Lemma \ref{lem:var_est}, we have
\begin{align*}
    R_{14} &= \frac{1}{n} \sum_{g \in [G]} (z_{[g]}^\top \Pi_{[g]}) (\Pi_{[g]}^\top z_{[g]}) + \frac{1}{n} \sum_{g \in [G]} (z_{[g]}^\top V_{[g]}) (V_{[g]}^\top z_{[g]}) + o_P(1) \\
    &= \frac{1}{n} \sum_{g \in [G]} (z_{[g]}^\top \Pi_{[g]}) (\Pi_{[g]}^\top z_{[g]}) + \frac{1}{n} \sum_{g \in [G]} (z_{[g]}^\top \tilde  V_{[g]}) (\tilde V_{[g]}^\top z_{[g]}) + o_P(1) \\
    &= \frac{1}{n} \sum_{g \in [G]} (z_{[g]}^\top \Pi_{[g]}) (\Pi_{[g]}^\top z_{[g]}) + \frac{1}{n} \sum_{g \in [G]} \mathbb E  (z_{[g]}^\top \tilde V_{[g]}) (\tilde V_{[g]}^\top z_{[g]}) + o_P(1) \\
    &= \Omega_z^{\Pi, \Pi} + \Omega_z^{\tilde V, \tilde V} + o_P(1),
\end{align*}
and
\begin{align*}
    R_{15} &= \frac{1}{n} \sum_{g \in [G]} (z_{[g]}^\top V_{[g]}) (e_{[g]}^\top z_{[g]}) + o_P(1) \\
    &= \frac{1}{n} \sum_{g \in [G]} (z_{[g]}^\top \tilde V_{[g]}) (\tilde e_{[g]}^\top z_{[g]}) + o_P(1) \\
    &= \frac{1}{n} \sum_{g \in [G]} \mathbb E (z_{[g]}^\top \tilde V_{[g]}) (\tilde e_{[g]}^\top z_{[g]}) + o_P(1) \\
    &= \Omega_z^{\tilde V, \tilde e} + o_P(1),
\end{align*}
and 
\begin{align*}
    R_{16} &= \frac{1}{n} \sum_{g \in [G]} (z_{[g]}^\top e_{[g]}) (V_{[g]}^\top z_{[g]}) + o_P(1) \\
    &= \frac{1}{n} \sum_{g \in [G]} (z_{[g]}^\top \tilde e_{[g]}) (\tilde V_{[g]}^\top z_{[g]}) + o_P(1) \\
    &= \frac{1}{n} \sum_{g \in [G]} \mathbb E (z_{[g]}^\top \tilde e_{[g]}) (\tilde V_{[g]}^\top z_{[g]}) + o_P(1) \\
    &= \Omega_z^{\tilde e, \tilde V} + o_P(1).
\end{align*}
We thus obtain
\begin{align*}
    \frac{1}{n} \tilde \Omega = \hat \Delta_1^2 \Omega_z^{\Pi, \Pi} + \hat \Delta_1^2 \Omega_z^{\tilde V, \tilde V} - \hat \Delta_1 \Omega_z^{\tilde V, \tilde e} - \hat \Delta_1 \Omega_z^{\tilde e, \tilde V} + \Omega_z^{\tilde e, \tilde e} + o_P(1),
\end{align*}
since $\hat \Delta_1 = O_P(1)$.

Combining the above results, we have
\begin{align*}
    \dot \Phi_1 & = \left((\frac{1}{\sqrt{n}}X^\top z) (\frac{1}{\tilde \lambda_n}\hat A_n) (\frac{1}{\sqrt{n}}z^\top X)\right)^{-2} \left((\frac{1}{\sqrt{n}}X^\top z) (\frac{1}{\tilde \lambda_n}\hat A_n) (\frac{1}{n}\tilde \Omega) (\frac{1}{\tilde \lambda_n}\hat A_n) (\frac{1}{\sqrt{n}}z^\top X)\right) \\
    &= \frac{\left( (\frac{1}{\sqrt{n}}X^\top z) (\frac{1}{\tilde \lambda_n} \hat A_n) (\frac{1}{\sqrt{n}} z^\top \tilde e) \right)^2}{\left((\frac{1}{\sqrt{n}}X^\top z) (\frac{1}{\tilde \lambda_n}\hat A_n) (\frac{1}{\sqrt{n}}z^\top X)\right)^4} \times \left((\frac{1}{\sqrt{n}}X^\top z) (\frac{1}{\tilde \lambda_n}\hat A_n) \Omega_z^{\Pi, \Pi} (\frac{1}{\tilde \lambda_n}\hat A_n) (\frac{1}{\sqrt{n}}z^\top X)\right) \\
    &+ \frac{\left( (\frac{1}{\sqrt{n}}X^\top z) (\frac{1}{\tilde \lambda_n} \hat A_n) (\frac{1}{\sqrt{n}} z^\top \tilde e) \right)^2}{\left((\frac{1}{\sqrt{n}}X^\top z) (\frac{1}{\tilde \lambda_n}\hat A_n) (\frac{1}{\sqrt{n}}z^\top X)\right)^4} \times \left((\frac{1}{\sqrt{n}}X^\top z) (\frac{1}{\tilde \lambda_n}\hat A_n) \Omega_z^{\tilde V, \tilde V} (\frac{1}{\tilde \lambda_n}\hat A_n) (\frac{1}{\sqrt{n}}z^\top X)\right) \\
    &- \frac{\left( (\frac{1}{\sqrt{n}}X^\top z) (\frac{1}{\tilde \lambda_n} \hat A_n) (\frac{1}{\sqrt{n}} z^\top \tilde e) \right)}{\left((\frac{1}{\sqrt{n}}X^\top z) (\frac{1}{\tilde \lambda_n}\hat A_n) (\frac{1}{\sqrt{n}}z^\top X)\right)^3} \times \left((\frac{1}{\sqrt{n}}X^\top z) (\frac{1}{\tilde \lambda_n}\hat A_n) \Omega_z^{\tilde V, \tilde e} (\frac{1}{\tilde \lambda_n}\hat A_n) (\frac{1}{\sqrt{n}}z^\top X)\right) \\
    &- \frac{\left( (\frac{1}{\sqrt{n}}X^\top z) (\frac{1}{\tilde \lambda_n} \hat A_n) (\frac{1}{\sqrt{n}} z^\top \tilde e) \right)}{\left((\frac{1}{\sqrt{n}}X^\top z) (\frac{1}{\tilde \lambda_n}\hat A_n) (\frac{1}{\sqrt{n}}z^\top X)\right)^3} \times \left((\frac{1}{\sqrt{n}}X^\top z) (\frac{1}{\tilde \lambda_n}\hat A_n) \Omega_z^{\tilde e, \tilde V} (\frac{1}{\tilde \lambda_n}\hat A_n) (\frac{1}{\sqrt{n}}z^\top X)\right) \\
    &+ \frac{1}{\left((\frac{1}{\sqrt{n}}X^\top z) (\frac{1}{\tilde \lambda_n}\hat A_n) (\frac{1}{\sqrt{n}}z^\top X)\right)^2} \times \left((\frac{1}{\sqrt{n}}X^\top z) (\frac{1}{\tilde \lambda_n}\hat A_n) \Omega_z^{\tilde e, \tilde e} (\frac{1}{\tilde \lambda_n}\hat A_n) (\frac{1}{\sqrt{n}}z^\top X)\right) + o_P(1),
\end{align*}
and thus $\dot \Phi_1 \convD \bar \Phi_1$, by the continuous mapping theorem, where
\begin{align*}
    \bar \Phi_1 &=  \frac{\left( (\zeta_{\tilde V} + \pi)^\top A \zeta_{\tilde e} \right)^2}{\left( (\zeta_{\tilde V} + \pi)^\top A (\zeta_{\tilde V} + \pi) \right)^4} \times \left( (\zeta_{\tilde V} + \pi)^\top A \Omega_z^{\Pi, \Pi} A (\zeta_{\tilde V}+\pi) \right) \\
    &+ \frac{\left( (\zeta_{\tilde V} + \pi)^\top A \zeta_{\tilde e} \right)^2}{\left( (\zeta_{\tilde V} + \pi)^\top A (\zeta_{\tilde V} + \pi) \right)^4} \times \left( (\zeta_{\tilde V} + \pi)^\top A \Omega_z^{\tilde V, \tilde V} A (\zeta_{\tilde V}+\pi) \right) \\
    &- \frac{\left( (\zeta_{\tilde V} + \pi)^\top A \zeta_{\tilde e} \right)}{\left( (\zeta_{\tilde V} + \pi)^\top A (\zeta_{\tilde V} + \pi) \right)^3} \times \left( (\zeta_{\tilde V} + \pi)^\top A \Omega_z^{\tilde V, \tilde e} A (\zeta_{\tilde V}+\pi) \right) \\
    &- \frac{\left( (\zeta_{\tilde V} + \pi)^\top A \zeta_{\tilde e} \right)}{\left( (\zeta_{\tilde V} + \pi)^\top A (\zeta_{\tilde V} + \pi) \right)^3} \times \left( (\zeta_{\tilde V} + \pi)^\top A \Omega_z^{\tilde e, \tilde V} A (\zeta_{\tilde V}+\pi) \right) \\
    &+ \frac{1}{\left( (\zeta_{\tilde V} + \pi)^\top A (\zeta_{\tilde V} + \pi) \right)^2} \times \left( (\zeta_{\tilde V} + \pi)^\top A \Omega_z^{\tilde e, \tilde e} A (\zeta_{\tilde V}+\pi) \right) \\
    &= \frac{1}{\left( (\zeta_{\tilde V} + \pi)^\top A (\zeta_{\tilde V} + \pi) \right)^4} \left\{ \left( (\zeta_{\tilde V} + \pi)^\top A \zeta_{\tilde e} \right)^2 \times \left( (\zeta_{\tilde V} + \pi)^\top A \Omega_z^{\Pi, \Pi} A (\zeta_{\tilde V}+\pi) \right) \right. \\
    &\left. + \left( (\zeta_{\tilde V} + \pi)^\top A \zeta_{\tilde e} \right)^2 \times \left( (\zeta_{\tilde V} + \pi)^\top A \Omega_z^{\tilde V, \tilde V} A (\zeta_{\tilde V}+\pi) \right) \right. \\
    &\left.- \left( (\zeta_{\tilde V} + \pi)^\top A \zeta_{\tilde e} \right) \times \left( (\zeta_{\tilde V} + \pi)^\top A (\zeta_{\tilde V} + \pi) \right) \times \left( (\zeta_{\tilde V} + \pi)^\top A \Omega_z^{\tilde V, \tilde e} A (\zeta_{\tilde V}+\pi) \right) \right.\\
    &\left.- \left( (\zeta_{\tilde V} + \pi)^\top A \zeta_{\tilde e} \right) \times \left( (\zeta_{\tilde V} + \pi)^\top A (\zeta_{\tilde V} + \pi) \right) \times \left( (\zeta_{\tilde V} + \pi)^\top A \Omega_z^{\tilde e, \tilde V} A (\zeta_{\tilde V}+\pi) \right) \right.\\
    &\left.+ \left( (\zeta_{\tilde V} + \pi)^\top A (\zeta_{\tilde V} + \pi) \right)^2 \times \left( (\zeta_{\tilde V} + \pi)^\top A \Omega_z^{\tilde e, \tilde e} A (\zeta_{\tilde V}+\pi) \right) \right\} \\
    & = \frac{ \left( (\zeta_{\tilde V} + \pi)^\top A \zeta_{\tilde e} \right)^2 \times \left( (\zeta_{\tilde V} + \pi)^\top A \Omega_z^{\Pi, \Pi} A (\zeta_{\tilde V}+\pi) \right)}{\left( (\zeta_{\tilde V} + \pi)^\top A (\zeta_{\tilde V} + \pi) \right)^4} \\
    & + \frac{ \begin{bmatrix}
  &      \begin{pmatrix}
        - \left( (\zeta_{\tilde V} + \pi)^\top A (\zeta_{\tilde V} + \pi) \right) \left( A(\zeta_{\tilde V} + \pi) \right) \\
        \left( (\zeta_{\tilde V} + \pi)^\top A \zeta_{\tilde e} \right) \left( A(\zeta_{\tilde V} + \pi) \right)
    \end{pmatrix}^\top
    \begin{pmatrix}
        \Omega_z^{\tilde e, \tilde e} & \Omega_z^{\tilde e, \tilde V} \\
        \Omega_z^{\tilde V, \tilde e} & \Omega_z^{\tilde V, \tilde V}
    \end{pmatrix} \\
& \times     \begin{pmatrix}
        - \left( (\zeta_{\tilde V} + \pi)^\top A (\zeta_{\tilde V} + \pi) \right) \left( A(\zeta_{\tilde V} + \pi) \right) \\
        \left( (\zeta_{\tilde V} + \pi)^\top A \zeta_{\tilde e} \right) \left( A (\zeta_{\tilde V} + \pi) \right)
    \end{pmatrix} 
    \end{bmatrix}} {\left( (\zeta_{\tilde V} + \pi)^\top A (\zeta_{\tilde V} + \pi) \right)^4},
\end{align*}
with
\begin{align*}
    \begin{pmatrix}
        \zeta_{\tilde e} \\ \zeta_{\tilde V} 
    \end{pmatrix} \stackrel{d}{=} \N \left( 0, \begin{pmatrix}
                \Omega_z^{\tilde e, \tilde e} & \Omega_z^{\tilde e, \tilde V} \\
                \Omega_z^{\tilde V, \tilde e} & \Omega_z^{\tilde V, \tilde V}
            \end{pmatrix}  \right).
\end{align*}

Finally, we note that 
\begin{align*}
    \left( (\zeta_{\tilde V} + \pi)^\top A \zeta_{\tilde e} \right)^2 \times \left( (\zeta_{\tilde V} + \pi)^\top A \Omega_z^{\Pi, \Pi} A (\zeta_{\tilde V}+\pi) \right) \geqslant 0,
\end{align*}
and
\begin{align*}
 \begin{bmatrix}
  &      \begin{pmatrix}
        - \left( (\zeta_{\tilde V} + \pi)^\top A (\zeta_{\tilde V} + \pi) \right) \left( A(\zeta_{\tilde V} + \pi) \right) \\
        \left( (\zeta_{\tilde V} + \pi)^\top A \zeta_{\tilde e} \right) \left( A(\zeta_{\tilde V} + \pi) \right)
    \end{pmatrix}^\top
    \begin{pmatrix}
        \Omega_z^{\tilde e, \tilde e} & \Omega_z^{\tilde e, \tilde V} \\
        \Omega_z^{\tilde V, \tilde e} & \Omega_z^{\tilde V, \tilde V}
    \end{pmatrix} \\
& \times     \begin{pmatrix}
        - \left( (\zeta_{\tilde V} + \pi)^\top A (\zeta_{\tilde V} + \pi) \right) \left( A(\zeta_{\tilde V} + \pi) \right) \\
        \left( (\zeta_{\tilde V} + \pi)^\top A \zeta_{\tilde e} \right) \left( A (\zeta_{\tilde V} + \pi) \right)
    \end{pmatrix} 
    \end{bmatrix} \geqslant  0,
\end{align*}
and the equalities hold if and only if 
\begin{gather*}
    \left( (\zeta_{\tilde V} + \pi)^\top A (\zeta_{\tilde V} + \pi) \right) \left( A(\zeta_{\tilde V} + \pi) \right) = 0, \\
    \left( (\zeta_{\tilde V} + \pi)^\top A \zeta_{\tilde e} \right) \left( A(\zeta_{\tilde V} + \pi) \right) = 0.
\end{gather*}
Given that $A$ is positive definite, the above two equalities hold if and only if $\zeta_{\tilde V} + \pi =0$, which has probability zero since $\zeta_{\tilde V}$ has a non-degenerate normal distribution. In addition, the denominator of $\bar \Phi_1$ is positive with probability one, which implies 
\begin{align*}
    1 / \dot \Phi_1^{1/2} = O_P(1).
\end{align*}
This concludes the proof. $\hfill\qedsymbol$

\subsection{Proof of Lemma \ref{lem:linear_form}}
By Lemma \ref{lem:betahat}, we have $\hat \beta \convP \beta$ and $(\hat \beta - \beta)^2 \Pi^\top \Pi / \sqrt{K} = o_P(1)$. 

For $T(\beta_0)$, by Lemma \ref{lem:var_est}, 
\begin{align*}
    \frac{\Psi}{ \widehat \Psi} \convP 1,
\end{align*}
which, by (\ref{eq:low_strong}), implies that
\begin{align*}
    \frac{\Phi_1}{ \widehat \Phi_1} \convP 1. 
\end{align*}
We have
\begin{align*}
    T(\beta_0) &= \frac{ (X^\top z \hat A_n z^\top X)^{-1}(X^\top z \hat A_n z^\top e)}{ \sqrt{\widehat \Phi_1}} + a_1 \delta + \left(\frac{d_n}{\sqrt{\widehat \Phi_1}} - a_1\right) \delta \\
    &= \frac{ (X^\top z \hat A_n z^\top X)^{-1}(X^\top z \hat A_n z^\top e)}{ \sqrt{\widehat \Phi_1}} + a_1 \delta + o_P(1),
\end{align*}
where the second equality holds because 
\begin{align*}
\frac{d_n}{\sqrt{\widehat \Phi_1}} - a_1 = \frac{d_n}{\sqrt{\Phi_1}} \frac{\sqrt{\Phi_1}}{\sqrt{\widehat \Phi_1}} - a_1 = o_p(1).   
\end{align*}
In addition, we have
\begin{align*}
    \frac{ (X^\top z \hat A_n z^\top X)^{-1}(X^\top z \hat A_n z^\top e)}{ \sqrt{\widehat \Phi_1}} &= \frac{(X^\top z \hat A_n z^\top X)^{-1}}{\sqrt{(X^\top z \hat A_n z^\top X)^{-2}}} \times \frac{(X^\top z \hat A_n z^\top e)}{\sqrt{\hat \Psi}} \\
    &= \sqrt{\frac{\Psi}{\hat \Psi}} \times \frac{(X^\top z \hat A_n z^\top e)}{\sqrt{\Psi}} \times (1 + o_P(1)) \\
    &= \frac{(\Pi^\top z A_n z^\top \tilde e)}{\sqrt{\Pi^\top z A_n \Omega A_n z^\top \Pi}} \times (1 + o_P(1)) \\
    &= \frac{(\Pi^\top z A_n z^\top \tilde e)}{\sqrt{\Pi^\top z A_n \Omega A_n z^\top \Pi}} + o_P(1),
\end{align*}
whence
\begin{align*}
    T(\beta_0) = \frac{1}{\sqrt{\Psi}} \sum_{g \in [G]} \acute \Pi_{[g]}^\top \tilde e_{[g]} + a_1 \delta + o_P(1).
\end{align*}

For $LM(\beta_0)$, if $\Pi^\top \Pi / \sqrt{K} \rightarrow \infty$, we have
\begin{align*}
    LM(\beta_0) &= \frac{X^\top (P - \bar P) e}{\sqrt{\hat \Sigma}} + a_2 \delta + \left(\frac{d_n (X^\top (P -\bar P) X) }{\sqrt{\hat \Sigma}} - a_2\right) \delta \\
    &= \frac{X^\top (P - \bar P) e}{\sqrt{\hat \Sigma}} + a_2 \delta + \left(\frac{d_n}{\sqrt{\Phi_2}} \times (1 + o_P(1)) - a_2\right) \delta \\
    &= \frac{X^\top (P - \bar P) e}{\sqrt{\hat \Sigma}} + a_2 \delta + o_P(1),
\end{align*}
where the second equality holds because by Lemma \ref{lem:var_est},
\begin{align*}
    \frac{\Sigma}{\hat \Sigma} \convP 1,
\end{align*}
which, by (\ref{eq:high_strong}), implies that
\begin{align*}
\frac{\Phi_2 (X^\top (P- \bar P)X)^2}{\hat \Sigma} = \frac{\Phi_2 (X^\top (P- \bar P)X)^2}{\Sigma} \frac{\Sigma}{\hat \Sigma} = \frac{ (X^\top (P- \bar P)X)^2}{ (\Pi^\top (P- \bar P)\Pi)^2} \frac{\Sigma}{\hat \Sigma} \convP 1.
\end{align*}
Alternatively, if $\Pi^\top \Pi / \sqrt{K} = O(1)$, we have
\begin{align*}
    LM(\beta_0) &= \frac{X^\top (P - \bar P) e}{\sqrt{\hat \Sigma}} + \frac{d_n \delta (X^\top (P -\bar P) X) }{\sqrt{\hat \Sigma}} \\
    &= \frac{X^\top (P - \bar P) e}{\sqrt{\hat \Sigma}} + o_P(1),
\end{align*}
where we use the fact that
\begin{align*}
    \frac{X^\top (P - \bar P) X}{\sqrt{\hat \Sigma}} = \frac{X^\top (P - \bar P) X}{\sqrt{ \Sigma}} \times \frac{\Sigma}{\hat \Sigma} = \frac{\Pi^\top (P - \bar P) \Pi + O_P(\sqrt{\Sigma})}{\sqrt{ \Sigma}} \times O_P(1) = O_P(1)
\end{align*}
by (\ref{eq:X(P-Pbar)X}), $\Pi^\top \Pi / \sqrt{K} = O(1)$, and 
\begin{align*}
    \frac{1}{\sqrt{\Phi_2}} = \sqrt{\frac{(\Pi^\top (P - \bar P) \Pi)^2}{\Sigma}} = O(\frac{\Pi^\top \Pi}{\sqrt{K}}) = O(1),
\end{align*} Assumption \ref{ass:local_alternative_and_covariance} implies that $a_2 = \lim_{n \rightarrow \infty} d_n / \sqrt{\Phi_2} = 0$ in this case. In either case, we obtain
\begin{align*}
    LM(\beta_0) = \frac{X^\top (P - \bar P) e}{\sqrt{\hat \Sigma}} + a_2 \delta + o_P(1).
\end{align*}
In addition, we have
\begin{align*}
    \frac{X^\top (P - \bar P) e}{\sqrt{\hat \Sigma}} & = \sqrt{\frac{\Sigma}{\hat \Sigma}} \times \frac{1}{\sqrt{\Sigma}} \left( \hat \Pi^\top \tilde e + \tilde V^\top (P - \bar P) \tilde e +  \tilde V^\top P_W \bar P \tilde e + \tilde V^\top \bar P P_W \tilde e - \tilde V^\top P_W \bar P P_W \tilde e \right) \\ 
    & = \sqrt{\frac{\Sigma}{\hat \Sigma}} \times \frac{1}{\sqrt{\Sigma}} \left( \hat \Pi ^\top \tilde e + \tilde V^\top (P - \bar P) \tilde e + o_P(\sqrt{\Sigma}) \right), \\
    &= \frac{1}{\sqrt{\Sigma}} \left( \hat \Pi ^\top \tilde e + \tilde V^\top (P - \bar P) \tilde e \right) + o_P(1),
\end{align*}
where the second equality holds by $\Sigma \geq C(\Pi^\top \Pi + K) \rightarrow \infty $ as shown in Lemmas \ref{lem:quad_form_1} and \ref{lem:linear_quad_form_1}, and the last equality holds by Lemma \ref{lem:var_est}. It follows that 
\begin{align*}
    LM(\beta_0) = \frac{1}{\sqrt{\Sigma}} \left( \sum_{g \in [G]} \hat \Pi_{[g]}^\top \tilde e_{[g]} + \sum_{g,h \in [G]^2, g \neq h} \tilde V_{[g]}^\top P_{[g,h]} \tilde e_{[h]} \right) + a_2 \delta + o_P(1).
\end{align*}

For $AR$, we have
\begin{align*}
    e^\top (P -\bar P) e - \tilde e^\top (P -\bar P) \tilde e = \tilde e^\top P_W \bar P \tilde e + \tilde e^\top \bar P P_W \tilde e - \tilde e^\top P_W \bar P P_W \tilde e = O_P(1)
\end{align*} 
by Lemma \ref{lem:quad_form_1}, and $\Upsilon \geq CK$ by (\ref{eq:Upsilon}). This implies that
\begin{align*}
    \frac{1}{\sqrt{\Upsilon}} \left(\sum_{g,h \in [G]^2, g \neq h} e_{[g]}^\top P_{[g,h]} e_{[h]} - \sum_{g,h \in [G]^2, g \neq h} \tilde e_{[g]}^\top P_{[g,h]} \tilde e_{[h]} \right) = O_P(K^{-1/2}) = o_P(1).
\end{align*}
We also have
\begin{align*}
    \hat e^\top (P -\bar P) \hat e - e^\top (P -\bar P) e &= (\hat \beta - \beta)^2 X^\top (P -\bar P) X - 2 (\hat \beta - \beta) X^\top (P -\bar P) e \\
    &= (\hat \beta - \beta)^2 \left(\Pi^\top (P- \bar P) \Pi + O_P(\sqrt{\Sigma}) \right) \\
    &- 2 (\hat \beta - \beta) \left( \hat \Pi ^\top \tilde e + \tilde V^\top (P - \bar P) \tilde e + o_P(\sqrt{\Sigma}) \right) \\
    &= (\hat \beta - \beta)^2 O_P(\Pi^\top \Pi + \sqrt{\Sigma}) + (\hat \beta - \beta) O_P(\sqrt{\Sigma}) \\
    &= \left( (\hat \beta - \beta)^2 \Pi^\top \Pi + (\hat \beta - \beta)^2 \sqrt{K} + (\hat \beta - \beta) \sqrt{K} \right) \times O_P(1)
\end{align*}
by (\ref{eq:X(P-Pbar)e}) and (\ref{eq:X(P-Pbar)X}), $\Pi^\top \Pi = O(\sqrt{K})$, and $\Sigma = O(\Pi^\top \Pi + K) = O(K)$. Then by $\hat \beta \convP \beta$ and $(\hat \beta - \beta)^2 \Pi^\top \Pi / \sqrt{K} = o_P(1)$, we have
\begin{align*}
    &\quad \frac{1}{\sqrt{\Upsilon}} \left(\sum_{g,h \in [G]^2, g \neq h} \hat e_{[g]}^\top P_{[g,h]} \hat e_{[h]} - \sum_{g,h \in [G]^2, g \neq h} e_{[g]}^\top P_{[g,h]} e_{[h]} \right) \\
    &= \left( \frac{(\hat \beta - \beta)^2 \Pi^\top \Pi}{\sqrt{K}} + (\hat \beta - \beta)^2 + (\hat \beta - \beta) \right) \times O_P(1) \\
    &= o_P(1).
\end{align*}
It follows that, by Lemma \ref{lem:var_est}, we have 
\begin{align*}
    AR &= \sqrt{\frac{\Upsilon}{\hat \Upsilon}} \times \frac{1}{\sqrt{\Upsilon}} \sum_{g,h \in [G]^2, g \neq h} \hat e_{[g]}^\top P_{[g,h]} \hat e_{[h]} \\
    &= \sqrt{\frac{\Upsilon}{\hat \Upsilon}} \times \frac{1}{\sqrt{\Upsilon}} \left(\sum_{g,h \in [G]^2, g \neq h} \hat e_{[g]}^\top P_{[g,h]} \hat e_{[h]} - \sum_{g,h \in [G]^2, g \neq h} e_{[g]}^\top P_{[g,h]} e_{[h]} \right) \\
    &+ \sqrt{\frac{\Upsilon}{\hat \Upsilon}} \times \frac{1}{\sqrt{\Upsilon}} \left(\sum_{g,h \in [G]^2, g \neq h} e_{[g]}^\top P_{[g,h]} e_{[h]} - \sum_{g,h \in [G]^2, g \neq h} \tilde e_{[g]}^\top P_{[g,h]} \tilde e_{[h]} \right) \\
    &+ \sqrt{\frac{\Upsilon}{\hat \Upsilon}} \times \frac{1}{\sqrt{\Upsilon}} \sum_{g,h \in [G]^2, g \neq h} \tilde e_{[g]}^\top P_{[g,h]} \tilde e_{[h]} \\
    &= \frac{1}{\sqrt{\Upsilon}} \sum_{g,h \in [G]^2, g \neq h} \tilde e_{[g]}^\top P_{[g,h]} \tilde e_{[h]} + o_P(1).
\end{align*} 
This concludes the proof. $\hfill\qedsymbol$

\subsection{Proof of Lemma \ref{lem:limit_dist}}

Note that by Assumption \ref{ass:reg_cluster_size}, we have $G\rightarrow \infty$ as $n \rightarrow \infty$. The proof follows the same steps as in \cite{Chao(2012)} to check the conditions for the martingale central limit theorem; see, for example, \cite{hall1980martingale}.

\vspace{5mm}
\noindent \textbf{Step 1: Construct martingale difference array.} 
Let 
\begin{align*}
    \rho_{1n} &= \text{cov} \left[\frac{1}{\sqrt{\Psi}} \sum_{g \in [G]} \acute \Pi_{[g]}^\top \tilde e_{[g]},  \frac{1}{\sqrt{\Sigma}} \left( \sum_{g \in [G]} \hat \Pi_{[g]}^\top \tilde e_{[g]} + \sum_{g,h \in [G]^2, g \neq h} \tilde V_{[g]}^\top P_{[g,h]} \tilde e_{[h]} \right)\right] \\
    &=\frac{1}{\sqrt{\Psi \Sigma}} \sum_{g \in [G]} \mathbb E \left[ \left( \acute \Pi_{[g]}^\top \tilde e_{[g]} \right) \left( \hat \Pi_{[g]}^\top \tilde e_{[g]} \right) \right],
\end{align*}
such that $\rho_1 = \lim_{n \rightarrow \infty} \rho_{1n}$, and 
\begin{align*}
    \rho_{2n} &= \text{cov} \left[\frac{1}{\sqrt{\Sigma}} \left( \sum_{g \in [G]} \hat \Pi_{[g]}^\top \tilde e_{[g]} + \sum_{g,h \in [G]^2, g \neq h} \tilde V_{[g]}^\top P_{[g,h]} \tilde e_{[h]} \right), \frac{1}{\sqrt{\Upsilon}} \sum_{g,h \in [G]^2, g \neq h} \tilde e_{[g]}^\top P_{[g,h]} \tilde e_{[h]} \right] \\
    &=\frac{2}{\sqrt{\Sigma \Upsilon}}  \sum_{g,h \in [G]^2, g \neq h} \mathbb E \left[ \left(\tilde V_{[g]}^\top P_{[g,h]} \tilde e_{[h]} \right) \left( \tilde e_{[g]}^\top P_{[g,h]} \tilde e_{[h]} \right) \right],
\end{align*} such that $\rho_2 = \lim_{n \rightarrow \infty} \rho_{2n}$, and note that
\begin{align*}
    \text{cov} \left[\frac{1}{\sqrt{\Psi}} \sum_{g \in [G]} \acute \Pi_{[g]}^\top \tilde e_{[g]}, \frac{1}{\sqrt{\Upsilon}} \sum_{g,h \in [G]^2, g \neq h} \tilde e_{[g]}^\top P_{[g,h]} \tilde e_{[h]}  \right] = 0.
\end{align*}
The assumptions for $\rho_1$ and $\rho_2$ in Theorem \ref{thm:limit_str_lcl} ensure that \begin{align*}
     \begin{pmatrix}
    1 &  \rho_{1n} & 0 \\
    \rho_{1n} & 1 & \rho_{2n} \\
    0 & \rho_{2n} & 1
\end{pmatrix}^{-1}
\end{align*}
exists for all $n$ large enough, and by the Slutsky theorem, the result would follow if
\begin{align*}
\begin{pmatrix}
    1 &  \rho_{1n} & 0 \\
    \rho_{1n} & 1 & \rho_{2n} \\
    0 & \rho_{2n} & 1
\end{pmatrix}^{-1/2} 
& \begin{pmatrix}
    \frac{1}{\sqrt{\Psi}} \sum_{g \in [G]} \acute \Pi_{[g]}^\top \tilde e_{[g]} \\
    \frac{1}{\sqrt{\Sigma}} \left( \sum_{g \in [G]} \hat \Pi_{[g]}^\top \tilde e_{[g]} + \sum_{g,h \in [G]^2, g \neq h} \tilde V_{[g]}^\top P_{[g,h]} \tilde e_{[h]} \right)  \\
    \frac{1}{\sqrt{\Upsilon}} \sum_{g,h \in [G]^2, g \neq h} \tilde e_{[g]}^\top P_{[g,h]} \tilde e_{[h]}
\end{pmatrix} \\
& \convD \N \left( \begin{pmatrix}
    0 \\
    0 \\
    0
\end{pmatrix}, \begin{pmatrix}
    1 &  0 & 0 \\
    0 & 1 & 0 \\
    0 & 0 & 1
\end{pmatrix} \right).
\end{align*}
Let $v = (v_1, v_2, v_3)^\top$ with $\left\Vert v \right\Vert_2 = 1$, and 
\begin{align*}
    c \equiv (c_1, c_2, c_3)^\top = \begin{pmatrix}
    1 &  \rho_{1n} & 0 \\
    \rho_{1n} & 1 & \rho_{2n} \\
    0 & \rho_{2n} & 1
\end{pmatrix}^{-1/2} (v_1, v_2, v_3)^\top,
\end{align*}
and note that $\left\Vert c \right\Vert_2$ is bounded for all $n$ large enough. By Cramer-Wald device, it suffices to show that
\begin{align*}
    c^\top \begin{pmatrix}
    \frac{1}{\sqrt{\Psi}} \sum_{g \in [G]} \acute \Pi_{[g]}^\top \tilde e_{[g]} \\
    \frac{1}{\sqrt{\Sigma}} \left( \sum_{g \in [G]} \hat \Pi_{[g]}^\top \tilde e_{[g]} + \sum_{g,h \in [G]^2, g \neq h} \tilde V_{[g]}^\top P_{[g,h]} \tilde e_{[h]} \right)  \\
    \frac{1}{\sqrt{\Upsilon}} \sum_{g,h \in [G]^2, g \neq h} \tilde e_{[g]}^\top P_{[g,h]} \tilde e_{[h]}
\end{pmatrix} \convD \N \left( 0, 1 \right).
\end{align*}

Denote
\begin{align*}
    M_n &= \frac{c_1}{\sqrt{\Psi}} \sum_{g \in [G]}\acute \Pi_{[g]}^\top \tilde e_{[g]} + \frac{c_2}{\sqrt{\Sigma}} \left( \sum_{g \in [G]} \hat \Pi_{[g]}^\top \tilde e_{[g]} + \sum_{g,h \in [G]^2, g \neq h} \tilde V_{[g]}^\top P_{[g,h]} \tilde e_{[h]} \right) \\
    & + \frac{c_3}{\sqrt{\Upsilon}} \sum_{g,h \in [G]^2, g \neq h} \tilde e_{[g]}^\top P_{[g,h]} \tilde e_{[h]},
\end{align*}
and note that $M_n$ can be written as sum of martingale difference array. To see this, define
\begin{align*}
    M_{1G} = \frac{c_1}{\sqrt{\Psi}} \acute \Pi_{[1]}^\top \tilde e_{[1]} + \frac{c_2}{\sqrt{\Sigma}} \hat \Pi_{[1]}^\top \tilde e_{[1]},
\end{align*}
and for $g \geq 2$ define
\begin{align*}
    M_{gG} &= \frac{c_1}{\sqrt{\Psi}} \acute \Pi_{[g]}^\top \tilde e_{[g]} + \frac{c_2}{\sqrt{\Sigma}} \hat \Pi_{[g]}^\top \tilde e_{[g]} + \frac{c_2}{\sqrt{\Sigma}} \sum_{h < g} \left(\tilde V_{[g]}^\top P_{[g,h]} \tilde e_{[h]} + \tilde V_{[h]}^\top P_{[h,g]} \tilde e_{[g]} \right) \\
    &+ \frac{c_3}{\sqrt{\Upsilon}} \sum_{h < g} \left(\tilde e_{[g]}^\top P_{[g,h]} \tilde e_{[h]} + \tilde e_{[h]}^\top P_{[h,g]} \tilde e_{[g]} \right),
\end{align*}
and we have $M_n = \sum_{g \in [G]} M_{gG}$. Further define $\eps_{[g]} = \left( \tilde e_{[g]}^\top, \tilde V_{[g]}^\top \right)^\top$ and the sequence of $\sigma$-fields $\mathcal F_{gG} = \{ \eps_{[1]}, \cdots, \eps_{[g]} \}$ such that $\mathcal F_{(g-1)G} \subset \mathcal F_{gG}$ with $\mathcal F_{0G} = \{ \emptyset, \Omega \}$. It is clear that $\{M_{gG} \}_{g=1}^G$ is a sequence of martingale difference array with respect to $\{\mathcal F_{gG} \}_{g=1}^G$. Note that
\begin{align*}
    \sum_{g \in [G]} \mathbb E (M_{gG}^2) = \mathbb E (M_n^2) = 1
\end{align*}
by the property of martingale difference array. 

In order to show
\begin{align*}
    \sum_{g \in [G]} M_{gG} \convD \N \left(0, 1 \right),
\end{align*}
it remains to check the Lindeberg's condition 
\begin{align*}
    \sum_{g \in [G]} \mathbb E \left(M_{gG}^2 \mathbf 1_{\{|M_{gG}| \geq c \}} \right) \rightarrow 0
\end{align*}
for every $c > 0$, and the stability condition 
\begin{align*}
    \sum_{g \in [G]} \mathbb E \left(M_{gG}^2| \mathcal F_{(g-1)G} \right) \convP \sum_{g \in [G]} \mathbb E (M_{gG}^2).
\end{align*}

\vspace{5mm}
\noindent \textbf{Step 2: Check Lindeberg's condition.} It suffices to show that
\begin{align*}
    \sum_{g \in [G]} \mathbb E (M_{gG}^{(2+\delta)}) \rightarrow 0
\end{align*}
for some $\delta > 0$, and we will verify that for $\delta = 2$ in what follows. We have
\begin{align*}
    \sum_{g \in G} \mathbb E (M_{gG}^{4}) &\leqslant  \frac{C}{\Psi^2} \sum_{g \in [G]} \mathbb E \left(\acute \Pi_{[g]}^\top \tilde e_{[g]} \right)^4 + \frac{C}{\Sigma^2} \sum_{g \in [G]}\mathbb E \left( \hat \Pi_{[g]}^\top \tilde e_{[g]} \right)^4 + \frac{C}{\Sigma^2} \sum_{g \in [G]} \mathbb E \left( \sum_{h < g} \tilde V_{[g]}^\top P_{[g,h]} \tilde e_{[h]} \right)^4 \\
    &+ \frac{C}{\Sigma^2} \sum_{g \in [G]} \mathbb E \left( \sum_{h < g} \tilde V_{[h]}^\top P_{[h,g]} \tilde e_{[g]} \right)^4 + \frac{C}{\Upsilon^2} \sum_{g \in [G]} \mathbb E \left( \sum_{h < g} \tilde e_{[h]}^\top P_{[h,g]} \tilde e_{[g]} \right)^4.
\end{align*}
For the first term, we have
\begin{align*}
    &\quad \frac{1}{\Psi^2} \sum_{g \in [G]} \mathbb E \left( \acute \Pi_{[g]}^\top \tilde e_{[g]} \right)^4 \\
    &\leqslant \frac{C}{\Psi^2} \sum_{g \in [G]}\left( \acute \Pi_{[g]}^\top \acute \Pi_{[g]} \right)^2 \\
    &\leqslant \frac{C\max_{1\leq g\leq G} \left\Vert \acute \Pi_{[g]} \right\Vert_2^2 \acute \Pi^\top \acute \Pi}{\left( \acute \Pi^\top \acute \Pi  \right)^2} \\
    &=o(1).
\end{align*} 
Here we use the fact that $\acute \Pi_{[g]} = z_{[g]} A_n z^\top \Pi$, so that 
\begin{align*}
    \frac{\max_{1\leq g\leq G} \left\Vert \acute \Pi_{[g]} \right\Vert_2^2}{\acute \Pi^\top \acute \Pi} &= \frac{\max_{1\leq g\leq G} \left\Vert \acute \Pi_{[g]} \right\Vert_2^2 / n}{\Pi^\top z A_n (z^\top z/n) A_n z^\top \Pi} \\
    &\leq \frac{C \max_{1\leq g\leq G} \Pi^\top z A_n (z_{[g]}^\top z_{[g]}/n) A_n z^\top \Pi}{\left\Vert A_n z^\top \Pi \right\Vert_2^2} \\
    &\leq \frac{C \max_{1\leq g\leq G} n_g \times \max_{i \in I_g, g \in [G]} \left\Vert z_{i,g} \right\Vert_2^2}{n} \\
    &= o(1).
\end{align*}
For the second term, we have
\begin{align*}
    &\quad \frac{1}{\Sigma^2} \sum_{g \in [G]}\mathbb E \left( \hat \Pi_{[g]}^\top \tilde e_{[g]} \right)^4 \\
    &\leqslant \frac{C}{\Sigma^2} \sum_{g \in [G]}\left( \hat \Pi_{[g]}^\top \hat \Pi_{[g]} \right)^2 \\
    &\leqslant \frac{C\max_{1\leq g\leq G} \left\Vert \hat \Pi_{[g]} \right\Vert_2^2 \hat \Pi^\top \hat \Pi}{\left(\Pi^\top \Pi + K \right)^2} \\
    &=o(1)
\end{align*}
as $K \rightarrow \infty$. 

For the third term, we have
\begin{align*}
    \frac{1}{\Sigma^2} \sum_{g \in [G]} \mathbb E \left( \sum_{h < g} \tilde e_{[h]}^\top P_{[h,g]} \tilde V_{[g]} \right)^4 &\leqslant \frac{C}{\Sigma^2} \sum_{g \in [G]} \sum_{h < g} \mathbb E \left( \tilde e_{[h]}^\top P_{[h,g]} \tilde V_{[g]} \right)^4 \\
    &+ \frac{C}{\Sigma^2} \sum_{g \in [G]} \sum_{h,k < g, h \neq k} \mathbb E \left( \tilde e_{[h]}^\top P_{[h,g]} \tilde V_{[g]} \right)^2 \left( \tilde e_{[k]}^\top P_{[k,g]} \tilde V_{[g]} \right)^2,
\end{align*}
where
\begin{align*}
    &\quad \frac{1}{\Sigma^2} \sum_{g \in [G]} \sum_{h < g} \mathbb E \left( \tilde e_{[h]}^\top P_{[h,g]} \tilde V_{[g]} \right)^4 \\
    &\leqslant \frac{1}{\Sigma^2} \sum_{g \in [G]} \sum_{h < g} \lambda_{\max}^2 \left( P_{[h,g]} P_{[g,h]} \right) \\
    &\leqslant \frac{C}{K^2} \sum_{g \in [G]} \sum_{h < g} \tr \left( P_{[h,g]} P_{[g,h]} \right) \\
    &=o(1),
\end{align*}
and
\begin{align*}
    &\frac{1}{\Sigma^2} \sum_{g \in [G]} \sum_{h,k < g, h \neq k} \mathbb E \left( \tilde e_{[h]}^\top P_{[h,g]} \tilde V_{[g]} \right)^2 \left( \tilde e_{[k]}^\top P_{[k,g]} \tilde V_{[g]} \right)^2 \\
    &\leqslant \frac{C}{\Sigma^2} \sum_{g \in [G]} \sum_{h,k < g, h \neq k} \lambda_{\max} \left( P_{[g,h]} P_{[h,g]} \right) \lambda_{\max} \left( P_{[g,k]} P_{[k,g]} \right)\\
    &\leqslant \frac{C}{\Sigma^2} \sum_{g \in [G]} \sum_{h,k < g, h \neq k} \tr \left( P_{[g,h]} P_{[h,g]} \right) \tr \left( P_{[g,k]} P_{[k,g]} \right) \\
    &\leqslant \frac{C}{K^2} \sum_{g \in [G]} \sum_{h < g} \tr \left( P_{[g,h]} P_{[h,g]} \right) \\
    &=o(1).
 \end{align*} 
The last two terms can be handled similarly.

\vspace{5mm}
\noindent \textbf{Step 3: Check stability condition.} The variance and conditional variance can be written as
\begin{align*}
    \mathbb E \left( M_{1G}^2 \right) = \mathbb E \left( M_{1G}^2 | \mathcal F_{0G} \right) = \frac{c_1^2}{\Psi}\acute \Pi_{[1]}^\top \Omega_1^{\tilde e, \tilde e} \acute \Pi_{[1]} + \frac{2c_1 c_2}{\sqrt{\Psi \Sigma}} \hat \Pi_{[1]}^\top \Omega_1^{\tilde e, \tilde e} \acute \Pi_{[1]} + \frac{c_2^2}{\Sigma}\hat \Pi_{[1]}^\top \Omega_1^{\tilde e, \tilde e} \hat \Pi_{[1]},
\end{align*}
and for $g \geq 2$,
\begin{align*}
    \mathbb E \left( M_{gG}^2 \right) &= \frac{c_1^2}{\Psi}\acute \Pi_{[g]}^\top \Omega_g^{\tilde e, \tilde e} \acute \Pi_{[g]} + \frac{2c_1 c_2}{\sqrt{\Psi \Sigma}} \hat \Pi_{[g]}^\top \Omega_g^{\tilde e, \tilde e} \acute \Pi_{[g]} + \frac{c_2^2}{\Sigma}\hat \Pi_{[g]}^\top \Omega_g^{\tilde e, \tilde e} \hat \Pi_{[g]} \\
    &+ \frac{c_2^2}{\Sigma} \sum_{h<g} \tr \left( \Omega_h^{\tilde e, \tilde e} P_{[h,g]} \Omega_g^{\tilde V, \tilde V} P_{[g,h]} \right) + \frac{c_2^2}{\Sigma} \sum_{h<g} \tr \left( \Omega_h^{\tilde V, \tilde V} P_{[h,g]} \Omega_g^{\tilde e, \tilde e} P_{[g,h]} \right) \\
    &+ \frac{4c_3^2}{\Upsilon} \sum_{h<g} \tr \left( \Omega_h^{\tilde e, \tilde e} P_{[h,g]} \Omega_g^{\tilde e, \tilde e} P_{[g,h]} \right) + \frac{2c_2^2}{\Sigma} \sum_{h<g} \tr \left( \Omega_h^{\tilde e, \tilde V} P_{[h,g]} \Omega_g^{\tilde e, \tilde V} P_{[g,h]} \right) \\
    &+ \frac{4c_2c_3}{\sqrt{\Sigma \Upsilon}} \sum_{h<g} \tr \left( \Omega_h^{\tilde e, \tilde e} P_{[h,g]} \Omega_g^{\tilde V, \tilde e} P_{[g,h]} \right) + \frac{4c_2c_3}{\sqrt{\Sigma \Upsilon}} \sum_{h<g} \tr \left( \Omega_h^{\tilde e, \tilde V} P_{[h,g]} \Omega_g^{\tilde e, \tilde e} P_{[g,h]} \right),
\end{align*}
and
\begin{align*}
    \mathbb E \left( M_{gG}^2 | \mathcal F_{(g-1)G} \right) &= \frac{c_1^2}{\Psi}\acute \Pi_{[g]}^\top \Omega_g^{\tilde e, \tilde e} \acute \Pi_{[g]} + \frac{2c_1 c_2}{\sqrt{\Psi \Sigma}} \hat \Pi_{[g]}^\top \Omega_g^{\tilde e, \tilde e} \acute \Pi_{[g]} + \frac{c_2^2}{\Sigma}\hat \Pi_{[g]}^\top \Omega_g^{\tilde e, \tilde e} \hat \Pi_{[g]} \\
    &+ \frac{2c_1 c_2}{\sqrt{\Psi \Sigma}} \sum_{h < g} \tilde e_{[h]}^\top P_{[h,g]} \Omega_g^{\tilde V, \tilde e} \acute \Pi_{[g]} + \frac{2c_1 c_2}{\sqrt{\Psi \Sigma}} \sum_{h < g} \tilde V_{[h]}^\top P_{[h,g]} \Omega_g^{\tilde e, \tilde e} \acute \Pi_{[g]} \\
    & + \frac{4c_1 c_3}{\sqrt{\Psi \Upsilon}} \sum_{h < g} \tilde e_{[h]}^\top P_{[h,g]} \Omega_g^{\tilde V, \tilde e} \acute \Pi_{[g]} + \frac{2c_1 c_2}{\sqrt{\Psi \Sigma}} \sum_{h < g} \tilde e_{[h]}^\top P_{[h,g]} \Omega_g^{\tilde V, \tilde e} \hat \Pi_{[g]} \\
    & + \frac{2c_1 c_2}{\sqrt{\Psi \Sigma}} \sum_{h < g} \tilde V_{[h]}^\top P_{[h,g]} \Omega_g^{\tilde e, \tilde e} \hat \Pi_{[g]} + \frac{4c_1 c_3}{\sqrt{\Psi \Upsilon}} \sum_{h < g} \tilde e_{[h]}^\top P_{[h,g]} \Omega_g^{\tilde V, \tilde e} \hat \Pi_{[g]} \\
    &+  \frac{c_2^2}{\Sigma} \sum_{h,k <g} \tilde e_{[h]}^\top P_{[h,g]} \Omega_g^{\tilde V, \tilde V} P_{[g, k]} \tilde e_{[k]} + \frac{c_2^2}{\Sigma} \sum_{h,k <g} \tilde V_{[h]}^\top P_{[h,g]} \Omega_g^{\tilde e, \tilde e} P_{[g, k]} \tilde V_{[k]} \\
    &+ \frac{4c_3^2}{\Upsilon} \sum_{h,k <g} \tilde e_{[h]}^\top P_{[h,g]} \Omega_g^{\tilde e, \tilde e} P_{[g, k]} \tilde e_{[k]} + \frac{2c_2^2}{\Sigma} \sum_{h, k < g} e_{[h]}^\top P_{[h,g]} \Omega_g^{\tilde V, \tilde e} P_{[g,k]} \tilde V_{[k]} \\
    &+ \frac{4c_2c_3}{\sqrt{\Sigma \Upsilon}} \sum_{h, k < g} V_{[h]}^\top P_{[h,g]} \Omega_g^{\tilde e, \tilde e} P_{[g,k]} \tilde e_{[k]} + \frac{4c_2c_3}{\sqrt{\Sigma \Upsilon}} \sum_{h, k < g} e_{[h]}^\top P_{[h,g]} \Omega_g^{\tilde V, \tilde e} P_{[g,k]} \tilde e_{[k]}.
\end{align*}
We thus obtain
\begin{align*}
    &\quad \mathbb E \left( M_{gG}^2 | \mathcal F_{(g-1)G} \right) - \mathbb E \left( M_{gG}^2 \right) \\
    & =\begin{Bmatrix}
      &  \frac{2c_1 c_2}{\sqrt{\Psi \Sigma}} \sum_{h < g} \tilde e_{[h]}^\top P_{[h,g]} \Omega_g^{\tilde V, \tilde e} \acute \Pi_{[g]} + \frac{2c_1 c_2}{\sqrt{\Psi \Sigma}} \sum_{h < g} \tilde V_{[h]}^\top P_{[h,g]} \Omega_g^{\tilde e, \tilde e} \acute \Pi_{[g]} + \frac{4c_1 c_3}{\sqrt{\Psi \Upsilon}} \sum_{h < g} \tilde e_{[h]}^\top P_{[h,g]} \Omega_g^{\tilde V, \tilde e} \acute \Pi_{[g]}  \\
    & +\frac{2c_1 c_2}{\sqrt{\Psi \Sigma}} \sum_{h < g} \tilde e_{[h]}^\top P_{[h,g]} \Omega_g^{\tilde V, \tilde e} \hat \Pi_{[g]} + \frac{2c_1 c_2}{\sqrt{\Psi \Sigma}} \sum_{h < g} \tilde V_{[h]}^\top P_{[h,g]} \Omega_g^{\tilde e, \tilde e} \hat \Pi_{[g]} + \frac{4c_1 c_3}{\sqrt{\Psi \Upsilon}} \sum_{h < g} \tilde e_{[h]}^\top P_{[h,g]} \Omega_g^{\tilde V, \tilde e} \hat \Pi_{[g]}
    \end{Bmatrix} \\
    & + \begin{Bmatrix}
      &  \frac{c_2^2}{\Sigma} \left[\sum_{h,k <g} \tilde e_{[h]}^\top P_{[h,g]} \Omega_g^{\tilde V, \tilde V} P_{[g, k]} \tilde e_{[k]} - \sum_{h<g} \tr \left( \Omega_h^{\tilde e, \tilde e} P_{[h,g]} \Omega_g^{\tilde V, \tilde V} P_{[g,h]} \right) \right]  \\
    &+ \frac{c_2^2}{\Sigma} \left[\sum_{h,k <g} \tilde V_{[h]}^\top P_{[h,g]} \Omega_g^{\tilde e, \tilde e} P_{[g, k]} \tilde V_{[k]} - \sum_{h<g} \tr \left( \Omega_h^{\tilde V, \tilde V} P_{[h,g]} \Omega_g^{\tilde e, \tilde e} P_{[g,h]} \right) \right] \\
    & +\frac{4c_3^2}{\Upsilon} \left[\sum_{h,k <g} \tilde e_{[h]}^\top P_{[h,g]} \Omega_g^{\tilde e, \tilde e} P_{[g, k]} \tilde e_{[k]} - \sum_{h<g} \tr \left( \Omega_h^{\tilde e, \tilde e} P_{[h,g]} \Omega_g^{\tilde e, \tilde e} P_{[g,h]} \right) \right]   \\
    &+\frac{2c_2^2}{\Sigma} \left[\sum_{h, k < g} e_{[h]}^\top P_{[h,g]} \Omega_g^{\tilde V, \tilde e} P_{[g,k]} \tilde V_{[k]} - \sum_{h<g} \tr \left( \Omega_h^{\tilde e, \tilde V} P_{[h,g]} \Omega_g^{\tilde e, \tilde V} P_{[g,h]} \right) \right]  \\
    &+ \frac{4c_2c_3}{\sqrt{\Sigma \Upsilon}} \left[\sum_{h, k < g} V_{[h]}^\top P_{[h,g]} \Omega_g^{\tilde e, \tilde e} P_{[g,k]} \tilde e_{[k]} - \sum_{h<g} \tr \left( \Omega_h^{\tilde e, \tilde e} P_{[h,g]} \Omega_g^{\tilde V, \tilde e} P_{[g,h]} \right) \right] \\
    &+  \frac{4c_2c_3}{\sqrt{\Sigma \Upsilon}} \left[ \sum_{h, k < g} e_{[h]}^\top P_{[h,g]} \Omega_g^{\tilde V, \tilde e} P_{[g,k]} \tilde e_{[k]} - \sum_{h<g} \tr \left( \Omega_h^{\tilde e, \tilde V} P_{[h,g]} \Omega_g^{\tilde e, \tilde e} P_{[g,h]} \right) \right] 
    \end{Bmatrix} \\
     &\equiv M_{gG}^{(1)} + M_{gG}^{(2)},
\end{align*}
and it suffices to show that \begin{align*}
    M_n^{(1)} &\equiv \sum_{g \in [G]} M_{gG}^{(1)} = o_P(1), \\
    M_n^{(2)} &\equiv \sum_{g \in [G]} M_{gG}^{(2)} = o_P(1).
\end{align*} 

Consider first $M_n^{(1)}$, we shall only compute the sum of the first term in $M_{gG}^{(1)}$, as the other terms can be handled similarly. Recall that $\tilde P$ is the block lower triangular matrix corresponding to $P - \bar P$, we have
\begin{align*}
    \mathbb E \left( \frac{2c_1 c_2}{\sqrt{\Psi \Sigma}} \sum_{g \in [G]}\sum_{h < g} \tilde e_{[h]}^\top P_{[h,g]} \Omega_g^{\tilde V, \tilde e} \acute \Pi_{[g]} \right) = 0,
\end{align*}
and
\begin{align*}
    \mathbb V \left( \frac{2c_1 c_2}{\sqrt{\Psi \Sigma}} \sum_{g \in [G]}\sum_{h < g} \tilde e_{[h]}^\top P_{[h,g]} \Omega_g^{\tilde V, \tilde e} \acute \Pi_{[g]} \right) &= \frac{4c_1^2 c_2^2}{\Psi \Sigma} \mathbb E \left(\tilde e^\top \tilde P^\top \Omega_{\tilde V, \tilde e} \acute \Pi \right)^2 \\
    &\leqslant \frac{C \lambda_{\max} \left(\tilde P \tilde P^\top \right) \lambda_{\max}\left(\Omega_{\tilde e, \tilde V} \Omega_{\tilde V, \tilde e} \right) \acute \Pi^\top \acute \Pi}{(\acute \Pi^\top \acute \Pi) K} \\ 
    &\leqslant \frac{C \lambda_{\max} \left(\tilde P \tilde P^\top \right)}{K} \\
    &\leqslant \frac{C\left\Vert \tilde P \tilde P^{\top} \right\Vert_F}{K} \\
    &= o(1),
\end{align*}
where we use Lemma \ref{lem:PandQ} in the last equality.

Now consider $M_n^{(2)}$, we shall only compute the sum of the first term in $M_{gG}^{(2)}$, as the other terms can be handled similarly. We have
\begin{align*}
    \mathbb E \left( \frac{c_2^2}{\Sigma} \sum_{g \in [G]}\left[\sum_{h,k <g} \tilde e_{[h]}^\top P_{[h,g]} \Omega_g^{\tilde V, \tilde V} P_{[g, k]} \tilde e_{[k]} - \sum_{h<g} \tr \left( \Omega_h^{\tilde e, \tilde e} P_{[h,g]} \Omega_g^{\tilde V, \tilde V} P_{[g,h]} \right) \right] \right) = 0,
\end{align*}
and
\begin{align*}
    &\quad \mathbb V \left( \frac{c_2^2}{\Sigma} \sum_{g \in [G]}\left[\sum_{h,k <g} \tilde V_{[h]}^\top P_{[h,g]} \Omega_g^{\tilde e, \tilde e} P_{[g, k]} \tilde V_{[k]} - \sum_{h<g} \tr \left( \Omega_h^{\tilde V, \tilde V} P_{[h,g]} \Omega_g^{\tilde e, \tilde e} P_{[g,h]} \right) \right] \right) \\
    &\leqslant \frac{C}{\Sigma^2}\mathbb E \left( \sum_{g \in [G]} \sum_{h<g} \left( \tilde V_{[h]}^\top P_{[h,g]} \Omega_g^{\tilde e, \tilde e} P_{[g, h]} \tilde V_{[h]} - \tr \left( \Omega_h^{\tilde V, \tilde V} P_{[h,g]} \Omega_g^{\tilde e, \tilde e} P_{[g,h]} \right) \right) \right)^2 \\
    &\quad + \frac{C}{\Sigma^2}\mathbb E \left( \sum_{g \in [G]} \sum_{h,k<g, h \neq k} \tilde V_{[h]}^\top P_{[h,g]} \Omega_g^{\tilde e, \tilde e} P_{[g, k]} \tilde V_{[k]} \right)^2.
\end{align*}
For the first term, we have
\begin{align*}
    &\quad \frac{1}{\Sigma^2}\mathbb E \left( \sum_{g \in [G]} \sum_{h<g} \left( \tilde V_{[h]}^\top P_{[h,g]} \Omega_g^{\tilde e, \tilde e} P_{[g, h]} \tilde V_{[h]} - \tr \left( \Omega_h^{\tilde V, \tilde V} P_{[h,g]} \Omega_g^{\tilde e, \tilde e} P_{[g,h]} \right) \right) \right)^2 \\
    &=\frac{1}{\Sigma^2}\mathbb E \left( \sum_{h \in [G]} \tilde V_{[h]}^\top \left( \sum_{g>h} P_{[h,g]} \Omega_g^{\tilde e, \tilde e} P_{[g, h]} \right) \tilde V_{[h]} - \tr \left( \Omega_h^{\tilde V, \tilde V} \left( \sum_{g>h} P_{[h,g]} \Omega_g^{\tilde e, \tilde e} P_{[g, h]} \right) \right) \right)^2 \\
    &\leqslant \frac{1}{\Sigma^2} \sum_{h \in [G]} \mathbb E \left( \tilde V_{[h]}^\top \left( \sum_{g>h} P_{[h,g]} \Omega_g^{\tilde e, \tilde e} P_{[g, h]} \right) \tilde V_{[h]} \right)^2 \\
    &\leqslant \frac{1}{\Sigma^2} \sum_{h \in [G]} \lambda_{\max}^2 \left(
    \left( \tilde P^\top \Omega_{\tilde e} \tilde P \right)_{[h,h]} \right) \\
    &\leqslant \frac{C}{\Sigma^2} \sum_{h \in [G]} \tr \left(
    \left( \tilde P^\top \Omega_{\tilde e} \tilde P \right)_{[h,h]} \right) \\
    &\leqslant \frac{C}{\Sigma^2} \tr \left(\tilde P^\top \Omega_{\tilde e} \tilde P \right) \\
    &\leqslant \frac{C}{K^2} \tr \left(\tilde P^\top \tilde P \right) \\
    &=o(1),
\end{align*}
where we use the fact that
\begin{align*}
    \tr \left(\tilde P^\top \tilde P \right)  = ||\tilde P||_F^2 \leq ||P||_F^2 = O(K).
\end{align*}
For the second term, we have
\begin{align*}
    &\quad \frac{1}{\Sigma^2} \mathbb E \left( \sum_{g \in [G]} \sum_{h,k<g, h \neq k} \tilde V_{[h]}^\top P_{[h,g]} \Omega_g^{\tilde e, \tilde e} P_{[g, k]} \tilde V_{[k]} \right)^2 \\
    &= \frac{1}{\Sigma^2} \mathbb E \left(\sum_{h,k \in [G]^2, h \neq k} \tilde V_{[h]}^\top \left( \sum_{g>h \vee k} P_{[h,g]} \Omega_g^{\tilde e, \tilde e} P_{[g, k]} \right) \tilde V_{[k]} \right)^2 \\
    &= \frac{1}{\Sigma^2} \mathbb E \left(\sum_{h,k \in [G]^2, h < k} \tilde V_{[h]}^\top \left( \sum_{g>k} P_{[h,g]} \Omega_g^{\tilde e, \tilde e} P_{[g, k]} \right) \tilde V_{[k]} \right.\\
    & \qquad \left.+ \sum_{h,k \in [G]^2, h > k} \tilde V_{[h]}^\top \left( \sum_{g>h} P_{[h,g]} \Omega_g^{\tilde e, \tilde e} P_{[g, k]} \right) \tilde V_{[k]} \right)^2 \\
    &\leqslant \frac{C}{\Sigma^2} \sum_{h,k \in [G]^2, h \neq k} \mathbb E \left( \tilde V_{[h]}^\top \left( \sum_{g>h \vee k} P_{[h,g]} \Omega_g^{\tilde e, \tilde e} P_{[g, k]} \right) \tilde V_{[k]} \right)^2 \\
    &\leqslant \frac{C}{\Sigma^2} \sum_{h,k \in [G]^2, h \neq k} \tr \left( \left( \sum_{g>h \vee k} P_{[h,g]} \Omega_g^{\tilde e, \tilde e} P_{[g, k]} \right) \left( \sum_{g>h \vee k} P_{[k,g]} \Omega_g^{\tilde e, \tilde e} P_{[g, h]} \right) \right) \\
    &\leqslant \frac{C}{\Sigma^2} \sum_{h,k \in [G]^2} \tr \left( \left( \tilde P^\top \Omega_{\tilde e} \tilde P \right)_{[h,k]} \left( \tilde P^\top \Omega_{\tilde e} \tilde P \right)_{[k,h]} \right) \\
    &\leqslant \frac{C}{K^2} \tr \left( \tilde P^\top \Omega_{\tilde e} \tilde P \tilde P^\top \Omega_{\tilde e} \tilde P \right) \\
    &\leqslant \frac{C \left\Vert \tilde P \tilde P^{\top} \right\Vert_F^2}{K^2} \\
    &=o(1),
\end{align*}
where we use Lemma \ref{lem:PandQ} in the last equality. This concludes the proof. $\hfill\qedsymbol$

\subsection{Proof of Lemma \ref{lem:cov_est}}

For the first result, for $j \in [d_z]$, let $v_j$ be the $d_z$-dimensional unit vector with $j$-th element one and other elements zero, and denote
$\bar z_j = z \Omega^{-1/2} v_j$, then  
\begin{align*}
    \bar z_j^\top \bar z_j &= v_j^\top (\Omega/n)^{-1/2} (z^\top z / n) (\Omega/n)^{-1/2} v_j = O(1), \\
    \max_{g \in [G]} \bar z_{j,[g]}^\top \bar z_{j,[g]} &= \max_{g \in [G]} v_j^\top (\Omega/n)^{-1/2} (z_{[g]}^\top z_{[g]} / n) (\Omega/n)^{-1/2} v_j = o(1).
\end{align*}
It follows that
\begin{align*}
    \left\vert\frac{1}{\sqrt{\Sigma}} \sum_{g \in [G]} \mathbb E\left[ \left(\bar z_{j,[g]}^\top \tilde e_{[g]} \right) \left( \hat \Pi_{[g]}^\top \tilde e_{[g]} \right) \right] \right\vert \leq \left( \sum_{g \in [G]} \mathbb E\left(\bar z_{j,[g]}^\top \tilde e_{[g]} \right)^2 \right)^{1/2} \left(\frac{1}{\Sigma} \sum_{g \in [G]} \mathbb E\left( \hat \Pi_{[g]}^\top \tilde e_{[g]} \right)^2 \right)^{1/2} = O(1),
\end{align*}
and as $d_z$ is fixed, we obtain
\begin{align} 
    \frac{1}{\sqrt{\Sigma}} \Omega^{-1/2}   \sum_{g \in [G]} \mathbb E\left[ \left( z_{[g]}^\top \tilde e_{[g]} \right) \left( \hat \Pi_{[g]}^\top \tilde e_{[g]} \right) \right] = O(1).
    \label{eq:zehatxhatehat}
\end{align}
In addition, let $\hat V = M_W (P -\bar P) V = Q \tilde V$, we have
\begin{align*}
    \frac{1}{\sqrt{\Sigma}} \Omega^{-1/2}   \sum_{g \in [G]} \left[ \left( z_{[g]}^\top \hat e_{[g]} \right) \left( \hat X_{[g]}^\top \hat e_{[g]} \right) \right] &= \frac{1}{\sqrt{\Sigma}} \Omega^{-1/2}   \sum_{g \in [G]} \left[ \left( z_{[g]}^\top \hat e_{[g]} \right) \left( \hat \Pi_{[g]}^\top \hat e_{[g]} \right) \right] \\
    &+ \frac{1}{\sqrt{\Sigma}} \Omega^{-1/2}   \sum_{g \in [G]} \left[ \left( z_{[g]}^\top \hat e_{[g]} \right) \left( \hat V_{[g]}^\top \hat e_{[g]} \right) \right].
\end{align*}
Thus, it suffices to show that, for any $j$,
\begin{align}
    &\frac{1}{\sqrt{\Sigma}} \sum_{g \in [G]} \left[ \left( \bar z_{j,[g]}^\top \hat e_{[g]} \right) \left( \hat \Pi_{[g]}^\top \hat e_{[g]} \right) \right] - \frac{1}{\sqrt{\Sigma}} \sum_{g \in [G]} \mathbb E\left[ \left(\bar z_{j,[g]}^\top \tilde e_{[g]} \right) \left( \hat \Pi_{[g]}^\top \tilde e_{[g]} \right) \right] = o_P(1), \label{eq:zehatxhatehat1}\\
    &\frac{1}{\sqrt{\Sigma}} \sum_{g \in [G]} \left[ \left(\bar z_{j,[g]}^\top \hat e_{[g]} \right) \left( \hat V_{[g]}^\top \hat e_{[g]} \right) \right] = o_P(1).\label{eq:zehatxhatehat2}
\end{align}

For \eqref{eq:zehatxhatehat1}, the left-hand side can be written as
\begin{align*}
    &\quad \frac{1}{\sqrt{ \Sigma}}  \left( \sum_{g \in [G]} \left( \bar z_{j,[g]}^\top \hat e_{[g]} \right) \left( \hat \Pi_{[g]}^\top \hat e_{[g]} \right) - \sum_{g \in [G]} \mathbb E \left(\bar z_{j,[g]}^\top \tilde e_{[g]} \right) \left( \hat \Pi_{[g]}^\top \tilde e_{[g]} \right) \right) \\
    &= \underbrace{\frac{1}{\sqrt{\Sigma}} \left( \sum_{g \in [G]} \left(\bar z_{j,[g]}^\top \tilde e_{[g]} \right) \left( \hat \Pi_{[g]}^\top \tilde e_{[g]} \right) - \sum_{g \in [G]} \mathbb E \left(\bar z_{j,[g]}^\top \tilde e_{[g]} \right) \left( \hat \Pi_{[g]}^\top \tilde e_{[g]} \right) \right)}_{R_{27}} \\
    &+ \underbrace{\frac{1}{\sqrt{ \Sigma}} \left( \sum_{g \in [G]} \left(\bar z_{j,[g]}^\top e_{[g]} \right) \left( \hat \Pi_{[g]}^\top e_{[g]} \right) - \sum_{g \in [G]} \left( \bar z_{j,[g]}^\top \tilde e_{[g]} \right) \left( \hat \Pi_{[g]}^\top \tilde e_{[g]} \right) \right)}_{R_{28}} \\
    &+ \underbrace{\frac{1}{\sqrt{\Sigma}} \left( \sum_{g \in [G]} \left(\bar z_{j,[g]}^\top \hat e_{[g]} \right) \left( \hat \Pi_{[g]}^\top \hat e_{[g]} \right) - \sum_{g \in [G]} \left(\bar z_{j,[g]}^\top e_{[g]} \right) \left( \hat \Pi_{[g]}^\top e_{[g]} \right) \right)}_{R_{29}}.
\end{align*}
For $R_{27}$, it has mean zero and 
\begin{align*}
    \mathbb V \left( R_{27} \right) &= \frac{1}{\Sigma} \mathbb E\left( \sum_{g \in [G]} \left(\bar z_{j,[g]}^\top \tilde e_{[g]} \right) \left( \hat \Pi_{[g]}^\top \tilde e_{[g]} \right) - \sum_{g \in [G]} \mathbb E \left(\bar z_{j,[g]}^\top \tilde e_{[g]} \right) \left( \hat \Pi_{[g]}^\top \tilde e_{[g]} \right) \right)^2 \\
    &\leqslant \frac{1}{ \Sigma}  \sum_{g \in [G]} \mathbb E\left( \left( \bar z_{j,[g]}^\top \tilde e_{[g]} \right) \left( \hat \Pi_{[g]}^\top \tilde e_{[g]} \right) \right)^2 \\
    &\leqslant \frac{C\max_{1\leq g\leq G} \left\Vert \hat \Pi_{[g]} \right\Vert_2^2 \bar z_j^\top \bar z_j}{( \Pi^\top \Pi + K) } \\
    &= o(1).
\end{align*}
For $R_{28}$, we have
\begin{align*}
    R_{28} &= \underbrace{\frac{1}{\sqrt{ \Sigma}} \sum_{g \in [G]} \left( \bar z_{j,[g]}^\top W_{[g]} \hat \gamma_{\tilde e} \right) \left( \hat \Pi_{[g]}^\top W_{[g]} \hat \gamma_{\tilde e} \right)}_{R_{28,1}} \\
    &- \underbrace{\frac{1}{\sqrt{ \Sigma}} \sum_{g \in [G]} \left( \bar z_{j,[g]}^\top W_{[g]} \hat \gamma_{\tilde e} \right) \left( \hat \Pi_{[g]}^\top \tilde e_{[g]} \right)}_{R_{28,2}} \\
    &- \underbrace{\frac{1}{\sqrt{ \Sigma}} \sum_{g \in [G]} \left( \bar z_{j,[g]}^\top \tilde e_{[g]} \right) \left( \hat \Pi_{[g]}^\top W_{[g]} \hat \gamma_{\tilde e} \right)}_{R_{28,3}},
\end{align*}
where
\begin{align*}
    \left| R_{28,1} \right| &\leqslant \max_{1 \leq g \leq G} \left\Vert W_{[g]} \hat \gamma_{\tilde e} \right\Vert_2^2 \times  \sqrt{\bar z_j^\top \bar z_j} \times \sqrt{\frac{\hat \Pi^\top \hat \Pi}{\Sigma}} = o_P(1), \\
    \left| R_{28,2} \right| &\leqslant \max_{1 \leq g \leq G} \left\Vert W_{[g]} \hat \gamma_{\tilde e} \right\Vert_2 \times  \sqrt{\bar z_j^\top \bar z_j} \times \sqrt{\frac{1}{\Sigma}\sum_{g \in [G]} \left( \hat \Pi_{[g]}^\top \tilde e_{[g]} \right)^2} = o_P(1), \\
    \left| R_{28,3} \right| &\leqslant \max_{1 \leq g \leq G} \left\Vert W_{[g]} \hat \gamma_{\tilde e} \right\Vert_2 \times  \sqrt{ \sum_{g \in [G]} \left( \bar z_{j,[g]}^\top \tilde e_{[g]} \right)^2} \times \sqrt{\frac{\hat \Pi^\top \hat \Pi}{\Sigma}} = o_P(1),
\end{align*}
whence $R_{28} = o_P(1)$. For $R_{29}$, we have
\begin{align*}
    R_{29} &= \hat \Delta^2 \times \underbrace{\frac{1}{\sqrt{\Sigma}} \sum_{g \in [G]} \left( \bar z_{j,[g]}^\top X_{[g]} \right) \left( \hat \Pi_{[g]}^\top X_{[g]} \right)}_{R_{29,1}} \\
    &- \hat \Delta \times \underbrace{\frac{1}{\sqrt{\Sigma}} \sum_{g \in [G]} \left( \bar z_{j,[g]}^\top X_{[g]} \right) \left( \hat \Pi_{[g]}^\top e_{[g]} \right)}_{R_{29,2}} \\
    &- \hat \Delta \times \underbrace{\frac{1}{\sqrt{ \Sigma}} \sum_{g \in [G]} \left( \bar z_{j,[g]}^\top e_{[g]} \right) \left( \hat \Pi_{[g]}^\top e_{[g]} \right)}_{R_{29,3}},
\end{align*}
where $\hat \Delta = o_P(1)$ and
\begin{align*}
    \left| R_{29,1} \right| &\leqslant \sqrt{\sum_{g \in [G]}\bar z_{j,[g]}^\top \bar z_{j,[g]} X_{[g]}^\top X_{[g]}} \times \sqrt{\frac{1}{\Sigma} \sum_{g \in [G]}\hat \Pi_{[g]}^\top \hat \Pi_{[g]} X_{[g]}^\top X_{[g]}} = O_P(1), \\
    \left| R_{29,2} \right| &\leqslant \sqrt{\sum_{g \in [G]}\bar z_{j,[g]}^\top \bar z_{j,[g]} X_{[g]}^\top X_{[g]}} \times \sqrt{\frac{1}{\Sigma} \sum_{g \in [G]}\hat \Pi_{[g]}^\top \hat \Pi_{[g]} e_{[g]}^\top e_{[g]}} = O_P(1), \\
    \left| R_{29,3} \right| &\leqslant \sqrt{\sum_{g \in [G]}\bar z_{j,[g]}^\top \bar z_{j,[g]} e_{[g]}^\top e_{[g]}} \times \sqrt{\frac{1}{\Sigma} \sum_{g \in [G]}\hat \Pi_{[g]}^\top \hat \Pi_{[g]} e_{[g]}^\top e_{[g]}} = O_P(1),
\end{align*}
since $\max_{g \in [G]}\mathbb E \left(X_{[g]}^\top X_{[g]} \right) = O(1)$ and $\max_{g \in [G]}\mathbb E \left(e_{[g]}^\top e_{[g]} \right) = O(1)$ by Lemma \ref{lem:4th_moment}. It follows that $R_{29} = o_P(1)$.

For \eqref{eq:zehatxhatehat2}, the left-hand side can be written as
\begin{align*}
    &\quad \frac{1}{\sqrt{\Sigma}} \sum_{g \in [G]} \left( \bar z_{j,[g]}^\top \hat e_{[g]} \right) \left( \hat V_{[g]}^\top \hat e_{[g]} \right) \\
    &= \frac{\hat \Delta^2}{\sqrt{\Sigma}} \sum_{g \in [G]} \left( \bar z_{j,[g]}^\top X_{[g]} \right) \left( \hat V_{[g]}^\top X_{[g]} \right) - \frac{\hat \Delta}{\sqrt{ \Sigma}} \sum_{g \in [G]} \left( \bar z_{j,[g]}^\top X_{[g]} \right) \left( \hat V_{[g]}^\top  e_{[g]} \right) \\
    &- \frac{\hat \Delta}{\sqrt{ \Sigma}} \sum_{g \in [G]} \left( \bar z_{j,[g]}^\top e_{[g]} \right) \left( \hat V_{[g]}^\top X_{[g]} \right) + \frac{1}{\sqrt{ \Sigma}} \sum_{g \in [G]} \left( \bar z_{j,[g]}^\top e_{[g]} \right) \left( \hat V_{[g]}^\top e_{[g]} \right).
\end{align*}
Here we only show that the last term is $o_P(1)$, which is most difficult since it does not involve $\hat \Delta$. We have
\begin{align*}
    \underbrace{\frac{1}{\sqrt{ \Sigma}} \sum_{g \in [G]} \left( \bar z_{j,[g]}^\top e_{[g]} \right) \left( \hat V_{[g]}^\top e_{[g]} \right)}_{R_{30}} &= \underbrace{\frac{1}{\sqrt{ \Sigma}} \sum_{g \in [G]} \left( \bar z_{j,[g]}^\top \tilde e_{[g]} \right) \left( \hat V_{[g]}^\top \tilde e_{[g]} \right)}_{R_{30,1}} \\
    &- \underbrace{\frac{1}{\sqrt{ \Sigma}} \sum_{g \in [G]} \left( \bar z_{j,[g]}^\top W_{[g]} \hat \gamma_{\tilde e} \right) \left( \hat V_{[g]}^\top \tilde e_{[g]} \right)}_{R_{30,2}} \\
    &- \underbrace{\frac{1}{\sqrt{\Sigma}} \sum_{g \in [G]} \left( \bar z_{j,[g]}^\top \tilde e_{[g]} \right) \left( \hat V_{[g]}^\top W_{[g]} \hat \gamma_{\tilde e} \right)}_{R_{30,3}}  \\
    &+ \underbrace{\frac{1}{\sqrt{\Sigma}} \sum_{g \in [G]} \left( \bar z_{j,[g]}^\top W_{[g]} \hat \gamma_{\tilde e} \right) \left( \hat V_{[g]}^\top W_{[g]} \hat \gamma_{\tilde e} \right)}_{R_{30,4}}.
\end{align*}
Note that
\begin{align*}
    \hat V_{[g]} &= \sum_{h \in [G], h \neq g} P_{[g,h]} \tilde V_{[h]} + \sum_{h \in [G]} P_{W,[g,h]} P_{[h,h]} \tilde V_{[h]} \\
    &+ \sum_{h \in [G]} P_{[g,g]} P_{W,[g,h]} \tilde V_{[h]} - \sum_{h \in [G]} \left(\sum_{k \in [G]} P_{W,[g,k]} P_{[k,k]} P_{W,[k,h]} \right) \tilde V_{[h]}.
\end{align*}
Therefore, we have
\begin{align*}
    \frac{1}{\Sigma} \sum_{g \in [G]} \left( \hat V_{[g]}^\top \tilde e_{[g]} \right)^2 &\leq \frac{C}{\Sigma} \sum_{g \in [G]} \left(\sum_{h \in [G], h \neq g} \tilde V_{[h]}^\top P_{[h,g]} \tilde e_{[g]} \right)^2 \\
    &+ \frac{C}{\Sigma} \sum_{g \in [G]} \left(\sum_{h \in [G]} \tilde V_{[h]}^\top  P_{[h,h]} P_{W,[h,g]} \tilde e_{[g]}  \right)^2 \\
    &+ \frac{C}{\Sigma} \sum_{g \in [G]} \left(\sum_{h \in [G]} \tilde V_{[h]}^\top  P_{W,[h,g]} P_{[g,g]} \tilde e_{[g]}  \right)^2 \\
    &+ \frac{C}{\Sigma} \sum_{g \in [G]} \left(\sum_{h \in [G]} \tilde V_{[h]}^\top \left(\sum_{k \in [G]} P_{W,[h,k]} P_{[k,k]} P_{W,[k,g]}  \right) \tilde e_{[g]}  \right)^2 \\
    &= \frac{C}{\Sigma} \sum_{g \in [G]} \left(\sum_{h \in [G], h \neq g} \tilde V_{[h]}^\top P_{[h,g]} \tilde e_{[g]} \right)^2 + o_P(1) \\
    &= \frac{C}{\Sigma} \sum_{g \in [G]} \mathbb E \left(\sum_{h \in [G], h \neq g} \tilde V_{[h]}^\top P_{[h,g]} \tilde e_{[g]} \right)^2 + o_P(1) \\
    &= O_P(1),
\end{align*}
by Lemma \ref{lem:quad_form_2} and the facts that 
\begin{align*}
    \frac{1}{\Sigma} \sum_{g \in [G]} \mathbb E \left(\sum_{h \in [G]} \tilde V_{[h]}^\top P_{[h,h]} P_{W,[h,g]} \tilde e_{[g]} \right)^2 &= \frac{1}{\Sigma} \sum_{g,h \in [G]^2} \mathbb E \left( \tilde V_{[h]}^\top P_{[h,h]} P_{W,[h,g]} \tilde e_{[g]} \right)^2 \\
    &\leqslant \frac{C\tr \left( P_W \bar P^2 P_W \right)}{\Sigma} \\
    &= o(1),
\end{align*} 
\begin{align*}
    \frac{1}{\Sigma} \sum_{g \in [G]} \mathbb E \left(\sum_{h \in [G]} \tilde V_{[h]}^\top  P_{W,[h,g]} P_{[g,g]} \tilde e_{[g]}  \right)^2 &= \frac{1}{\Sigma} \sum_{g, h \in [G]^2} \mathbb E \left( \tilde V_{[h]}^\top  P_{W,[h,g]} P_{[g,g]} \tilde e_{[g]}  \right)^2 \\
    &\leqslant \frac{C\tr \left( \bar P P_W^2 \bar P \right)}{\Sigma} \\
    &= o(1),
\end{align*}
and
\begin{align*}
&    \frac{1}{\Sigma} \sum_{g \in [G]} \left(\sum_{h \in [G]} \tilde V_{[h]}^\top \left(\sum_{k \in [G]} P_{W,[h,k]} P_{[k,k]} P_{W,[k,g]}  \right) \tilde e_{[g]}  \right)^2 \\
& = \frac{1}{\Sigma} \sum_{g,h \in [G]^2} \left( \tilde V_{[h]}^\top \left(\sum_{k \in [G]} P_{W,[h,k]} P_{[k,k]} P_{W,[k,g]}  \right) \tilde e_{[g]}  \right)^2 \\
    &\leqslant \frac{C\tr \left( P_W \bar P P_W^2 \bar P P_W \right)}{\Sigma} \\
    &= o(1).
\end{align*}

This implies 
\begin{align*}
    \left| R_{30,2} \right| &\leqslant \max_{1 \leq g \leq G} \left\Vert W_{[g]} \hat \gamma_{\tilde e} \right\Vert_2 \times  \sqrt{\bar z_j^\top \bar z_j} \times \sqrt{\frac{1}{\Sigma}\sum_{g \in [G]} \left( \hat V_{[g]}^\top \tilde e_{[g]} \right)^2} = o_P(1).
\end{align*}
In addition, we have
\begin{align*}
    \left| R_{30,3} \right| &\leqslant \max_{1 \leq g \leq G} \left\Vert W_{[g]} \hat \gamma_{\tilde e} \right\Vert_2 \times  \sqrt{\sum_{g \in [G]}\left( \bar z_{j,[g]}^\top \tilde e_{[g]} \right)^2} \times \sqrt{\frac{\tilde V^\top Q^2 \tilde V}{\Sigma}} = o_P(1), \\
    \left| R_{30,4} \right| &\leqslant \max_{1 \leq g \leq G} \left\Vert W_{[g]} \hat \gamma_{\tilde e} \right\Vert_2^2 \times  \sqrt{\bar z_j^\top \bar z_j} \times \sqrt{\frac{\tilde V^\top Q^2 \tilde V}{\Sigma}} = o_P(1).
\end{align*}

It follows that
\begin{align*}
    R_{30} = R_{30,1} + o_P(1),
\end{align*}
and for $R_{30,1}$, note that similar to the proof above, we also have
\begin{align*}
    R_{30,1} = \frac{1}{\sqrt{ \Sigma}} \sum_{g,h \in [G]^2 g \neq h} \left( \bar z_{j,[g]}^\top \tilde e_{[g]} \right) \left( \tilde V_{[h]}^\top P_{[h,g]} \tilde e_{[g]} \right) + o_P(1),
\end{align*}
where
\begin{align*}
    & \mathbb E \left( \frac{1}{\sqrt{\Sigma}} \sum_{g,h \in [G]^2 g \neq h} \left( z_{[g]}^\top \tilde e_{[g]} \right) \left( \tilde V_{[h]}^\top P_{[h,g]} \tilde e_{[g]} \right) \right) = 0,
\end{align*}
and
\begin{align*}
    &\quad \mathbb V \left( \frac{1}{\sqrt{ \Sigma}} \sum_{g,h \in [G]^2 g \neq h} \left( \bar z_{j,[g]}^\top \tilde e_{[g]} \right) \left( \tilde V_{[h]}^\top P_{[h,g]} \tilde e_{[g]} \right) \right) \\
    &= \frac{1}{\Sigma} \mathbb E \left(\sum_{g,h \in [G]^2 g \neq h} \left( \bar z_{j,[g]}^\top \tilde e_{[g]} \right) \left( \tilde V_{[h]}^\top P_{[h,g]} \tilde e_{[g]} \right) \right)^2 \\
    &\leqslant \frac{C}{ \Sigma} \mathbb E \left( \sum_{g,h \in [G]^2 g \neq h} \bar z_{j,[g]}^\top \left( \tilde e_{[g]} \tilde e_{[g]}^\top - \Omega_g^{\tilde e, \tilde e} \right) P_{[g,h]} \tilde V_{[h]} \right)^2 \\
    &+ \frac{C}{\Sigma} \mathbb E \left( \sum_{g,h \in [G]^2 g \neq h} \bar z_{j,[g]}^\top \Omega_g^{\tilde e, \tilde e} P_{[g,h]} \tilde V_{[h]} \right)^2 \\
    &\leq \frac{C}{\Sigma}  \sum_{g,h \in [G]^2 g \neq h} \mathbb E \left( \bar z_{j,[g]}^\top \tilde e_{[g]} \tilde e_{[g]}^\top  P_{[g,h]} \tilde V_{[h]} \right)^2 \\
    &+ \frac{C}{ \Sigma} \mathbb E \left( \bar z_j^\top \Omega_{\tilde e} (P - \bar P) \tilde V \right)^2 \\
    &\leq \frac{C \max_{g \in [G]} \left\Vert \bar z_{j,[g]} \right\Vert_2^2}{ \Sigma} \sum_{g,h \in [G]^2 g \neq h} \tr \left( P_{[g,h]} P_{[h,g]} \right) + \frac{C \bar z_j^\top \bar z_j}{\Sigma} \\
    &= o(1).
\end{align*}

For the second result, we have
\begin{align}\label{eq:rho2}
    &\quad \frac{1}{\sqrt{\Sigma \Upsilon}} \left[\sum_{g,h \in [G]^2, g \neq h} \left(X_{[g]}^\top P_{[g,h]} \hat e_{[h]} \right) \left( \hat e_{[g]}^\top P_{[g,h]} \hat e_{[h]} \right)  \right. \notag \\
    &\left. \qquad\qquad\quad -  \sum_{g,h \in [G]^2, g \neq h} \mathbb E \left(\tilde V_{[g]}^\top P_{[g,h]} \tilde e_{[h]} \right) \left( \tilde e_{[g]}^\top P_{[g,h]} \tilde e_{[h]} \right) \right] \notag \\
    &= \frac{1}{\sqrt{\Sigma \Upsilon}} \sum_{g,h \in [G]^2, g \neq h} \left(\Pi_{[g]}^\top P_{[g,h]} \hat e_{[h]} \right) \left( \hat e_{[g]}^\top P_{[g,h]} \hat e_{[h]} \right) \notag \\
    &+ \frac{1}{\sqrt{\Sigma \Upsilon}} \left[\sum_{g,h \in [G]^2, g \neq h} \left(V_{[g]}^\top P_{[g,h]} \hat e_{[h]} \right) \left( \hat e_{[g]}^\top P_{[g,h]} \hat e_{[h]} \right)  \right. \notag \\
    &\left. \qquad\qquad\quad-  \sum_{g,h \in [G]^2, g \neq h} \mathbb E \left(\tilde V_{[g]}^\top P_{[g,h]} \tilde e_{[h]} \right) \left( \tilde e_{[g]}^\top P_{[g,h]} \tilde e_{[h]} \right) \right].
\end{align}

The first term on the RHS of \eqref{eq:rho2} can be written as
\begin{align*}
    &\quad \frac{1}{\sqrt{\Sigma \Upsilon}} \sum_{g,h \in [G]^2, g \neq h} \left(\Pi_{[g]}^\top P_{[g,h]} \hat e_{[h]} \right) \left( \hat e_{[g]}^\top P_{[g,h]} \hat e_{[h]} \right) \\
    & = \underbrace{\frac{1}{\sqrt{\Sigma \Upsilon}} \sum_{g,h \in [G]^2, g \neq h} \left(\Pi_{[g]}^\top P_{[g,h]} \tilde e_{[h]} \right) \left( \tilde e_{[g]}^\top P_{[g,h]} \tilde e_{[h]} \right)}_{R_{31}}\\
    &+ \underbrace{\frac{1}{\sqrt{\Sigma \Upsilon}} \left( \sum_{g,h \in [G]^2, g \neq h} \left(\Pi_{[g]}^\top P_{[g,h]} e_{[h]} \right) \left(e_{[g]}^\top P_{[g,h]} e_{[h]} \right) -  \sum_{g,h \in [G]^2, g \neq h} \left(\Pi_{[g]}^\top P_{[g,h]} \tilde e_{[h]} \right) \left( \tilde e_{[g]}^\top P_{[g,h]} \tilde e_{[h]} \right) \right)}_{R_{32}} \\
    &+ \underbrace{\frac{1}{\sqrt{\Sigma \Upsilon}} \left( \sum_{g,h \in [G]^2, g \neq h} \left(\Pi_{[g]}^\top P_{[g,h]} \hat e_{[h]} \right) \left(\hat e_{[g]}^\top P_{[g,h]} \hat e_{[h]} \right) -  \sum_{g,h \in [G]^2, g \neq h} \left(\Pi_{[g]}^\top P_{[g,h]} e_{[h]} \right) \left(e_{[g]}^\top P_{[g,h]} e_{[h]} \right) \right)}_{R_{33}}.
\end{align*}
By using the same argument as in the proof of Lemma \ref{lem:linear_quad_form_2}, we can show that $R_{31} = o_P(1)$ and $R_{32} = o_P(1)$. For $R_{33}$, we have
\begin{align*}
    R_{33} &= - \hat \Delta^3 \times \underbrace{\frac{1}{\sqrt{\Sigma \Upsilon}} \sum_{g,h \in [G]^2, g \neq h} \left(\Pi_{[g]}^\top P_{[g,h]} X_{[h]} \right) \left(X_{[g]}^\top P_{[g,h]} X_{[h]} \right)}_{R_{33,1}} \\
    &+ \hat \Delta^2 \times \underbrace{\frac{1}{\sqrt{\Sigma \Upsilon}} \sum_{g,h \in [G]^2, g \neq h} \left(\Pi_{[g]}^\top P_{[g,h]}e_{[h]} \right) \left(X_{[g]}^\top P_{[g,h]} X_{[h]} \right)}_{R_{33,2}} \\
    &+ \hat \Delta^2 \times \underbrace{\frac{1}{\sqrt{\Sigma \Upsilon}} \sum_{g,h \in [G]^2, g \neq h} \left(\Pi_{[g]}^\top P_{[g,h]}X_{[h]} \right) \left(e_{[g]}^\top P_{[g,h]} X_{[h]} \right)}_{R_{33,3}} \\
    &+ \hat \Delta^2 \times \underbrace{\frac{1}{\sqrt{\Sigma \Upsilon}} \sum_{g,h \in [G]^2, g \neq h} \left(\Pi_{[g]}^\top P_{[g,h]}X_{[h]} \right) \left(X_{[g]}^\top P_{[g,h]} e_{[h]} \right)}_{R_{33,4}} \\
    &- \hat \Delta \times \underbrace{\frac{1}{\sqrt{\Sigma \Upsilon}} \sum_{g,h \in [G]^2, g \neq h} \left(\Pi_{[g]}^\top P_{[g,h]} X_{[h]} \right) \left(e_{[g]}^\top P_{[g,h]} e_{[h]} \right)}_{R_{33,5}} \\
    &- \hat \Delta \times \underbrace{\frac{1}{\sqrt{\Sigma \Upsilon}} \sum_{g,h \in [G]^2, g \neq h} \left(\Pi_{[g]}^\top P_{[g,h]} e_{[h]} \right) \left(X_{[g]}^\top P_{[g,h]} e_{[h]} \right)}_{R_{33,6}} \\
    &- \hat \Delta \times \underbrace{\frac{1}{\sqrt{\Sigma \Upsilon}} \sum_{g,h \in [G]^2, g \neq h} \left(\Pi_{[g]}^\top P_{[g,h]} e_{[h]} \right) \left(e_{[g]}^\top P_{[g,h]} X_{[h]} \right)}_{R_{33,7}},
\end{align*}
and by using a similar argument as in the proof of Lemma \ref{lem:linear_quad_form_2}, we can show that
\begin{align*}
    R_{33, i} = O_P(1), \quad i=1, \dots, 7,
\end{align*}
which implies that $R_{33} = o_P(1)$. 

The second term on the RHS of \eqref{eq:rho2} can be written as
\begin{align*}
    &\quad \frac{1}{\sqrt{\Sigma \Upsilon}} \left(\sum_{g,h \in [G]^2, g \neq h} \left(V_{[g]}^\top P_{[g,h]} \hat e_{[h]} \right) \left( \hat e_{[g]}^\top P_{[g,h]} \hat e_{[h]} \right)  -  \sum_{g,h \in [G]^2, g \neq h} \mathbb E \left(\tilde V_{[g]}^\top P_{[g,h]} \tilde e_{[h]} \right) \left( \tilde e_{[g]}^\top P_{[g,h]} \tilde e_{[h]} \right) \right) \\
    &= \underbrace{\frac{1}{\sqrt{\Sigma \Upsilon}} \left(\sum_{g,h \in [G]^2, g \neq h} \left(\tilde V_{[g]}^\top P_{[g,h]} \tilde e_{[h]} \right) \left( \tilde e_{[g]}^\top P_{[g,h]} \tilde e_{[h]} \right)  -  \sum_{g,h \in [G]^2, g \neq h} \mathbb E \left(\tilde V_{[g]}^\top P_{[g,h]} \tilde e_{[h]} \right) \left( \tilde e_{[g]}^\top P_{[g,h]} \tilde e_{[h]} \right) \right)}_{R_{34}} \\
    &+ \underbrace{\frac{1}{\sqrt{\Sigma \Upsilon}} \left(\sum_{g,h \in [G]^2, g \neq h} \left(V_{[g]}^\top P_{[g,h]} e_{[h]} \right) \left( e_{[g]}^\top P_{[g,h]} e_{[h]} \right)  -  \sum_{g,h \in [G]^2, g \neq h} \left(\tilde V_{[g]}^\top P_{[g,h]} \tilde e_{[h]} \right) \left( \tilde e_{[g]}^\top P_{[g,h]} \tilde e_{[h]} \right) \right)}_{R_{35}} \\
    &+ \underbrace{\frac{1}{\sqrt{\Sigma \Upsilon}} \left(\sum_{g,h \in [G]^2, g \neq h} \left(V_{[g]}^\top P_{[g,h]} \hat e_{[h]} \right) \left( \hat e_{[g]}^\top P_{[g,h]} \hat e_{[h]} \right)  -  \sum_{g,h \in [G]^2, g \neq h} \left(V_{[g]}^\top P_{[g,h]} e_{[h]} \right) \left(e_{[g]}^\top P_{[g,h]} e_{[h]} \right) \right)}_{R_{36}}.
\end{align*}
By using the same argument as in the proof of Lemma \ref{lem:linear_quad_form_2}, we can show that $R_{34} = o_P(1)$ and $R_{35} = o_P(1)$. In addition, we can show that $R_{36} = o_P(1)$ as in the proof for $R_{33}$. This concludes the proof.  $\hfill\qedsymbol$

\subsection{Proof of Lemma \ref{lem:nuisance_parameter}}

If Assumptions \ref{ass:reg}--\ref{ass:local_alternative_and_covariance} hold, then by Lemma \ref{lem:betahat}, we have $\hat \beta \convP \beta$. Consider first $\hat \rho_1$. It suffices to show that $\hat \rho_1 - \rho_{1n} = o_P(1)$, and we note that $\hat \rho_1$ can be written as
\begin{align*}
    \hat \rho_1 &= \frac{1}{\sqrt{\hat \Psi \hat \Sigma}} \sum_{g \in [G]} \left[ \left( \acute X_{[g]}^\top \hat e_{[g]} \right) \left( \hat X_{[g]}^\top \hat e_{[g]} \right) \right] \\
    &= \frac{1}{\sqrt{\hat \Psi \hat \Sigma}} \sum_{g \in [G]} \left[ \left( (z \hat A_n z^\top X)_{[g]}^\top \hat e_{[g]} \right) \left( \hat X_{[g]}^\top \hat e_{[g]} \right) \right] \\
    &= \frac{1}{\sqrt{\hat \Psi \hat \Sigma}} X^\top z \hat A_n \sum_{g \in [G]} \left[ \left( z_{[g]}^\top \hat e_{[g]} \right) \left( \hat X_{[g]}^\top \hat e_{[g]} \right) \right] \\
    &= \sqrt{\frac{\Sigma}{\hat \Sigma}} \times \frac{1}{\sqrt{X^\top z \hat A_n \hat \Omega \hat A_n z^\top X}} X^\top z \hat A_n \hat \Omega^{1/2}  \\
    &\times \hat \Omega^{-1/2} \Omega^{1/2} \times \frac{1}{\sqrt{\Sigma}} \Omega^{-1/2}   \sum_{g \in [G]} \left[ \left( z_{[g]}^\top \hat e_{[g]} \right) \left( \hat X_{[g]}^\top \hat e_{[g]} \right) \right].
\end{align*}
By Lemma \ref{lem:var_est}, we have
\begin{align*}
    \sqrt{\frac{\Sigma}{\hat \Sigma}} &= 1 + o_P(1), \\
     \hat \Omega^{-1/2} \Omega^{1/2} &= I_{d_z} + o_P(1),
\end{align*}
and by (\ref{eq:low_id}) and (\ref{eq:low_An}) we have
\begin{align*}
    &\quad \frac{1}{\sqrt{X^\top z \hat A_n \hat \Omega \hat A_n z^\top X}} X^\top z \hat A_n \hat \Omega^{1/2} \\
    &= \frac{1}{\sqrt{(X^\top z/r_n) (\hat A_n/\lambda_n) (\hat \Omega/n) (\hat A_n/\lambda_n) (z^\top X/r_n)}} (X^\top z/r_n) (\hat A_n/\lambda_n) (\hat \Omega/n)^{1/2} \\ 
    &= \frac{1}{\sqrt{(\Pi^\top z/r_n) ( A_n/\lambda_n) (\Omega/n) (A_n/\lambda_n) (z^\top \Pi/r_n)}} (\Pi^\top z/r_n) ( A_n/\lambda_n) (\Omega/n)^{1/2} + o_P(1) \\
    &= \frac{1}{\sqrt{\Pi^\top z A_n \Omega A_n z^\top \Pi}} \Pi^\top z A_n \Omega^{1/2} + o_P(1).
\end{align*}
Further recall that
\begin{align*}
    \rho_{1n} &= \frac{1}{\sqrt{\Psi \Sigma}} \sum_{g \in [G]} \mathbb E \left[ \left( \acute \Pi_{[g]}^\top \tilde e_{[g]} \right) \left( \hat \Pi_{[g]}^\top \tilde e_{[g]} \right) \right] \\
    &= \frac{1}{\sqrt{\Pi^\top z A_n \Omega A_n z^\top \Pi}} \Pi^\top z A_n \Omega^{1/2} \times \frac{1}{\sqrt{\Sigma}} \Omega^{-1/2}   \sum_{g \in [G]} \mathbb E\left[ \left( z_{[g]}^\top \tilde e_{[g]} \right) \left( \hat \Pi_{[g]}^\top \tilde e_{[g]} \right) \right]. 
\end{align*}
The consistency of $\hat \rho_1$ then follows by (\ref{eq:zehatxhatehat}) and Lemma \ref{lem:cov_est}.

Next, consider $\hat \rho_2$. It suffices to show that $\hat \rho_2 - \rho_{2n} = o_P(1)$, we have
\begin{align*}
    \hat \rho_2 - \rho_{2n} &= \frac{2}{\sqrt{\Sigma \Upsilon}} \left[\sum_{g,h \in [G]^2, g \neq h} \left(X_{[g]}^\top P_{[g,h]} \hat e_{[h]} \right) \left( \hat e_{[g]}^\top P_{[g,h]} \hat e_{[h]} \right)  \right.\\
    &\left. \qquad\qquad\quad -  \sum_{g,h \in [G]^2, g \neq h} \mathbb E \left(\tilde V_{[g]}^\top P_{[g,h]} \tilde e_{[h]} \right) \left( \tilde e_{[g]}^\top P_{[g,h]} \tilde e_{[h]} \right) \right] \\ 
    &+\left(\sqrt{\frac{\Sigma \Upsilon}{\hat \Sigma \hat \Upsilon}} - 1 \right) \times \left( \frac{2}{\sqrt{\Sigma \Upsilon}} \sum_{g,h \in [G]^2, g \neq h} \left(X_{[g]}^\top P_{[g,h]} \hat e_{[h]} \right) \left( \hat e_{[g]}^\top P_{[g,h]} \hat e_{[h]} \right) \right).
\end{align*}
The consistency of $\hat \rho_2$ then follows by Lemmas \ref{lem:var_est} and \ref{lem:cov_est}.

Lastly, consider $\hat \alpha_1$ and $\hat \alpha_2$. If Assumptions \ref{ass:reg}-\ref{ass:high} hold with $\Pi^\top \Pi / \sqrt{K} \rightarrow \infty$, then by (\ref{eq:low_strong}), (\ref{eq:high_strong}) and Lemma \ref{lem:var_est}, we have $\widehat \Phi_1 / \Phi_1  \convP 1$ and $\widehat \Phi_2 / \Phi_2 \convP 1$. Therefore, if the assumptions for $d_n$ in Assumption \ref{ass:local_alternative_and_covariance} hold, then by the continuous mapping theorem, we have $\hat \alpha_1 \convP \alpha_1$ and $\hat \alpha_2 \convP \alpha_2$. Alternatively, 
if Assumption \ref{ass:reg}-\ref{ass:high} hold with $\Pi^\top \Pi / \sqrt{K} = O(1)$, we have $a_2 = 0$, so that $\alpha_2 = 0$ and $\alpha_1 = 1$. By Lemma \ref{lem:var_est}, we have $\hat \Sigma / \Sigma \convP 1$, and note that $\Sigma / \Gamma_{\tilde V, \tilde e} \rightarrow 1$ since $\Pi^\top \Pi / K \rightarrow 0$. With $\Gamma_{\tilde V, \tilde e}/K \rightarrow \Gamma_{22} > 0$, we can show that $\widehat \Phi_2 \convD \bar \Phi_2$ for some random variable $\bar \Phi_2$ such that $\bar \Phi_2 > 0$ with probability one, as in the proof of Step 3 of Lemma \ref{lem:betahat}. Combining this with the fact that $\widehat \Phi_1 \convP 0$ by (\ref{eq:low_strong}) and Lemma \ref{lem:var_est}, we have $\hat \alpha_2 \convP 0$ and $\hat \alpha_1 \convP 1$, by the continuous mapping theorem. This concludes the proof. $\hfill\qedsymbol$

\section{Proofs of Main Results} \label{sec:proof_thm}

\subsection{Proof of Proposition \ref{prop:limit}}

The proof follows exactly the same lines as in the proof of Lemma 2.2. in \cite{LWZ24} and is thus omitted. $\hfill\qedsymbol$

\subsection{Proof of Theorem \ref{thm:limit_str_lcl}}

The result follows from Lemma \ref{lem:linear_form}, Lemma \ref{lem:limit_dist} and the Slutsky theorem. $\hfill\qedsymbol$

\subsection{Proof of Theorem \ref{thm:opt_test_asypeff}}

Consider the set $\mathcal{M}'$ of data generating processes $m$ that satisfy the weak convergence result (\ref{eq:weak_convergence}) pointwise for all $\delta\in\Re$.
It is straightforward to see that $\mathcal{M}\subset\mathcal{M}'$. As a result, the test class $\mathfrak{C}$ under consideration is a subset of an augmented class $\mathfrak{C}'$ of $\phi_n$ satisfying that
\begin{align}
    \lim_{n \rightarrow \infty}\mathbb E \left[\phi_n  \right] \leqslant \alpha & \text{\quad for all\ }m\in\mathcal{M}', \delta=0, \label{eq:asymp_valid0}\\
    \liminf_{n \rightarrow \infty}\mathbb E \left[\phi_n  \right] \geqslant \alpha & \text{\quad for all\ }m\in\mathcal{M}', \delta\neq0. \label{eq:asymp_unbiased0}
\end{align}
We also note that the oracle version of the combination test, $\phi^{o}_n$ in (\ref{eq:opttest_oracle}), can be understood as taking the weak convergence result (\ref{eq:weak_convergence}) as the starting point and is simply the UMPU test in the limiting problem (under known $a_1(\alpha_1),a_2(\alpha_2),\rho_1,\rho_2$), evaluated at the sample analgoues (Wald, LM, and AR statistics). By construction, $\phi^{o}_n$ satisfies (\ref{eq:asymp_valid0}) and (\ref{eq:asymp_unbiased0}), so $\phi^{o}_n\in\mathfrak{C}'$. Furthermore, by a direct application of Theorem 1 in \cite{mueller2011}, it follows that for any $\delta_1\neq 0$ and any $\phi_n\in\mathfrak{C}'$,
\begin{align*}
    \lim_{n \rightarrow \infty}\mathbb E \left[\phi_n  \right] 
    \leqslant 
    \lim_{n \rightarrow \infty}\mathbb E \left[\phi^{o}_{n}  \right]
    \quad\text{for all }m\in\mathcal{M}', \delta=\delta_1,
\end{align*}
which subsequently implies that for any $\delta_1\neq 0$ and any $\phi_n\in\mathfrak{C}$,
\begin{align*}
    \lim_{n \rightarrow \infty}\mathbb E \left[\phi_n  \right] 
    \leqslant 
    \lim_{n \rightarrow \infty}\mathbb E \left[\phi^{o}_{n}  \right]
    \quad\text{for all }m\in\mathcal{M}, \delta=\delta_1.
\end{align*}
Lastly, note that $\mathbb E\left[\phi^*_n\right]$ is continuous in $(\hat\alpha_1,\hat\alpha_2,\hat\rho_1,\hat\rho_2)$, which are consistent for $(\alpha_1,\alpha_2,\rho_1,\rho_2)$ under the assumptions of Theorem \ref{thm:limit_str_lcl} by Lemma \ref{lem:nuisance_parameter}. By the continuous mapping theorem, we thus have that, for any $\delta$, $\mathbb E\left[\phi^*_n\right]=\mathbb E\left[\phi^{o}_{n}\right]+o_P(1)$. Therefore, we have $\phi^*_n$ satisfies (\ref{eq:asymp_valid}) and (\ref{eq:asymp_unbiased}), and thus $\phi^*_n\in\mathfrak{C}$. In addition, for any $\delta_1\neq 0$, and any $\phi_n\in\mathfrak{C}$,
\begin{align*}
    \lim_{n \rightarrow \infty}\mathbb E \left[\phi_n  \right] \leqslant \lim_{n \rightarrow \infty}\mathbb E \left[\phi^{o}_{n}  \right]= \lim_{n \rightarrow \infty} \mathbb E \left[\phi^*_n  \right] \text{\quad for all\ }m\in\mathcal{M}, \delta=\delta_1. 
\end{align*}

For the second part, it is straightforward to check that $\tilde \phi_n\in \mathfrak{C}$. Also, note that the comparison of local asymptotic power of $\phi^*_n$ and $\tilde \phi_n$ can be reduced to the comparison of the non-centrality parameters for $\phi^o_n$ and $\tilde \phi_n$ in their limiting distributions, which are given by $(a^\top V^{-1}a) \delta^2$ and $a_1^2 \delta^2$, respectively, where \begin{align*}
    a = \begin{pmatrix}
    a_1 \\
    a_2 \\
    0
\end{pmatrix}, \quad V = \begin{pmatrix}
    1 &  \rho_1 & 0 \\
    \rho_1 & 1 & \rho_2 \\
    0 & \rho_2 & 1
\end{pmatrix}.
\end{align*}
Direct calculation yields
\begin{align*}
    (a^\top V^{-1}a) \delta^2 - a_1^2 \delta^2 = \frac{\delta^2}{1-\rho_1^2-\rho_2^2} \left( a_2 - \rho_1 a_1\right)^2 \geqslant 0,
\end{align*}
and since $\rho_1^2 + \rho_2^2 < 1$ and $\delta \neq 0$, we obtain the desired result.  $\hfill\qedsymbol$

\subsection{Proof of Theorem \ref{thm:fixed_alter}}

We shall distinguish between the two cases: (i) $\Pi^\top \Pi / \sqrt{K}$ is diverging and (ii) $\Pi^\top \Pi / \sqrt{K}$ is bounded, and argue along the appropriate subsequence as in Step 2 and 3 in the proof of Lemma \ref{lem:betahat}.

For the first case, we have
\begin{align*}
    T(\beta_0) &= \frac{ (X^\top z \hat A_n z^\top X)^{-1}(X^\top z \hat A_n z^\top e)}{ \sqrt{\widehat \Phi_1}} + \frac{\delta}{\sqrt{\widehat \Phi_1}}, \\
    LM(\beta_0) &= \frac{X^\top (P - \bar P) e}{\sqrt{\hat \Sigma}} + \frac{\delta (X^\top (P -\bar P) X) }{\sqrt{\hat \Sigma}}, \\
    AR &= \frac{\hat e^\top (P - \bar P) \hat e}{\sqrt{\hat \Upsilon}}
\end{align*}
under the fixed alternative. On the event $X^\top (P - \bar P)X > 0$, we can write
\begin{align*}
    \frac{\delta (X^\top (P -\bar P) X) }{\sqrt{\hat \Sigma}} = \frac{\delta}{\sqrt{(X^\top (P -\bar P) X)^{-1} \hat \Sigma (X^\top (P -\bar P) X)^{-1}}} = \frac{\delta}{\sqrt{\widehat \Phi_2}}.
\end{align*}
By repeating the proof of Lemma \ref{lem:linear_form} for the other terms in the above expression beside $\delta/\sqrt{\widehat \Phi_1}$ and $\delta/\sqrt{\widehat \Phi_2}$ (which do not depend on $a_1$ and $a_2$), we can write
\begin{align*}
    \begin{pmatrix}
    T (\beta_0) \\
    LM(\beta_0 ) \\  
    AR
    \end{pmatrix}
    = 
    \begin{pmatrix}
    \delta / \sqrt{\widehat \Phi_1}  \\
    \delta / \sqrt{\widehat \Phi_2} \\  
    0
    \end{pmatrix} + R_n,
\end{align*}
where $R_n = \Op(1)$. Recall the definition $(\hat \omega_1,\hat \omega_2,\hat \omega_3)'$:
\begin{align*}
    \begin{pmatrix}
    \hat \omega_1 \\
    \hat \omega_2 \\  
    \hat \omega_3
    \end{pmatrix}
    = \frac{1}{\sqrt{\hat b_1^2 + \hat b_2^2 + \hat b_3^2}} 
    \times \begin{pmatrix}
    1 & \hat \rho_1 & 0  \\
    \hat \rho_1 & 1 & \hat \rho_2  \\
    0 & \hat \rho_2 & 1
    \end{pmatrix}^{-1/2} 
    \begin{pmatrix}
            \hat b_1 \\
            \hat b_2 \\
            \hat b_3
    \end{pmatrix}, \begin{pmatrix}
    \hat b_1 \\
    \hat b_2 \\
    \hat b_3        
    \end{pmatrix}
    = \begin{pmatrix}
    1 & \hat \rho_1 & 0 \\
    \hat \rho_1 & 1 & \hat \rho_2 \\
    0 & \hat \rho_2 & 1
    \end{pmatrix}^{-1/2} 
    \begin{pmatrix}
            \hat \alpha_1 \\
            \hat \alpha_2 \\
            0
    \end{pmatrix},
\end{align*}
and note that $(\hat \omega_1,\hat \omega_2,\hat \omega_3)' = \Op(1)$. We have
\begin{align*}
    0 &\leq 1-\mathbb E \left[ \phi^*_n \right] \\
    &= \mathbb P \left( \left( \hat \omega_1 T (\beta_0) + \hat \omega_2 LM(\beta_0 ) + \hat \omega_3 AR \right)^2 < \mathbb C_\alpha \right) \\
    &\leq \mathbb P \left( \left( \hat \omega_1 T (\beta_0) + \hat \omega_2 LM(\beta_0 ) + \hat \omega_3 AR \right)^2 < \mathbb C_\alpha, X^\top (P - \bar P)X > 0 \right) + \mathbb P \left( X^\top (P - \bar P)X \leqslant 0 \right) \\
    &\leq \mathbb P \left( \left( \delta \sqrt{\frac{1}{\widehat \Phi_1} + \frac{1}{\widehat \Phi_2}} \times \sqrt{\hat \alpha^\top \hat V^{-1} \hat \alpha} + \Op(1) \right)^2 < \mathbb C_\alpha  \right) + \mathbb P \left( X^\top (P - \bar P)X \leqslant 0 \right) ,
\end{align*}
where
\begin{align*}
    \hat \alpha \equiv \begin{pmatrix}
    \hat \alpha_1 \\
    \hat \alpha_2 \\
    0
\end{pmatrix}, \quad \hat V \equiv \begin{pmatrix}
    1 &  \hat \rho_1 & 0 \\
    \hat \rho_1 & 1 & \hat \rho_2 \\
    0 & \hat \rho_2 & 1
\end{pmatrix},\quad \begin{pmatrix}
    \delta / \sqrt{\widehat \Phi_1}  \\
    \delta / \sqrt{\widehat \Phi_2} \\  
    0
    \end{pmatrix} = \delta \sqrt{\frac{1}{\widehat \Phi_1} + \frac{1}{\widehat \Phi_2}} \times 
    \begin{pmatrix}
            \hat \alpha_1 \\
            \hat \alpha_2 \\
            0
    \end{pmatrix},
\end{align*}
and we used the fact that $\hat \omega_1 \times \hat \alpha_1 + \hat \omega_2 \times \hat \alpha_2 + \hat \omega_3 \times 0 = \sqrt{\hat \alpha^\top \hat V^{-1} \hat \alpha}$. In addition, we have $\widehat \Phi_1/ \Phi_1 \convP 1$ and $\widehat \Phi_2/ \Phi_2 \convP 1$ by (\ref{eq:low_strong}), (\ref{eq:high_strong}) and Lemma \ref{lem:var_est}, whence $ \widehat \Phi_1 = \op(1)$ and $\widehat \Phi_2 = \op(1)$ by (\ref{eq:Phi1}) and (\ref{eq:Phi2}).  Therefore, we have
\begin{align*}
    \delta \sqrt{\frac{1}{\widehat \Phi_1} + \frac{1}{\widehat \Phi_2}} \times \sqrt{\hat \alpha^\top \hat V^{-1} \hat \alpha} \convP \infty (-\infty), \quad \text{for}\ \delta > 0\ (\delta<0), 
\end{align*}
where we use the fact that $\left\Vert \hat \alpha \right\Vert_2 = 1$ by construction, $\hat V^{-1} \convP V^{-1}$ where
\begin{align*}
    V \equiv \begin{pmatrix}
    1 & \rho_1 & 0 \\
    \rho_1 & 1 & \rho_2 \\
    0 & \rho_2 & 1
\end{pmatrix},
\end{align*}
by Lemma \ref{lem:nuisance_parameter}, whence $\hat \alpha^\top \hat V^{-1} \hat \alpha$ can be bounded away from zero with probability approaching one. Together with the fact that $X^\top (P - \bar P)X > 0$ with probability approaching one by (\ref{eq:high_strong}) and $\Pi^\top \Pi \rightarrow \infty$, this implies that $\lim_{n \rightarrow \infty} \mathbb E \left[ \phi^*_n \right] = 1$. 

For the second case, by Lemma \ref{lem:nuisance_parameter}, we have
$\hat \rho_1 \convP \rho_1$ and $\hat \rho_2 \convP \rho_2$, and note that $\rho_1 = 0$ in this case since $\Pi^\top \Pi / K \rightarrow 0$. In addition, similar to the proof of Step 3 of Lemma \ref{lem:betahat}, it can be shown that $\widehat \Phi_2 \convD \bar \Phi_2$ for some random variable $\bar \Phi_2$ such that $\bar \Phi_2 > 0$ with probability one, and since $\widehat \Phi_1 \convP 0$ by (\ref{eq:low_strong}), (\ref{eq:Phi1}) and Lemma \ref{lem:var_est}, it follows that $\hat \alpha_1 \convP 1$ and $\hat \alpha_2 \convP 0$, by the continuous mapping theorem. Therefore, we have $\hat \omega_1 \convP 1$, $\hat \omega_2 \convP 0$ and $\hat \omega_3 \convP 0$. Finally, as in the proof for the first case above, we have $T(\beta_0) =\delta/\sqrt{\widehat \Phi_1} + \Op(1)$ where $\widehat \Phi_1 = \op(1)$, $LM(\beta_0) = \Op(1)$ and $AR = \Op(1)$. It follows that $\lim_{n \rightarrow \infty} \mathbb E \left[ \phi^*_n \right] = 1$. This concludes the proof. $\hfill\qedsymbol$

\subsection{Proof of Theorem \ref{thm:low_weak_scalar}}

We shall argue along the appropriate subsequence as in Step 4 of the proof of Lemma \ref{lem:betahat}.  Suppose we are under the local alternative that $\beta - \beta_0 = \delta d_n$. We have
\begin{align*}
    \hat \rho_1 &= \frac{1}{\sqrt{\hat \Psi \hat \Sigma}} \sum_{g \in [G]} \left[ \left( \acute X_{[g]}^\top \hat e_{[g]} \right) \left( \hat X_{[g]}^\top \hat e_{[g]} \right) \right] \\
    &= \sqrt{\frac{\Sigma}{\hat \Sigma}} \times \frac{1}{\sqrt{X^\top z \hat A_n \hat \Omega \hat A_n z^\top X}} X^\top z \hat A_n \sqrt{\hat \Omega} \\
    &\times \sqrt{\frac{\Omega}{\hat \Omega}} \times \frac{1}{\sqrt{\Omega \Sigma}}  \sum_{g \in [G]} \left[ \left( z_{[g]}^\top \hat e_{[g]} \right) \left( \hat X_{[g]}^\top \hat e_{[g]} \right) \right],
\end{align*}
and note the important fact that
\begin{align*}
    \left( \frac{1}{\sqrt{X^\top z \hat A_n \hat \Omega \hat A_n z^\top X}} X^\top z \hat A_n \sqrt{\hat \Omega}  \right)^2 = 1
\end{align*}
when $d_z = 1$, and thus 
\begin{align*}
    \hat \rho_1^2 = \frac{\Sigma}{\hat \Sigma} \times \frac{\Omega}{\hat \Omega} \times \left( \frac{1}{\sqrt{\Omega \Sigma}}  \sum_{g \in [G]} \left[ \left( z_{[g]}^\top \hat e_{[g]} \right) \left( \hat X_{[g]}^\top \hat e_{[g]} \right) \right] \right)^2.
\end{align*}
By Lemmas \ref{lem:var_est} and \ref{lem:cov_est}, we have $\hat \rho_1^2 \convP \bar \rho_1^2$. Together with the fact that $\hat \rho_2^2 \convP \bar \rho_2^2$ by Lemmas \ref{lem:var_est} and \ref{lem:cov_est}, this implies that $\hat \rho_1^2 + \hat \rho_2^2 < 1$ with probability approaching one. On that event, direct calculation yields
\begin{align*}
     &\quad \hat \omega_1 T (\beta_0) + \hat \omega_2 LM(\beta_0 ) + \hat \omega_3 AR \\
     &= \frac{T(\beta_0)\hat \alpha_1(1-\hat \rho_2^2) - T(\beta_0)\hat \rho_1 \hat \alpha_2 + LM(\beta_0) (\hat \alpha_2-\hat \rho_1 \hat \alpha_1) + AR \hat \rho_2(\hat \rho_1 \hat \alpha_1-\hat \alpha_2)}{\sqrt{1-\hat \rho_1^2 - \hat \rho_2^2} \sqrt{\hat \alpha_1^2(1-\hat \rho_2^2) - 2 \hat \rho_1 \hat \alpha_1 \hat \alpha_2 + \hat \alpha_2^2}}.
\end{align*}
We shall analyze each term in turn. To begin with, we note that, similar to the proof of Step 4 of Lemma \ref{lem:betahat}, it can be shown that $\widehat \Phi_1 \convD \bar \Phi_1$ for some random variable $\bar \Phi_1$ such that $\bar \Phi_1 > 0$ with probability one, and since $\widehat \Phi_2 \convP 0$ by (\ref{eq:Phi2}), (\ref{eq:high_strong}) and Lemma \ref{lem:var_est}, it follows that $\hat \alpha_1 \convP 0$ and $\hat \alpha_2 \convP 1$, by the continuous mapping theorem. Note also that
\begin{align*}
    T(\beta_0) &= \frac{(X^\top z \hat A_n z^\top X)^{-1}}{\sqrt{(X^\top z \hat A_n z^\top X)^{-2}}} \times \frac{1}{\sqrt{X^\top z \hat A_n \hat \Omega \hat A_n z^\top X}} X^\top z \hat A_n \sqrt{\hat \Omega} \\
    &\times \sqrt{\frac{\Omega}{\hat \Omega}} \times \frac{1}{\sqrt{\Omega}} \sum_{g \in [G]} z_{[g]}^\top \tilde e_{[g]} + \frac{d_n}{\sqrt{\widehat \Phi_1}} \delta \\
    &= \Op(1),
\end{align*}
and thus
\begin{align*}
    T(\beta_0)\hat \alpha_1(1-\hat \rho_2^2) = o_P(1).
\end{align*}
Next, we note that
\begin{align*}
    T(\beta_0)\hat \rho_1 &= \frac{(X^\top z \hat A_n z^\top X)^{-1}}{\sqrt{(X^\top z \hat A_n z^\top X)^{-2}}} \times \sqrt{\frac{\Sigma}{\hat \Sigma}} \times \frac{\Omega}{\hat \Omega} \\
    &\times \frac{1}{\sqrt{\Omega}} \sum_{g \in [G]} z_{[g]}^\top \tilde e_{[g]} \times \frac{1}{\sqrt{\Omega \Sigma}}  \sum_{g \in [G]} \left[ \left( z_{[g]}^\top \hat e_{[g]} \right) \left( \hat X_{[g]}^\top \hat e_{[g]} \right) \right] + o_P(1).
\end{align*}
In addition, we can show that
\begin{align*}
    \frac{1}{\sqrt{n}} z^\top X \convD \N \left( \pi, \Omega_z^{\tilde V, \tilde V} \right),
\end{align*}
which, combining with the fact that $\hat A_n/ \lambda_n \convP A$, implies that
\begin{align*}
    \frac{(X^\top z \hat A_n z^\top X)^{-1}}{\sqrt{(X^\top z \hat A_n z^\top X)^{-2}}} = 1 + o_P(1).
\end{align*}
If follows that, by Lemmas \ref{lem:var_est} and \ref{lem:cov_est},
\begin{align*}
    T(\beta_0)\hat \rho_1 \hat \alpha_2 = \bar\rho_1 \times \frac{1}{\sqrt{\Omega}} \sum_{g \in [G]} z_{[g]}^\top \tilde e_{[g]} + o_P(1).
\end{align*}
Next, we note that
\begin{align*}
    LM(\beta_0) &= \frac{X^\top (P - \bar P) e}{\sqrt{\hat \Sigma}} + a \delta + o_P(1) \\
    &= \frac{1}{\sqrt{\Sigma}} \left( \sum_{g \in [G]} \hat \Pi_{[g]}^\top \tilde e_{[g]} + \sum_{g,h \in [G]^2, g \neq h} \tilde V_{[g]}^\top P_{[g,h]} \tilde e_{[h]} \right) + a \delta + o_P(1),
\end{align*}
where the first equality is by (\ref{eq:high_strong}) and Lemma \ref{lem:var_est}, and the second equality is by Lemmas \ref{lem:quad_form_1} and \ref{lem:var_est}. It follows that
\begin{align*}
    LM(\beta_0) (\hat \alpha_2-\hat \rho_1 \hat \alpha_1) = \frac{1}{\sqrt{\Sigma}} \left( \sum_{g \in [G]} \hat \Pi_{[g]}^\top \tilde e_{[g]} + \sum_{g,h \in [G]^2, g \neq h} \tilde V_{[g]}^\top P_{[g,h]} \tilde e_{[h]} \right) + a \delta + o_P(1).
\end{align*}
Next, we note that
\begin{align*}
    AR &= \frac{1}{\sqrt{\widehat \Upsilon}} \left( \hat e^\top (P - \bar P) \hat e - e^\top (P - \bar P) e \right) + \frac{1}{\sqrt{\widehat \Upsilon}} \left( e^\top (P - \bar P) e - \tilde e^\top (P - \bar P) \tilde e \right) + \frac{1}{\sqrt{\widehat \Upsilon}}  \tilde e^\top (P - \bar P) \tilde e \\
    &= \frac{1}{\sqrt{\Upsilon}} \sum_{g,h \in [G]^2, g \neq h} \tilde e_{[g]}^\top P_{[g,h]} \tilde e_{[h]} + o_P(1).
\end{align*}
by (\ref{eq:X(P-Pbar)e}), (\ref{eq:X(P-Pbar)X}), and Lemmas \ref{lem:quad_form_1} and \ref{lem:var_est}. It follows that
\begin{align*}
    AR \hat \rho_2(\hat \rho_1 \hat \alpha_1-\hat \alpha_2) = -\bar \rho_2 \times  \frac{1}{\sqrt{\Upsilon}} \sum_{g,h \in [G]^2, g \neq h} \tilde e_{[g]}^\top P_{[g,h]} \tilde e_{[h]} + o_P(1)
\end{align*}
by Lemma \ref{lem:cov_est}. Finally, we have
\begin{align*}
    &1-\hat \rho_1^2 - \hat \rho_2^2 \convP 1-  \bar \rho_1^2 - \bar \rho_2^2, \\
    &\hat \alpha_1^2(1-\hat \rho_2^2) - 2 \hat \rho_1 \hat \alpha_1 \hat \alpha_2 + \hat \alpha_2^2 \convP 1.
\end{align*}
Combining all the results, we have
\begin{align*}
    &\quad \hat \omega_1 T (\beta_0) + \hat \omega_2 LM(\beta_0 ) + \hat \omega_3 AR \\
    &= \frac{-\bar \rho_1}{\sqrt{1-\bar \rho_1^2-\bar \rho_2^2}} \times \frac{1}{\sqrt{\Omega}} \sum_{g \in [G]} z_{[g]}^\top \tilde e_{[g]} \\
    &+ \frac{1}{\sqrt{1-\bar \rho_1^2-\bar \rho_2^2}} \times \left( \frac{1}{\sqrt{\Sigma}} \left( \sum_{g \in [G]} \hat \Pi_{[g]}^\top \tilde e_{[g]} + \sum_{g,h \in [G]^2, g \neq h} \tilde V_{[g]}^\top P_{[g,h]} \tilde e_{[h]} \right) + a \delta \right) \\
    &+ \frac{-\bar \rho_2}{\sqrt{1-\bar \rho_1^2-\bar \rho_2^2}} \times \frac{1}{\sqrt{\Upsilon}} \sum_{g,h \in [G]^2, g \neq h} \tilde e_{[g]}^\top P_{[g,h]} \tilde e_{[h]} + o_P(1).
\end{align*}

Furthermore, we note that in the proof of Lemma \ref{lem:limit_dist}, we only require $\acute \Pi$ to satisfy
\begin{align*}
    \frac{1}{\Psi} \sum_{g \in [G]} \acute \Pi_{[g]}^\top \acute \Pi_{[g]} = O(1), \quad \frac{1}{\Psi} \max_{g \in [G]}  \acute \Pi_{[g]}^\top \acute \Pi_{[g]} = o(1),
\end{align*}
which is guaranteed by Assumptions \ref{ass:reg} and \ref{ass:low_id}. We also have
\begin{align*}
    \frac{1}{\Omega} \sum_{g \in [G]} z_{[g]}^\top z_{[g]} = O(1), \quad \frac{1}{\Omega} \max_{g \in [G]} z_{[g]}^\top z_{[g]} = o(1),
\end{align*}
by Assumption \ref{ass:reg} alone. Therefore, we can replace $\frac{1}{\sqrt{\Psi}} \sum_{g=1}^G \acute \Pi_{[g]}^\top \tilde e_{[g]} $ with $\frac{1}{\sqrt{\Omega}} \sum_{g \in [G]} z_{[g]}^\top \tilde e_{[g]}$ in the proof of Lemma \ref{lem:limit_dist}. Note that in this way, Assumption \ref{ass:low_id} is no longer needed. We thus can follow the same argument in the proof of Lemma \ref{lem:limit_dist} and obtain that 
\begin{align*}
    \begin{pmatrix}
        \frac{1}{\sqrt{\Omega}} \sum_{g \in [G]} z_{[g]}^\top \tilde e_{[g]} \\
        \frac{1}{\sqrt{\Sigma}} \left( \sum_{g \in [G]} \hat \Pi_{[g]}^\top \tilde e_{[g]} + \sum_{g,h \in [G]^2, g \neq h} \tilde V_{[g]}^\top P_{[g,h]} \tilde e_{[h]} \right) + a \delta \\
        \frac{1}{\sqrt{\Upsilon}} \sum_{g,h \in [G]^2, g \neq h} \tilde e_{[g]}^\top P_{[g,h]} \tilde e_{[h]}
    \end{pmatrix} \convD \N \left( \begin{pmatrix}
    0 \\
    a \delta \\
    0
\end{pmatrix}, \begin{pmatrix}
    1 &  \bar \rho_1 & 0 \\
    \bar \rho_1 & 1 & \bar\rho_2 \\
    0 & \bar \rho_2 & 1
\end{pmatrix} \right).
\end{align*}
The desired result then follows.

Next, suppose we are under the fixed alternative. From the proof above, we have
\begin{align*}
    &\quad \hat \omega_1 T (\beta_0) + \hat \omega_2 LM(\beta_0 ) + \hat \omega_3 AR \\
    &= \frac{(1+ o_P(1))}{\sqrt{1-\bar \rho_1^2- \bar \rho_2^2}} LM(\beta_0) + O_P(1) \\
    &= \frac{(1+ o_P(1))}{\sqrt{1-\bar \rho_1^2-\bar \rho_2^2}} \times \left( \frac{\delta}{\sqrt{\widehat \Phi_2}} \times (1 + o_P(1)) + O_P(1)\right) + O_P(1)
\end{align*}
by (\ref{eq:high_strong}) and Lemma \ref{lem:var_est}, and note that $\widehat \Phi_2 = o_P(1)$ by (\ref{eq:Phi2}) and Lemma \ref{lem:var_est}. If follows that
\begin{align*}
    \left( \hat \omega_1 T (\beta_0) + \hat \omega_2 LM(\beta_0 ) + \hat \omega_3 AR \right)^2 \convP \infty,
\end{align*}
and the desired result follows. This concludes the proof. $\hfill\qedsymbol$

\subsection{Proof of Theorem \ref{thm:low_weak_multi}}

Under the local alternative, similar to the proof of Theorem \ref{thm:low_weak_scalar}, we have
\begin{align*}
    T(\beta_0) &= \frac{(X^\top z \hat A_n z^\top X)^{-1}}{\sqrt{(X^\top z \hat A_n z^\top X)^{-2}}} \times \frac{1}{\sqrt{X^\top z \hat A_n \hat \Omega \hat A_n z^\top X}} X^\top z \hat A_n \hat \Omega^{1/2} \\
    &\quad \times \hat \Omega^{-1/2} \Omega^{1/2} \times \Omega^{-1/2} \sum_{g \in [G]} z_{[g]}^\top \tilde e_{[g]} + \frac{d_n}{\sqrt{\widehat \Phi_1}} \delta = \Op(1),\\
        LM(\beta_0)& = \frac{1}{\sqrt{\Sigma}} \left( \sum_{g \in [G]} \hat \Pi_{[g]}^\top \tilde e_{[g]} + \sum_{g,h \in [G]^2, g \neq h} \tilde V_{[g]}^\top P_{[g,h]} \tilde e_{[h]} \right) + a \delta + \op(1),\\
        AR & = \frac{1}{\sqrt{\Upsilon}} \sum_{g,h \in [G]^2, g \neq h} \tilde e_{[g]}^\top P_{[g,h]} \tilde e_{[h]} + \op(1).
\end{align*}
By repeating the proof of Lemma \ref{lem:limit_dist} and ignoring $\frac{1}{\sqrt{\Psi}} \sum_{g=1}^G \acute \Pi_{[g]}^\top \tilde e_{[g]} $ (and thus Assumption \ref{ass:low_id} is not needed), we have
\begin{align*}
    \begin{pmatrix}
    LM(\beta_0) \\
    AR
\end{pmatrix} \convD \N \left( \begin{pmatrix}
    a \delta \\
    0
\end{pmatrix}, \begin{pmatrix}
    1 & \rho \\
    \rho & 1
\end{pmatrix} \right),
\end{align*}
by the Slutsky theorem. For $\hat \rho_1$, we have
\begin{align*}
    \hat \rho_1 &= \frac{1}{\sqrt{\hat \Psi \hat \Sigma}} \sum_{g \in [G]} \left[ \left( \acute X_{[g]}^\top \hat e_{[g]} \right) \left( \hat X_{[g]}^\top \hat e_{[g]} \right) \right] \\
    &= \sqrt{\frac{\Sigma}{\hat \Sigma}} \times \frac{1}{\sqrt{X^\top z \hat A_n \hat \Omega \hat A_n z^\top X}} X^\top z \hat A_n \hat \Omega^{1/2} \\
    &\times \hat \Omega^{-1/2} \Omega^{1/2} \times \frac{1}{\sqrt{\Sigma}} \Omega^{-1/2}   \sum_{g \in [G]} \left[ \left( z_{[g]}^\top \hat e_{[g]} \right) \left( \hat X_{[g]}^\top \hat e_{[g]} \right) \right],
\end{align*}
where
\begin{align*}
    \frac{1}{\sqrt{X^\top z \hat A_n \hat \Omega \hat A_n z^\top X}} X^\top z \hat A_n \hat \Omega^{1/2} = O_P(1),
\end{align*}
and by Lemma \ref{lem:cov_est},
\begin{align*}
    \frac{1}{\sqrt{\Sigma}} \Omega^{-1/2}  \sum_{g \in [G]} \left[ \left( z_{[g]}^\top \hat e_{[g]} \right) \left( \hat X_{[g]}^\top \hat e_{[g]} \right) \right] = \frac{1}{\sqrt{\Sigma}} \Omega^{-1/2}  \sum_{g \in [G]} \mathbb E\left[ \left( z_{[g]}^\top \tilde e_{[g]} \right) \left( \hat \Pi_{[g]}^\top \tilde e_{[g]} \right) \right] + o_P(1).
\end{align*}
Further note that the first term on the right-hand side of the above display is $o(1)$ since $\Pi^\top \Pi / K \rightarrow 0$, so that $\hat \rho_1 \convP 0$ by Lemma \ref{lem:var_est}. In addition, we have $\hat \rho_2 \convP \rho$ by Lemmas \ref{lem:var_est} and \ref{lem:cov_est}. Finally, similar to the proof of Step 4 of Lemma \ref{lem:betahat}, it can be shown that $\widehat \Phi_1 \convD \bar \Phi_1$ for some random variable $\bar \Phi_1$ such that $\bar \Phi_1 > 0$ with probability one, and since $\widehat \Phi_2 \convP 0$ by (\ref{eq:Phi2}), (\ref{eq:high_strong}) and Lemma \ref{lem:var_est}, it follows that $\hat \alpha_1 \convP 0$ and $\hat \alpha_2 \convP 1$, by the continuous mapping theorem. This implies that $\hat \omega_1 \convP 0$, $\omega_2 \convP 1/\sqrt{1-\rho^2}$ and $\omega_3 \convP -\rho/\sqrt{1-\rho^2}$, whence  
\begin{align*}
& \quad \hat \omega_1  T(\beta_0) + \hat \omega_2 LM(\beta_0) +  \hat \omega_3 AR\\
    & = \frac{1}{\sqrt{1-\rho^2}}LM(\beta_0) - \frac{\rho}{\sqrt{1-\rho^2}}AR + \op(1) \\
    & \convD \frac{1}{\sqrt{1-\rho^2}} \N_1 - \frac{\rho}{\sqrt{1-\rho^2}} \N_2,
\end{align*}
where $\N_1$ and $\N_2$ are defined in Theorem \ref{thm:low_weak_multi}.

Under the fixed alternative, similar to the proof of Theorem \ref{thm:low_weak_scalar}, we have $\lim_{n \rightarrow \infty} \mathbb E \left[ \phi^*_n \right] = 1$. This concludes the proof. $\hfill\qedsymbol$

 \section{Additional Simulations} \label{sec:simu_add}

\begin{figure}[H]
    \centering
    \includegraphics[width=1\textwidth]
    {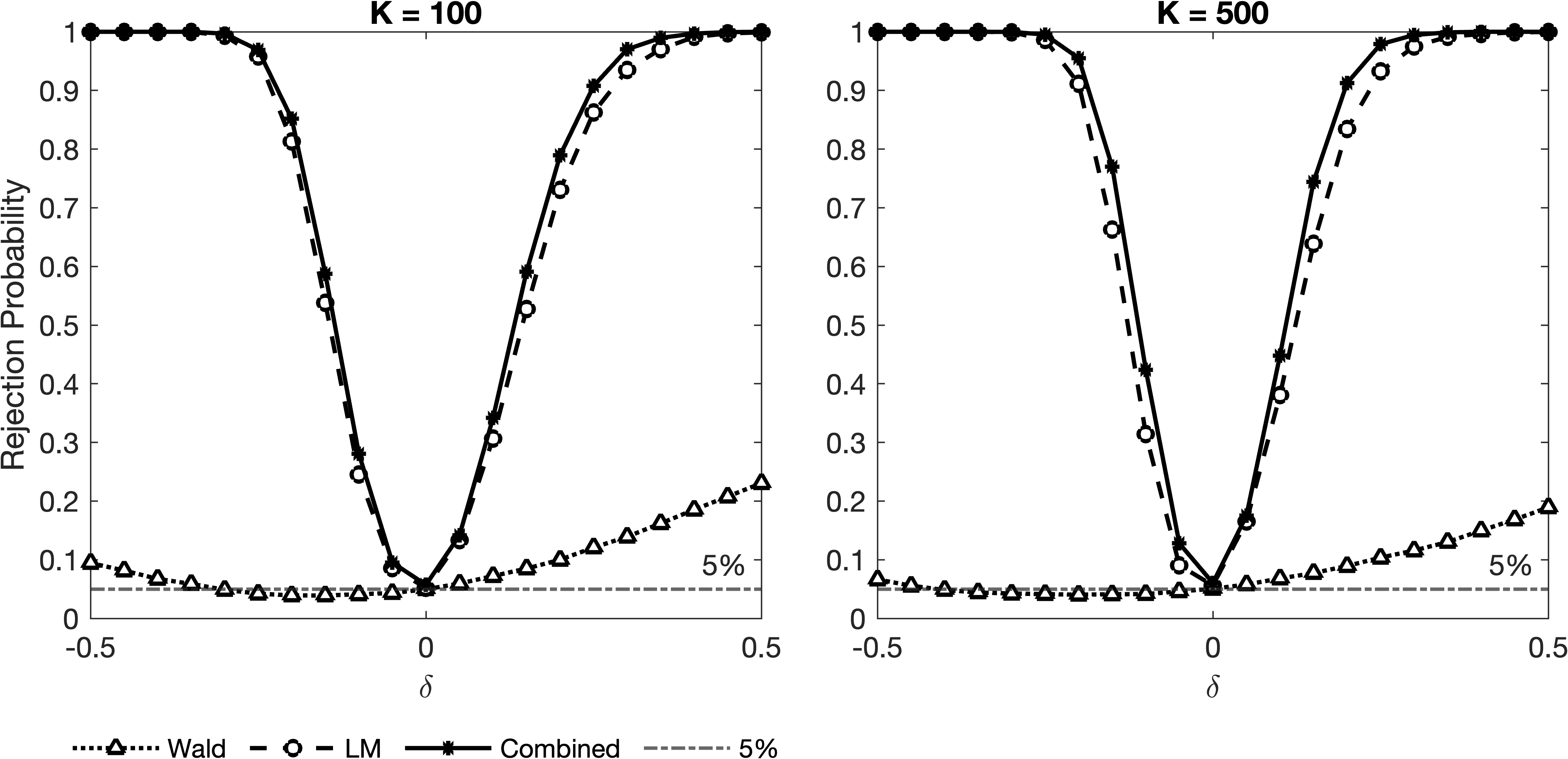}
    \caption{Power Curve of the combination, Wald, and LM tests.}
    \floatfoot{\doublespacing \textit{Notes}: This figure displays the power curves for our combination test $\phi_n^*$ along with those for the component Wald and LM tests, at different values of $K$ (the dimension of the many IVs), $\psi=30$ (the identification strength of the many IVs is relatively strong), and $\phi=0$ (the identification strength of the one-dimensional IV is weak). The horizontal axis represents the deviations in the parameter of interest from the maintained hypothesis, that is, we are interested in testing $\mathcal{H}_0: \beta = \beta_0 \  \text{against}\  \mathcal{H}_{1}:  \beta \neq \beta_0$, and $\delta = \beta - \beta_0$. See Section \ref{sec: simu} in the main text for a detailed description of the simulation setup. All results are based on $5,000$ simulations.}
    \label{fig:res_lw}
\end{figure}

In this section, we present some additional simulation results to illustrate the effect of weak low-dimensional IVs. 
We set $\psi = 30$ so that the identification strength of many IVs remains relatively strong, and $\phi = 0$ so that the identification strength of the one-dimensional IV is rather weak. Figure \ref{fig:res_lw} displays the power curves for $K = 100$ and $K = 500$, respectively, which can be regarded as extensions of Panels A and B of Figure \ref{fig:simulations-stacked} in the main text. Overall, the dominant performance of our combination test remains robust to different strengths of one-dimensional IV. In addition, we note that although the one-dimensional IV is relatively weak, the Wald test still provides a nontrivial gain in power for our combination test (as seen from the noticeable gaps between the power curves of $\phi^*_n$ and the jackknife LM test). This gain arises from its correlation with the jackknife LM statistic, in line with the theoretical result stated in Theorem \ref{thm:low_weak_scalar}.



\section{Additional Empirical Applications} \label{sec:emp_add}

\begin{figure}[H]
    \centering
    \includegraphics[width=0.9\textwidth]
    {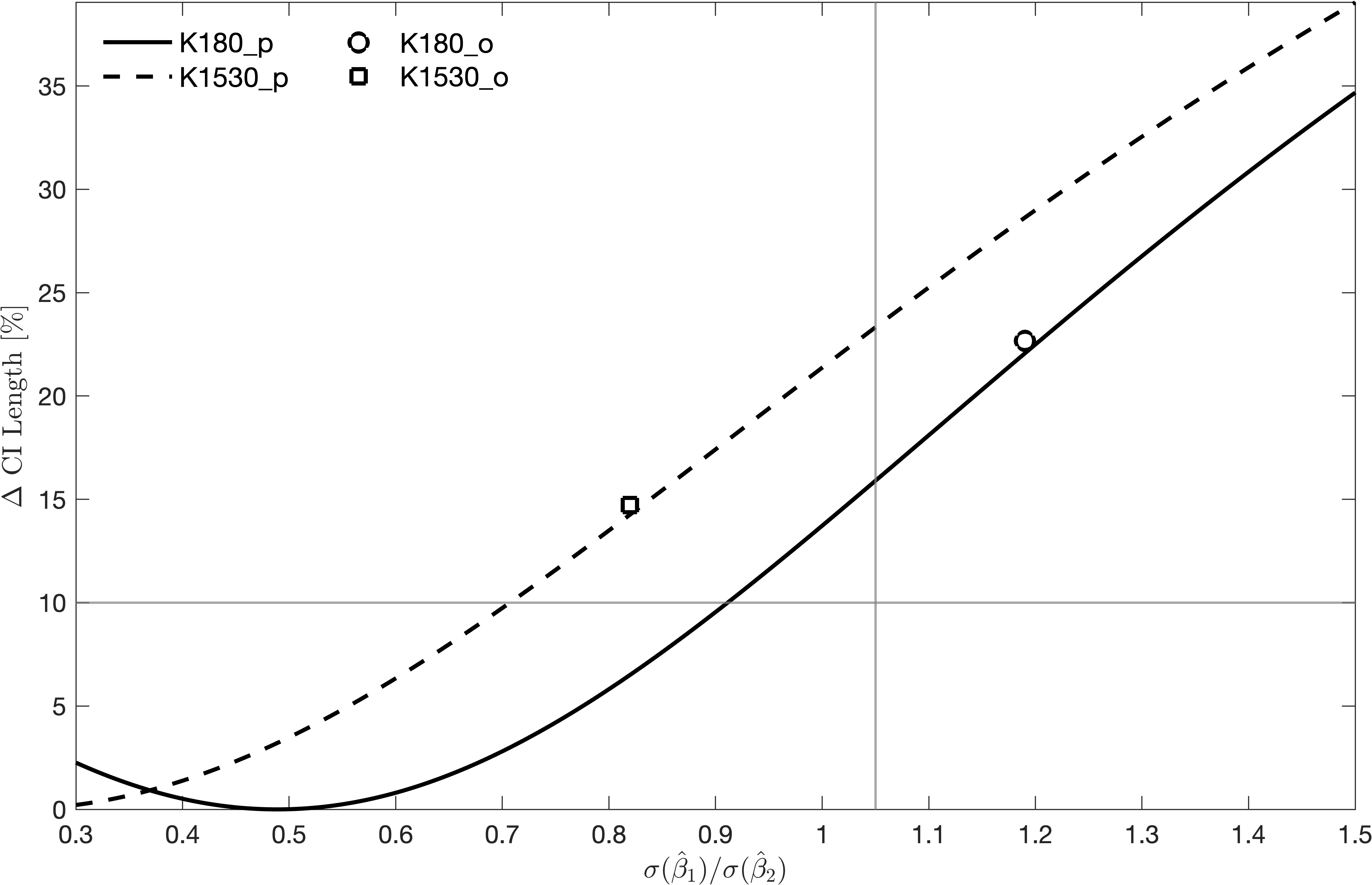}
    \caption{Realized percentage reduction in confidence interval length: Return to education}
    \floatfoot{\doublespacing \textit{Notes}: This figure shows, for each specification in the return to education example, the observed percentage decrease in confidence interval length (Combined CI versus Wald CI, as in Table \ref{tab:res_ak91}, and indicated by ``o" in figure legends) plotted as a point against the standard error ratio ($\hat \sigma(\hat \beta_1) / \hat \sigma(\hat \beta_2)$ in Table \ref{tab:res_ak91}). Also shown is the theoretical lower bound for the reduction (indicated by ``p" in figure legends), analogous to Figure \ref{fig:CI_theory} in the main text, but now computed using the specification-specific estimate $\hat\rho_1$, as reported in Table \ref{tab:res_ak91}. Here, ``K180" refers to the specifications with $K=180$ instruments, and ``K1530" refers to the specifications with $K=1,530$ instruments. The horizontal axis is the ratio of standard deviations (errors) of $\hat \beta_1$ and $\hat \beta_2$. The vertical axis is the reduction in the length of confidence interval in percentage points. As a final remark, note that the actual numerical values of the relevant quantities in Table \ref{tab:res_ak91}, rather than the rounded values shown there, are used to produce Figure \ref{fig:res_ak91}.}
    \label{fig:res_ak91}
\end{figure}

In this section, we consider the return to education application, using the dataset of \cite{Angrist-Krueger(1991)}. In this application, the outcome variable is the log weekly wages and the endogenous variable is the years of schooling. We follow \cite{MS22} and \cite{LWZ24} to consider two specifications with $180$ and $1,530$ instruments. The set of $180$ instruments consists of $30$ quarter and year of birth interactions (QoB--YoB) and $150$ quarter and place of birth interactions (QoB-PoB). The set of $1,530$ instruments includes all interactions among QoB-YoB-PoB. The quantitative implications obtained from Table \ref{tab:res_ak91} and Figure \ref{fig:res_ak91} are in line with the discussion in Section \ref{sec: illu} of the main text. Our confidence intervals are roughly $23\%$ and $15\%$ shorter, respectively, than the Wald CIs in the specifications with $K=180$ and $K=1,530$. Here, we note that the actual reduction in the length of the confidence interval can be greater than $10\%$ even when the standard error ratio is much lower than $1.05$ (in the specification with $K=1,530$), again indicating the conservativeness of our rule of thumb. However, we reiterate that, although the low-dimensional IVs (the three QoB IVs, and therefore $\hat \beta_1$ and the Wald confidence interval) are identical across the two specifications ($K=180$ and $1,530$) by construction, we are not attempting to obtain improved inference based on pooling statistics across specifications, and our theoretical results do not justify such an approach.

\newpage

\newpage

\begin{table}[H]
\centering
    \caption{Point estimates and confidence intervals: Return to education.}
\label{tab:res_ak91}
\begin{tabular}{l
                S
                S}
\hline\hline
 & \multicolumn{1}{c}{$K=180$}
 & \multicolumn{1}{c}{$K=1,530$} \\
\hline
$\hat \rho_1$
& 0.489 & 0.236 \\
$\hat \rho_2$
& -0.170 & -0.200 \\
$\hat \sigma(\hat \beta_1) / \hat \sigma(\hat \beta_2)$
& 1.190 & 0.820 \\
$\hat \beta_1$
& 0.098 & 0.098 \\
Wald CI
& {(0.059, 0.138)} & {(0.059, 0.138)} \\
$\hat \beta_2$
& 0.099 & 0.084 \\
LM CI
& {(0.066, 0.132)} & {(0.036, 0.133)} \\
$\hat \beta^*$
& 0.097 & 0.093 \\
Comb. CI
& {(0.066, 0.127)} & {(0.059, 0.126)} \\
\hline\hline
\end{tabular}
\floatfoot{\doublespacing \textit{Notes}: This table reports the estimation and inference results for the return to education example using the \cite{Angrist-Krueger(1991)} dataset, shown separately for specifications with $K=180$ instruments and $K=1,530$ instruments. The IV set in the column labeled ``$K=180$" consists of $30$ quarter and year of birth interactions (QoB--YoB) and $150$ quarter and place of birth interactions (QoB--PoB), while the IV set in the column with ``$K=1,530$" includes the full set of interactions among QoB--YoB--PoB. See Appendix D in \cite{LWZ24} for a more detailed description of data and this empirical application. The point estimates are obtained from the standard two-stage least squares (TSLS) estimator with the three-dimensional QoB instruments, $\hat\beta_1$, and, in addition, from the leave-one-cluster-out estimator, $\hat\beta_2$, which makes use of many IVs. Wald CI and LM CI denote the confidence intervals based on $\hat \beta_1$ and $\hat \beta_2$, respectively. The estimator $\hat\beta^*$ is the combined estimator for $\beta$, defined in Section \ref{sec: efficiency_gain} of the main text. It is essentially the midpoint of the confidence interval in (\ref{eq:impliedCI}), which is obtained from our combination test and labeled as ``Comb. CI" in the table. In addition, $\hat\rho_1$ and $\hat\rho_2$ denote estimates of the asymptotic correlation between the Wald and LM statistics, and between the LM and AR statistics, respectively. Finally, $\hat \sigma(\hat \beta_1) / \hat \sigma(\hat \beta_2)$ denotes the ratio of standard errors of $\hat \beta_1$ and $\hat \beta_2$. All displayed numbers are rounded to three decimal places.}
\end{table}

\newpage

\end{document}